# The BigBOSS Experiment

D. Schlegel[p], F. Abdalla[z], T. Abraham[r], C. Ahn[g], C. Allende Prieto[j], J. Annis[h],
E. Aubourg[a], M. Azzaro[i], S. Bailey[p]. C. Baltay[ii], C. Baugh[f], C. Bebek[p], S. Becerril[i],
M. Blanton[s], A. Bolton[gg], B. Bromley[gg], R. Cahn[p], P.-H. Carton[c],
J. L. Cervantes-Cota[jj,ll], Y. Chu[ff], M. Cortês[p,x], K. Dawson[gg], A. Dey[r], M. Dickinson[r],
H. T. Diehl[h], P. Doel[z], A. Ealet[d], J. Edelstein[x], D. Eppelle[c], S. Escoffier[d], A. Evrard[cc],
L. Faccioli[p,x], C. Frenk[f], M. Geha[ii], D. Gerdes[cc], P. Gondolo[gg], A. Gonzalez-Arroyo[m],
B. Grossan[x], T. Heckman[n], H. Heetderks[x], S. Ho[p], K. Honscheid[u], D. Huterer[cc],
O. Ilbert[o], I. Ivans[gg], P. Jelinsky[x], Y. Jing[v], D. Joyce[r], R. Kennedy[p], S. Kent[h],
D. Kieda[gg], A. Kim[p], C. Kim[g], J.-P. Kneib[o], X. Kong[ff], A. Kosowsky[dd], K. Krishnan[g],
O. Lahav[z], M. Lampton[x], S. LeBohec[gg], V. Le Brun[o], M. Levi[p], C. Li[v], M. Liang[r],
H. Lim[g], W. Lin[v], E. Linder[g,x], W. Lorenzon[cc], A. de la Macorra[kk,ll], Ch. Magneville[c],
R. Malina[o], C. Marinoni[d], V. Martinez[t], S. Majewski[hh], T. Matheson[r], R. McCloskey[r],
P. McDonald[p,b], T. McKay[cc], J. McMahon[cc], B. Menard[n], J. Miralda-Escudé[l],
M. Modjaz[s], A. Montero-Dorta[i], I. Morales[i], N. Mostek[p,x], J. Newman[dd], R. Nichol[ee],
P. Nugent[p,x], K. Olsen[r], N. Padmanabhan[ii], N. Palanque-Delabrouille[c], I. Park[g],
J. Peacock[aa], W. Percival[ee], S. Perlmutter[p,x], C. Peroux[o], P. Petitjean[k], F. Prada[i],
E. Prieto[o], J. Prochaska[y], K. Reil[w], C. Rockosi[y], N. Roe[p], E. Rollinde[k], A. Roodman[w],
N. Ross[p], G. Rudnick[bb], V. Ruhlmann-Kleider[c], J. Sanchez[i], D. Sawyer[r], C. Schimd[o],
M. Schubnell[cc], R. Scoccimaro[s], U. Seljak[g,p,x], H. Seo[x], E. Sheldon[b], M. Sholl[x],
R. Shulte-Ladbeck[dd], A. Slosar[b], D. S. Smith[u], G. Smoot[g,p,x], W. Springer[gg], A. Stril[p],
A. S. Szalay[n], C. Tao[d], G. Tarlé[cc], E. Taylor[x], A. Tilquin[d], J. Tinker[s], F. Valdes[r],
J. Wang[ff], T. Wang[ff], B. A. Weaver[s], D. Weinberg[u], M. White[p,x], M. Wood-Vasey[dd],
J. Yang[g], X. Yang[v]. Ch. Yèche[c], N. Zakamska[n], A. Zentner[dd], C. Zhai[ff], P. Zhang[v]

[a] Astrophysique, Particules et Cosmologie Laboratoire (APC), Paris
[b] Brookhaven National Laboratory
[c] CEA/IRFU, Saclay
[d] Centre de Physique des Particules de Marseille
[e] Centre de Physique Theorique, Université de Marseille
[f] Durham University
[g] WCU Institute for the Early Universe at Ewha Womans University, Seoul, Korea
[h] Fermi National Accelerator Laboratory
[i] Instituto de Astrofísica de Andalucía (IAA-CSIC), Granada, Spain
[j] Instituto de Astrofísica de Canarias, Tenerife, Spain
[jj] Instituto Nacional de Investigaciones Nucleares (ININ), Mexico
[kk] Instituto Avanzado de Cosmologia, Mexico
[k] Institut d'Astrophysique (IAP), Paris
[l] ICREA / Institut de Ciències del Cosmos, Universitat de Barcelona - IEEC, Barcelona, Spain
[m] Instituto de Física Teórica, (UAM-CSIC), Madrid, Spain
[n] Johns Hopkins University
[o] Laboratoire d'Astrophysique de Marseille
[p] Lawrence Berkeley National Laboratory

[r] National Optical Astronomy Observatory
[s] New York University
[t] Observatori Astronomic, Universitat de Valencia, Spain
[u] The Ohio State University
[v] Shanghai Astronomical Observatory
[w] SLAC National Accelerator Laboratory
[ll] Universidad Nacional Autonoma de Mexico (UNAM),Mexico
[x] University of California, Berkeley
[y] University of California, Santa Cruz/Lick Observatory
[z] University College, London
[aa] University of Edinburgh
[bb] University of Kansas
[cc] University of Michigan
[dd] University of Pittsburgh
[ee] University of Portsmouth
[ff] University of Science and Technology of China
[gg] University of Utah
[hh] University of Virginia
[ii] Yale University

# EXECUTIVE SUMMARY

BigBOSS is a Stage IV ground-based dark energy experiment to study baryon acoustic oscillations (BAO) and the growth of structure with a wide-area galaxy and quasar redshift survey. It has been conditionally accepted by NOAO in response to a call for major new instrumentation and a high-impact science program for the 4-m Mayall telescope at Kitt Peak. The BigBOSS instrument is a robotically-actuated, fiber-fed spectrograph capable of taking 5000 simultaneous spectra over a wavelength range from 340 nm to 1060 nm, with a resolution $R = \lambda/\Delta\lambda = 3000-4800$. The focal plane is located at prime focus, where a new optical corrector will provide an impressive 3 degree diameter field of view. The BigBOSS collaboration will also deliver a spectroscopic pipeline and data management system to reduce and archive all data for public access. BigBOSS builds upon the SDSS-III/BOSS project, re-using many elements of the BOSS instrument and computing pipeline designs.

The BigBOSS Key Project is a 14,000 square degree survey that will be carried out using 500 nights over five years. Using data from imaging surveys that are already underway, spectroscopic targets are selected that trace the underlying dark matter distribution. In particular, targets include luminous red galaxies (LRGs) up to $z = 1.0$, extending the BOSS LRG survey in both redshift and survey area. To probe the universe out to even higher redshift, BigBOSS will target bright [OII] emission line galaxies (ELGs) up to $z = 1.7$. In total, approximately 20 million galaxy redshifts are obtained to measure the BAO feature, trace the matter power spectrum at smaller scales, and detect redshift space distortions. BigBOSS represents at least an order of magnitude improvement over BOSS both in the co-moving volume it probes and the number of galaxies it will map. In addition to the cosmological constraints coming from the galaxy survey, BigBOSS will provide additional constraints on early dark energy and on the curvature of the universe by measuring the Ly-$\alpha$ forest in the spectra of over 600,000 $2.2 < z < 3.5$ quasars.

The BigBOSS Mayall survey will provide a significant advance in our understanding of the expansion history of the universe and dark energy. BigBOSS will achieve sub-percent accuracy on the BAO standard ruler: 0.4% for $0.5 < z < 1.0$ and 0.6% for $1.0 < z < 1.7$. This precision on the expansion history of the universe is a powerful probe of the nature of dark energy, as quantified by the Dark Energy Task Force figure of merit (DETF FOM), which measures the combined precision on the dark energy equation of state, $w$, and its evolution with redshift $w_a$. BigBOSS galaxy BAO measurements achieve a DETF FOM of 95 with Planck priors only. This is a factor of three over the DETF FOM of all Stage III galaxy BAO measurements combined with Planck, clearly satisfying the DETF criteria for a Stage IV experiment.

Expanding the analysis to a broadband power analysis, including the Ly-$\alpha$ forest in BigBOSS quasar spectra together with the galaxies, BigBOSS achieves a FOM of 395 with Planck plus Stage III priors. This FOM is based on conservative assumptions for the analysis of broad band power ($k_{max} = 0.15$), and could grow to over 600 if current work allows us to push the analysis to higher wave numbers ($k_{max} = 0.3$). In addition, BigBOSS will place significant constraints on theories of modified gravity and inflation, and will measure the sum of neutrino masses with an error of 0.024 eV, sufficient to make the first detection of neutrino mass for the normal mass hierarchy at $2\sigma$ significance, and rule out the the inverted mass hierarchy at 95% CL. Alternately, the inverted mass hierarchy could be detected with $4\sigma$ significance.

BigBOSS will also enable an unprecedented multi-object spectroscopic capability for the U.S. community through an existing NOAO facility. Community-led P.I. programs using the BigBOSS spectrographs will be offered through the traditional time allocation process administered by NOAO. Additionally, during nights assigned to the BigBOSS Key Project, between 10% and 20% of the fibers per observation (i.e., representing a total of between 5 and 10 million spectra over the duration of the BigBOSS survey), will be made available for community use for "synchronous science" observing programs. Much as with SDSS, a rich variety of projects will be enabled with the legacy data from the BigBOSS survey and PI programs.

BigBOSS is complementary to the imaging surveys that are underway or planned for this decade. The BigBOSS survey will cover much of the PanSTARRS-1 and PTF survey areas in the Northern hemisphere. It will also overlap with significant coverage regions of the imminent Dark Energy Survey and future LSST survey projects. The overlap with DES and LSST in particular will provide spectra for photometric redshift training useful for the weak lensing and galaxy cluster measurements and will enable a host of other scientific goals.

We have determined that the BigBOSS instrument is optically compatible with the Blanco telescope at CTIO, the 4-m "twin" to the Mayall telescope at KPNO. An additional 10,000 square degree BAO survey could be carried out from the Blanco, for a total survey area of 24,000 square degrees (subject, of course, to invitation and approval from NOAO). From CTIO, the BigBOSS instrument would then have concurrent access to the same sky that LSST will be mapping, and the prospects for synergistic programs are enormous.

This report is based on the BigBOSS proposal submission to NOAO in October 2010, and reflects the project status at that time with minor updates.

# Contents

# 1 Introduction

## 1.1 BigBOSS in Context

On November 18, 2009, NOAO announced an opportunity to "pursue a large science program with the Mayall 4-meter telescope on Kitt Peak and to develop a major observing capability" for the National Observatory[1]. In particular, the Call encouraged proposals that would enable (and pursue) "large, high-impact science programs and [improve] the capabilities [within] the U.S. System of ground-based optical and near-IR telescopes". In response to this call, the BigBOSS Collaboration submitted a proposal for a highly multiplexed, wide-field fiber-fed spectrograph for the prime focus of the Mayall 4-m Telescope. The Big-BOSS spectrograph focal plane has 5000 robotically actuated fibers covering a 3° diameter field of view. The fibers feed ten identical spectrographs, each covering the wavelength range from 340 to 1060 nm with a resolution $R = \lambda/\Delta\lambda = 3000 - 4800$. This instrument is designed to enable a Key Project addressing fundamental questions in cosmology and will provide the NOAO community with a significant new observational resource. The proposal was reviewed by an external, non-advocate committee chaired by Brian Schmidt, which concluded that BigBOSS would be "a highly effective use of the Mayall Telescope in the period of 2016-2020, and the resulting survey would be one of the telescope's major scientific contributions during its lifetime."

With the discovery that the bulk of gravitating matter in the universe is in a "dark" form [Zwicky, 1933; Rubin, Ford & Thonnard, 1980], and the even more startling discovery that the universal expansion is *accelerating* [Riess et al., 1998; Perlmutter et al., 1999], we have had to come to terms with the fact that 96% of the energy density of the universe is contained in some hitherto undetected (and unsuspected!) form. Over the last decade, there has been a growing realization that understanding these new components of the universe (i.e., the dark matter and dark energy) requires fundamentally new physics. Numerous ideas have been advanced to explain the acceleration and predict its redshift evolution [e.g., see review by Frieman et al., 2008]. Nevertheless, despite intense efforts over the last decade since its discovery, there is still no consensus as to the nature of dark energy. Our understanding is still limited by a lack of data, specifically by our limited knowledge of the expansion rate and growth of structure as a function of redshift. The field looks to astronomical observations for guidance.

It is therefore hardly surprising that numerous recent community-based reviews have recommended that a major undertaking of the astronomy and physics communities be focused on constraining the equation of state of dark energy, or more generally the accurate measure of the Universe's expansion history. These include *Connecting Quarks to the Cosmos* (Committee on the Physics of the Universe 2003); the Report of the Dark Energy Task Force (DETF; Albrecht et al. 2006); the Report of the High Energy and Particle Astrophysics (HEPAP) Particle Astrophysics Scientific Assessment Group (PASAG; Ritz et al. 2009); *New Worlds, New Horizons in Astronomy & Astrophysics*, and the Report of the Committee for a the Astro2010 Decadal Survey of Astronomy and Astrophysics (Blandford et al. 2010, http://www.nap.edu/catalog/12951.html).

---

[1]See "Announcement of Opportunity for Large Science Programs Providing Observing Capabilities for the Mayall 4m Telescope", http://www.noao.edu/kpno/largescience.html

BigBOSS will obtain observational constraints that will bear on three of the four "science frontier" questions identified by the Astro2010 Cosmology and Fundamental Physics Panel of the Decadal Survey: Why is the universe accelerating? What is dark matter? What are the properties of neutrinos? Indeed, the BigBOSS project was recommended for substantial immediate R&D support by the PASAG report. The second highest ground-based priority from the Astro2010 Decadal Survey was the creation of a funding line within the NSF to support a "Mid-Scale Innovations" program, and it used BigBOSS as a "compelling" example for support. This choice was the result of the Decadal Survey's Program Prioritization panels reviewing 29 mid-scale projects and recommending BigBOSS "very highly".

## 1.2 The BigBOSS Cosmology Program

BigBOSS on the 4-m Mayall Telescope at the Kitt Peak National Observatory will enable (1) a cosmological investigation of unprecedented scale and scientific value; and (2) a unique spectroscopic survey capability for NOAO's user community.

The legacy of the NOAO telescopes includes fundamental advances in cosmological exploration, namely the discovery of dark matter in galaxies [e.g., Rubin, 1983; Rubin, Ford & Thonnard, 1980; Rubin et al., 1985], the pioneering use of supernovae as standard candles [e.g., Phillips, 1993; Hamuy et al., 1995, 1996; Phillips et al., 1999], and the resulting unexpected discovery of the cosmological acceleration [e.g., Riess et al., 1998; Schmidt et al., 1998; Garnavich et al., 1998; Perlmutter et al., 1999; Wood-Vasey et al., 2007; Miknaitis et al., 2007]. As mentioned, this last discovery, in particular, has revolutionized cosmology and focused efforts on characterizing the acceleration and understanding the "dark energy" that is driving it.

BigBOSS will equip the Mayall telescope for the next phase in this endeavor of cosmological discovery to measure the geometry of the universe and characterize its accelerating expansion with unprecedented accuracy. As described in more detail in the following chapters, the BigBOSS Key Project will use 500 dark/gray nights spread over five years to undertake a redshift survey of approximately 20 million galaxies up to a redshift $z = 1.7$ over 14,000 deg$^2$ to directly measure the baryon acoustic oscillation (BAO) scale. BigBOSS represents at least one order of magnitude increase over BOSS both in co-moving volume and number of galaxies, and will measure the Hubble parameter and angular diameter distance to sub-percent accuracies.

The BigBOSS experiment galaxy BAO measurement alone provides a factor of three improvement in the Dark Energy Task Force figure of merit (DETF FOM) over all Stage III BAO experiments - BOSS, WiggleZ, HETDEX - combined. The full power of BigBOSS extends this with the inclusion of Lyman-$\alpha$ forest data from QSOs and the full galaxy power spectrum analysis. Making very conservative assumptions ($k_{max} = 0.15$), BigBOSS combined with Stage III DETF priors will achieve a FOM of 395, and this will increase to over 600 if current work allows us to extend this analysis to higher wave numbers ($k_{max} = 0.3$). Exactly how far these analyses can be pushed is a very active area of research especially among BigBOSS collaborators who are also working on the BOSS Stage III experiment, with results expected in the next few years.

The redshift survey used to measure the baryon acoustic scale can also address other problems in cosmology. In particular, the survey will yield constraints on the neutrino mass, inflationary models, and modified gravity. Importantly, BigBOSS could make the first detection of the absolute neutrino mass scale, at $4\sigma$ significance for the inverted mass hierarchy and at $2\sigma$ significance for the normal mass hierarchy.

The survey will also be used for photometric redshift calibration of DES and LSST. BigBOSS will follow the model of the Sloan Digital Sky Survey by making publicly available both its reduced and calibrated survey data and high-order data products, since these will enable a host of studies related to galaxy evolution and large scale structure.

## 1.3 BigBOSS and the NOAO Community

BigBOSS will greatly enhance the NOAO community's ability to successfully undertake large spectroscopic surveys of high astrophysical impact. Beyond its impact on cosmology, BigBOSS is an instrument of remarkable astrophysical grasp and fills an important hole in the U.S. System. Wide-field multi-object spectroscopy has been called out as a desired capability on 4-m class telescopes by various reports, including "The Ground-Based O/IR National Observatory: A Roadmap to 2020", the report of the Future of NOAO Committee; "Renewing Small Telescopes for Astronomical Research" (ReSTAR); "Report of the First Workshop on the Ground-Based O/IR System" (2000). Most recently, the Report of the Astro2010 Decadal Survey's Panel on Optical and Infrared Astronomy from the Ground identified massively multiplexed spectroscopy as an essential capability for addressing many astrophysical questions.[2]

Currently, the only wide-field ($\geq 1°$) spectroscopic capabilities available publicly to the U.S. astronomical community are the Hydra spectrographs on the 3.5-m WIYN and 4-m Blanco telescopes, and Hectospec on the MMT (see Table 1.1). BigBOSS therefore fills an important need for the U.S. community, by providing a capability of unprecedented scientific reach, and represents an order-of-magnitude gain over our current ability to undertake large spectroscopic surveys. With the exception of LAMOST (which is located at a poorer site, limited by design to certain observing modes, and unavailable to the U.S. community), BigBOSS is unmatched in its ability to efficiently undertake wide-field spectroscopic surveys. Covering a significant fraction of the sky spectroscopically requires an instrument with a very large field of view ($> 1$ deg$^2$), which BigBOSS provides.

The addition of BigBOSS to the U.S. System is very timely. BigBOSS will provide the much-needed spectroscopic follow-up for imaging surveys, with a $3°$ field of view that is comparable to PanSTARRS ($3°$), Dark Energy Survey ($2.2°$) and LSST ($3.5°$). With its ability to obtain 5000 spectra over a $3°$ diameter field with a single observation, BigBOSS will enable the U.S. astronomical community to undertake revolutionary studies of astronomical objects. Such studies include: surveying the kinematics and chemical properties of stars in the Milky Way to understand our Galaxy's structure, dynamical and chemical history; mapping the evolution of large scale structure in the galaxy distribution over the last 8 billion years; extending studies of galaxy evolution of the scope undertaken by the Sloan Digital Sky Survey to higher redshift; carrying out large scale surveys for identifying and studying rare populations (e.g., high redshift QSOs, bright lensed galaxies, low metallicity stars, very cool white dwarfs); probing the structure of the intergalactic media along the lines of sight to background galaxies and quasars; measuring the dynamics, stellar populations, and star-formation properties within and around low-redshift galaxy clusters; kinematics of stars in open clusters and moving groups; mapping the ionization structure and kinematics of gas in the interstellar medium; and much, much more.

---

[2]From the Report of the Panel on OIR Astronomy from the Ground: "Massively multiplexed optical/NIR spectrographs and spectroscopic surveys on 4 to 8-m telescopes to map large-scale structure for the study of dark energy and cosmology, measure the evolution of galaxies across redshift and environment using spectral diagnostics, and to study the chemical and dynamical history of the Milky Way with large spectroscopic samples of stars. Several SFPs identified the need for surveys at least an order or magnitude larger than those currently underway; such surveys require new instrumentation for either fully or highly dedicated facilities as well as large survey teams. BigBOSS and HETDEX are compelling examples of next generation projects in this category."

Table 1.1: Existing & Planned Optical Multi-Object Spectrographs

| Telescope/ Spectrograph | Aper. (m) | Slit/ Fiber | Resolution $\lambda/\Delta\lambda$ | $\eta^1$ | FOV sq.deg. | $N_{max}^2$ | $\epsilon^3$ | $\mu^4$ |
|---|---|---|---|---|---|---|---|---|
| SDSS/BOSS | 2.5 | f | 1560-2650 | 0.25 | 7.07 | 1000 | 1.000 | 1.00 |
| WIYN/Hydra* | 3.5 | f | 700-22000 | 0.05 | 0.78 | 100 | 0.04 | 0.04 |
| Mayall/MARS* | 3.9 | s | 1000 | 0.40 | 0.005 | 25 | 0.1 | 0.003 |
| Mayall/KOSMOS* | 3.9 | s | 2000 | 0.40 | 0.014 | 60 | 0.2 | 0.008 |
| Blanco/Hydra* | 3.9 | f | 700-50000 | 0.05 | 0.35 | 138 | 0.07 | 0.024 |
| AAT/AAOmega | 3.9 | f | 1200-10000 | 0.25 | 3.14 | 392 | 1.0 | 1.08 |
| LAMOST | 4.0 | f | 1000 | 0.10 | 19.6 | 4000 | 4.1 | 2.84 |
| Magellan/IMACS* | 6.5 | s | 2000-20000 | 0.30 | 0.20 | 600 | 4.9 | 0.23 |
| MMT/Hectospec* | 6.5 | f | 1000 | 0.24 | 0.78 | 300 | 1.9 | 0.72 |
| MMT/Hectochelle* | 6.5 | f | 32000 | 0.10 | 0.78 | 300 | 0.8 | 0.30 |
| MMT/Binospec* | 6.5 | s | 1000-3000 | 0.40 | 0.07 | 150 | 1.6 | 0.10 |
| Gemini/GMOS* | 8.1 | s | 600-3700 | 0.40 | 0.008 | 100 | 1.7 | 0.02 |
| VLT/VIMOS | 8.2 | s | 180-2520 | 0.30 | 0.06 | 750 | 9.7 | 0.11 |
| Subaru/FOCAS | 8.3 | s | 250-2000 | 0.30 | 0.01 | 50 | 0.7 | 0.02 |
| LBT/MODS* | 8.4 | s | 130-1730 | 0.40 | 0.01 | 20 | 0.4 | 0.03 |
| HET/LRS | 9.2 | s | 550-1300 | 0.40 | 0.004 | 13 | 0.3 | 0.01 |
| HET/VIRUS-W | 9.2 | f | 550-1300 | 0.18 | 0.10 | 500? | 4.9 | 0.14 |
| Keck/LRIS* | 10.0 | s | 300-3000 | 0.35 | 0.01 | 20 | 0.4 | 0.03 |
| Keck/DEIMOS* | 10.0 | s | 1700-4800 | 0.35 | 0.022 | 80 | 1.8 | 0.07 |
| **Mayall/BigBOSS*** | **3.9** | **f** | **3000-4800** | **0.30** | **7.07** | **5000** | **14.6** | **2.92** |

[1] Optical throughput of spectrograph and telescope.
[2] Maximum number of objects which can be simultaneously observed.
[3] **Survey efficiency** relative to SDSS-III/BOSS, defined as (Aperture)$^2$ × $N_{obj}$ × $\eta$.
[4] **Sky mapping efficiency** relative to SDSS-III/BOSS, defined as (Aperture)$^2$ × FOV × $\eta$.
* Public access available to the US Community through NOAO and/or NASA.

The tools for planning observations and reducing the data are deliverables with the instrument, allowing NOAO users to plan and execute their observations. BigBOSS will therefore benefit the community in three basic ways: (1) the data from the BigBOSS survey will be made available publicly to the astronomical community through an archive; (2) fibers (between 10 and 20%) could be made available to the community during the regular BigBOSS survey for targets of opportunity or community-proposed science targets; (3) the instrument and observing system will be made available for use by the community through NOAO's proposal process, for both large surveys and short (i.e., few-night) PI-led programs. This brings many of the benefits of the SDSS-I, SDSS-II and SDSS-III more directly to the NOAO community, and on a grander scale.

## 1.4   Organization of this Document

This document is organized as follows. This chapter introduces the NOAO Large Science Call and the BigBOSS concept, and briefly describes the impact of BigBOSS on the spectroscopic capabilities available to the U.S. community. Chapter 2 describes in detail the motivation behind the dark energy Key Project to be undertaken with BigBOSS. Chapter 3 outlines, by way of a few examples, the broad range of science enabled by BigBOSS, and the various ways in which the NOAO community may choose to utilize the BigBOSS instrument. Chapter 4 describes the plan for target selection for the Key Project. Details of the BigBOSS instrument are presented in Chapter 5. The BigBOSS survey strategy and operations plan is described in Chapter 6, and the plan for data management is described in Chapter 7.

# 2  Key Science Project

## 2.1  BigBOSS and the Investigation of Dark Energy

We now know that our rather complete understanding of the fundamental interactions of matter is limited to 4% of the energy composition of the universe. Some 23% is composed of dark matter, presumably yet-to-be-discovered elementary particles, and the remaining 73% is not matter at all. That 73% – the dark energy –might be due to a uniform and unchanging energy density described by a cosmological constant, albeit with a value minuscule by comparison with what would be expected on dimensional grounds, or alternatively it might be variable in time and space. A third possibility is that the observation of dark energy is due to a failure of General Relativity. Establishing any of these explanations would cause a dramatic change in our understanding of the universe as a whole.

Among the four primary techniques for studying dark energy identified by the Dark Energy Task Force [DETF; Albrecht et al., 2006], measurement of baryon acoustic oscillations (BAO) was singled out as having the fewest astrophysical uncertainties. BAO uses only the redshift and angular locations of galaxies; the brightness and shapes of the galaxies are irrelevant. What is measured is the two-point correlation as a function of the distance between galaxies. We know that there is an enhancement at a co-moving distance of 150 Mpc (100 Mpc/$h$) as a relic of waves that propagated in the electron-photon-baryon plasma until recombination at a redshift of 1087. However, to make this technique competitive requires an enormous number of redshifts and thus a substantial investment in a new, more capable instrument and significant allocation of the telescope time.

Measuring the apparent size of this 150-Mpc standard ruler at various redshifts yields measurements of the Hubble parameter $H(z)$ and the angular-diameter distance $d_A(z)$. From these it is straightforward to tightly constrain the dark energy density $\Omega_{DE}$ and the dark energy equation of state $w(z) = p(z)/\rho(z)$, the ratio of its pressure to its energy density. In particular, these measurements would enable us to rule out a cosmological constant as the source of the accelerating expansion if $w$ is sufficiently different from $-1$.

A first measurement of BAO was achieved in 2005 [Eisenstein et al., 2005], using the spectroscopic survey of SDSS. A sample of 47,000 luminous red galaxies (LRGs) in the range $0.16 < z < 0.47$ showed a peak in the two-point correlation function with 3-$\sigma$ significance. A similar level of detection was achieved the same year by the 2dF galaxy redshift survey [Cole et al., 2005]. Another SDSS observation was made using photometric-$z$ measurements with a much larger sample, 600,000 galaxies [Padmanabhan et al., 2007]. Using photometric redshifts (photo-$z$'s) degrades the measurements, especially that of $H(z)$, and thus requires many more galaxies. The SDSS Data Release 7 result [Percival et al., 2010] uses nearly 900,000 galaxies (including all spectroscopic SDSS galaxies, not just LRGs, and including 2dFGRS galaxies) to obtain measurements of $[d_A^2/H]^{1/3}$ with a precision of 3%.

The next step in BAO measurement is the BOSS experiment [Schlegel, White, and Eisenstein, 2009], currently in progress. It is part of the SDSS-III program and will collect spectra of 1.5 million LRGs out to $z = 0.7$. In addition, BOSS will use QSOs as sources to detect the distribution of neutral hydrogen along the line of sight. These Ly-$\alpha$ forest measurements will supplement the LRG measurements and extend the range of redshift that can be studied.

DETF established a nomenclature for dark energy experiments, which has been adopted generally. Stage I dark energy experiments are those that were completed at the time of the DETF report, May 2006. Stage II experiments were those underway at the time of the DETF report. Stage III were near-term, medium cost experiments, such as the BOSS and DES experiments, and were expected to improve by at least a factor of three in the DETF FOM over Stage II. Stage IV experiments as defined by the DETF are major, future experiments that would provide at least another factor of three improvement in FOM over Stage III. BigBOSS has the science reach of a Stage IV experiment, in the same category as major projects like LSST or WFIRST, but at a much more modest cost. BigBOSS will measure baryon acoustic oscillations, extending the maximum redshift probed using galaxies from $z \approx 0.7$ for BOSS out to $z \approx 1.7$, increasing by an order of magnitude both the volume probed and the number of galaxies mapped.

BigBOSS will measure much more than the BAO signal. The three-dimensional galaxy power spectrum, which is the Fourier transform of the two-point correlation function, encodes information about the initial source of fluctuations and the expansion history, constituents, and structure of the universe. While the remarkable measurements of the cosmic microwave background by COBE, WMAP, and now Planck, give a two-dimensional snapshot of the universe at the moment of recombination of electrons and nuclei to form atoms, the tomographic measurements of the three-dimensional power spectrum provide a motion picture of the evolution of universe.

DETF provided a single figure of merit, which can be used to compare the capability of various combinations of experiments to probe dark energy. Using the simple parameterization

$$w(a) = w_0 + (1 - a)w_a, \tag{2.1}$$

a figure of merit can be defined as the reciprocal of the area of an error ellipse in the $w_0 - w_a$ plane. A conventional normalization takes for the figure of merit the square root of the determinant of the $2 \times 2$ Fisher matrix for $w_0$ and $w_a$.

As we show below, BigBOSS will dramatically increase our understanding of cosmology and of dark energy in particular. At a minimum, the galaxy measurement of BAO will triple the figure of merit from all Stage III galaxy BAO experiments (BOSS, WiggleZ and HETDEX) combined, using only the Planck CMB results as a prior in both cases. But the potential from additional BigBOSS measurements is very much greater. Redshift space distortions (RSD), described below, provide additional information on the basic cosmological parameters. Exploitation of the full power spectrum, also described below, provides additional information, which could provide a further factor of three to six in the DETF FOM over galaxy BAO alone Exactly how far this can take us will depend on the range of scales over which we can understand the gravitational non-linearity and the galaxy bias. Detailed estimates are provided in subsequent sections.

BigBOSS will have an impact on cosmology beyond just the issue of dark energy. By measuring the power spectrum across a range of $z$, it will obtain measurements and limits on the primordial power spectrum that can be compared with the predictions of inflationary theory. Similarly, it will search for primordial non-Gaussianity, which would be a signature of either an unusual class of inflation models or an additional source of primordial fluctuations. BigBOSS will also address a fundamental question of particle physics, probing the absolute scale of neutrino masses and potentially ruling out the inverted mass hierarchy. Comparable sensitivity is extremely hard to obtain in nuclear or particle physics experiments.

Altogether, BigBOSS will provide a remarkably broad program in fundamental science.

## 2.2   Overview of the BigBOSS BAO Survey

The BigBOSS instrument features 5000 robotically actuated fibers located at the Mayall prime focus, feeding ten three-arm spectrographs. This new instrument will enable a massively parallel wide-field spectroscopic survey using 500 nights over five years at the Mayall telescope and will also provide the NOAO community with a powerful new capability to carry out large-scale spectroscopic observations. An overview of the BigBOSS instrument is given in Table 2.1; more details can be found in Chapter 5.

Table 2.1: BigBOSS instrument overview.

| Parameter | Value | units |
|---|---|---|
| Configuration | Prime Focus | |
| Focal plane diameter | 0.95 | m |
| Linear field of view diameter | 3 | degrees |
| Slew & settle time (3 deg move) | < 1 | minute |
| Number of fibers | 5000 | |
| Fiber density | 713 | per square degree |
| Focal plane plate scale | 82.6 | $\mu$m arcsec$^{-1}$ |
| Fiber center-to-center spacing | 12 | mm |
| Fiber actuator throw diameter | 3 | arcmin |
| Fiber diameter | 1.45 | arcsec |
| Wavelength coverage | 340 - 1060 | nm |
| Resolution | 3000 - 4800 | $\lambda/\Delta\lambda$ |
| Re-positioning speed | < 1 | minute |
| Re-positioning accuracy | < 5 | $\mu m$ |

The BigBOSS key science project is a wide-area survey designed to map the large-scale structure of the universe over the redshift range $0.2 < z < 1.7$ using emission-line galaxies (ELGs) and luminous red galaxies (LRGs). ELG redshifts are determined through detection of the [OII] doublet at a rest-frame wavelength of 3727 Å (the doublet nature of the line makes it identifiable even when it is the only feature present), while LRG redshifts are determined using the 4000 Å Balmer break feature. LRGs are more highly biased tracers of the dark matter halos and thus better suited for measuring the BAO feature, while the less strongly biased ELGs can be detected to higher redshifts and are better suited for measuring the early growth of structure and redshift space distortions. The use of two different galaxy populations with overlapping redshift distributions provides a check on the intrinsic systematic errors.

In addition, quasars (QSOs) at high redshift ($z > 2.2$) will be used as backlights to probe large-scale structure delineated via Ly-$\alpha$ absorption, a technique that is currently being pioneered by the BOSS experiment. Each QSO provides many measurements of the matter distribution along its path, so the small fraction of fibers allocated to QSOs still provides a well-sampled map of large scale structure at very early times, constraining models of early dark energy and providing strong constraints on curvature.

We have carried out simulations of the BigBOSS survey with numerous variants to determine the optimal use of the available survey time. Generally speaking, the most powerful survey is the one that covers the largest area, with the constraint that the co-moving target density be greater than about $1 \times 10^{-4} (h/\text{Mpc})^3$. The lower limit on the density results from our desire to use reconstruction to partially correct for the erasure of structure due to non-linearity, as described by [Eisenstein et al., 2007].

The BigBOSS targets are selected from photometric imaging surveys that are currently in progress and are expected to be completed and available in time for BigBOSS, such as the Palomar Transient Factory (PTF), PanSTARRS-I and WISE. As described in detail in Chapter 4, ELG and LRG targets are defined by selection algorithms based on colors and magnitudes. LRG targets are chosen to complement the ongoing 10,000 deg² BOSS survey with an additional 4,000 deg² and with galaxies at higher redshift. The first observation of any patch of sky (out of an eventual five) will target a higher density of QSO targets (250 deg⁻²) covering the redshift range from $z = 0$ to $z = 3.5$ (quasars at $z > 3.5$, which are still valuable on a per-object basis, will also be targeted, but they are too sparse to make a big overall contribution). The QSO spectra collected during the first pass will be analyzed, and a subset of QSO targets (denoted "Ly-$\alpha$ QSOs") will be selected at a target density of about 50 deg⁻² with high purity in the redshift range $2.2 < z < 3.5$. This strategy requires at least a day between the first and subsequent passes (see Chapter 4 for detailed description).

In this document we present a baseline survey as a demonstration of the scientific reach that will be possible with BigBOSS. While it has been developed with sufficient detail to provide a credible demonstration of the strong scientific potential of BigBOSS, we fully expect it to be improved as planning continues. The full justification of this baseline survey feasibility, informed by exposure time calculations, completeness estimates, and weather simulations, can be found in Chapter 6.

Further optimization of survey parameters will be possible, folding together the projected instrument sensitivity, the available targets that can be selected from photometric survey data, the allocation of fibers to various categories of objects for different exposure times, projections of expected weather, airmass and seeing conditions, and tradeoffs between survey depth, breadth and completeness. A number of improvements in efficiency can be anticipated. For example, the robotically actuated targeting is intrinsically flexible, and it will even be possible to modify the survey strategy as it progresses, in response to initial results and new information from other experiments.

For the baseline BigBOSS survey we assume 500 dark/grey nights of observing at the Mayall telescope, distributed over five years (with emphasis on the first four), excluding the monsoon months. The average observing time per night during this period is 9.5 hours, defined to include the period when the sun is at least 18° below the horizon (astronomical twilight). Based on weather and seeing records at the Mayall we project that 62% of this time will be useful for astronomical observations, for a total of 2945 hours.

The BigBOSS survey configuration described in Table 2.2 is composed of a densely packed set of $\approx 10,000$ spectroscopic observations over 14,000 deg². Since the field of view available in each observation is $7.0 \deg^2$, the observations are highly overlapping. Each area of sky is covered on average by five independent observations ("Mean # of observations"). A survey planning tool has been used to optimize the observing plan over the nominal footprint, taking into account airmass, seeing, extinction and sky brightness. The details are described in Chapter 6; the essential result is that BigBOSS can cover all 14,000 square degrees in 500 nights, with the galaxy and QSO densities in Table 2.2.

Table 2.2: BigBOSS survey overview.

| Parameter | Value | units |
|---|---|---|
| Survey area | 14,000 | sq. degrees |
| Focal plane area | 7 | sq. degrees |
| Fibers per exposure | 5000 | |
| Fiber density | 714 | per sq. degree |
| Exposures in survey | 10,000 | |
| Mean # of observations | 5 | per area |
| Max. target density | 3570 | per sq. degree |
| Number of nights | 500 | |
| Fraction clear | 0.62 | |
| Useful observing | 2945 | hours |
| Ave. time per pointing | 88 | minutes |
| Overhead per exposure | 1 | minute |
| Ave. exposure per tile | 16.6 | minutes |
| ELG min. [OII] flux | $0.9 \times 10^{-16}$ | ergs/s/cm$^2$ |
| ELG exposures per target | 1 | |
| ELG mean exposure time | 16.6 | minutes |
| ELG fiber allocation[1] | 0.64 | |
| ELG target density | 2335 | per sq. degree |
| ELG fiber completeness | 0.80 | |
| ELG target selection efficiency | 0.65 | |
| ELG redshift measurement efficiency | 0.9 | |
| ELG redshifts | 1092 | per sq. degree |
| Total ELGs | 15,302,200 | |
| LRG exposures per target | 2 | |
| LRG mean exposure time | 33.2 | minutes |
| LRG fiber allocation[1] | 0.2 | |
| LRG target density | 356 | per sq degree |
| LRG fiber completeness | 0.80 | |
| LRG target selection efficiency | 0.90 | |
| LRG redshift measurement efficiency | 0.95 | |
| LRG redshifts | 244 | per sq. degree |
| Total LRGs | 3,409,000 | |
| Total Galaxies | 18,711,200 | |
| QSO exposures per target | 5 | |
| QSO mean exposure time | 83 | minutes |
| Ly-$\alpha$ QSO fiber allocation[1] | 0.14 | |
| Ly-$\alpha$ QSO target density | 65 | per sq. degree |
| Ly-$\alpha$ QSO fiber completeness | 0.80 | |
| Ly-$\alpha$ QSO redshift measurement efficiency | 0.9 | |
| Ly-$\alpha$ QSOs | 47 | per sq. degree |
| Total Ly-$\alpha$ QSOs | 630,000 | |

[1]Before fiber completeness and averaged over the survey. See Table 6.2.

This survey configuration, combined with the flexibility of the fiber actuator system, gives BigBOSS some ability to balance exposure times and target number densities. A single exposure on the sky will be 1000 seconds, dynamically adjusted upwards in worse-than-median observing conditions. This exposure time is sufficient to determine the redshift for 65% of the ELG targets from the prominent [OII] doublet. The doublet is detected at $8\sigma$ for a line flux of $0.9 \times 10^{-16} \mathrm{ergs/s/cm}^2$. This threshold is conservative and our galaxy counts do not include the slightly fainter ELGs, which will be detected at lower significance. LRGs are each observed twice, for a total exposure time of 2000 seconds, and QSOs are each observed five times, for a total exposure time of 5000 seconds. The exposure times for galaxy spectra are generated using the exposure time calculator (ETC) described in Appendix A. Wherever possible, the ETC draws on measured data from the BOSS survey, such as the sky glow out to 10000Å at moderate resolution and instrumental throughputs. We also incorporate the seeing (average 1.1 arcsec) and atmospheric extinction measured from the Mayall. By basing our ETC on measured data from these sites and instrumentation, we have confidence that our projected exposure times are reasonable.

To estimate the total number of objects that will be surveyed we must take into account several sources of inefficiency, which we list in Table 2.2. First, we find that in the proposed configuration and proposed target densities we achieve about 80% fiber completeness (the percentage of all potential targets that actually have their full set of exposures completed). Second, some fraction of the selected objects will either (1) not have bright enough spectral features required to attain a redshift, (2) lie outside our redshift range of interest, or (3) are the incorrect type of object. We detail these effects in Chapter 4 and encapsulate these effects into an overall "target selection efficiency". We expect this factor to be most important for ELGs and QSOs where the efficiency critically depends on photometric selection techniques. Third, some fraction of otherwise properly selected objects will have non-detectable or low quality redshifts from pipeline software ("redshift measurement efficiency"). This effect is largest for the ELGs, where some fraction of the targets inevitably will have [OII] doublets that are lost due to bright sky emission lines. However, these last two sources of inefficiency are less important for LRGs, which we expect to target and detect very reliably.

Also listed in Table 2.2 is the fraction of available fibers used for each target class ("fiber allocation"; see also Table 6.2). Note that in this baseline survey, on average 80% of the fibers are utilized in each exposure, leaving a substantial number of unused fibers. We reserve some fraction of the unused fibers for calibration (i.e., sky and standard stars). Although the number of calibration fibers is to be determined, a significant fraction (between 10% and 20% of the 5000 fibers in each BigBOSS pointing, corresponding to between 5 and 10 million spectra over the course of the 14,000 $\deg^2$ survey) can be made available for ancillary science targets (see Chapters 3 and 6). A complete breakdown of the fiber allocation scheme is presented in Table 6.2.

This survey will yield the galaxy density distribution in $z$ displayed in Table 2.3 and the QSO density distribution shown in Table 2.4.

Table 2.3: Expected galaxy density distributions and resulting signal power to shot-noise power ratio. The BOSS LRG sample is included. $\bar{n}P_\star \equiv \bar{n}P(k = 0.14\ h\mathrm{Mpc}^{-1}, \mu = 0.6)$ represents the signal-to-noise ratio for a typical BAO-scale, redshift-space Fourier mode (see text – $\bar{n}P$ is summed over the two types of galaxy, given their biases).

| $z$ | $dn/dz_{LRG}$ (sq. deg.)$^{-1}$ | $dn/dz_{ELG}$ (sq. deg.)$^{-1}$ | $dn/dV_{LRG}$ ($10^{-4}h^3\mathrm{Mpc}^{-3}$) | $dn/dV_{ELG}$ ($10^{-4}h^3\mathrm{Mpc}^{-3}$) | $nP_\star$ |
|---|---|---|---|---|---|
| 0.15 | 47 | 247 | 2.80 | 14.73 | 8.32 |
| 0.25 | 117 | 148 | 2.80 | 3.55 | 4.53 |
| 0.35 | 209 | 69 | 2.82 | 0.93 | 3.67 |
| 0.45 | 314 | 120 | 2.84 | 1.08 | 3.76 |
| 0.55 | 426 | 429 | 2.86 | 2.88 | 4.42 |
| 0.65 | 443 | 888 | 2.36 | 4.73 | 4.49 |
| 0.75 | 533 | 1359 | 2.37 | 6.04 | 4.96 |
| 0.85 | 541 | 1712 | 2.08 | 6.58 | 4.79 |
| 0.95 | 435 | 1654 | 1.49 | 5.65 | 3.74 |
| 1.05 | 289 | 1284 | 0.90 | 3.98 | 2.44 |
| 1.15 | 104 | 941 | 0.30 | 2.70 | 1.28 |
| 1.25 | 0 | 680 | 0.00 | 1.83 | 0.63 |
| 1.35 | 0 | 582 | 0.00 | 1.48 | 0.50 |
| 1.45 | 0 | 630 | 0.00 | 1.53 | 0.52 |
| 1.55 | 0 | 592 | 0.00 | 1.39 | 0.46 |
| 1.65 | 0 | 424 | 0.00 | 0.97 | 0.32 |

Table 2.4: Expected QSO density, obtained by rescaling the BOSS QSO distribution to the BigBOSS target density.

| $z_{med}$ | 1.85 | 1.95 | 2.05 | 2.15 | 2.25 | 2.35 |
|---|---|---|---|---|---|---|
| $dn/dz_{QSO}$ | 5.51 | 7.54 | 12.0 | 39.6 | 74.1 | 68.7 |
| $z_{med}$ | 2.45 | 2.55 | 2.65 | 2.75 | 2.85 | 2.95 |
| $dn/dz_{QSO}$ | 54.7 | 43.7 | 34.1 | 25.0 | 22.3 | 20.0 |
| $z_{med}$ | 3.05 | 3.15 | 3.25 | 3.35 | 3.45 | 3.55 |
| $dn/dz_{QSO}$ | 19.1 | 16.2 | 13.7 | 8.52 | 4.55 | 3.62 |

## 2.3 Galaxies as Tracers of Cosmic Structure

### 2.3.1 The Galaxy Power Spectrum

Large portions of the cosmological information in BigBOSS is attainable with the reduction of 3-D maps to 2-D statistics. The fundamental statistic measured by BigBOSS will be the correlation function of galaxies, $\hat{\xi}_g(\mathbf{r}) = \langle \delta_g(\mathbf{x}) \delta_g(\mathbf{x}+\mathbf{r}) \rangle$, or its Fourier transform, the power spectrum, $\hat{P}_g(\mathbf{k})$, where $\delta_g(\mathbf{x})$ is the fluctuation in galaxy density at redshift-space position $\mathbf{x}$. In the large-scale limit, in the standard cosmological model, the relationship between $\delta_g$ and mass density fluctuations $\delta_m$ is described by the well-motivated [Kaiser, 1987; McDonald & Roy, 2009] linear bias plus shot-noise model, written here in Fourier space:

$$\delta_g(\mathbf{k}) = (b + f\mu^2)\delta_m(\mathbf{k}) + \epsilon \qquad (2.2)$$

where $\mathbf{k}$ is the wavevector, $\mu$ is the cosine of the angle between $\mathbf{k}$ and the observer's line of sight, $b = b(z)$ is linear bias, an *a priori* unknown parameter, with unknown redshift dependence, that is determined by galaxy formation physics, $f(z) = d\ln D/d\ln a$, where $D(z)$ is the linear growth factor (in the linear regime, $\delta_m(z) = D(z)\delta_i$ where $\delta_i$ is the initial perturbation) and $a = (1 + z)^{-1}$ is the expansion factor, and $\epsilon$ is an approximately white (spatially uncorrelated) noise variable. In this model the power spectrum of galaxies is related to the power spectrum of mass by

$$\hat{P}_g(\mathbf{k}) = P_g(\mathbf{k}) + P_N = (b + f\mu^2)^2 P_m(k) + \bar{n}^{-1} \qquad (2.3)$$

where $\bar{n}$ is the mean number density of galaxies – the noise power spectrum is $P_N = \bar{n}^{-1}$ because galaxies are to a good approximation Poisson distributed around the biased large-scale density field. This model is well-motivated in the large-scale ($k \to 0$) limit, but inevitably breaks down on small, non-linear scales. We can hope to detect deviations from this model on large scales signaling a non-standard cosmology, e.g., if the initial fluctuations are non-Gaussian, or the gravity theory is not GR.

Estimating how well BigBOSS can measure the power spectrum in the linear regime is straightforward because the field is very close to Gaussian (e.g., Feldman, Kaiser & Peacock [1994]). The key quantity is the signal-to-noise ratio per Fourier mode,

$$\frac{S}{N}(k) = \frac{P_g(k)}{P_g(k) + \bar{n}^{-1}} = \frac{\bar{n}P_g(k)}{1 + \bar{n}P_g(k)} \qquad (2.4)$$

We see that the S/N per mode goes to 1 for high galaxy density (low noise) and diminishes with decreasing density, with the key quantity governing the transition being $nP_g(k)$. The error on $P_g(k)$ averaged over many independent modes in some Fourier space volume element $d^3k$ ("band") is given by

$$\frac{\sigma_{P_g(k)}}{P_g(k)} = \frac{1}{\sqrt{N}}\frac{N}{S}(k); \qquad N = V\frac{d^3k}{2(2\pi)^3}, \qquad (2.5)$$

where the extra $(1/2)$ is the result of $\delta(x)$ being real. The errors on different bands are approximately uncorrelated (as long as the bands are not too narrow relative to the size of the survey). Fig. 2.1 shows what power spectrum measurements from the BigBOSS survey described in Sec. 2.2 will look like. The expected uncertainties in the power spectrum as a function of $k$ for various bands in $z$ are shown in Fig. 2.2. For all of our projections, we use the figure of merit defined by the JDEM Science Working Group (hereafter FoMSWG) in taking the values of the parameters of the fiducial cosmology to be the $\Lambda$CDM model of WMAP5 displayed in Table 2.5.

It is clear from Eq. 2.3 that any feature, like the BAO feature, that appears in the large-scale matter power spectrum will also appear in the galaxy power spectrum. In §2.3.2 we will isolate and discuss the BAO distance scale information in $P_g(k)$ and in §2.3.3 we will isolate the redshift-space distortions, where $f(z)$ is measured using the $\mu$ dependence of Eq. 2.3. Considerable cosmological information resides in the broadband power spectrum apart from the acoustic oscillation features and redshift-space distortions. The geometry of the Universe can be probed because the observable coordinates are redshift and angle, while Eq. 2.3 is implicitly written in comoving coordinates (the Alcock-Paczynski test is one aspect of this geometry effect). The low-$k$ slope measures the primordial perturbations power law index, carrying information on inflation and non-Gaussianity; the turnover depends on the redshift of matter-radiation equality, constraining the matter density and extra relativistic degrees of freedom such as additional neutrino species; the higher-$k$ tail provides leverage on the primordial power index tilt and running, neutrino masses, and dark matter properties.

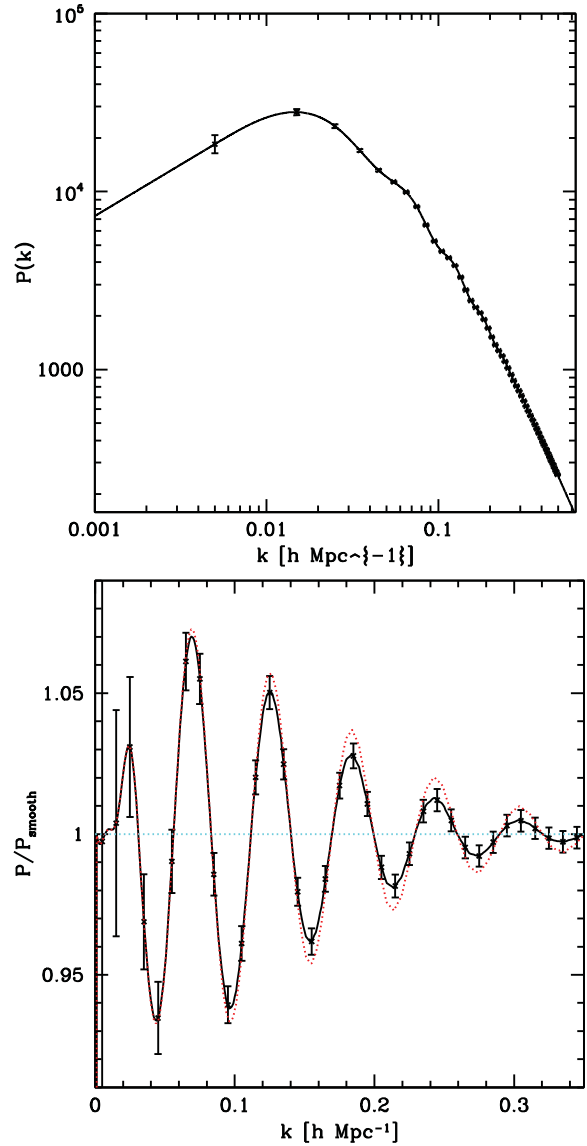

Figure 2.1: Simulated data showing the power spectrum for a bin $0.5 < z < 0.9$, with $\Delta k = 0.01h$ Mpc$^{-1}$. The upper panel shows the absolute power spectrum while the lower panel shows the same divided by a smooth, wiggle-free power spectrum to emphasize the BAO signal. The red dotted is linear, while the solid black line includes smearing due to non-linear effects, corrected by reconstruction.

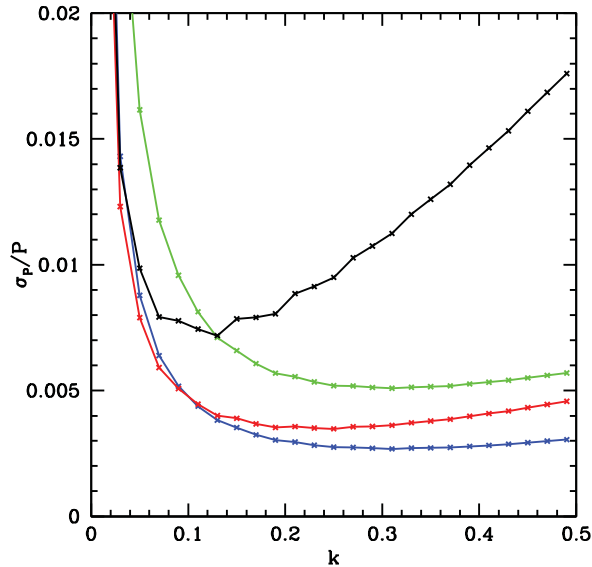

Figure 2.2: Expected fractional uncertainties in the power spectrum for bins of $\Delta k = 0.02h^{-1}$ Mpc (indicated by points). The curves represent different redshift bins $z < 0.5$ - green, $0.5 < z < 0.9$ - blue, $0.9 < z < 1.3$ - red, $z > 1.3$ - black. The gravitational growth function is best constrained on scales $k > 0.05h^{-1}$ Mpc (smaller than 200 Mpc).

As we will show, constraints on cosmological parameters based on RSD or the full broadband power are much more powerful than those derived from just the BAO features. The BAO measurement is generally emphasized because the BAO feature has a unique form that cannot be corrupted by any known systematics. With increasing time, non-linearities in gravitational evolution and bias lead to the breakdown of Eq. 2.3 on increasingly large scales (small $k$); however, these effects generally only smear the BAO feature and degrade the statistical errors; they do not lead to any significant bias in the distance scale measurement [Seo et al., 2010; Montesano, Sanchez, & Phleps, 2010; Jeong & Komatsu, 2009]. To fully exploit the information in the broadband power spectrum non-linearities must be modeled carefully. This will generally be possible up to some $k_{max}$, beyond which it cannot be done reliably and the power spectrum can no longer be used to constrain cosmological parameters.

Table 2.5: Values of the cosmological parameters used in calculations. Here $k_\star = 0.05$ Mpc$^{-1}$. Using $\omega_M + \omega_B + \omega_{\rm rad} + \omega_k = h^2$, the value of the Hubble constant is 71.9 kms$^{-1}$Mpc$^{-1}$. Our model has $\sigma_8 = 0.791$.

| | |
|---|---|
| $n_S$ | 0.963 |
| $\omega_M$ | 0.13263 |
| $\omega_B$ | 0.02273 |
| $\omega_k$ | 0 |
| $\omega_{\rm DE}$ | 0.3843 |
| $\ln \Delta_\zeta^2(k_\star) \equiv \ln A_s$ | $-19.9628$ |

Modeling issues for nonlinearities (including scale-dependent bias) are common to both RSD and more general uses of the broad-band power and can be treated by marginalization over nuisance parameters [Schulz & White, 2006; Padmanabhan & White, 2009],the halo model, higher order perturbation theory, or a low order polynomial or Padé approximant in $k$ [Schulz & White, 2006; Padmanabhan & White, 2009; Cresswell & Percival, 2009; White, 2005; McDonald, 2006; McDonald & Roy, 2009].

### 2.3.2 BAO Features in the Power Spectrum

Initial fluctuations in the matter density provided sources for "acoustic" waves that propagated in the photon-electron-baryon plasma of the early universe (see, for example, [Eisenstein & Hu, 1998]). Before the wave stops propagating at the epoch of recombination, it travels a co-moving distance $s \approx 150$ Mpc, which can be computed quite precisely from known cosmological parameters. An excess of matter is left both at the source of the wave and at the surface of the sphere of radius $s$. The sources are distributed randomly but the pattern of separation at a distance $s$ is visible in the two-point correlation function $\xi(r)$.

The distance $s$ provides a standard rule. Viewed transversely, the 150 Mpc meter stick subtends an angle $\theta$ such that

$$s = (1+z)d_A(z)\theta \qquad (2.6)$$

where $d_A(z)$ is the angular-diameter distance to an object at redshift $z$. Viewed along the line of sight, a peak in the two-point correlation function will be present when $\Delta z$ is such that

$$\frac{\Delta z}{H(z)} \approx s \qquad (2.7)$$

The observation of the peak in the two-point correlation function thus provides a means of measuring both $H(z)$ and $d_A(z)$.

Although we observe only luminous galaxies and not the dominant dark matter, fluctuations in the galaxy density track the fluctuations in matter on large scales, as described by Eq. 2.3. In looking for the signal of baryon acoustic oscillations we do not need to know the galaxy bias nor must we even assume a strictly linear growth of structure, since it is the location of the peak in the two-point correlation function and not its amplitude that is of interest.

We have used the Fisher matrix code of [Seo & Eisenstein, 2007] to make predictions for the effectiveness of BigBOSS galaxy BAO measurements. The smearing of the BAO feature by non-linear effects is taken into account by including the BAO-signal suppression factor [Seo & Eisenstein, 2007]

$$\exp\left(-\frac{1}{2}k_\perp^2 \Sigma_\perp^2 - \frac{1}{2}k_\parallel^2 \Sigma_\parallel^2\right). \tag{2.8}$$

The parameters $\Sigma_\perp$ and $\Sigma_\parallel$ are given by

$$\Sigma_\perp = \Sigma_0 D(z), \tag{2.9}$$
$$\Sigma_\parallel = \Sigma_\perp (1+f), \tag{2.10}$$

where $D(z)$ is the growth function, normalized to $(1+z)^{-1}$ for large $z$ and where $f = d\ln D/d\ln a$. The scale for non-linear erasure of the BAO feature (related to the typical Lagrangian displacement scale) is given by [Seo & Eisenstein, 2007; Eisenstein, Seo & White, 2007]:

$$\Sigma_0 = 12.4 h^{-1}\,\mathrm{Mpc} \times (\sigma_8/0.9)\ . \tag{2.11}$$

The damping effects of non-linear clustering can be partially corrected by the process of "reconstruction" [Eisenstein et al., 2007]. We assume that the level of reconstruction scales with the number density as in White [2010].

Our calculations are based on the expected numbers of galaxies BigBOSS will obtain as described in Sec. 2.2 and Sec. 4. These numbers are displayed in Table 2.3.

For the galaxy distributions in Table 2.3 we have taken the bias of the LRGs to be 1.7 and that of the ELGs to be 0.84 at $z = 0$. We assume the power spectrum amplitude for each type of galaxy, $i$, is roughly constant with redshift, i.e.,

$$P_i(z,k) \sim P_i(0,k) = b_i(0)^2 P_m(0,k). \tag{2.12}$$

Note that in Table 2.3 we list $\bar{n}P(k = 0.14\ h\mathrm{Mpc}^{-1}, \mu = 0.6)$, rather than the traditional $\bar{n}P(k = 0.2\ h\mathrm{Mpc}^{-1}, \mu = 0.0)$. Recall that the motivation for targeting $\bar{n}P = 1$ is that it can be shown that this gives the optimum survey volume/density for fixed total number of galaxies, however, $P$ is not the same for all $\mathbf{k}$ modes, so one is left with the question of which $\mathbf{k}$ to choose if a quick representation of typical S/N is desired. We find by experimentation that setting $\bar{n}P(k = 0.14\ h\mathrm{Mpc}^{-1}, \mu = 0.6) = 1$ will generally produce an optimal BAO distance measurement in the volume/density trade-off at fixed total number, i.e., this definition gives a better picture of where a survey, considering all different $\mathbf{k}$, falls in the density vs. area trade-off than does $\bar{n}P(k = 0.2\ h\mathrm{Mpc}^{-1}, \mu = 0.0)$.

We work with bins of $\Delta z = 0.1$ and in each bin we determine the expected fractional uncertainty in $d_A(z)$ and $H(z)$. The results are shown in Table 2.6.

### 2.3.3 Redshift-space Distortions

Under the standard assumption of isotropy, $\xi$ is a function only of the magnitude of $\mathbf{r}$, not its direction; correspondingly the power spectrum depends solely on the magnitude of $\mathbf{k}$. In fact, the redshift of a galaxy is affected by its peculiar velocity as well as by the Hubble flow.

Table 2.6: Predicted fractional uncertainties for BigBOSS in $D$ and $H$ in bins of $z$ for 14k square degrees assuming the galaxy density distribution given in Table 2.3.

| $z$ | $\sigma(d_A/s)/(d_A/s)$ | $\sigma(Hs)/Hs$ |
|---|---|---|
| 0.05 | 0.0657 | 0.1261 |
| 0.15 | 0.0285 | 0.0520 |
| 0.25 | 0.0201 | 0.0357 |
| 0.35 | 0.0156 | 0.0276 |
| 0.45 | 0.0125 | 0.0221 |
| 0.55 | 0.0102 | 0.0180 |
| 0.65 | 0.0089 | 0.0156 |
| 0.75 | 0.0078 | 0.0137 |
| 0.85 | 0.0073 | 0.0126 |
| 0.95 | 0.0073 | 0.0123 |
| 1.05 | 0.0080 | 0.0128 |
| 1.15 | 0.0103 | 0.0153 |
| 1.25 | 0.0154 | 0.0203 |
| 1.35 | 0.0172 | 0.0224 |
| 1.45 | 0.0164 | 0.0213 |
| 1.55 | 0.0172 | 0.0223 |
| 1.65 | 0.0222 | 0.0282 |
| $< 0.5$ | 0.0083 | 0.0148 |
| $0.5 - 1.0$ | 0.0036 | 0.0063 |
| $> 1.0$ | 0.0049 | 0.0070 |

The peculiar velocity has both a random component and a component that is a response to the gravitational field of the non-uniform distribution $\delta$ itself. The magnitude of the latter is proportional to the rate of growth of the perturbation, $\dot{D}$. As first shown by Kaiser [1987], this leads to the $f(z)\mu^2$ term in in Eq. 2.3.

It is possible to show that, to a very good approximation, $f$ can be written as

$$f = \Omega_m(a)^\gamma, \tag{2.13}$$

where $\Omega_m(a)$ is the fraction of the energy density at scale factor $a$ that is due to matter and $\gamma$ is a constant, the gravitational growth index. If General Relativity holds, $\gamma$ is very near $6/11$, with only a slight dependence on the equation of state of dark energy. In alternatives to General Relativity, $\gamma$ can differ from this value by as much as $0.1 - 0.2$.

Observations to date have already demonstrated the feasibility of tracing the amplitude of the distortion field in the two point clustering statistics and have already led to preliminary estimates of $f(a)$ at various redshifts, although not yet at a useful level of precision [Guzzo et al., 2008]. For example, using Eq. (2.13) current constraints on the growth index $\gamma$ have an accuracy of about 40%, which is still far from the precision required to reject at least the two modified gravity models most discussed in literature i.e. DGP and $f(R)$ models. Nonetheless, it has been shown that large (nearly all-sky) and deep ($0 < z < 2$) galaxy redshift surveys will constrain the amplitude of the distortion field to the precision needed to discriminate distinctive departures from general relativity on cosmological length scales [White et al., 2009; McDonald & Seljak, 2009; Guzzo et al., 2008; Percival & White, 2009; Simpson & Peacock, 2010].

BigBOSS, covering an order of magnitude larger volume than present-day surveys, will bring uncertainties on $\sigma_8(z)f(z)$ (the amplitude of the power spectrum times the growth rate, i.e., what is really measured is $f^2(z)P_m(z,k)$) down to a few percent in the whole redshift window $0.5 < z < 1.6$. See Fig. 2.3. This accuracy will result in unprecedented constraints on both cosmological and gravitational parameters. Note that, when computing constraints using broadband power, we continue to use the Seo & Eisenstein [2007] BAO wiggle damping factor to damp the full broadband power signal (specifically, we divide the overall noise in the measurement, i.e., the Fisher matrix derivative, by this factor). This method has not been tested for this purpose but should give at least a rough approximation to the inevitable loss of information to non-linearities. $k_{max}$ then represents the scale where systematic errors in estimating the effect of non-linearities become too large to use the measurement at all. We use the same reconstruction factor as for BAO.

There are some potential sources of systematics that might bias redshift distortions estimators and prevent BigBOSS estimates from being effectively "data-limited" i.e. from exploiting the whole information contained in the data. Orbital motions of galaxies within virialized structures scatter galaxy redshifts along the line-of-sight creating 'Fingers of God' and thereby erasing spatial information on small scales. This non-linear effect can be phenomenologically modeled and disentangled from the aspect of interest, i.e. linear bulk motions. A common procedure assumes that non-linear random velocities can be treated as an independent nuisance parameter, which can be marginalized (e.g., Verde et al. [2002]). A more complete modeling of the distortion pattern seems required if we are to reach the forecasted level of precision (see, e.g, Scoccimarro [2004]; Percival & White [2009]; McDonald & Roy [2009]; Tinker, Weinberg, & Zheng [2006]; Tinker [2007]; Heavens, Matarrese, &

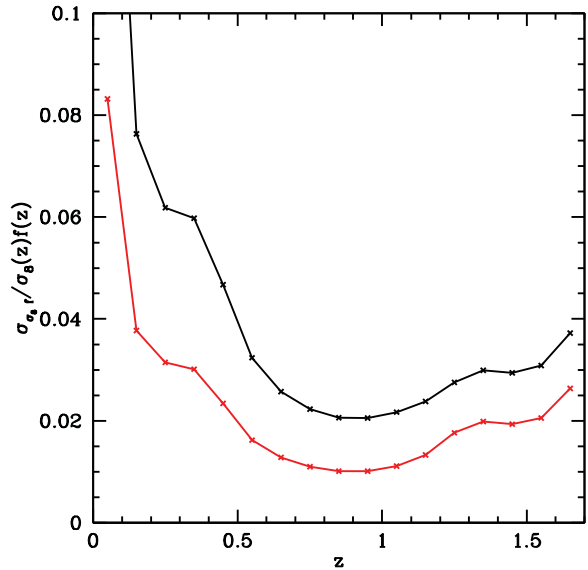

Figure 2.3: Constraints, derived from redshift-space distortions, on $\sigma_8(z)f(z)$ (amplitude of the power times growth rate), for bins of $\Delta z = 0.1$, for $k_{max} = 0.1$ or $0.2$ $h\mathrm{Mpc}^{-1}$ (upper and lower lines, respectively).

Verde [1998]). Also galaxy biasing, especially if it manifests with non-linear and/or scale-dependent features, must be efficiently removed in order not to contaminate cosmological interpretation. The possibilities currently explored range from combining different probes to define bias-independent estimators [Song & Percival, 2009; Zhang et al., 2007], to exploiting independent biasing estimates extracted from higher than two-point clustering statistics [Verde et al., 2002; Marinoni et al., 2005].

### 2.3.4  Galaxy-lensing Cross-correlation

The next 10 years will see dramatic advances in weak lensing observations due to new facilities coming on line. The first of these will be Pan-STARRS and Hyper Suprime-Cam in the Northern hemisphere, and Dark Energy Survey in the Southern hemisphere. LSST will follow with a deeper map over 70% of the sky. These surveys are essentially 2D in nature, relying on photometric redshifts to break the sky into a series of "shells" to obtain distance information. The BigBOSS Key Project provides an opportunity to improve our constraints from these surveys by providing information in 3D.

These new surveys will provide a measure of the growth of structure by comparing the lensing signal from sources and lenses at different distances and factoring out the contribution due to the geometry. This standard technique is called shear-shear correlations, and the full power of combining such lensing maps with galaxy surveys are only now being explored (e.g., [Bernstein, G.M., 2009]) but a key role is played by spectroscopic observations which can help to pin down the redshift distribution. Shear-shear weak lensing analysis also suffers from systematic error such as spurious signals induced by distortions and intrinsic alignments. In contract, weak lensing analysis around point sources such as galaxies or clusters eliminates most of these systematic issues. If the redshift of these lens galaxies is known one can express the lensing signal in terms of physical surface mass density as a function of transverse separation, instead of dimensionless shear as a function of angular separation in shear-shear analysis, thereby eliminating one projection along the line of sight. BigBOSS will provide the redshifts for such an application. In addition BigBOSS will measure velocity-density correlations, which probe the logarithmic growth of the structure $f = d \ln D / d \ln a$. The combination of lensing and velocity correlations allow another determination of galaxy bias, and hence the reconstruction of the three-dimensional dark matter field. These represent powerful approaches to measuring the growth of structure that will strengthen the science reach of Pan-STARRS, Hyper Suprime-Cam as well as DES and LSST in their areas of overlap.

Additionally, a joint analysis of weak lensing maps with BigBOSS provides a test of General Relativity. Galaxy clustering, weak lensing and redshift-space distortions can be combined to derive a quantity that does not depend on either galaxy bias or the amplitude of clustering. This quantity, $E_g$, is extremely sensitive to modifications in the gravity sector [Zhang et al., 2007]. Such an analysis has already been used to eliminate a specific alternative to dark matter called TeVeS [Reyes et al., 2010]. BigBOSS combined with LSST should reduce these errors substantially, providing a detailed probe of gravity on cosmological scales.

## 2.4   The Ly-$\alpha$ Forest Flux as Tracer of Cosmic Structure

### 2.4.1   Ly-$\alpha$ Forest

The Ly$\alpha$ forest is the collection of absorption lines blueward of the Ly$\alpha$ emission line (at 1215.67Å rest-frame wavelength) observed in the spectra of distant quasars. The absorption is caused by the neutral hydrogen along the line of sight to the quasar in question. Since the neutral hydrogen traces cosmological density fields, measuring the Ly-$\alpha$ forest can provide the statistical properties of the Universe at the high redshift and thus constrain cosmology.

The atmosphere becomes increasingly opaque at wavelengths blueward of 3500Å and ground-based observations of the Ly-$\alpha$ forest are limited to the quasars at redshifts of $z > 1.8$. In principle there is no upper limit to the redshift of the quasars used, but the decreasing apparent brightness of quasars means that the relative contribution to a BAO measurement becomes small above $z \sim 3.5$. Moreover, in each quasar, the readily available forest lies between the rest frame Ly-$\alpha$ and Ly-$\beta$ emissions, where absorption is dominated by the Ly-$\alpha$ absorption. Therefore, a sample of quasars limited to quasar redshift $z_q \lesssim 3.5$ can be used to measure the density fields in the redshift range between $z \gtrsim 1.9$ and $z \lesssim 3.4$. The Ly-$\alpha$ forest is a unique probe of cosmology at those redshifts[3].

The gas in the intergalactic-medium is in photoionization equilibrium with a nearly uniform ionizing background. Since recombination is a two-body process and since the ionization fraction is nearly unity, the number density of neutral hydrogen atoms is roughly proportional to the square of the density of ionized hydrogen. In practice, the exponent is closer to 1.8 because the gas in the denser environment is also hotter and the recombination coefficient has a non-negligible temperature dependence. The flux measured in the forest is therefore approximately given by

$$f(\lambda) = C(\lambda)e^{-\tau(z)} = C(\lambda)e^{A(1+\delta)^{\beta}}, \tag{2.14}$$

where $f$ is the flux measured by the spectrograph, $C$ is the unabsorbed quasar continuum, $A \sim 0.1$ and $\beta \sim 1.8$ are constants and we will use $\tau$ to denote optical depth throughout. This approach is usually referred to as the fluctuating Gunn-Peterson (FGPA) approximation [Weinberg, Katz, & Hernquist, 1998].

---

[3]Obtaining a sufficiently dense sample of galaxies at those redshifts would require a prohibitive amount of observation time . Proposed 21cm intensity mapping experiments (e.g., Chang et al. [2008]) might be able to measure BAO in this redshift range, with a telescope greater than 100m in diameter, but it would require an even larger telescope to resolve the smaller scales probed by the Ly-$\alpha$ forest. Furthermore, this attempt would be fundamentally limited by the fact that the observed DLAs which dominate the total neutral density are much more sparse and non-linear than the IGM probed by the Ly-$\alpha$ forest.

### 2.4.2 BAO with Ly-$\alpha$ Forest

The possibility of measuring BAO through Ly-$\alpha$ forest was first discussed by White [2003]. McDonald & Eisenstein [2007] performed calculations that, when applied to BOSS, show that it should be able to measure the BAO distance scale at $z \sim 2.5$ to $\sim 1.5\%$ precision using the Ly-$\alpha$ forest. In Slosar et al. [2009], cosmological simulations were populated with neutral hydrogen gas according to FGPA approximation and the ability of BOSS to measure BAO through Ly-$\alpha$ forest was confirmed. This work has been further extended with even larger simulations in White et al. [2009].

BAO has not been seen yet in the Ly-$\alpha$ forest; however, the signal must inevitably be there (as it would be in any tracer of the large-scale density field), and there are several reasons to be optimistic that potential systematic effects can be kept under control. The most important tool that we have is the knowledge that the only signal that corresponds to real three-dimensional fluctuations in the optical depth to the Ly-$\alpha$ transition will in fact correlate across different quasars. Other effects, such as continuum fluctuations, will be a source of noise, but these will average out in the cross-correlated pairs. Moreover, any systematic that could correlate across quasars in the rest-frame will correlate equally for neighboring as well as widely separated quasars. Therefore, any correlations between fluxes in neighboring quasars above that present in widely separated quasars is unlikely to come from sources other than the real fluctuations in the optical depth.

An important caveat is that fluctuations in the optical depth can also be associated with non-gravitational processes. For example, if the photo-ionization field is modified by large-scale fluctuations in quasar number density, the optical depth field would follow. Such processes can introduce large-scale fluctuations. However, these effects generally will not produce a sharp feature that could be mistaken for the BAO peak. The smooth contributions to the two-point function can be modelled and marginalized over using techniques that are very similar to those employed by galaxy surveys. In fact, we have measured the 1D power in SDSS quasar spectra on the BAO scale (at the time of McDonald et al. [2006] but unpublished), so we know that there is not a lot of extra power in the data beyond the standard model that we include in our Fisher matrix projections. Therefore the statistical error projections should be robust, and we only need to rely on the usual argument that BAO measurements are not systematically sensitive to broad-band effects, because the BAO feature is sharp and well-localised in configuration space. Hence it can be detected well even in the present of large and potentially unknown background, as long as these are slowly varying, an assumption which is valid for all known contaminants.

### 2.4.3 Small Scale Power Spectrum with the Ly-$\alpha$ Forest

As discussed above, our ability to measure the large-scale fluctuations from the Ly-$\alpha$ forest beyond BAO has not yet been demonstrated. However, the small scale fluctuations in the Ly-$\alpha$ forest are much better understood. On small scales (100kpc - 10Mpc), the photo-ionization background can be assumed to be constant. Since the average line of sight through the Ly-$\alpha$ forest probes a typical rather than an overdense region (as galaxies do), so the fluctuations are in the weakly non-linear regime where hydrodynamic simulations are known to be reliable. There are, however, uncertainities associated with our understanding of the inter-galactic medium (IGM). For example, the thermal pressure prevents cosmic baryons from gravitationally collapsing on scales smaller than Jean's scale, setting a characteristic filtering scale, which depends on the thermal history of the IGM. Such uncertainties can be marginalised over, but higher-order correlators often break these degeneracies and provide useful cross-checks[Mandelbaum, McDonald, Seljak & Cen, 2003].

It has already been explicitly demonstrated that this approach works, using just 3000 quasars from the SDSS in McDonald et al. [2006]. The statistic of choice in this case is the one-dimensional flux power-spectrum on small scales (where continuum fluctuations can be assumed to be negligible). These measurements are very competitive at constraining the amplitude and slope of the matter power spectrum at wave-vectors corresponding to scales around 1 Mpc. These can in turn be used to put constraints on neutrino masses and inflationary parameters. Six hundred thousand BigBOSS quasar sightlines will provide not only an unprecedented opportunity to control systematics, but will also lead to some very competitive constraints as discussed in the Sections 2.5.2 and 2.5.3.

## 2.5 Cosmological Constraints from BigBOSS

### 2.5.1 Combined Constraints on Dark Energy

The galaxy-BAO measurement from BigBOSS provides the most tried and tested constraints. From the uncertainties in $d_A$ and $H$ in Table 2.6, we can derive uncertainties in $w_0$ and $w_a$, and thus the DETF figure of merit. The results are shown in Table 2.7 and Figure 2.4. BigBOSS will improve over *all* Stage III galaxy BAO by a factor of three. The bar chart in Figure 2.5 provides a further illustration of the improvement gains in cosmological parameter estimation that result from the proposed BigBOSS Key Project over current uncertainties [from Komatsu et al., 2010]. In addition to the gains made using the well-tested galaxy-BAO, the figure also shows the added benefits of using Ly-$\alpha$ BAO and the full broad-band power spectrum analysis.

To combine the galaxy-BAO constraints with those from other techniques, we simply add together the Fisher matrices containing the information about the cosmological parameters for all the $z$-bins and for all techniques considered. The results are shown in Table 2.7, which shows that the galaxy BAO constraints from BigBOSS alone will approximately triple the DETF FoM for all Stage III galaxy BAO results combined, and that the potential for BigBOSS is very much greater when the Ly-$\alpha$ forest and full broadband galaxy power are used. The results indicated by $k_{max} = 0.15, 0.3$ use the full broadband spectrum as observed in angle and redshift coordinates, and not just the linear theory redshift-space angular dependence; this includes (but is not limited to) the Alcock-Paczynski effect. In all cases we use "propagator limited" galaxy information, i.e., signal power is suppressed by Seo & Eisenstein [2007]-like Gaussian damping factors (after reconstruction) (e.g., our $k_{max} = 0.3 h^{-1}$Mpc case is *not* assuming linear theory is perfect out to this $k$). The coverage is 14k square degrees with $dn/dz$ as predicted for BigBOSS in Table 2.3.

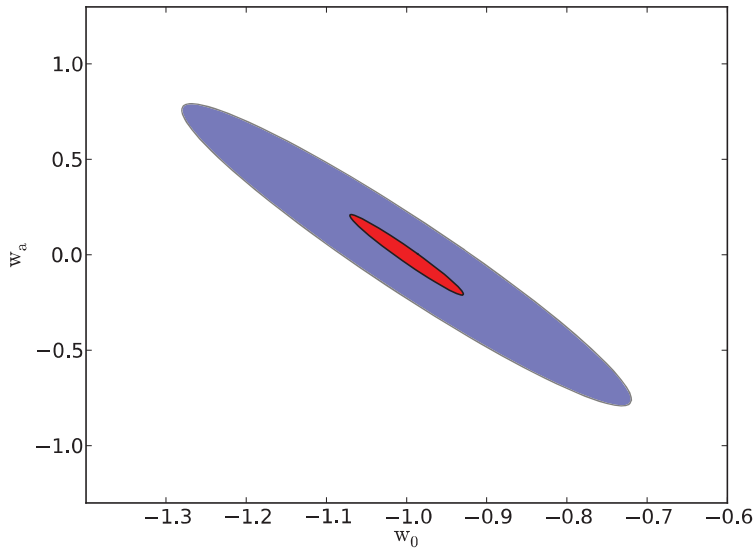

Figure 2.4: Contours at 68% confidence in the $w_0$-$w_a$ plane. The inner (red) ellipse uses BigBOSS galaxies and Ly-$\alpha$ broadband power spectrum ($k_{max} = 0.15$) over 14,000 sq. degrees together with Planck. The outer ellipse is Stage III BAO (BOSS, WiggleZ, and HETDEX) with Planck.

As pointed out by Simpson & Peacock [2010], measuring redshift-space distortions while neglecting geometric effects, and combining it with a BAO measurement treated as an independent constraint, is not self-consistent. This approach could underestimate the errors, i.e., the errors using all information could in principle be larger than those of the BAO+RSD case. However, our results show that including all the information self-consistently does improve the results significantly over the BAO+RSD case. This is not inconsistent with Simpson & Peacock [2010], who focused specifically on distinguishing dark energy from modified gravity. Since there is no reason to believe that using RSD alone is significantly more reliable than using all of the information available, we take this as the BigBOSS projection, if one can go beyond BAO.

### 2.5.2 Neutrino Properties

The effects of neutrinos in cosmology are well understood. They decouple from the cosmic plasma when the temperature of the Universe is about 1 MeV, just before electron-positron annihilation. While ultra-relativistic, they behave as extra radiation (albeit not electromagnetically coupled) with a temperature equal to $(4/11)^{1/3}$ of the temperature of the cosmic microwave background. As the universe expands and cools, they become non-relativistic and ultimately behave as additional dark matter.

Table 2.7: Figures of merit achieved with the BigBOSS 14,000 deg$^2$ survey. The Stage III BAO includes BOSS, HETDEX, and WiggleZ. All Stage III adds SNe and WL using the FOMSWG Fisher matrices.

| Experiments | DETF FoM |
|---|---|
| *Galaxy BAO:* | |
| Planck + StageIII BAO | 31 |
| Planck + BigBOSS BAO | 102 |
| Planck + all StageIII + BigBOSS BAO | 178 |
| | |
| *Galaxies and Lyα BAO:* | |
| Planck + BigBOSS | 136 |
| Planck + All StageIII + BigBOSS | 217 |
| | |
| *Galaxy broadband power spectrum and Lyα BAO:* | |
| Planck + All StageIII + BigBOSS ($k_{max} = 0.15$) | 324 |
| Planck + All StageIII + BigBOSS ($k_{max} = 0.30$) | 544 |

Neutrinos have two important effects in the early universe. First, as an additional radiation component they affect the timing of the epoch of matter-radiation equality. And second, the process of neutrinos becoming non-relativistic imprints a characteristic scale in the power spectra of fluctuations. This is termed the 'free-streaming scale' and is roughly equal to the distance a typical neutrino has traveled while it is relativistic. Fluctuations on smaller scales are suppressed by a non-negligible amount, of the order of a few percent. This allows us to put limits on the neutrino masses.

In principle, if multiple mass eigenstates are present, the cosmological measurements should be sensitive to individual mass states, but in practice they are strongly degenerate and the only quantity cosmology really measures is the sum of neutrino masses. The present data are already very competitive with laboratory experiments, with the most aggressive data combinations giving $\sum m_\nu < 0.17$eV (at 95% c.l., see [Seljak, Slosar and McDonald, 2006]). This result is already better than the expected sensitivity of experiments like KATRIN[Drexlin et al, 2005], which should set a limit of about 0.2eV for mass eigenstates associated with the electron neutrino. A more conservative recent analysis of cosmological data put a limit on the sum of neutrino masses of about 0.3eV (at 95% c.l., see e.g. [Thomas, Abdalla & Lahav, 2010; Vikhlinin et al., 2010]).

Global fits over cosmological parameters including the sum of neutrino masses and time-varying dark energy were carried out by Stril, Cahn, & Linder [2010] for realizations similar to the BigBOSS galaxy BAO survey. Because one of the splittings of the squares of the neutrino mass is about $2.43 \times 10^{-3}$ eV$^2$[KamLAND Collaboration, 2005], we know at least one neutrino has a mass of at least 0.05 eV. If neutrinos have an inverted mass hierarchy, the minimum sum of the neutrino masses is roughly twice this (since the other splitting is considerably smaller). In Table. 2.8 we show forecasts for BigBOSS together with priors from the Planck satellite. When all the information is combined, the BigBOSS $1 - \sigma$ limits obtained on the sum of neutrino masses should be about $\sigma(\Sigma\, m_\nu) = 0.024$ eV (integrating in k-space up to $k_{\max} = 0.1h/\mathrm{Mpc}$ for galaxies, and marginalizing over parameters of the IGM gas model for Ly-$\alpha$, as described in Sec. 2.4.1, on which we obtain the $1 - \sigma$ limits: $\langle F \rangle = 0.80 \pm 0.0033$, $T_0 = 2.0 \pm 0.033 \, [\times 10^4 \, \mathrm{K}]$, and $\gamma - 1 = 0.50 \pm 0.035$).

Thus BigBOSS should be able to make a 2-sigma detection of the sum of the neutrino masses, even in the case of a normal mass hierarchy, and it could rule out the inverted mass hierarchy at a similar confidence level. Either of these would be a major result with important repercussions for particle physics as well as cosmology. In Figure 2.6 we illustrate the reach of various combinations of ground-based and cosmological probes. Note that we have used a conservative upper limit of integration of information for the galaxies $(0.1h/\mathrm{Mpc})$ and that increasing this limit quickly brings the experiment into the regime of guaranteed detection. Therefore, with improved understanding of galaxy bias, its scale dependence and stochasticity, BigBOSS will be an even more powerful probe of neutrino masses than what we quote here.

The other parameter relevant for neutrino physics is the effective number of neutrino species $N_\nu$, which parametrizes the energy density attributed to any non-electromagnetically interacting ultrarelativistic species (including e.g. axions) in units of the equivalent of one neutrino species that fully decouples before electron-positron annihilation. The value for the standard cosmological model is $N_\nu = 3.04$[4]. The detection of any discrepancy from the expected value would be a major result, indicating either a new particle, non-standard neutrino physics or some other exotic new physics. Our forecasts for this parameter are also shown in Table 2.8. Again we see that the effective number of neutrino species will be measured to an accuracy much better than unity, providing strong constraints on the alternative models involving extra sterile neutrinos, axions or partly-thermalised species.

We conclude that BigBOSS will be an excellent probe of neutrino masses and will beautifully complement laboratory experiments. If neutrino physics has more complexity to it, as recently indicated by the MiniBooNE experiment [MiniBooNE Collaboration, 2010] confirming previous LSND anomalies[Athanassopoulos et al., 1996], BigBOSS might be able to shed new light on its features.

### 2.5.3   Constraints on Inflation Models

The inflationary paradigm is to date the best contender for the origin of the fluctuations of primordial density, which seeded the large-scale structure we observe today. In its simplest formulation it predicts perturbations in the initial distribution that are very nearly scale-independent and Gaussian-distributed about the mean.

Arguments of symmetry together with predictions for nearly scale-invariant fluctuations lead us to parameterize the primordial spectrum as a function of $k$ through the spectral

---

[4]The small increase with respect to $N_\nu = 3$ is due to the fact that some neutrinos are still coupled at the onset of electron-positron annihilation.

Table 2.8: Constraints on the sum of neutrino masses and number of effective neutrino species from BigBOSS forecasts in combination with Planck satellite information. The base model is a Planck prior with BAO information from BigBOSS galaxies and Ly-$\alpha$ forest (LyaF). We then add successively broadband power spectrum (nBAO) information from galaxies, Ly-$\alpha$, and taken together. In the case where broadband galaxy power is used, $k_{max} = 0.1 \, h/\text{Mpc}$ for the broadband power, however, we continue to use the BAO information from higher $k$. When all information is combined the BigBOSS potential for detection of a non-null sum of neutrinos masses is around $\Sigma\, m_\nu = 0.05 \pm 0.024$ (eV).

|  | $\Sigma\, m_\nu$ [eV] | $\Sigma\, N_\nu$ |
|---|---|---|
| Fiducial values | 0.05 | 3.04 |
| $\sigma-$ Planck+BAO(LyaF+galaxies) | 0.094 | 0.18 |
| $\sigma-$ Planck+BAO(LyaF+galaxies)+nBAO(galaxies) | 0.039 | 0.097 |
| $\sigma-$ Planck+BAO(LyaF+galaxies)+nBAO(LyaF) | 0.031 | 0.056 |
| $\sigma-$ Planck+BAO(LyaF+galaxies)+nBAO(galaxies+LyaF) | 0.024 | 0.056 |

index or tilt

$$n_{\rm S}(k) = \frac{d\ln P}{d\ln k}.\tag{2.15}$$

If we take $k_0$ as a reference scale, the primordial power spectrum can be expanded as

$$P(k) = P(k_0)(k/k_0)^{n_{\rm S}(k_0)+\frac{1}{2}\alpha\ln(k/k_0)}\tag{2.16}$$

where $\alpha = dn_{\rm S}/d\ln k$ at $k_0$. If there is no "running" of the spectral index, the primordial power spectrum is a pure power law.

The Harrison-Zel'dovich primordial spectrum has $n_{\rm S} = 1$, while inflation predicts slight deviations from unity. Ruling out $n_{\rm S} = 1$ at a significant level of confidence would strengthen the case for inflation. The WMAP7 result is $0.963 \pm 0.014$ [Komatsu et al., 2010]. The current limits on running of the spectral index, obtained by the WMAP team, are $-0.061 < dn_{\rm S}/d\ln k < 0.017$ (95% CL) . The Ly-$\alpha$ forest, because it is in the regime of linearity for a wide range of $k$, is an excellent complementary probe of $n_s$ and $\alpha$.

In Table. 2.9 we present forecasts on inflationary observables obtained with Fisher-matrix formalism applied to the power spectrum obtained from BigBOSS quasars and galaxies, combined with Planck priors and existing higher resolution QSO spectra. We marginalize over intergalactic medium nuisance parameters, the mean absorption level $\langle F \rangle$, and $T_0$ and $\gamma$ in the temperature-density relation. The maximum k considered is $k_{max} = 0.1 h/\text{Mpc}$ for galaxies. For the Ly-$\alpha$ forest we include essentially all information, up to $k_{max} = 5 h/\text{Mpc}$, although the smallest scale information is absorbed into constraining the IGM model parameters. Note that this limit is only relevant to the information we assume from $\sim 100$ existing high resolution spectra. BigBOSS does not have sufficient resolution to probe radial $k$ this high, and in the transverse direction we are limited to a much smaller $k$ set by the Nyquist frequency corresponding to the typical separation between lines of sight (i.e., we really only have fully 3D information for $k \quad 0.15 \, h \, \text{Mpc}^{-1}$). The marginalized $1-\sigma$ limits on these are $\langle F \rangle = 0.80 \pm 0.0057$, $T_0 = 2.0 \pm 0.025 \, [\times 10^4 \text{K}]$, and $\gamma - 1 = 0.50 \pm 0.074$. In Fig.2.7 we present projected constraints in the $n_S - \alpha$ plane, Fisher matrix ellipse contours for the same experimental realization.

Table 2.9: Constraints on inflationary observables obtained combining BigBOSS with Planck satellite information. As a baseline model we consider a Planck prior with BAO information from BigBOSS galaxies and Ly-$\alpha$ forest (LyaF), and then in conjunction with broadband power spectrum information (nBAO) from galaxies, Ly-$\alpha$ , and combined. In the case where broadband galaxy power is used, $k_{max} = 0.1$ h/Mpc for the broadband power, however, we continue to use the BAO information from higher $k$.

| | $n_S$ | $\alpha_S$ |
|---|---|---|
| Fiducial | 0.963 | 0.00 |
| $\sigma-$ Planck+BAO(LyaF+galaxies) | 0.0026 | 0.0071 |
| $\sigma-$ Planck+BAO(LyaF+galaxies)+nBAO(galaxies) | 0.0025 | 0.0068 |
| $\sigma-$ Planck+BAO(LyaF+galaxies)+nBAO(LyaF) | 0.0023 | 0.0027 |
| $\sigma-$ Planck+BAO(LyaF+galaxies)+nBAO(galaxies+LyaF) | 0.0022 | 0.0024 |

These are impressive results. In standard slow-rolling inflationary models, the running of the spectral index is of the order $O((1 - n_s)^2) \sim 1 \times 10^{-3}$ if $n_s \sim 0.96$. This means that BigBOSS will start to approach the region of guaranteed detection in minimal inflationary models. Detection of the running of the spectral index would be a confirmation of inflationary *prediction* and thus considerably strengthen the observational evidence for inflation.

The power spectrum, or the two-point correlation function, has traditionally been the statistic of choice in cosmological observations. This is because many theories predict that the initial seeds are nearly Gaussian distributed. Thus in the linear regime, expected to be valid on the largest scales in LSS and on almost all scales in CMB, one expects the fluctuations to be nearly Gaussian and all the information is contained in their two-point function statistic. In the non-linear regime non-Gaussianity develops and higher order correlations become non-vanishing, created by the nonlinear gravitational evolution.

Primordial non-Gaussianity, determined by the higher order correlations present in the linear regime, is complementary to the information contained in the power spectrum, since it probes aspects of physics during inflation that cannot be probed otherwise. For example, while a single field slow-roll inflation with canonical kinetic energy and adiabatic vacuum predicts very small amount of non-Gaussianity, violation of any of these conditions may lead to large non-Gaussianity. Many of these models predict the non-Gaussianity of local type, $\Phi = \phi + f_{nl}\phi^2$, where $\Phi$ is the gravitational potential in the matter era and $\phi$ is the corresponding primordial Gaussian case. Alternatives to inflation based on bounce also tend to predict large non-Gaussianity of this type. A detection of primordial non-Gaussianity would rule out the simplest model of inflation. Conversely, a non-detection at a level of $f_{nl} < 1$ would rule out many of its alternatives.

Until recently, the most powerful method to place limits on $f_{nl}$ was based on the bispectrum of cosmic microwave background (CMB), with the latest WMAP constraint giving one-sigma error of around 20 on $f_{nl}$ [Senatore, Smith & Zaldarriaga, 2010]. With a better angular resolution one can sample more modes and the error should be improved to about 5 with the higher angular resolution Planck satellite [Cooray, Sarkar & Serra, 2008].

An alternative approach using clustering of biased tracers of structure on very large scales has recently been proposed [Dalal et al., 2008]. It was shown that the non-Gaussianity leads to a unique scale dependence of the large-scale bias, one that increases strongly towards the large scales, and whose amplitude scales with the bias of the tracer relative to the dark matter. One can therefore place the limits on $f_{nl}$ by comparing the scale dependence of the power spectrum of the biased tracer to the one expected in cosmological models under the assumption of a scale independent bias. A first application of this method has been presented using the large-scale clustering of quasar and luminous red galaxies (LRG) galaxy data from the Sloan Digital Sky Survey (SDSS) [Slosar et al., 2008]. The result, a non-detection with one sigma error of about 25, is comparable to the latest CMB constraints from WMAP, suggesting this is a competitive method compared to the bispectrum from CMB and should be pursued further.

Based on the size and volume of the BigBOSS survey one could in principle expect to reduce the current errors by a factor of 5-10 [McDonald, 2008]. However, to achieve this the galaxies measured in BigBOSS must have sufficiently large bias, since only for biased tracers is the non-Gaussian scale-dependent clustering is revealed. The current projections suggest the galaxies will have bias of around 2 at $1 < z < 2$, suggesting this may be possible. In this case the expected one-sigma error in $f_{nl}$ is about 5. One way to further improve the errors is by combining two tracers of LSS, one with a high bias and one with a low bias: in this case it may possible to cancel sampling variance, which is the dominant source of error on large scales [Seljak, 2009], but due to low number density this will have to include an additional tracer of structure, potentially combining with the LSST and DES data.

More detailed studies of halo mass distribution of BOSS galaxies, combined with numerical simulations of non-Gaussian models [Desjacques, Seljak & Iliev, 2009], are needed to provide a better answer to the question what the ultimate reach of BigBOSS for non-Gaussianity studies is, but it seems likely that the limits will be at least comparable to the best limits from CMB and possibly much better than that.

### 2.5.4   Modified Gravity

While BAO measure the change in geometric scales with cosmic expansion history, the power spectrum as a whole probes the growth history of structure in the universe. The power spectrum amplitude depends on the growth factor $D(z)^2$ and redshift space distortions are sensitive to the growth rate $f(z)$. Of course, a change in the normalization of the mass power spectrum is degenerate with a change in bias, so we generally assume that only $dD/d \ln a$ can be measured, although it may be possible to put constraints on the bias using higher order statistics or lensing. In this modified gravity section we take a simplified approach and assume a concrete model for bias evolution with one parameter over which we marginalise, namely $b(z) = b_0/D(z)$ (in other sections of this document we marginalize over completely free bias parameters in each $\Delta z = 0.1$ bin). Within the framework of general relativity, growth factor information essentially repeats the information in the expansion history and distance measurements; that is, there is a one-to-one correspondence between expansion and growth.

This implies that measuring both expansion and growth (redshift-space distortions),

through the BAO scale and the overall power spectrum, enables a test of Einstein gravity. One model independent parameterization of the growth deviation from general relativity is the gravitational growth index $\gamma$ [Linder, 2005; Linder & Cahn, 2007; Linder, 2008]. This characterizes the growth rate as

$$f(z) = \Omega_m(z)^\gamma. \tag{2.17}$$

Note that at high redshift, as matter domination tightens, $\Omega_m(z)$ is close to unity and the uncertainty on $\gamma$ increases. Thus this test of gravity is essentially a low redshift test (although not so low that nonlinearities cloud interpretation). This is further strengthened by growth being a continuing process, so small differences in the rate are amplified as growth persists into the late universe. Stril, Cahn, & Linder [2010] found that an experiment similar to BigBOSS is capable of determining $\gamma$ to within 0.04 (7%), simultaneously with fitting the expansion history, neutrino mass, and other cosmological parameters.

However, most extensions to gravity modify the physics in time- and scale-dependent ways, so another approach to exploring gravity uses combinations of the potentials $\psi$ and $\phi$ entering the metric. One can think of these as characterizing the gravity connecting the Newtonian potential and the density field, i.e. $\nabla^2 \phi = 4\pi G a^2 \delta \rho_m$, and the gravity connecting the potential and the velocity field, $\nabla \psi = -\dot{v}$. In general relativity these are the same: $\psi = \phi$, but they can differ in many other theories of gravity. The two connecting relationships most closely tied to observations are given by

$$-k^2(\phi + \psi) = 8\pi G_N a^2 \bar{\rho}_m \Delta_m \times \mathcal{G} \tag{2.18}$$
$$-k^2 \psi = 8\pi G_N a^2 \bar{\rho}_m \Delta_m \times \mathcal{V}, \tag{2.19}$$

where $\bar{\rho}_m \Delta_m$ is the gauge invariant matter density perturbation and $G_N$ is Newton's constant. In general relativity, the time- and scale-dependent functions $\mathcal{G}$ and $\mathcal{V}$ are identically unity. While the cosmic microwave background and to a large extent weak gravitational lensing are sensitive to $\mathcal{G}$, strong growth probes such as the galaxy power spectrum can constrain $\mathcal{V}$, thus breaking the degeneracy between the metric ingredients. Seeing the two parameters separately is a critical step toward identifying the class of modification to gravity.

Gravity beyond general relativity tends to give scale dependent effects, and indeed this is a major clue to distinguishing gravitational modifications from galaxy bias properties. A wide field survey that provides accurate growth measurements over a wide range of scales gives an important lever arm in wavenumber $k$. BigBOSS covers the key redshift range for growth of $z \approx 0.7 - 1.4$ and covers up to 14,000 deg$^2$. The detailed spectroscopy delivers information on the velocity field, providing measurements of the growth rate $dD/d \ln a$ in addition to the growth factor $D(a) = (\delta\rho/\rho)_a/(\delta\rho/\rho)_{\text{init}}$ (with the caveat that we are assuming perfectly known bias evolution in this modified gravity section).

Daniel & Linder [2010] demonstrate that the BigBOSS science design makes it a major experiment for testing gravity. Figure 2.8 reveals that BigBOSS can determine each gravity function and is an important complement to next generation Planck CMB and high redshift supernovae distance measurements. Even allowing for both time (redshift) and scale (wave mode) dependence, BigBOSS delivers results a factor of 10-100 times better than current constraints.

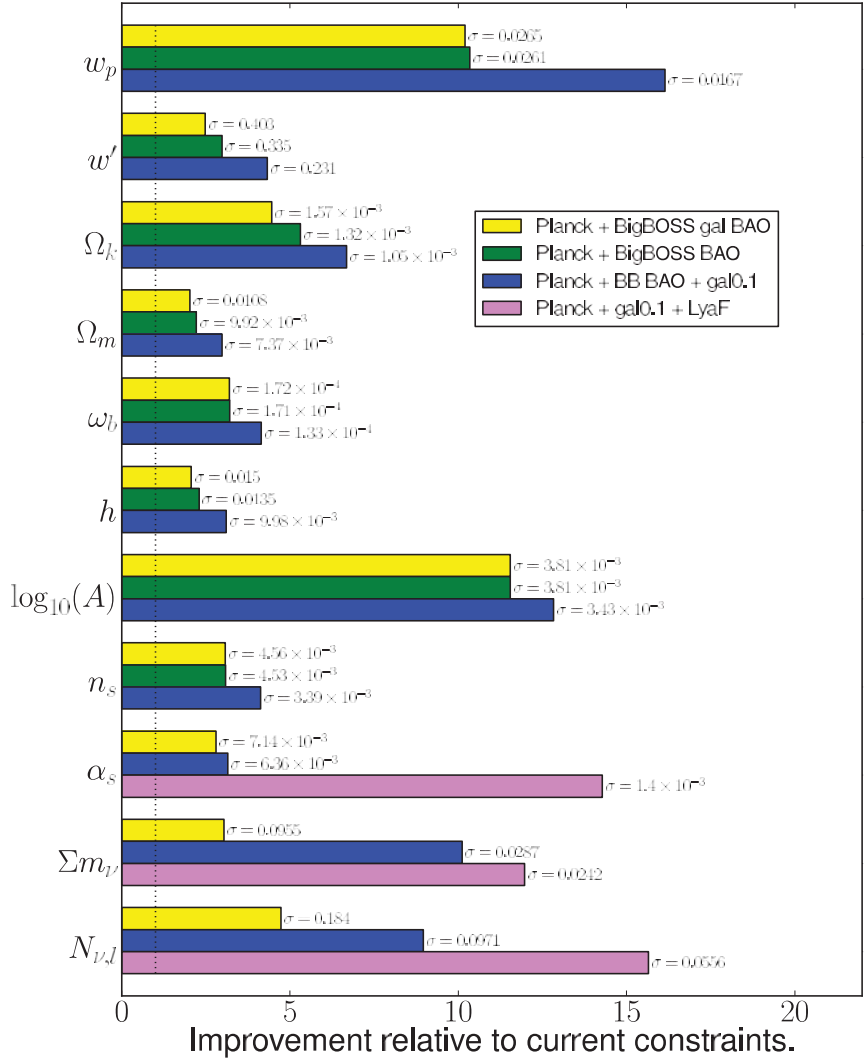

Figure 2.5: Summary of the fundamental parameter constraint projections resulting from the BigBOSS Key Project. The length of a bar is the measurement accuracy (i.e., $1/\sigma$), relative (dotted line = 1) to current constraints from Komatsu et al. [2010] using WMAP 7 year, BAO [Percival et al., 2010], and $H_0$ [Riess et al., 2009] constraints. All include Planck priors, with colors indicating: yellow: BAO from BigBOSS galaxies; green: BAO from BigBOSS galaxies and Ly-$\alpha$; blue: as green, plus all BigBOSS galaxy clustering information at $k < 0.1h$ Mpc$^{-1}$; and violet: as blue, but including Ly-$\alpha$ information at all scales. Parameters through $n_s$ are normalized to the Komatsu et al. [2010] OWCDM model (non-flat, constant $w$), except for the $w'^{\parallel}$ normalization, which is simply set to 1. The normalization for the sum of neutrino masses is 0.29, half the Komatsu et al. [2010] 95% upper limit for the $\Lambda$CDM+neutrinos model. The running spectra index, $\alpha_s$, and extra radiation, $N_\nu$, cases are normalized to the Komatsu et al. (2010) "$\Lambda$CDM plus this parameter" constraints.

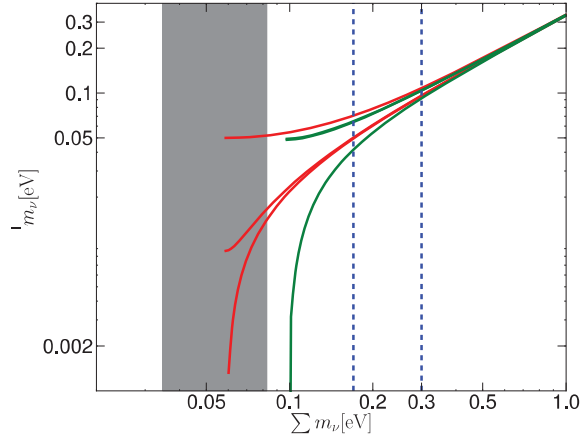

Figure 2.6: This plot shows the masses of individual neutrino mass eigenstates as a function of the sum of the masses for normal (red) and inverted hierarchies (green). We have taken the mass splittings to be $\Delta m_{21}^2 = 7.58 \times 10^{-5} \text{eV}^2$ (KAMLAND experiment[KamLAND Collaboration, 2005]) and $|\Delta m_{32}^2| = 2.43 \times 10^{-3} \text{eV}^2$ (MINOS experiment[MINOS Collaboration, 2008]). Vertical dotted lines represent typical limits from cosmology at 95% c.l. [Seljak, Slosar and McDonald, 2006; Thomas, Abdalla & Lahav, 2010; Vikhlinin et al., 2010]. The grey band shows the $\pm 1\sigma$ error of 0.024 eV on the sum of the neutrino masses as forecast from BigBOSS, centered on the minimum allowed value of 0.058 eV from current measurements of neutrino mixing.

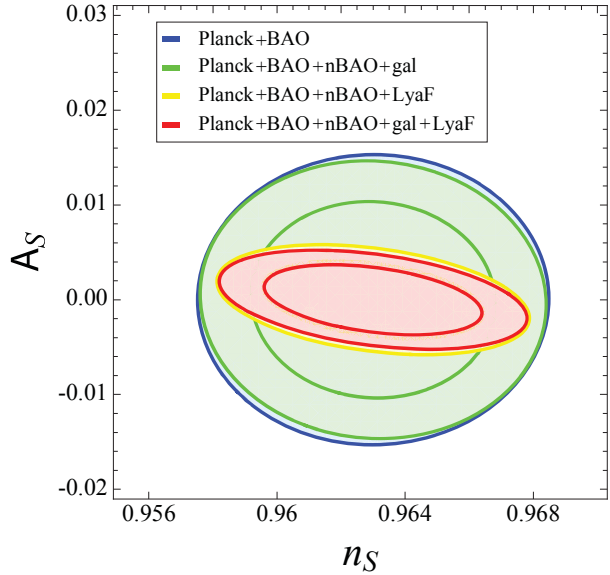

Figure 2.7: $1 - \sigma$ and $2 - \sigma$ constraints in the $n_S - \alpha$ plane obtained with broadband power information (nBAO) from BigBOSS's QSO + galaxies, combined with Planck satellite forecasts and BAO-scale constraints from the Ly-$\alpha$ forest plus galaxies.

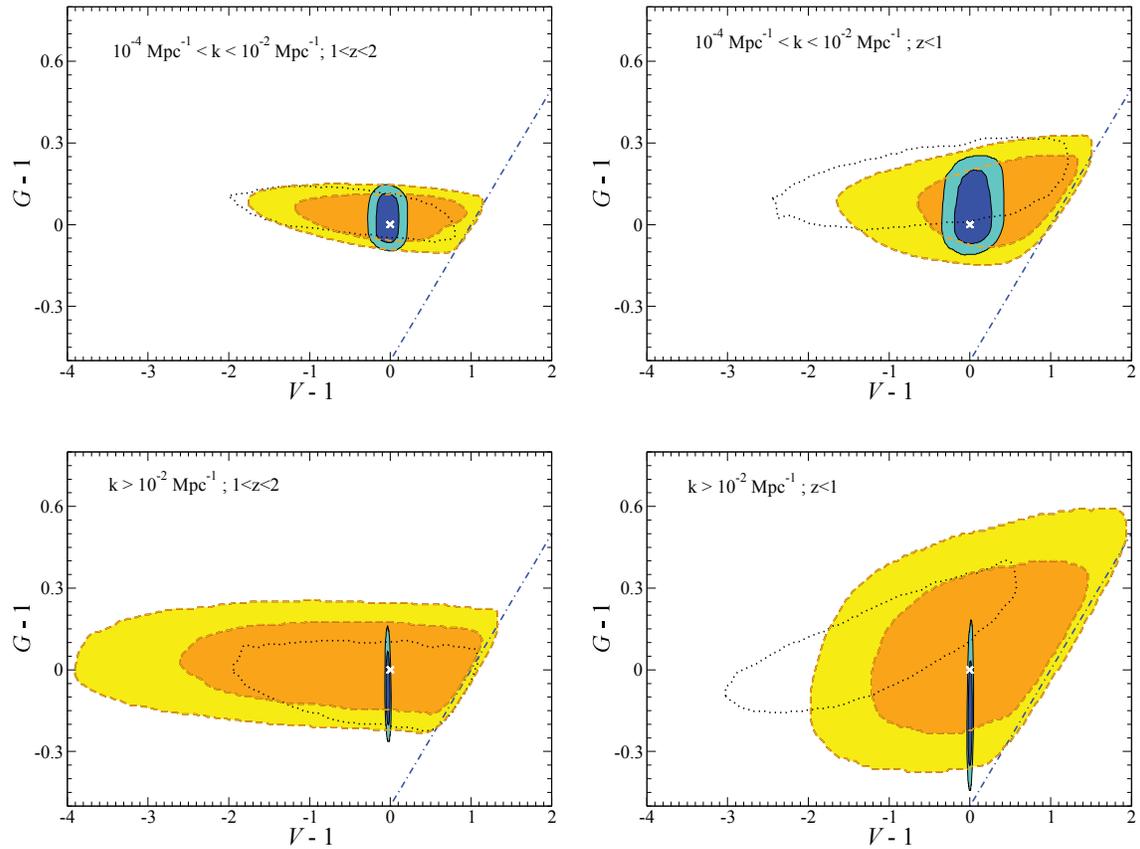

Figure 2.8: 68% and 95% cl constraints on $\mathcal{V}-1$ and $\mathcal{G}-1$ are plotted for the two redshift and two wavenumber bins using mock future BigBOSS, Planck, and supernova data (solid blue contours; dashed yellow contours show the degradation without BigBOSS). The dotted contours recreate the 95% CL current data contours, including galaxy and weak lensing information. The anticipated improvement in constraints is a factor of 10-100 in area. The x's denote the fiducial GR values. From Daniel & Linder [2010]

# 3  Community Science with BigBOSS

## 3.1  Overview

The primary motivation for the BigBOSS survey and multi-object spectrograph is to measure and constrain cosmological parameters, in order to understand the nature of "dark energy". However, the data collected will enable a multitude of astrophysical science projects beyond this core cosmology survey, and the instrument itself will provide an unparalleled spectroscopic facility to the community. In this section, we describe a few of the many other scientific investigations that will be enabled by the BigBOSS survey and instrument.

The survey strategy for the Key Project is detailed in Chapter 6, but we review the key points here. Our proposed survey will cover roughly 14,000 deg$^2$ of sky, with each point in the survey footprint covered by at least 5 BigBOSS tiles. The Survey will primarily target emission line galaxies ($t_{exp}$=16.6min), luminous red galaxies ($t_{exp}$=33.2min), and QSOs ($t_{exp}$=83min) (see Table 2.2). During survey observations, roughly 80% of the fibers will be placed on the primary targets, and a to-be-determined fraction of the remaining 20% will be devoted to calibration (i.e., standard stars and sky). The remaining fibers (somewhere between 10% and 20% of the total 5000) will be available for "synchronous" science targets from outside the Key Project (e.g., see §3.8.2 & § 6.3). The survey will also repeatedly target several (4-6) Calibration Fields and sample these regions more densely and with deeper observations (e.g., see §6.1.2).

BigBOSS "community science" programs can be divided into three categories. First, the legacy spectroscopic database from the core BigBOSS survey will enable multiple projects, especially in the domain of extragalactic astronomy and galaxy evolution. Second, between 10% and 20% of the 5,000 BigBOSS spectrograph fibers will be unallocated to cosmology or calibration targets in any given tiling. Hence these fibers furnish an opportunity to schedule "synchronous observing" programs in parallel with the BigBOSS Key Project. Over the course of the full survey (nearly 10,000 tiles), this represents up to many million independent spectroscopic target observations available to the NOAO community for synchronous scheduling. Finally, the BigBOSS instrument will be available for dedicated use by PI-led programs through the regular NOAO time allocation process; many of the science projects described in this chapter may be proposed for execution in this PI-led mode.

It is important to note that the topics covered in this chapter are *by no means* an exhaustive list of the astrophysical grasp of BigBOSS. Instead, the cases presented here are meant to illustrate the ability of BigBOSS to address numerous astrophysical questions beyond the cosmological survey it is designed to execute. The BigBOSS project collaboration is committed to the full integration of input from the astronomical community into our survey planning and operations, so as to maximize the joint scientific return between the proposed cosmology programs and broader astronomical science. To this end, we will engage the community by way of a BigBOSS science planning workshop, as described in Section 3.10.

The breakdown of this chapter is as follows. Sections 3.2 through 3.7 present several science programs across a broad range of astronomy that can be carried out using BigBOSS on the Mayall. In Section 3.8, we discuss the three modes by which the NOAO user community can undertake these or other programs. In Section 3.9 we discuss the role BigBOSS can play in supporting other ground- and space-based experiments. Finally, in Section 3.10, we present the rationale for a Community Workshop that we propose in order to integrate the NOAO community into the use of this extraordinary capability.

## 3.2  Galaxy Evolution

The spectroscopic legacy data from the BigBOSS survey will be particularly powerful for a wide range of galaxy-evolution studies. In this section, we list a subset of the possible community-science projects in this area. These research opportunities will benefit both from the core BigBOSS survey data set, and from "calibration fields" that we will observe in order to properly characterize the BigBOSS survey performance and accurately measure the sampling and completeness functions. We will define 4 to 6 of these calibration fields, at least two of which can be targeted at any time of the year (see Chapter 6). These fields will be targeted at least once during each BigBOSS run, and over the period of the survey will build up a total area of $\approx$30-40 deg$^2$ which is densely sampled with deep spectroscopy. By carefully choosing the calibration fields to lie in regions which have wide-area multi-wavelength and archival spectroscopic coverage (e.g.: selected PS1 calibration fields; overlapping with M31; the best studied portions of the Sagittarius stream; the 9 deg$^2$ ND-WFS Boötes field; the 2 deg$^2$ COSMOS field; the SXDF and UDS fields etc.), these fields will be invaluable for many independent science programs (e.g., galaxy evolution, Galactic structure, etc.) and will have high legacy value. Thanks to repeat visits these calibration fields will cover a total area of $\approx$30-40 deg$^2$, with a much higher spatial sampling, and with the possibility of hitting targets multiple times to increase the effective survey depth.

### 3.2.1  The Evolution of Massive Galaxies: Star Formation, Merging, and AGN

The evolution in the number density, luminosity, and stellar mass of $L > L_\star$ galaxies provides a very strong constraint on galaxy evolution models [De Lucia & Blaizot, 2007], as merging and star formation result in different shape evolutions of the luminosity function at the bright end [Bower et al., 2006; Croton et al., 2006]. To accurately measure the luminosity function (LF) and its evolution significantly above $L_\star$ requires a survey with a large volume and with precise redshifts. Large volume need to be surveyed, since the most massive galaxies are rare and current studies suffer greatly from sampling uncertainties and cosmic variance. At the same time precise redshifts are necessary as redshift errors - and their corresponding error in luminosity - can significantly distort the shape and normalization of the exponential tail of the LF [Marchesini et al., 2007]. Current studies using medium band filters can achieve the desired redshift accuracy but they require large time allocations to obtain the photometric data over the requisite survey areas [van Dokkum et al., 2009].

The total BigBOSS Key Project survey area will contain roughly 880,000 galaxies with $M > 3 \times 10^{11} M_\odot$ at $z < 1.0$. BigBOSS will therefore yield orders of magnitude better measurements of the LRG LF compared to current studies [e.g., Brown et al., 2007; Faber et al., 2007], and will explore the relatively unstudied redshift range between the local and $z \sim 1$ studies, where the evolution in the massive galaxy population is rapid. BigBOSS will also characterize the ELG luminosity function to fainter magnitudes than for LRGs. Taken together these LFs will constrain the evolutionary channels by which ELGs merge and form LRGs.

BigBOSS will also deliver the best measurement of evolution in the [OII] luminosity function from redshift $z = 1.5$ to $z = 0.7$, where the color-selection for emission line galaxies is most effective. [OII] may be used as a measure of star-formation rate [e.g., Kewley, Geller & Jansen, 2004]. A statistical correction for the significant extinction can be determined using the Balmer decrement calculated from higher order Balmer lines. This will allow the measurement of a statistical [OII] extinction as a function of [OII] luminosity, stellar mass, and broad-band SED shape. BigBOSS is likely to detect H$\beta$ with similar significance to [OII] for most ELGs, and so should be able to adequately calibrate the [OII]-SFR relation for the sample. For lower-luminosity emitters the higher order Balmer lines will be statistically detected by stacking the spectra as a function of [OII] and H$\beta$ luminosity, stellar mass, and SED shape.

BigBOSS will also provide an extremely precise measurement of the evolution of average AGN activity from $z = 1.3$ to the present. AGN can be identified from the presence of high ionization lines (e.g., [NeV], large [OIII]/H$\beta$, or MgII emission), or the relative strength of the [OII] emission [e.g., Yan et al., 2006; Montero-Dorta et al., 2009]. Correlating AGN activity with other galaxy properties (e.g. color, star-formation rate, stellar mass, etc., derived in combination in multiwavelength photometric surveys) can further quantify the role of AGN in suppression of star formation. These studies are specifically enabled by the large wavelength range and high throughput of BigBOSS.

### 3.2.2 Strong Lensing and Galaxy Structure

The SDSS has led to the discovery of the largest single sample of confirmed strong galaxy-galaxy gravitational lens galaxies [Bolton et al., 2006, 2008a] through the detection of two redshifts in single spectra (see Figure 3.1). This sample of lenses has provided a measurement of the mass-density structure of elliptical galaxies and its dependence on galaxy mass [Koopmans et al., 2006; Bolton et al., 2008b]. Scaling from the known incidence of strong gravitational lenses identified by SDSS, the BigBOSS spectroscopic database should provide 10,000–20,000 new lenses through this discovery channel, which can be followed up at high spatial resolution with space telescope observations [Bolton et al., 2006], adaptive optics-aided imaging [Marshall et al., 2007], or integral-field spectroscopy [Bolton & Burles, 2007]. With this large number of lenses it will be possible to "stack" the lenses in bins of redshift and halo mass. This will open a unique path to measure the mass-dependent redshift evolution of galaxy mass-density profiles and dark-matter fractions. In addition, the large number of spectra to be obtained by BigBOSS makes it likely that several multiple-redshift lenses will be found such as the double Einstein ring system [Gavazzi et al., 2008], with additional applications to precise galaxy-structure measurement and cosmography.

### 3.2.3 A Blind Spectroscopic Survey

A small fraction of the BigBOSS survey fibers, partially overlapping with those allocated to the measurement of the night sky foreground, can be devoted to a blind spectroscopic survey of the sky. This will yield the largest ever full optical-band blind spectroscopic survey (covering $\approx 0.4 - 1$ deg$^2$) and will provide a fundamentally different census of the universe than that resulting from targeted observations, including discovery of extremely high equivalent-width emission line galaxies. By virtue of these spectra being distributed uniformly throughout the BigBOSS area, this survey will be unaffected by the cosmic variance that plagues narrower field blind spectroscopic surveys. The large wavelength coverage of BigBOSS will furthermore permit identification of multiple significant emission lines across a wide range of redshifts (from low-redshift H-$\alpha$ to high-redshift Ly-$\alpha$ - see Fig. 3.2), complementing narrow-band, longslit- or IFU-based programs that concentrate on small wavelength (and hence redshift) windows [e.g., Rhoads et al., 2000; Martin & Sawicki, 2004; Hill et al., 2008].

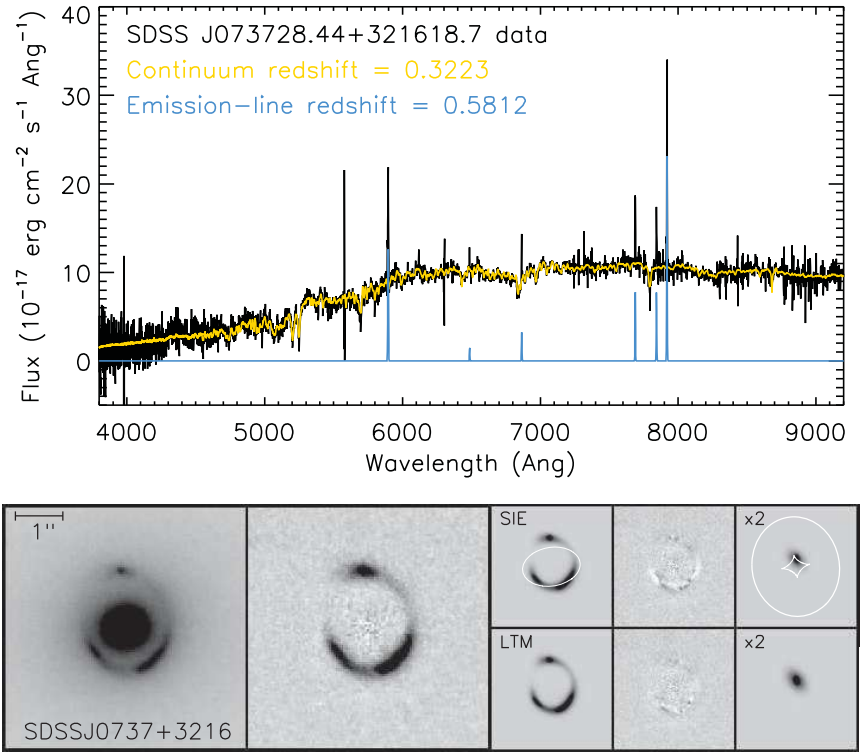

Figure 3.1: *Top:* SDSS discovery spectrum of the gravitational lens SDSS J0737+3216, showing a continuum redshift of 0.32 and an emission-line redshift of 0.58 in the same line of sight. *Bottom: Hubble Space Telescope* image and lens models of SDSS J0737+3216 from the SLACS Survey (Figure from [Bolton et al., 2008a].)

### 3.2.4 Stacked Spectra as a Function of Photometric Properties

Although the individual galaxy spectra from the core BigBOSS survey will be of relatively low signal-to-noise ratio (SNR), the vast number of such spectra will permit the creation of extremely high SNR stacked spectra [e.g., Eisenstein et al., 2003; Schiavon et al., 2006; Cimatti et al., 2008] as a function of redshift, color, and luminosity. These high SNR stacks will enable the measurement of precision abundances, low-level emission lines, detailed SFHs, and average velocity dispersions for much of the sample. With the combination of BigBOSS resolution and wavelength coverage, this library of stacked spectra will be a definitive resource for the study of physical galaxy properties and their evolution.

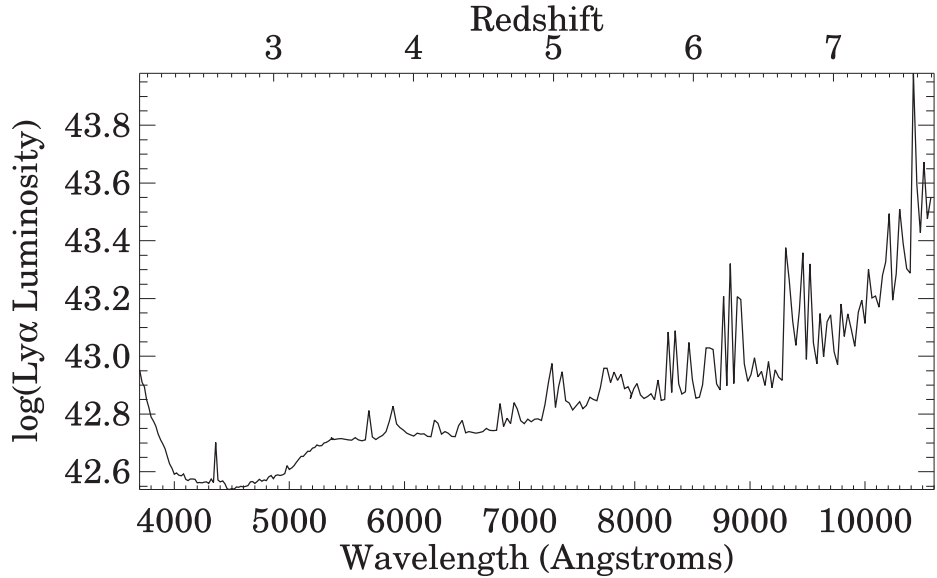

Figure 3.2: The expected 8-$\sigma$ detection limit for Ly$\alpha$ luminosity as a function of wavelength and redshift. The emission line is modeled as a Gaussian of width 70 km/s observed using BigBOSS for 5000 seconds split up into 5 separate exposures. The large wavelength range of this BigBOSS survey would sample the most luminous emitters over a very wide redshift range and thereby complement other narrow- band imaging or small-field IFU studies of the emission line galaxy population.

### 3.2.5 A Deep, Magnitude-Limited Sub-Survey

The calibration fields in particular will be used to construct a magnitude-limited sample of galaxies that goes significantly deeper than the main survey, potentially as deep as $i = 22.0$ to 22.5 with no color cut. It is well known that color selection techniques, such as those used in the main cosmology survey miss important elements of the galaxy population (e.g. [Franx et al., 2003]). A magnitude-limited survey will therefore be very important for more general galaxy-evolution studies by quantifying the biases associated with the color selection of the main survey.

In addition, this sample will be extremely valuable for measuring the evolution of the galaxy luminosity function to fainter luminosities, for correlating SFR with physical galaxy properties well below $L_\star$, and for calibrating the spectroscopic redshift distribution of the faint blue galaxy population used in weak gravitational lensing surveys. Precise knowledge of this distribution is essential for all applications of weak lensing to astrophysics and cosmology [e.g., van Waerbeke et al., 2006]]. The full BigBOSS survey will suppress the systematics of cosmic variance in this measurement to a greater extent than any other survey, while the deeper calibration-field component will push further down the luminosity func-tion. We note that all BigBOSS faint-galaxy redshift distributions can be boot-strapped
to the distributions of even fainter galaxies using the spatial cross-correlation calibration
technique of Newman [2008].

### 3.2.6  Galaxy Clusters

Over the course of its operations, BigBOSS will be able to obtain spectroscopic redshifts for
the brightest cluster galaxies of a large number of photometrically selected galaxy cluster
candidates, such as the $\sim 20,000$ clusters expected to be delivered by the Red-Sequence
Cluster Survey 2 [RCS-2;  Yee et al., 2007], or the many thousands of clusters expected
to be yielded by future Sunyaev-Zeldovich (SZ) surveys and by the Planck satellite [e.g.,
Chamballu et al., 2010]. The SZ cluster programs are especially promising as cosmological
probes, since they are directly sensitive to the intra-cluster medium and are not dependent
on virialization or galaxy properties within the cluster. For these surveys, however, no
redshift estimate is available from the SZ determination itself. The core BigBOSS survey
will likely target the brightest cluster galaxy (BCG) in many of these clusters at $z <$
$0.7$, providing critical redshift information that will maximize the scientific yield of cluster
surveys.

The main BigBOSS survey will obtain many thousands of these brightest-cluster-galaxy
redshifts. However, depending on the tiling strategies, it will be hard for the main survey to
sample more than $\sim 1$ galaxy per square arcminute, which implies that it will be very dif-
ficult to get more than a couple of redshifts for each cluster. In contrast, the higher spatial
sampling of the deeper calibration survey has the potential of getting many more redshifts
per cluster. In addition to providing secure spectroscopic confirmation of the clusters, this
will also allow dynamical masses to be calculated, thanks to the 70km/s resolution of the
spectrographs. Scaling from the number of clusters in the IRAC shallow survey [Eisenhardt
et al., 2008] we expect that the calibration fields contain $\sim 1000 - 1200$ galaxy clusters
at $z < 1$ and $\sim 450 - 550$ at $z > 1$. A well-characterized set of clusters over such a wide
redshift range will enable studies of the large-scale regions around galaxy clusters, the trend
in galaxy populations surrounding these intersections of the cosmic web, and a useful cross-
calibration for other cluster mass measurements.

### 3.2.7  The Brightest High-z Galaxies: Giants and Lensed Normal Galaxies

The large area of the BigBOSS instrument and survey will permit spectroscopy of rare,
bright color-selected Lyman-break galaxy (LBG) candidates, to be confirmed via the Ly-$\alpha$
emission line at redshifts $z \sim 3 - 5$. The small handful of known extremely bright LBGs
have mostly been discovered either serendipitously [e.g., Yee et al., 1996], or by systematic
searches that have resulted in tens of objects [e.g.,  Kubo et al., 2009]. BigBOSS offers the
possibility of systematically identifying several hundred such objects across the sky. Many
of these will be strongly lensed by intervening galaxies, groups, and clusters and therefore
will have substantially lower intrinsic luminosities that have been amplified by gravitational
lensing. There are only a handful of such objects currently known, which severely limits
their potential for detailed studies of the physical conditions in the "normal" star-forming population at $z > 1.5$. Deeper follow-up ground-based spectroscopy of a large sample of these objects identified by BigBOSS will permit high SNR survey-scale studies of the IGM and star-formation in young galaxies at high redshifts [e.g., Pettini et al., 2000]. From space, JWST will yield access to deep MIR spectroscopy for these rare objects, giving a direct view on the dust content and ionization sources in these distant galaxies [e.g., Papovich et al., 2009].

### 3.2.8    Correlation of Galaxy Properties with Environment

The local environment in which galaxies reside is known to be one of the key controlling factors for their formation, properties, and evolution [e.g., Dressler, 1980; Blanton et al., 2005; Cooper et al., 2006]. Using highly sampled data from the calibration fields, the BigBOSS project will permit extremely detailed studies of the empirical correlations between galaxy properties (color, luminosity, stellar mass) and environments [e.g., Cooper et al., 2008]. Even in the smaller calibration-field subset of the BigBOSS survey, there will be sufficient numbers to split galaxies into multiple bins in redshift, stellar mass, and environmental density simultaneously. The data on the individual $z < 0.5$ objects will be extremely good, even with the short exposure times.

A recent particularly promising approach has been to decompose galaxies into central and satellite galaxies based on their luminosities and clustering [e.g., Yang et al., 2007]. This approach has yielded new insights, as it appears that central and satellite galaxies have different properties at a fixed stellar and halo mass. Understanding this apparent dichotomy is key to understanding the physical role of environment in galaxy evolution. Locally this has been most successfully attempted with SDSS [e.g. Pasquali et al., 2009] by taking advantage of the extensive spectroscopic coverage, but the redshift evolution in the central vs. satellite population has not been explored due to the lack of sufficient sampling over a large area. The deep, magnitude-limited BigBOSS calibration-field survey would have the requisite spatial sampling, thus enabling group and satellite catalogs to be constructed at redshifts $z < 1$ and opening up a new avenue for environmental studies beyond the local universe.

## 3.3    AGN Science

The BigBOSS Key Project will provide spectroscopic observations of AGN and QSO samples of unprecedented size, enabling detailed studies of these populations. In this section, we describe some of the potential scientific projects that could make use of this sample.

### 3.3.1    Global Quasar Census

With the completion of two major quasar surveys, the 2dF QSO Redshift survey [2QZ; Croom et al., 2004] and the SDSS Quasar Survey [Schneider et al., 2010], the number of *spectroscopically confirmed* quasars stands a little over $10^5$ objects. Using these samples, great strides have been made in measuring the global properties of the quasar population, and its evolution with redshift.

In particular, the bright end of the optical quasar luminosity function (QLF) at all redshifts to $z \sim 6$ is now well understood [Richards et al., 2006], and first evidence for optical AGN downsizing at $z$ 2.5 has been seen [Croom et al., 2009]. The evolution of the clustering of the brightest optically selected quasars is also now reasonably measured at redshifts $z \leq 2.2$ [Croom et al., 2005; Ross et al., 2009] and $z > 3.5$ [Shen et al., 2007], although the $z > 3.5$ measurements are hampered by low-number statistics even in the final SDSS dataset. Clustering measurements constrain the mass of the host dark matter haloes that the observed quasar population inhabits; current measurements indicate that this mass is $M_{\rm DMH} \sim 2 \times 10^{12} h^{-1} M_{\odot}$, at all redshifts. The *combination* of the QLF and clustering measurements yield estimates of quasar lifetimes, $t_q$ [e.g. Martini & Weinberg, 2001; Haiman & Hui, 2001; Shen, 2009; Shankar et al., 2010]. The measured value of $t_q$ (and how it depends on host halo mass, environment and duty cycle of the central engine super-massive black hole [SMBH]), while currently still relatively poorly constrained, is a key discriminator between alternative scenarios represented in suites of semi-analytic and N-body simulation models [e.g. Booth & Schaye, 2009; Bonoli et al., 2009].

Due to the evolution of the QLF and the flux-limited nature of most quasar samples, there is a strong correlation between redshift and luminosity in current quasar samples, making it difficult to isolate luminosity dependence of clustering from redshift dependence: the so-called "luminosity-redshift" ($L - z$) degeneracy. This affect is especially acute at redshifts $z \sim 3$, at the height of "quasar epoch". Furthermore, once selection effects are taken into account, and the sample is divided into redshift, luminosity, or another physical parameter [Shen et al., 2009], even datasets of 100,000 quasars can only comprise a few thousand objects in each bin of interest, leading to a low S/N measurement, e.g., when pair-counting at small-scales in clustering measurements [e.g., Myers et al., 2007; Ross et al., 2009]. The dataset from the currently on-going SDSS-III:BOSS Quasar Survey will begin to address some of these issues, but only at redshifts with $2.2 < z < 3.5$ and only for objects $18.0 \leq i$ 21.5. Although BOSS observes quasars fainter than $i = 21.5$, the relatively low completeness for these objects will most likely lead to them not being used in any global statistical analysis.

BigBOSS has the capacity to completely revolutionize the measurements of global quasar properties such as the QLF and clustering, decisively testing and ruling out sets of models. With the efficient QSO target selection discussed in Section 4 below, BigBOSS will deliver a data set of $10^6$ spectroscopically confirmed luminous AGN, over all redshifts up to $z \sim 6$. This is illustrated in Figure 3.3, where the $L - z$ plane is comprehensively filled, with the dynamic range at any given redshift 5 magnitudes.

### 3.3.2   Dual Super-Massive Black Holes

A wealth of observations have shown that galaxy mergers are common and that nearly all galaxies host a central super-massive black hole (SMBH); consequently, some galaxies must host two SMBHs as the result of recent mergers. These are known as "dual SMBHs" for the first $\sim 100$ Myr after the merger when they are at separations $\gtrsim 1$ kpc [Begelman, Blandford & Rees, 1980; Milosavljević & Merritt, 2001]. These dual-SMBH systems are an important testing ground for theories of galaxy formation and evolution. For example, simulations predict that quasar feedback in mergers can have extreme effects on star formation [Springel, Di Matteo & Hernquist, 2005] and that the core-cusp division in nuclear stellar distributions may be caused by the scouring effects of dual SMBHs [Milosavljević et al., 2002; Lauer et al., 2007]]. A statistical study of dual SMBHs and their host galaxies would thus have important implications for theories of galaxy formation and SMBH growth. It could also place interesting constraints on the source population of future gravitational-wave experiments such as *LISA*.

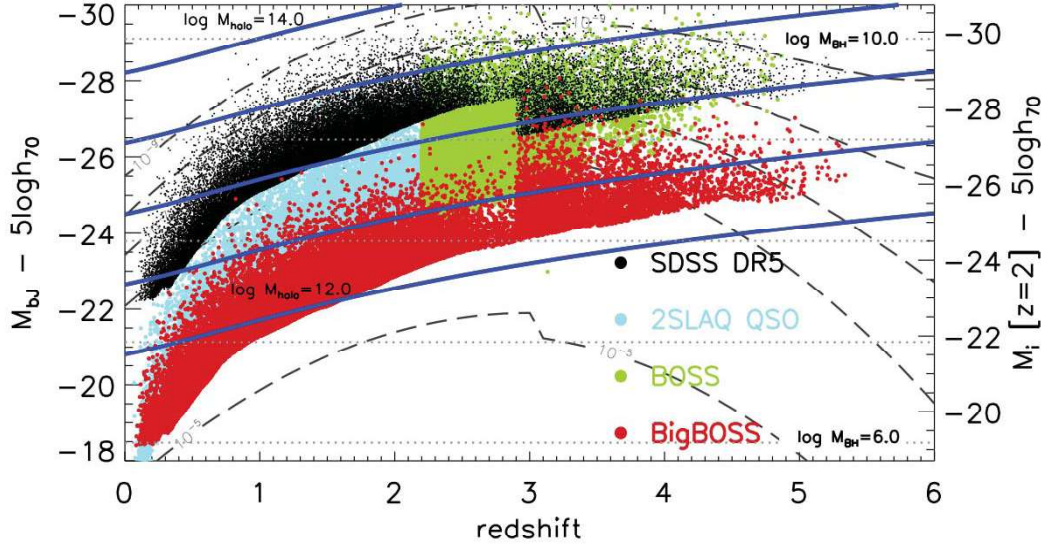

Figure 3.3: The Quasar $L - z$ plane for recent quasar surveys, including the SDSS (black points), 2SLAQ QSO (cyan), BOSS (green), and the projected BigBOSS data (red). The projected BigBOSS data come from assuming a significantly fainter sample than SDSS (see Section 4 for fuller details on quasar target selection), but it should be noted that the selection function present in the SDSS data, e.g. at $z \sim 2.7$, will not be present in the BigBOSS data. Note also that there are 105,000 objects in the SDSS DR7 catalog, but only $\approx 9,000$ faint, $g \leq 21.85$ low-redshift QSOs in hand from the 2SLAQ Survey [Croom et al., 2009]. Dashed lines indicate the evolution of the host halo space density, as inferred from the observed quasar luminosity function. Horizontal dotted lines show the corresponding black hole mass, while the solid lines are halo virial predictions. This figure is kindly modified from the original form in [Croton, 2009], where the reader is referred to for further details.

However, identifying dual SMBHs has so far been difficult. They are observationally identifiable when sufficient gas accretes onto them to power dual active galactic nuclei (AGN), and a handful of dual AGN have been spatially resolved in radio, infrared, optical, and/or X-ray images of nearby galaxies [Komossa et al., 2003; Rodriguez et al., 2006; Bianchi et al., 2008].

A promising new approach to building such a statistical sample is to select dual AGN candidates as galaxies with double-peaked narrow AGN emission lines in their optical spectra, as identified using standard line-ratio diagnostics [Baldwin, Phillips, & Terlevich, 1981; Kewley et al., 2006], While double-peaked *broad* lines can arise from AGN accretion disks [Eracleous et al., 1997], double-peaked *narrow* lines generally only arise from two accreting

SMBHs at separations $\gtrsim 1$ kpc, each with its own narrow-line region (NLR). Spatially resolved detections of the two NLRs are necessary to distinguish this situation from other sources of double-peaked emission, such as outflows within a single NLR. Two such dual AGN were identified in DEEP2, at redshifts $z = 0.6$ and $z = 0.7$ [Gerke et al., 2007; Comerford et al., 2009]. More recently 271 galaxies, at $0.1 < z < 0.6$, with double-peaked narrow AGN emission lines have been identified in the Sloan Digital Sky Survey (SDSS) as dual AGN candidates [Smith et al., 2010; Wang et al., 2009; Liu et al., 2010].

BigBOSS spectroscopy has the potential to yield $\gtrsim 20{,}000$ dual SMBH candidates at redshifts $0 < z < 2$ and with line-of-sight velocity differences of $\gtrsim 100$ km s$^{-1}$ between the SMBHs. Follow-up slit spectroscopy of these candidates could spatially resolve the two narrow-line regions and thereby confirm $\gtrsim 10{,}000$ of these. This would vastly increase the number of known dual SMBHs, extend the known population to $z > 1$, and thus enable unique evolutionary studies of SMBH merging.

## 3.4   IGM Science

BigBOSS provides the exciting possibility of performing tomographic studies of the intergalactic medium using bright background quasars as light sources. For bright ($I$   18) quasars, BigBOSS spectroscopy can yield measurements of metal absorption systems over a large range in redshift (e.g., MgII from $0.33 < z < 2.55$, CIV from $1.4 < z < 5.4$). At redshifts $z$   1 where BigBOSS will also well sample the galaxy population, comparative tomographic studies of the metal rich gas and the bright galaxy distribution will be possible on unprecedented scales. In the regions of the Calibration Fields, such studies can be extended to fainter quasars and to smaller spatial and angular scales. The BigBOSS Key Project will also generate a large dataset of Damped Ly-$\alpha$ (DLA) and Broad Absorption Line (BAL) objects that will provide key insights into the topology and feedback mechanisms of the IGM and cosmic web at high redshifts.

Recently, much progress has been made in studying the abundance and nature of high column density damped Ly-$\alpha$ (DLA) systems (with $N_{HI} > 10^{20.3}$ cm$^{-2}$; e.g., [Wolfe et al., 2005; Prochaska & Wolfe, 2009]), their relation to the universal distribution of HI gas, and the complex interaction between the IGM and galactic star formation. Also, catalogs, some with base timelines of decades or more, [Gibson et al., 2009; Allen et al., 2010] are now available that study the broad absorption lines (BALs) that are occasionally seen in the spectra of quasars. These BALs are extreme events, intrinsic to the quasar, and traditionally associated with winds originating from the central accretion disk.

The fraction of DLAs and BALs is typically $\sim 5\%$ and $\sim 10-15$ of the total, $z > 2.2$ quasar population, respectively. Hence, a survey such as BigBOSS, as opposed to SDSS (or even SDSS-III:BOSS), is required to make statistically significant progress. The BigBOSS Key Project will observe more than 600,000 high-$z$ quasar sight-lines over 14,000 deg$^2$, yielding more than 33,000 DLA systems and 100,000 BAL QSOs. Using the DLAs and the galaxy/AGN distribution, BigBOSS can map out the interlinked cosmic web over an unprecedented volume. The resulting large sample of DLAs can provide the statistics to quantify the number density of high column-density absorbers for comparison to the luminous galaxy population. The BAL QSO sample can form the basis for detailed temporal monitoring of the absorbers and follow-up studies of the environments of the BAL QSOs, with the goal of understanding the impact of the BAL phenomenon on the IGM.

## 3.5 Galactic Archaeology

In $\Lambda$CDM galaxy formation models, galaxies are assembled through the accretion and merging of smaller systems. Since dynamical relaxation times for some galactic components (e.g., the thick disk and halo) are typically longer than the age of the universe, the accretion histories of these components are encoded as substructure in the phase space distribution of stars. Even in cases where the dynamical history is erased, chemical tagging can be used to identify stars belonging to common merger remnants. Mapping the photometric, kinematic and chemical distribution of stars in a given galaxy can thus reveal its fossil accretion history and serve as a critical test of galaxy formation theory. The comparison of such observations with detailed quantitative predictions of the $\Lambda$CDM model of galaxy formation and evolution have ushered in a new age of "near-field cosmology".

With its wide field of view and high multiplex advantage, BigBOSS is well suited to undertake ambitious archaeological surveys of the Milky Way and the Andromeda Galaxy with the goal of understanding their origin.

### 3.5.1 The Milky Way

The history of star formation, chemical enrichment, hierarchical galaxy mass assembly, and dynamical evolution can be resolved in greater detail in our own Milky Way than in any other galaxy. The fossil record of this history is encoded in the detailed phase-space distribution and chemical composition of Galactic stars. The BigBOSS instrument offers the potential to map radial velocities and elemental abundances for a larger number of stars over a larger volume of our galaxy than any other ongoing or planned survey.

Surveys such as the SDSS/SEGUE project have revealed a wealth of kinematical substructures within the stellar populations of our Galaxy, in particular in the thick disk and stellar halo. In the southern hemisphere, at brighter magnitudes ($V < 12$) and covering shorter distances, the ongoing Radial Velocity Experiment (RAVE) is measuring radial velocities for $10^6$ stars, and will be followed by HERMES, a high-resolution spectroscopic survey of a sample of a similar size down to $V < 14$. The forthcoming APOGEE survey of the SDSS-III project will, in turn, extend these studies into the dust-obscured regions of the Milky Way by observing in the $H$ band and penetrating for the first time into distant parts of the disk and bulge, with rich information on up to 15 chemical elements for $10^5$ stars.

The era of near-field cosmology that these surveys have initiated will culminate with the release of data from the Gaia ESA cornerstone mission, which is set to launch in 2012 and to begin public data releases in 2014, with a final catalog expected in 2018. Gaia will measure parallaxes, proper motions, and photometry for an unbiased sample of $10^9$ stars down to $V < 20$, along with radial velocities for $10^8$ stars down to $V < 17$. These samples are essentially complete down to their magnitude limits and allow a full 6D phase-space mapping of the stars included. Gaia, however, leaves a conspicuous gap in coverage: there will be roughly $9 \times 10^8$ stars with photometry and astrometry but no radial velocities between $17 < V < 20$.

The accuracy of the radial velocities to be provided by the Gaia Radial Velocity Spectrometer (RVS) decreases between a few km s$^{-1}$ at $V \sim 15$ to about 15 km s$^{-1}$ at $V \sim 17$. The accuracy of the parallaxes and proper motions ($\mu$) to be provided by Gaia are 21 $\mu$as and 11 $\mu$as yr$^{-1}$, respectively, at $V = 15$, 90 $\mu$as and 50 $\mu$as yr$^{-1}$ at $V = 18$, and 275 $\mu$as and 145 $\mu$as yr$^{-1}$ at $V = 20$. For a star located at $d = 5$ kpc and moving perpendicular to the line of sight at $v = 100$ km s$^{-1}$, a value representative of the velocity dispersion in the Galactic halo, these figures are equivalent to uncertainties of $\sigma(v) \simeq v\sqrt{\sigma^2(\mu)/\mu^2 + \sigma^2(d)/d^2} = 10, 45$, and 140 km s$^{-1}$ at $V = 15$, 18, and 20, respectively. At the same distance, but for a disk star moving at 50 km s$^{-1}$, the uncertainties, dominated by the relative errors in the parallaxes, are accordingly reduced to 5, 23 and 70 km s$^{-1}$. Thus, this is roughly the radial velocity accuracy needed to match the astrometry from Gaia. From numerical experiments using scaled spectra from the CALSPEC (HST) library for P177D (a $V = 13.5$ G0 dwarf) and VB8 ($V = 16.8$, M7), and the exposure time calculator described in Appendix A (see Fig. 3.4), we find that with an exposure time of 600 s BigBOSS can easily match the quality of Gaia's astrometry.

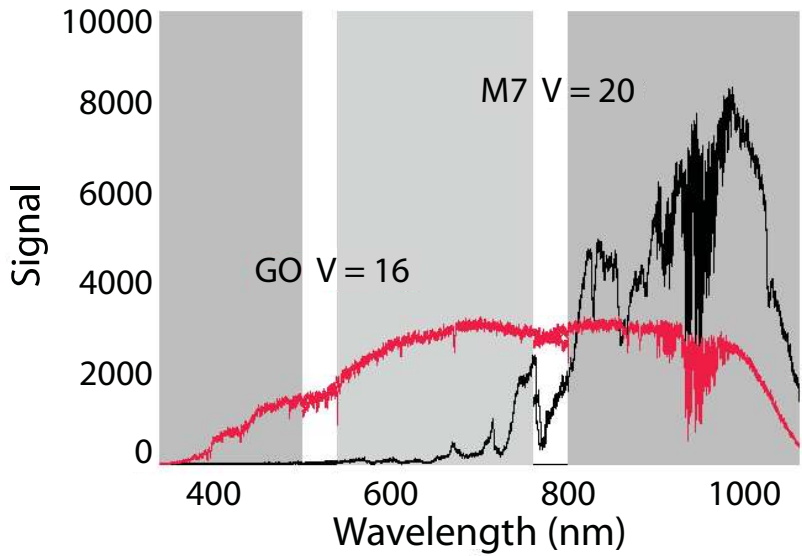

Figure 3.4: Simulated observations for dwarf stars with spectral types G5 (scaled HST spectrum of P177D) and M7 (scaled spectrum of VB8). The shadowed areas show the wavelength coverage of each BigBOSS arm, with the overlapping windows in white. Note that the telluric absorption near 900 nm has been excluded from the Monte-Carlo experiments. In practice these features can be used to set the zero point of the wavelength scale, avoiding systematic errors in the derived radial velocities.

Similarly, Gaia RVS spectra will only be capable of constraining the metal content of targets brighter than $V \sim 14$, due to the poor signal-to-noise ratio of the data for fainter sources. In synchronous observing mode, BigBOSS will be able to determine the overall stellar metallicity three magnitudes deeper than Gaia, down to $V \sim 17$. In a PI observing mode, the BigBOSS instrument could determine metallicities of Gaia stars to even deeper limits.

With its sensitivity, multiplex capability, field-of-view, and spectral resolution, the Big-BOSS instrument is the only project capable of complementing Gaia, a 1 billion-dollar mission, and fully realizing the potential of that survey. More broadly, the BigBOSS instrument will be the only one able to provide large-scale spectroscopic follow-up of Galactic stellar targets from deep, wide-field imaging surveys planned for the near future, such as the US-led Dark Energy Survey (DES) and the Large Synoptic Survey Telescope (LSST).

### 3.5.2   M31

The BigBOSS spectrograph on the Mayall telescope is well suited to perform a wide-field spectroscopic survey of our nearest massive neighbor, the Andromeda Galaxy (M31). The size of the BigBOSS spectroscopic field of view relative to the scale of M31 on the sky can be seen in Figure 3.5.

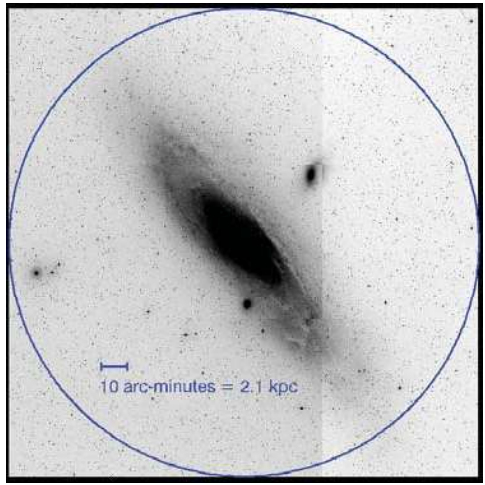

Figure 3.5: Digitized Sky Survey mosaic of the Andromeda Galaxy (M31), with the 3°-diameter BigBOSS spectroscopic field of view overlayed.

A wealth of substructure in the form of tidal streams and faint dwarf galaxy satellites has been discovered in recent photometric surveys of massive spiral galaxies in the local universe (e.g., MW: [Belokurov et al., 2006]; M31: [Ferguson et al., 2002]; nearby spiral galaxies: [Martinez-Delgado et al., 2010]). Much of this work is focused the Milky Way and M31 galaxies, where it possible to resolve individual stars and thus reach extremely low surface brightness levels. The limiting surface brightness for resolved star studies is set not only by the photometric depth, but can also be improved by minimizing the contamination of foreground/background sources in the sample. For example, using a spectroscopically selected sample of resolved stars, the surface brightness profile of M31 has been measured down to $\mu_V = 35$ mag sq arcsec [Guhathakurta et al., 2005]. A spectroscopic survey for substructure based on individual stars is motivated by two goals. First, spectroscopy provides an extra phase space dimension (two spatial and one velocity) in which to search for structure in 6D phase space. A second, and equally important, motivation for spectroscopy is to provide better discrimination between foreground and background objects based on velocity and chemistry, thus pushing the search for substructure to significantly fainter surface brightness limits than possible with photometry alone.

There are compelling advantages to studying the galactic structure of M31. It is sufficiently nearby that individual red giant branch stars are easily resolved and have an apparent magnitude of $r \sim 21$ at the mean distance of M31, providing many thousands of stellar spectroscopic targets per square degree. It is sufficiently far away that surveying a large portion of the M31 halo requires covering a few hundred square degrees, as compared to the full sky coverage needed to understand the Milky Way. An on-going photometric survey of M31, the Pandas survey, has discovered many new tidal streams and dwarf galaxies, and provides an good catalog for spectroscopic target selection. Existing spectroscopic coverage of M31 is limited is deep pencil beams through the M31 disk and halo. These data have provided and tantalizing look at the kinematic and chemical structure of M31. A full mapping of the kinematics and abundances of the M31 halo would be a tremendous step forward compared to these pencil beam surveys through the halo that currently exist.

The BigBOSS spectrograph would enable an unprecedented wide-field spectroscopic survey of M31. The number of M31 stars accessible to spectroscopy with the BigBOSS+Mayall telescope is between 1000-5000 M31 RGB stars per square degree for projected radii 50 kpc and beyond in M31. This is well matched to the number of BigBOSS fibers. With a few hundred pointings at reasonable exposure times, the BigBOSS instrument could map the resolved stellar population of M31 to impressive depth and completeness, with a spectral resolution delivering precise radial velocities and substantial chemical-composition sensitivity.

## 3.6  Stellar Science

Numerous Galactic stellar science programs with targets distributed sparsely on the sky would be well suited to execution in the BigBOSS "synchronous science" mode that we describe in detail in Section 3.8.2 below. In this section we describe some of the particular scientific projects that fall into this category.

### 3.6.1  Blue Horizontal Branch Stars

Blue horizontal branch stars in the halo of the Milky Way are important probes of the stellar density profile and total mass of the galaxy because their distances can be determined accurately ($\sim 10\%$, [Clewley et al., 2004]) out to 50 kpc for $r = 19.5$. Candidates can be selected photometrically, but require spectroscopic confirmation at sufficient S/N to separate then from the less luminous field blue straggler population. Using efficient color cuts that select BHB candidates based on their hot effective temperatures and large Balmer jumps, the density of candidates is a few per square degree. BHB candidates selected in this manner were observed under an SDSS ancillary spectroscopic program (similar to the BigBOSS synchronous science mode), and over the 8032 square degrees of the survey 14,366 BHB candidates were allocated fibers, about 8 per 7 square degree SDSS field. This SDSS sample was used in [Xue et al., 2008] to make one of the best available determinations of the mass of our Galaxy. For comparison, the SEGUE project, a survey of Milky Way stars using the SDSS survey telescope and spectrographs [Yanny et al., 2009], targeted BHB candidates at high priority, and allocated fibers to all such candidates in each of its 200 pointings. The final number of BHB candidate spectra in the same color-selection region as used for the SDSS ancillary program is 2930, a density not much larger than that from the SDSS BHB sample despite the fact that SEGUE had access to all 640 fibers in the SDSS field. For the BHBs, and for other similar sparse classes of targets, the survey area is the primary limitation on the size of the sample.

### 3.6.2   White Dwarf Stars

The white dwarf catalog of [Eisenstein et al., 2006] was assembled through the SDSS ancillary science observing mode, and has doubled the number of spectroscopically confirmed white dwarfs. This catalog has been used for statistical studies of white dwarf stellar physics as well as the star formation history and age of the Galactic stellar populations [Harris et al., 2006]. In addition, the catalog includes rare sub-classes like the "ultra-cool" white dwarfs that can be used to determine ages for the oldest stars in the Galactic disk and halo [Gates et al., 2004; Kilic et al., 2006], magnetic [Schmidt et al., 2003] and pulsating [Mukadam et al., 2004] white dwarfs, and white dwarf plus M-star binary systems [Silvestri et al., 2006]. These last systems can be further searched with time-resolved spectroscopy for compact, post common-envelope binaries in order to better understand the formation mechanism of these systems which are the precursors of cataclysmic variables X-ray binaries and possibly SN Ia. A similar program executed during the BigBOSS survey would represent an order-of-magnitude increase in statistical power and discovery potential of white dwarf science.

### 3.6.3   Other Stellar Samples

A BigBOSS synchronous observing program would also allow for efficient spectroscopy of luminous red giant stars at distances greater than 50 kpc from the Galactic center. There are only a few such stars per square degree, and they can be reliably selected from multicolor photometry [Morrison et al., 2001; Majewski et al., 2003; Yanny et al., 2009]. These rare stars are one of the few kinematic tracers available at large Galactocentric distance to measure the total mass of the Galaxy and its dark matter halo [e.g. Battaglia et al., 2005].

The BigBOSS instrument could also be employed for a spectroscopic search of color- or objective prism-selected candidates for the most metal-poor, chemically primitive stars in the Galaxy [Beers et al., 1985; Christlieb et al., 2001]. As observational upper limits are extended, searches for these and other population-extreme objects become harder to execute as a stand-alone programs, but they are ideally suited to execution in synchronous mode with BigBOSS.

## 3.7   BigBOSS and the Transient Sky

The BigBOSS survey has the potential to spectroscopically characterize the transient sky, especially if its mapping footprint on the sky is preceded closely in time by an imaging survey with a time-domain/transient-detection component. The likely imaging data that would be used for targeting this mode would be drawn from the Palomar Transient Factory and/or the PanSTARRS surveys; a small fraction of BigBOSS fibers could be assigned to targets selected due to their variability in these imaging surveys. The bulk of these objects are likely to be AGN or variable stars. Nevertheless, by targeting essentially all the variable targets above some relatively bright magnitude and variability amplitude, BigBOSS can provide a near-complete spectroscopic census of the transient sources and thus holds the potential to discover and physically characterize some hitherto unknown variable class of objects. Other transient science may be best pursued as part of a separate, PI program; we provide an example in §3.7.1.

### 3.7.1 LSST Supernovae Survey

It is estimated that LSST will discover and obtain well-measured light curves for 300,000 SNe Ia through an intensive campaign which covers 300 square degrees of sky with a more frequent cadence than the standard LSST field [Abell et al., 2009]. These fields are likely to be largely equatorial to ensure access by the maximum number of facilities. Obtaining spectra (and spectroscopic redshifts) for the host galaxies of this large crop of SNe would make a variety of new science possible and old science statistically relevant. In particular, host galaxy spectra will be needed to sharpen the utility of these SNe for precision cosmology, as described in the SN Ia Science White Paper submitted for the 2010 Decadal Review [Howell et al., 2009].

With existing spectrographs on 10m telescopes, covering such a large area of sky would require between 5,000 (VLT/VIMOS) and 17,000 (Keck / DEIMOS) separate pointings on the sky, with exposure times of at least an hour per pointing to obtain redshifts for the majority of SN hosts (which should extend to redshifts of $z \sim 1$, with host magnitudes commonly reaching $r \sim 23 - 24$). It has therefore generally been assumed that LSST supernova studies (which can provide a useful probe of dark energy) will have to rely on photometric redshift estimates. However, the greater uncertainty and unknown systematics of supernova photo-z's (as compared to spectroscopic redshifts) may significantly degrade the utility of LSST SNe.

However, obtaining redshifts for the hosts of a majority of these supernovae would be feasible with the BigBOSS instrument. Three hundred square degrees of sky would require only 43 BigBOSS pointings to cover. As the effective collecting area of the Mayall is 15% that of one of the Keck telescopes, if BigBOSS has a throughput comparable to that of the DEIMOS spectrograph, given the comparable resolution, spectra with equivalent signal-to-noise to a 1 hour DEIMOS spectrum could be obtained in 6.7 hours' observing time with BigBOSS. Hence, covering the full LSST SN survey region would require $\sim 40$ clear nights at Kitt Peak. This would provide only $\sim 200,000$ redshifts, however; a multiple-coverage scheme (e.g. with bright objects observed only once and fainter targets multiple times) would be necessary to obtain redshifts of all 300,000 LSST SNe Ia, requiring perhaps $\sim 60$ nights total. This is well within the range of past NOAO survey programs.

Fibers not placed on the hosts of SNe Ia could be placed at positions of other LSST SNe and transients, allowing an exploration of a much wider set of objects and providing crucial redshift measurements for the new (or rare) and astrophysically interesting classes of transients that LSST will undoubtedly discover, such as Macronovae, GRB orphan afterglows, tidal disruption flares, and fallback SNe [Abell et al., 2009; Rau et al., 2009]. These new classes of objects might appear photometrically similar to other objects without a distance estimate. Since their intrinsic properties are not well known, conventional photo-z estimates cannot be used, and thus, an independent secure spectroscopic host redshift will be needed to distinguish between competing possibilities.

## 3.8 Community Participation in BigBOSS Science

As described at the start of this Chapter, we envision three ways for the NOAO user community to benefit from the BigBOSS Project: by using the archive resulting from the Key Project; by proposing synchronous observing programs; and by proposing to use the instrument for their own science. We discuss each of these below. Table 3.1 provides a summary of the possible breakdown of science projects described above among these three observing modes. Once again, we stress that this is *only a subset* of the possible science impact of the BigBOSS survey and instrument beyond its core cosmology mission.

### 3.8.1 Archival Science

The BigBOSS Key Project will result in a database of more than 19 million spectra of galaxies and QSOs and upto 9.8 million ancillary targets (see Table 2.2). While many of the spectra may be of low signal-to-noise ratio, they may nevertheless enable astrophysical studies of unprecedented scope. A few examples of such archival studies have already been discussed in this Chapter (e.g., see Sections 3.2 and 3.3), but many more are feasible.

We propose to create a data archive in collaboration with the NOAO Science Data Management group, along the lines of the archive some of our team members have developed for the SDSS projects (see Section 7 for further details). The data and catalogs will be released in a pre-defined cadence that provides for a limited period of collaboration access, followed by full public release of all raw and processed data and associated catalogs. Observations made in the synchronous observing mode or PI-led observations will be made available as requested. The SDSS archive has a remarkable legacy of usage by the non-SDSS community, and we envision a similar level of benefit flowing from the BigBOSS survey.

### 3.8.2 Synchronous Science

The baseline BigBOSS survey, as described in Chapters 2 and 6, will be able to place approximately 80% of its fibers on the ELG, LRG and QSO targets that reach the survey goals in redshift coverage and exposure time. While work is continuing on optimizing the BigBOSS survey strategy, the baseline survey demonstrates that some fraction of the fibers in any tiling, 500-1000 fibers ($\approx 70 - 150 \ \mathrm{deg}^{-2}$), will be un-filled by targets from the main BigBOSS survey. We propose that these fibers, totaling between 5,000,000 and 10,000,000 observations over the course of the survey, be made available to other science programs proposed by the astronomical community at large, to be executed synchronously with the BigBOSS survey. Since this mode of observation is unique to BigBOSS, we describe it here in some detail.

The opportunity to submit target objects to such synchronous programs that use the un-allocated fraction of BigBOSS fibers would provide the astronomical community with a completely unique resource: access to a five year observing program that will cover 14,000 $\mathrm{deg}^2$ and can reach S/N 10 $\mathrm{\AA}^{-1}$ at 4000 Å for a point source $g$ magnitude of 19.5 [5]. This is a

---

[5]as calculated using the BigBOSS ETC (see Appendix A)

Table 3.1: **Science Impact of BigBOSS**

**1. Key Project**

- constraints on cosmological parameters ($H(z)$, $D_A(z)$, $\Omega_k$, $w_0$, $w_a$, etc.) [§2.5]
- constraints on growth parameter $D(z)$ [§2.3.3,§2.3.4]
- constraints on neutrino masses $\Sigma m_\nu$ [§2.5.2]
- constraints on inflation $\alpha$, $n_s$ [§2.5.3]

**2. Community-Led Archival Science from Key Project**

- galaxy evolution studies (based on stacking analysis, or on individual spectra of galaxies in the calibration / Pilot Survey fields) [§3.2]
    - evolution of age, mass, metallicity with redshift
    - environmental effects (only from calibration fields)

- large scale structure
    - redshifts for galaxy cluster candidates [e.g., § 3.2.6]
    - clustering of galaxies as function of redshift, type, etc. [§ 3.2.8]
    - analysis of halo occupation distribution as function of type, etc.
    - QSO clustering

- tomography of IGM and correlation with galaxy distribution [§3.4]
- structure of Galactic halo (from calibration stars, failed QSO candidates, etc.)
- rare objects (lenses, galaxies, AGN, stars) [§3.2.2,§3.2.7,§3.3,§3.6]
- *Your Idea Here*

**3. Community-Led Synchronous Science from Key Project**

- structure and dynamics of the Galactic Halo and thick disk [§3.5.1 ]
- AGN surveys, identification, evolution, clustering [§3.3]
- nearby galaxy surveys
- identification of sources from NASA sky surveys (e.g., WISE, GALEX, Spitzer, Herschel, Euclid, etc.)
- rare objects (white dwarfs, metal poor stars, high velocity stars, weird AGN, blue compact galaxies, TOO events, etc.) [e.g., §3.3,§3.6]
- *Your Idea Here*

**4. Community-Led PI Projects and Other Surveys**

- structure and dynamics of M31 halo [§3.5.2]
- detailed studies of Milky Way streams [§3.5.1]
- studies of open clusters and moving groups
- LSST Supernova Survey [§3.7]
- galaxy evolution studies [§3.2]
- ISM tomography
- IGM tomography [§3.4]
- dynamics of intracluster planetary nebulae
- studies of stellar activity in star forming regions
- *Your Idea Here*

capability uniquely well matched to programs with targets that are sparsely distributed over the sky, or that require a spectroscopic search for rare but scientifically valuable objects.

To implement these synchronous observing programs led by the community, we propose that NOAO solicit proposals and evaluate them for scientific and technical merit through its regular time allocation process. The successful investigators would then submit target lists to the BigBOSS team for each year's observing. After the location of the BigBOSS pointings are determined based on the optimal tiling of the survey targets, the same code that assigns the BigBOSS fibers would then assign the unused fibers to objects from the synchronous program target lists. With the pointing center already established, the only remaining freedom in placing the unused fibers is the 181 arcsec diameter of the patrol region of the actuators. These programs would be subject to the same bright magnitude cutoff as the BigBOSS targets in order to prevent excessive scattered light from contaminating neighboring fibers, and would be observed in units of the standard BigBOSS 16.6 minute exposures.

The most straightforward kind of synchronous program to implement would be those that can submit large lists of targets all at the same priority, expecting to obtain spectra for only the fraction of the input catalog that can be allocated to the unfilled fibers. A useful analogy is the *Hubble Space Telescope* Snapshot Survey program, where successful proposers submit a list of un-prioritized targets and a fraction of those targets are observed as scheduling of the other approved programs allows.

In contrast, synchronous programs that have very few targets per pointing would be included at high priority in the BigBOSS survey tiling optimization. Programs of this kind have the potential to create samples of many thousands of objects for targets as sparse as one per square degree, making it possible to characterize the populations of very rare and poorly understood classes of objects that are currently limited by the rate at which samples can be collected using single-object spectroscopy.

While the allocation of a fraction of the unused fibers to this second category of high-priority synchronous programs would probably not impact the BigBOSS survey efficiency appreciably, assigning *all* of the leftover fibers to such programs would likely impact survey efficiency adversely. In order to support this mode, we will carry out detailed tiling simulations to determine a threshold for additional constraints from the synchronous programs. The multi-year duration of the BigBOSS survey is an advantage here, as this more complicated but potentially very productive class of synchronous programs could be implemented in the second and later years of the survey, after the BigBOSS team has a season's experience with the tiling and fiber allocation.

### 3.8.3 PI Science

Since BigBOSS is a facility instrument for the Mayall telescope, it will be open to the community for P.I.-led science programs. Proposals will be accepted through the regular and survey time allocation processes administered by NOAO. Since BigBOSS is primarily a survey instrument, we envision that the bulk of programs proposed by the community will be large spectroscopic surveys. The large surveys of the Galactic halo and thick disk and the halo of M31 (Sections 3.5.1 and 3.5.2), many types of galaxy-evolution studies (§ 3.2), and the LSST SN follow-up survey (§ 3.7.1) would likely fall into this category.

Since we will be developing the software to process, analyze and archive the data for the Key Project, we propose to work with the NOAO Science Data Management group to make this end-to-end reduction pipeline available for the PI science programs. Since the Key Project will have a well-defined observing strategy and protocols, any observations that follow these protocols should be able to make use of the standard pipeline. We will work with NOAO to ensure that users have full access to the pipeline, but envision NOAO providing the necessary support of the pipeline for their users.

### 3.8.4 Community Support and Deliverables

As illustrated by the SDSS, the community value of large spectroscopic survey programs is immeasurably enhanced by the timely distribution of high-level data products and corresponding documentation. The details of the data products that we will release to the community, and the infrastructure with which this distribution will be carried out, are given in Section 7.5 of this document. In brief, we will distribute target lists, targeting photometry, survey window functions, reduced spectra, and derived spectroscopic parameters including redshifts and object classifications. The primary means of distribution will be via an online database interface, in the manner developed for the SDSS. All survey data will be released to the public at semi-regular intervals, the dates of which will be determined and advertised at the outset of the survey. Details of the survey data as distributed to the public will be posted to a website that will be available for public viewing simultaneously with the public data releases.

Community targets submitted for synchronous observation as described above will be calibrated and extracted to the 1D spectrum level by the main BigBOSS survey pipeline. These spectra will be made available to synchronous proposers with relatively short turnaround, and distributed via password-protected web or FTP directories. (We anticipate that during the hardware and software commissioning phase, there will be a somewhat longer delay in providing reduced synchronous-program data, during which we will verify the correct operation of our observing modes, pipeline analysis, and distribution system.)

In order to support the dedicated use of the BigBOSS spectrograph for PI-led programs, we will deliver a functioning suite of targeting and observation-planning software to NOAO. We will also release all of our spectral data-analysis software under an open-source license, and will consult with PI users in order to maximize their scientific return from their data. Since we anticipate that PI programs may operate in different regimes and have different scientific data-analysis requirements, it is anticipated that PI-led programs will contribute some software-development effort as necessary to meet their specific program goals, to the extent that they differ from the analysis requirements of the core BigBOSS survey.

## 3.9 Synergies with Ground and Space Missions

### 3.9.1 BigBOSS, DES and LSST

The massive spectroscopic survey provided by BigBOSS will provide a unique and important complement to direct-imaging science projects currently being planned. We focus here on surveys using the Dark Energy Camera (DECam) on the 4m Blanco Telescope, as well as on future imaging with the Large Synoptic Survey Telescope (LSST). BigBOSS spectroscopy will complement ongoing imaging surveys undertaken in the northern hemisphere (e.g., PanSTARRS, PTF, etc.) and future space-based all-sky surveys in similar ways.

Although both DECam and LSST will be located in the southern hemisphere, their planned surveys will include significant overlap with the baseline BigBOSS survey footprint. The LSST survey footprint (see figure 2.1 in the LSST Science Book, [Abell et al., 2009]) extends to a declination limit of a minimum of $+3^{\circ\|}$across the entire sky, with an additional northern extension to $\approx+30^{\circ\|}$(the "northern Ecliptic region") at right ascensions between RA = 60 and $120^{\circ\|}$($4$-$8^h$). The Dark Energy Survey (DES, the cornerstone DECam survey) will cover an equatorial strip from RA = $-50^{\circ\|}$to $+55^{\circ\|}$($-3.3^h$ to $+3.7^h$, overlapping with SDSS Stripe 82), and an additional region between RA = $30^{\circ\|}$and $55^{\circ\|}$($2^h$ and $3.7^h$) which lies north of DEC$=-15^{\circ\|}$(Jim Annis, personal communication). The BigBOSS spectroscopy within these regions will be a rich complement to these surveys.

One of the most significant impacts of BigBOSS upon future photometric surveys will be in the spectroscopic calibration of photometric redshifts, which represents one of the dominant systematic uncertainties in the weak-lensing cosmology and galaxy dark-matter halo measurements to be made from imaging data. BigBOSS will provide direct calibration redshifts for objects in the survey overlap regions, greatly increasing the number of redshifts available to develop and tune techniques. At fainter magnitudes (to $i = 24$ for DES and $i = 25$ for LSST), photometric redshift calibrations are likely to be based on cross-correlation techniques [Newman, 2008], which rely on the fact even the faintest galaxies at a given redshift will cluster with the brightest galaxies at that $z$ (precisely those galaxies which are easiest to obtain redshifts for and that BigBOSS will target). By combining measurements of the clustering of a photo-$z$-selected sample with members of a spectroscopic sample, as a function of the spectroscopic $z$, the actual redshift distribution of the photometric sample can be reconstructed accurately. For current deep survey samples, which consist of tens of thousands of objects in a few square degrees, the dominant error in cross-correlation estimates is due to finite-field-size effects [Matthews & Newman , 2010]. However, for surveys covering $> \sim$ 100 square degrees, these effects become negligible. A survey of $\sim 10,000$ objects (galaxies, QSOs, or absorption systems) per unit $z$ covering at least 100 square degrees is sufficient for LSST calibration purposes; in combination with existing SDSS and BOSS data at lower redshifts, BigBOSS will provide sufficient coverage at redshifts up to 2–3 (beyond which calibration requirements are much less stringent, so smaller samples are needed), almost single-handedly solving the photometric redshift calibration problem for both DES and LSST.

The synergies between BigBOSS and deep photometric surveys extend much further, encompassing many of the topics already discussed in Sections 3.2–3.7 above, but with the added power of deeper photometric data. Amongst other things, BigBOSS will measure redshifts for the dominant galaxies in cluster candidates from these surveys, enabling improved dark energy constraints; provide spectroscopic classifications for variable objects (e.g., via SN host redshifts, AGN typing, stellar typing, spectroscopic binary identification, etc.), as discussed in §3.7; improve dwarf/giant separation for Galactic structure studies; provide reddening measurements from stellar extinction; yield AGN classifications and redshifts for determining demographics, luminosity functions, evolution, etc.; measure redshifts of both foreground and background galaxies in strong gravitational lenses; perform metal line tomography of the intergalactic medium along the lines of sight to bright QSOs; etc. In turn, these surveys will add to the BigBOSS dataset both by providing morphological information and photometry extending beyond the wavelength range covered by BigBOSS (e.g., the LSST $y$ band). Even though only a fraction of the BigBOSS survey area will overlap DES and LSST coverage, much of the science yield may be obtained by using a limited common area to calibrate statistics to be applied to the full samples.

### 3.9.2  BigBOSS and NASA Missions

With its ability to efficiently deliver spectroscopy for large numbers of targets spread over large areas of sky, BigBOSS can provide a unique resource in support of space missions which map the sky. For example, the recent *Wide-Field Infrared Survey Explorer* (*WISE*) mission will map the entire sky and result in a large number of unidentified infrared galaxies. Folding a subset of these in as synchronous science targets with the Key Project can yield a relatively unbiased and very large sample of redshifts for *WISE* targets.

With the Astro2010 Decadal Survey recommending a significant augmentation to the NASA Explorer Program, it is likely that there will be future missions that will yield multiwavelength maps of large regions of the sky. BigBOSS can provide spectroscopic data for significant samples in support of these missions.

## 3.10  Community Science Workshop

In order to maximize the broader impact of the BigBOSS survey and instrument within the astronomical community, we propose to hold a BigBOSS community science workshop in advance of the start of survey operations. The purpose of this workshop would be three-fold. First, the workshop will convey the details of BigBOSS to a wide professional audience, and allow for extensive back-and-forth discussion between interested astronomers and the BigBOSS project team. Second, the workshop will provide an opportunity for interested researchers to meet and form collaboration networks based upon common interests in Big-BOSS data and capabilities. Third, the workshop will give our team the best possible overview of the range of interest in BigBOSS and the supplementary and PI proposals that are likely to be forthcoming, thereby allowing us to optimally strategize for commissioning, operations, and data reduction. We propose that this workshop be held in fall 2011, with ample advance notice and publicity. As the use of BigBOSS develops over subsequent years, additional workshops or conferences may be organized in order to continue the momentum that we hope to build in support of BigBOSS as a community resource.

# 4 Target Selection

## 4.1 Summary of Targeting Requirements

The primary science goal for the BigBOSS survey is to measure with high precision the baryon acoustic feature imprinted on the large-scale structure of the universe, as well as the distortions of galaxy clustering due to redshift-space effects. The survey will achieve this science goal through spectroscopic observations of three distinct classes of extragalactic sources: luminous red galaxies (LRGs), star-forming emission line galaxies (ELGs), and quasi-stellar objects (QSOs). Each of these categories will require a different set of selection techniques to provide sufficiently large samples of spectroscopic targets from available photometric data. Further, to ensure high efficiency, the methods used must select objects with spectral features that will produce a reliable redshift or Ly-$\alpha$ forest measurement within the BigBOSS wavelength range.

The targeting requirements for each of the BigBOSS targets is summarized in Table 4.1. The requirements table includes the spectral feature of interest, the desired redshift range, volume density, and the projected areal density summarized from the following target selection discussion. The volume density values in the table are the minimum required densities; we expect to achieve substantially higher target densities over much of the redshift range. Figure 4.1 shows the target redshift distributions resulting from the strategies discussed in this chapter. The distributions have been smoothed to show the general shape and redshift range of the target samples, and the total areal densities are scaled to match the values in Table 4.1. These distributions represent the raw galaxy populations that would be targeted by BigBOSS *before* inefficiencies (such as placing a fiber on the target or measuring a spectral feature of sufficient signal-to-noise) are taken into account.

Table 4.1: Summary of the minimum galaxy sample requirements for BigBOSS key science.

| Target | Feature | Redshift | Min. Vol. Density $h^{-3}$ Mpc$^3$ | Areal Density per deg$^{-2}$ |
|--------|---------|----------|-----------------------------------|------------------------------|
| LRG | 4000Å break | 0.6< $z$ <1.0 | $3 \times 10^{-4}$ | 350 |
| ELG | [OII] emission | 0.7< $z$ <1.7 | $1 \times 10^{-4}$ | 2300 |
| QSO | Ly-$\alpha$ forest | 2.2< $z$ <3.5 | $1 \times 10^{-5}$ | 65 |

The lowest redshift sample of BigBOSS targets will be composed of LRGs. These luminous, massive galaxies ceased star formation more than a billion years before the time of observation, and therefore have evolved, red composite spectral energy distributions (SEDs). The BOSS survey is targeting these objects to $z < 0.6$ using SDSS *gri* colors and measuring spectroscopic redshifts using the prominent 4000Å break continuum feature. BigBOSS will measure the same feature but extending to $z < 1.0$; as a result, other selection techniques will be required. In particular, we will select LRGs using the prominent 1.6$\mu$m (restframe) "bump". This feature corresponds to the peak of LRG SEDs and provides a strong correlation between optical/near-infrared (NIR) color and redshifts at $z < 1$. We will use 3.4$\mu$m photometry from the space-based Wide-Field Infrared Survey Explorer (*WISE*)

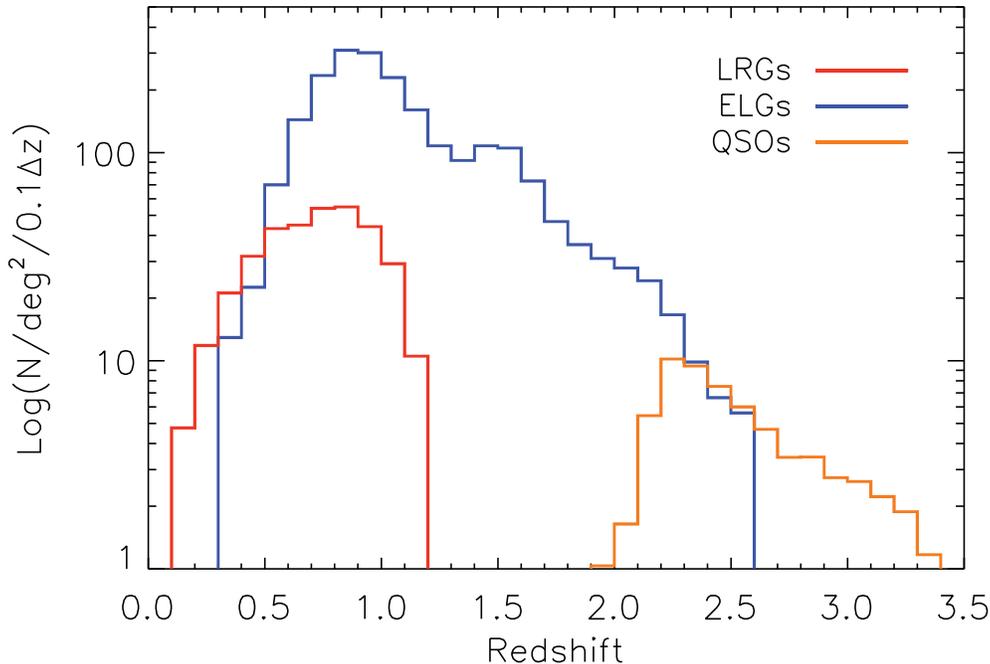

Figure 4.1: The general BigBOSS target redshift distributions summarized from the target-ing discussion and scaled to match the areal density values in Table 4.1. The distributions represent the expected underlying galaxy populations for the BigBOSS targets *before* losses (such as targeting efficiency and fiber completeness) are taken into account. In the case of the LRGs, the target distribution is subsampled to meet the volume density requirement.

to efficiently select LRGs in the redshift range of $0.6 < z < 1.0$.

The majority of the spectroscopic redshift measurements for BigBOSS will come from emission-line galaxies at redshifts $0.7 < z < 1.7$. These galaxies possess high star formation rates, and therefore they exhibit strong emission lines from ionized HII regions around massive stars as well as SEDs with a relatively blue continuum spectrum shape. One of the most prominent features of ELG spectra is line emission from the [OII] 3727Å doublet, which consists of a pair of emission lines separated in wavelength by 2.8Å. The spacing of this doublet provides a unique signature, allowing for definitive line identification and secure redshift measurements even if it is the only feature observed. The doublet feature is a key tool for measuring redshifts in several smaller-area spectroscopic surveys (such as DEEP2 and WiggleZ); as a result, the properties of strongly line-emitting galaxies have been well studied. The goal of the BigBOSS ELG target selection will be to provide a large sample of ELGs with sufficient [OII] line flux to obtain a redshift detection and measurement out to $z < 1.7$. We will use forthcoming large-area optical photometric datasets to select these targets.

The highest-redshift target sample will consist of QSOs. We will be able to measure large-scale structure using the foreground neutral-hydrogen absorption systems which make up the Ly-$\alpha$ forest at $2.2 < z < 3.5$. Unfortunately, QSOs have SEDs and magnitudes very similar to blue stars, which generally leads to inefficient and incomplete targeting for spectroscopic samples. BOSS selects quasars with a combination of techniques that depend critically on SDSS $u$-band measurements, resulting in 40 targets deg$^{-2}$; roughly 50% of those targets prove to be QSOs at an appropriate redshift for Ly-$\alpha$ absorption studies. BigBOSS will achieve twice the BOSS target density by using variability information from time-series photometric surveys and allocating a portion of the BigBOSS survey to high-redshift QSO identification. BigBOSS will provide redshifts for QSOs over a wide redshift range ($z < 3.5$), but only obtain long exposures on confirmed $z > 2.2$ quasars in order to measure the Ly-$\alpha$ forest.

For the purposes of the following targeting discussion, we define a few terms which we will use to describe the effectiveness of our selection techniques.

- *Completeness*: The fraction of objects selected out of all the available objects of the desired class (e.g., the fraction of all $0.6 < z < 1$ LRGs with $r < 22.5$ that are selected by our targeting techniques).

- *Redshift window efficiency*: The fraction of selected objects which lie within the desired redshift range and are of the correct object type.

- *Spectral feature efficiency*: The fraction of selected objects that possess a spectral feature that could potentially be detected at high enough significance for redshift measurement. For absorption-line measurements (e.g. LRG redshifts and QSO Ly-$\alpha$ studies), this will be primarily determined by the optical magnitude of the target. For ELGs, the spectral feature efficiency will be determined by what fraction of objects have sufficiently great [OII] flux to potentially yield a secure detection and redshift; this fraction is only loosely correlated with optical continuum magnitude.

- *Target selection efficiency*: As defined in §2.2, this is the product of the redshift window efficiency and spectral feature efficiency.

- *Redshift measurement efficiency*: The fraction of objects which possess a potentially identifiable spectral feature for which a secure redshift measurement is obtained.

These measures are separate from considerations of what fraction of objects selected as targets are actually placed on fibers, defined at *fiber completeness* and discussed in Chapter 6.

In this chapter, we will show that the combination of time-series photometry in forthcoming wide-area surveys and simple color selections can achieve all the target density requirements of Table 4.1. We also provide expected redshift distributions of the targeted samples based upon tests of strawman target selection algorithms. Elements of several other sections have relevance here. For instance, Chapter 5 details the design of the BigBOSS instrument, which informs a spectral simulator presented in Appendix A. This detailed spectral simulator aids in the design of the targeting strategy (such as magnitude limits), calculates exposure times, and estimates redshift measurement efficiencies. Given the expected target densities and exposure times, the overall survey strategy is developed in Chapter 6. Included in the survey strategy is an optimized method to tile the sky that maximizes the area covered and number of target redshifts obtained, while minimizing the overall time required for the survey. Chapter 6 also outlines a strategy for fiber allocation and calculates the total usage of available fiber exposure times. The main results of these chapters are summarized in Table 2.2 and are used to calculate the DETF Figure of Merit in Table 2.7.

## 4.2 Photometric Surveys

Selecting extragalactic sources for BigBOSS will require the use of imaging data for targeting. Therefore, the success of the survey is predicated on the availability of photometry over the entire BigBOSS footprint to sufficient depth to achieve our target number density requirements (after taking selection efficiency into account). Large area surveys with more than 10,000 deg$^2$ of extragalactic sky coverage are rare. However, there are several ongoing surveys in the Northern Hemisphere that will deliver multiband photometry in the BigBOSS footprint within the next few years. These forthcoming datasets will serve as the backbone for BigBOSS target selection. We describe these surveys in more detail below.

### 4.2.1 SDSS

The Sloan Digital Sky Survey [Abazajian et al., 2009] has served as an excellent photometric data source for wide-field studies. SDSS includes multiband ($ugriz$) photometry which can efficiently separate a wide variety of stellar and extragalactic sources using their optical spectral energy distributions (SEDs). The $5\sigma$ magnitude depths for the SDSS $ugriz$ bands are 22.0, 22.2, 22.2, 21.3, and 20.5, respectively. SDSS covers a 10,000 deg$^2$ footprint with contiguous coverage over the North Galactic Cap and partial coverage of the South Galactic Cap. The BOSS survey is designed to take advantage of this photometry, targeting both LRGs and Ly-$\alpha$ QSOs selected using SDSS imaging.

The main SDSS photometric sample will largely not be deep enough to be useful for spectroscopic targeting in BigBOSS. However, we can use the well characterized properties of SDSS spectrophotometry to help calibrate the spectroscopic properties of BigBOSS. For example, the relative spectral calibration of SDSS F-stars can readily be used by BigBOSS to calibrate relative throughputs and to monitor variable sky transmission. Further, the results of the BOSS QSO survey and variability studies in the deeper Stripe 82 will inform the BigBOSS QSO target selection and reduce the number of stellar contaminants in our quasar survey. SDSS photometry and spectroscopy will provide a well-tested data source to calibrate with and compare samples against throughout the BigBOSS survey.

### 4.2.2 PanSTARRS

The PanSTARRS $3\pi$ survey [PanSTARRS website, 2010] is a transient-sensitive survey designed to observe 30,000 deg$^2$ of sky over 12 epochs in each of the five $grizy$ survey filters. The multiband photometry generated from the co-added exposures will reach depths that exceed that of SDSS and will serve as a source database for BigBOSS target selection. PanSTARRS has been designed to be a staged experiment, with additional telescopes scheduled to come online in the next decade. However, only the first of those telescopes (PS1) is currently taking survey data and will accomplish 360 seconds of total exposure time in three years of operation. Upon completion of this survey, we expect that the PanSTARRS co-added data will be released for public consumption and use for spectroscopic followup.

Additional targeting information could also come from PS1 time-domain photometry, but the public availability of the time variability information is uncertain at this time.

### 4.2.3   Palomar Transient Factory

The Palomar Transient Factory (PTF) [Law et al., 2009] is a photometric survey designed to find transients over 12,000 deg$^2$ in the Northern Hemisphere. PTF is using the 1.2m Oschin Telescope at Palomar Observatory with the CFH12K camera to conduct this survey. Thus far, PTF has focused on obtaining Mould $R$ band photometry with a nominal 5 day cadence and 60 seconds of exposure time, as well as shallower coverage in the $g$'-band. Four years of survey operations will yield a total of three hours' exposure time in $R$ over the entire survey footprint. We project that the $R$-band depth of the final co-added data will be $\sim 0.5$ magnitudes fainter than PS1 $r$ and therefore more valuable to our ELG target selection (see Figure 4.1). The PTF collaboration will be releasing data within two years of observation. LBNL is a member of this collaboration.

### 4.2.4   Ground-based Photometric Error Model

Our strawman plan for BigBOSS ELG target selection will focus on the co-added $gi$ bands from the PS1 survey and the co-added $R$ band from PTF. Since neither PS1 nor PTF have completed their surveys, we must model the photometric errors that match the depths expected from each survey. The error model can then be applied to synthetic magnitudes generated from galaxy SED templates convolved with the PTF and PS1 filter bands to reasonably represent the photometric quality of the surveys. The photometric signal to noise ratio for various telescope parameters is modeled with the equation

$$\sigma_m = 10^{(-0.4m_{AB} + 0.4m_{site})} \times \left[\frac{t}{\pi\omega^2}\right]^{1/2}, \tag{4.1}$$

where $m_{AB}$ is the source magnitude, $m_{site}$ is the site-dependent sensitivity, $t$ is the total exposure time, and $\omega$ is the FWHM of the source in arcseconds. Each filter band an independent value of $m_{site}$ which is solved for from the survey-reported $5\sigma$ depth shown in Table 4.2 . Figure 4.2 shows the photometric error versus source magnitude for the $gri$ bands from PS1 and $R$ band from PTF. For these estimates, we use a mean galaxy half light radius of 0.3″‖to represent the extended ELG galaxy objects observed at high redshift.

Table 4.2: Assumed PTF and PS1 parameters for the photometric error model in Eq. 4.1.

| Survey | band | $m_{site}$ | t (s) | $\omega$ (arcsec) | $m_{AB}$ (5$\sigma$) |
|--------|------|-----------|-------|-------------------|---------------------|
| PS1    | $g$  | 22.85     | 360   | 1.3               | 23.4                |
| PS1    | $r$  | 22.53     | 360   | 1.3               | 23.0                |
| PTF    | $R$  | 21.55     | 10800 | 2.0               | 23.5                |
| PS1    | $i$  | 22.05     | 360   | 1.0               | 22.7                |

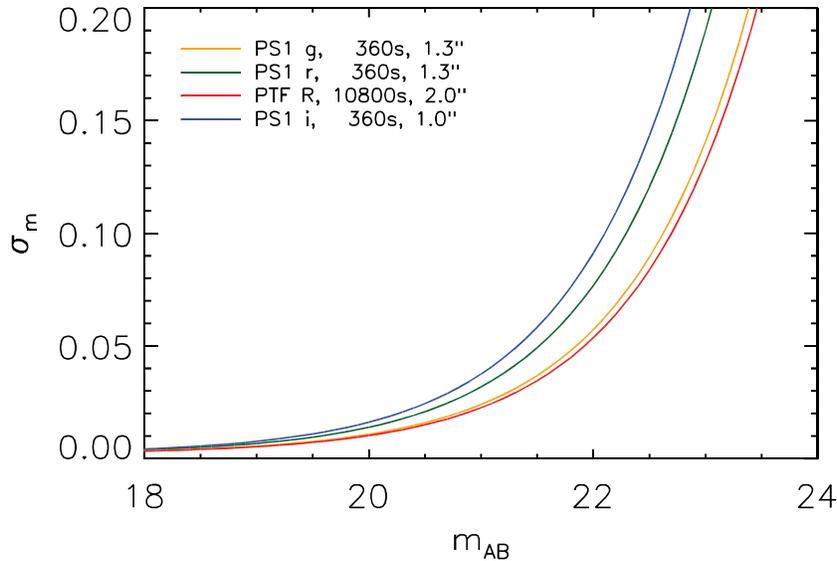

Figure 4.2: Assumed magnitude errors for the Palomar Transient Factory and PanSTARRS $3\pi$ survey.

### 4.2.5 WISE

Ground-based photometry will not always be optimal for selecting all targets of interest. In such cases, we can additionally make use of space-based surveys, which can obtain deep imaging at infrared wavelengths much more efficiently. The experiment of greatest utility for BigBOSS is the *WISE* (Wide-field Infrared Survey Explorer) satellite, which conducted an all-sky survey at wavelengths of 3.4, 4.6, 12 and $22\mu$m [Wright et al., 2010]. In the course of the extended mission, completed in early 2011, 99.99% of the sky was imaged at least 16 times, with an average depth of 32 exposures.

The key *WISE* channel for BigBOSS is $3.4\mu$ m, which will go the deepest for galaxy populations of interest. In this document, we use $100\mu$Jy (18.9 mag AB) as a conservative estimate of the *WISE* limit, only reaching the confusion noise limit of 63 $\mu$Jy (19.4 mag AB) in the deepest regions. In fact, the extended WISE mission achieved much better coverage (approximately twice) than the nominal mission coverage assumed here. The final *WISE* public data release is scheduled to occur in March 2012, providing ample time for optimizing BigBOSS target selection.

### 4.2.6 Other Imaging Surveys

For reference, we list below other wide field imaging surveys which, if available and well documented by the time of the BigBOSS survey, could be used to help define our targets.

### A. U-band Surveys

The South Galactic Cap U-band Sky Survey [SCUSS website, 2010] is a joint project amongst the Chinese Academy of Sciences, its National Astronomical Observatories unit, and Steward Observatory, with observations planned to begin in September 2010. Using a mosaic of four 4K×4K CCDs covering a one square degree field, the survey plans to observe a 3,700 square degree field within the South Galactic cap using the 90-inch (2.3m) Bok telescope at Kitt Peak (belonging to Steward Observatory). Given the expected survey exposure time of 5 minutes per field, the limiting magnitude reached is estimated to be $u \sim 23$ ($5\sigma$).

A complimentary survey to SCUSS could also be performed in the Northern Hemisphere. A collaboration of French and Canadian astronomers have proposed a $u$-band CFHT Survey which would cover a minimum of 5,000 square degrees in the Northern extragalactic sky. A Pilot Survey, which should start in 2011, will observe the ∼800 sq. degree region covered by the CFHT Red Cluster Sequence-2 (RCS2) survey with MegaCam. Upon completion of this Pilot Survey, the CFHT $u$-band survey could then continue to partially cover the SDSS and PS1 footprint. The exposure times are expected to be about 10 minutes per field and the limiting magnitude will reach roughly $u \sim 24$ ($5\sigma$).

## B. DES

The Dark Energy Survey (DES) [Abbott et al., 2005] will use a new wide-field camera for the 4-meter Blanco telescope at CTIO, the Dark Energy Camera (DECam), to probe dark energy via a wide-area photometric survey (as well as a smaller-area survey focused on detecting type Ia supernovae; as the latter will only cover ∼ 40 square degrees, it is of little relevance for BigBOSS). The camera is scheduled to be installed in 2011. In total, DES will cover 5000 square degrees, primarily in the Southern sky, over the course of 525 nights of observations over five years. The survey will deliver $griz$ imaging is expected to deliver $5\sigma$ (point source, $0.9''\|$seeing) limiting magnitudes of $g = 26.1$, $r = 25.6$, $i = 25.8$, $z = 25.4$, considerably deeper than BigBOSS requirements. The DES footprint is expected to have ∼ 500 deg$^2$ overlap with the BigBOSS footprint, primarily in the equatorial SDSS Stripe 82 region.

## C. LSST

The proposed Large Synoptic Survey Telescope (LSST) [Ivezic et al., 2008] will conduct a deep, 6-band ($ugrizy$) photometric survey covering over 20,000 square degrees (primarily in the Southern sky) focused principally on studies of dark energy. By combining a large field of view camera (observing 9.6 square degrees at a time) with a large-aperture (8.4-meter diameter) telescope, LSST is designed to rapidly survey the sky to deep depths. This will enable studies of faint transients and asteroids as well as yielding extremely deep co-added images by combining roughly 1000 observations of each area of sky over 10 years. For the main survey, a single visit to each field will yield $5\sigma$ magnitude limits of $u = 23.9$, $g = 25.0$, $r = 24.7$, $i = 24.0$, $z = 23.3$, and $y = 22.1$; co-added depths will reach 26.3, 27.5, 27.7, 27.0, 26.2, and 24.9, respectively. Each patch of sky will be visited about 1000 times in ten years with a camera that covers 9.6 square degree field of view. The main survey will also extend well into the Northern Hemisphere (Dec< +33 for 2.2 airmass limit) to cover the entire Ecliptic plane. Therefore, we expect that there will be significant overlap between the BigBOSS footprint and LSST, perhaps as large as 6,000 deg$^2$. Once LSST starts survey operations in 2018, inclusion of their photometry from the first year of operations could rapidly improve target selection for BigBOSS in the overlapping area.

## 4.3   Luminous Red Galaxies

### 4.3.1   Target Properties

The largest volume surveys of large-scale structure to date have targeted the highest mass galaxies in the $z < 1$ universe, a population commonly known as luminous red galaxies (LRGs) [Eisenstein et al., 2001]. These objects are luminous and red in the restframe optical bands due to their high stellar mass and lack of ongoing star formation. They are commonly found in massive galaxy clusters today, and therefore they exhibit strong clustering and a relatively high large scale structure bias ([Eisenstein et al., 2005], [Ho et al., 2009], [Kazin et al., 2010]). Because of their strong 4000Å breaks and the correlation between their apparent magnitudes and luminosity distance, LRGs at $z < 0.6$ can be selected efficiently and their redshifts estimated based on SDSS-depth photometry [Padmanabhan et al., 2007], while the strong absorption features around the break allow redshifts to be identified definitively in spectra of modest signal-to-noise. They have therefore formed the cornerstone of the BOSS spectroscopic redshift survey.

Surveying LRGs at higher redshifts is beneficial for studying cosmology as their strong biasing to the underlying dark matter halos leads to a greater power spectrum amplitude, aiding BAO measurements. However, LRGs are increasingly difficult to select at higher redshifts as the 4000Å break passes into the $i$ band (at $z \sim 0.75$) and imaging at longer wavelengths (e.g. $z, J, H$, or $K$-band) is required to estimate LRG redshifts. At sufficiently high redshifts, an additional difficulty is that LRGs will be less common simply due to galaxy evolution. At these early times before their star formation has ceased, they will have bluer restframe SEDs, lower stellar masses, and weaker absorption breaks than local LRGs ([Faber et al., 2007], [Brown et al., 2007]). Only a small subset of the massive red galaxy population was in place as early as $z \sim 2$ ([Daddi et al., 2005], [López-Corredoira, 2010]).

At $z < 0.55$, the BOSS LRG sample selection yields a number density above $3 \times 10^{-4}$ galaxies per $h^{-3}$ Mpc$^3$, sufficient to achieve the BigBOSS science goals. Therefore, at lower redshifts, we will either use existing BOSS spectroscopic samples or apply the BOSS target selection in regions not yet covered. The BOSS selection will yield 119 LRGs per deg$^2$. At higher redshifts, however, we require different selection techniques, taking advantage of near-infrared imaging from space. The remainder of this section will focus on the strategy we will use in this domain.

### 4.3.2   Selection Technique

The spectral energy distributions of cool stars exhibit a local maximum at a wavelength of roughly $1.6\mu m$, corresponding to a local minimum in the opacity of H$^{-\parallel}$ions [John, 1988]. This feature, commonly referred to as the "$1.6\mu m$ bump" dominates the near-infrared spectra of stellar populations with ages above $\sim 10$ Myr, and represents the global peak in $f_\nu$ for populations older than $\sim 500$ Myr [Sawicki, 2002]. Since they possess few young stars, luminous red galaxies at $z \sim 0.5 - 1$ will therefore exhibit relatively large near-infrared to optical flux ratios at wavelengths of $\sim 2 - 4\mu m$.

The lowest-wavelength channel in *WISE*, centered at $3.4\mu m$, is nearly optimal for selecting these objects as it overlaps the bump at redshift $z \sim 1$. The infrared-to-optical flux ratio of LRGs rises monotonically with redshift as $z$ approaches 1, then will decline beyond $z \sim 1.1$. As a consequence of both the increased rarity of LRGs and the greater luminosity distance, LRGs at $z > 1$ are uncommon at the magnitudes BigBOSS will survey. A simple cut in $r$ - [3.4$\mu$m] color should therefore select LRGs effectively while adding in information from more optical bands can help in rejecting non-LRGs. *WISE* data is particularly well-suited for this application, as its survey depth was designed specifically to be able to detect $L_{*\|}$red-sequence galaxies to $z = 1$; LRGs are generally significantly brighter than this limit.

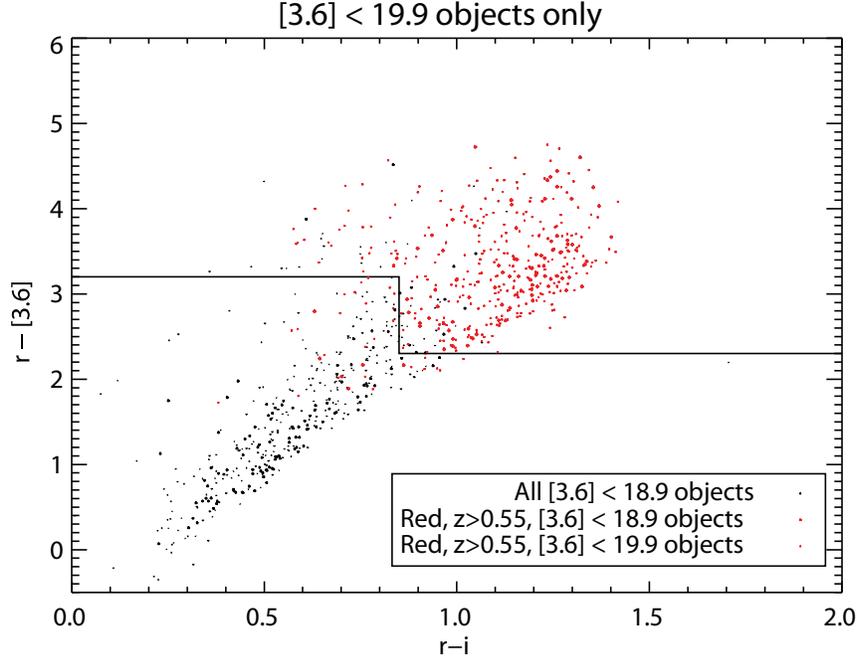

Figure 4.3: An optical/near-infrared color-color diagram for galaxies observed by the CFHT Legacy Survey, *Spitzer* IRAC, and the DEEP2 Galaxy Redshift Survey. In this and below figures, $r$ indicates CFHT LS $r$-band magnitude, $i$ indicates CFHT LS $i$, and [3.6] indicates IRAC 3.6$\mu$m AB magnitude. Galaxies with LRG-like SEDs at $z > 0.55$ are indicated by red points; those with 3.6$\mu$m magnitudes brighter than 18.9 (a conservative estimate of the *WISE* 5-$\sigma$ detection depth in the [3.4$\mu$m] band) are indicated by larger symbols.

To test selection techniques, we have employed publicly-released data from the AEGIS survey [Davis et al., 2007], which incorporates pan-chromatic imaging and spectroscopy from the DEEP2 Galaxy Redshift Survey [Davis et al., 2003]. In particular, we use optical catalogs derived from CFHT Legacy Survey data [Gwyn, 2008], NIR imaging catalogs from *Spitzer* IRAC [Barmby et al., 2008], and redshifts and restframe colors from DEEP2. All magnitudes used are on the AB system. In our tests, we use IRAC 3.6$\mu$m magnitudes as a proxy for *WISE* 3.4$\mu$m photometry and hereafter refer to the 3.6$\mu$m band; actual BigBOSS target selection will be optimized using *WISE* itself. At $z < 1.25$, 3.6$\mu$m lies on the long-wavelength side of the bump, so the measured IRAC flux should be lower than 3.4$\mu$m flux for a given galaxy; as a consequence, estimates from [3.6$\mu$m] $< 18.9$ or $< 19.4$ sample sizes from this analysis will be conservative. As seen in Figures 4.3, 4.4, and 4.5, galaxies with red restframe colors (restframe $U - B > 0.9$) at redshift $z > 0.55$ are almost entirely confined to a limited region in an optical/near-infrared color-color plot. A strawman LRG selection criterion is shown by the solid lines in this figure.

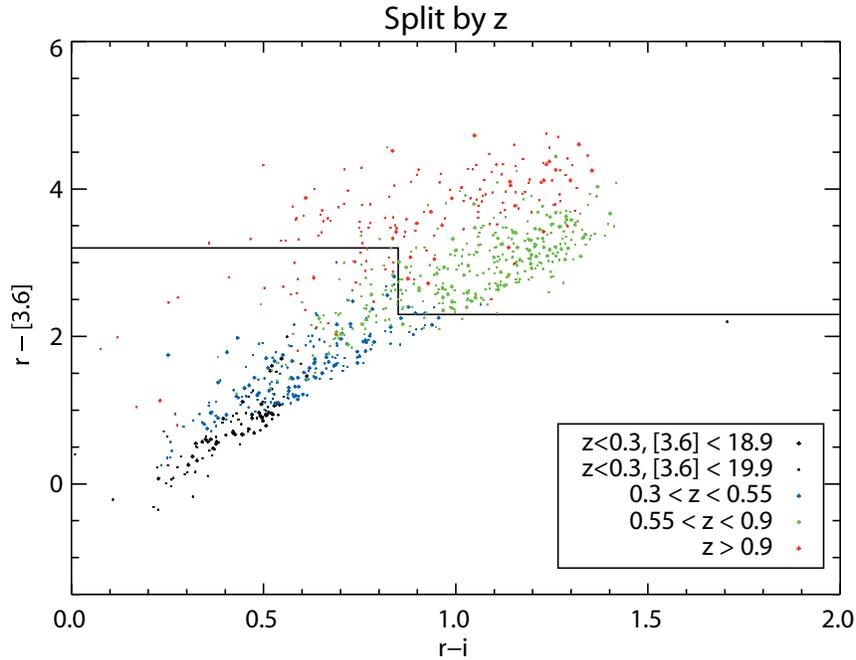

Figure 4.4: As Figure 4.3, with objects color-coded according to their redshift and symbol sizes determined by $[3.6\mu m]$ magnitude.

Since the $1.6\mu m$ bump is present in all but the youngest stellar populations, a pure cut in infrared-to-optical ratio (or equivalently $r - [3.6\mu m]$ color) is effective at selecting objects in the target redshift range, but roughly 15% of the selected objects will be bluer than LRGs. By making the selection cut dependent on an optical color (both $g - r$ and $r - i$ have been tested and prove to be equally effective), these blue interlopers can be partially rejected; even a crudely optimized box (as shown in Figs. 4.3 – 4.5) improves the LRG redshift window efficiency to 90%.

### 4.3.3 Sample Properties

There are 420 objects per square degree within the depicted selection box with $[3.6\mu m] <$ 18.9 (a conservative limit), or 1120 with $[3.6\mu m] < 19.4$; we adopt these as two fiducial scenarios for BigBOSS LRG samples. As these target densities are based on a 0.4 square degree region within the Extended Groth Strip, these source densities are subject to sample (or "cosmic") variance as well as Poisson uncertainty; they are uncertain at the 10-15% level as a result.

We use DEEP2 redshifts to estimate the redshift distributions we will obtain from our $z > 0.55$ LRG target selection, though given the limited area covered by DEEP2, CFHT LS, and IRAC, both sample/cosmic variance and Poisson variance are large within small $0.1\Delta z$ bins. We consider two scenarios: a shallow sample selected to have $[3.6\mu m] < 18.9$ (a conservative estimate of the *WISE* survey depth) and $i < 21.5$; and a deeper sample with $[3.6\mu m] < 19.4$ (a more optimistic estimate) and $i < 22$. These samples yield 380 and 670 targets per square degree, respectively.

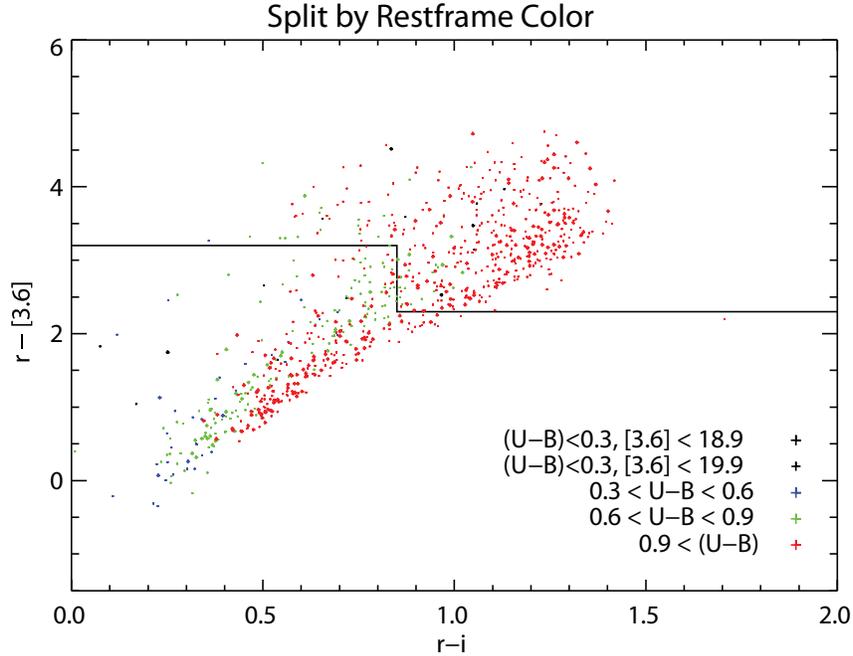

Figure 4.5: As Figure 4.3, with objects color-coded according to their restframe $U - B$ color. Objects with $U - B > 0.9$ generally have spectral energy distributions similar to LRGs.

In Figure 4.6, we plot the redshift distributions of the resulting samples, along with the overall redshift distribution of all galaxies in our LRG selection box and the number density goal of $3 \times 10^{-4}$ objects per $h^{-3}$ Mpc$^3$. Both of these samples are larger than the LRG population assumed in Section 2.2. However, as seen in the figure, we have more than enough targets at $z < 0.8$ and will downsample at those redshifts accordingly. The apparent magnitude of LRGs is strongly correlated with their redshift as they are on the exponential tail of the luminosity function, allowing us to sculpt the LRG redshift distribution efficiently. Even using conservative assumptions about the depth of *WISE* photometry, we find that we can select a sufficiently large sample of LRGs to meet BigBOSS survey goals.

Our spectral feature efficiency for $z > 0.55$ LRGs will primarily be a function of optical magnitude, a fact that will strongly affect the signal-to-noise we achieve in the spectrum of a given galaxy and determine whether or not we can detect absorption lines. We therefore will only target *WISE*-selected LRGs down to some $r$ or $i$-band magnitude limit, which will correspond to a limit in spectral signal-to-noise.

Figure 4.7 shows the effect changing this limiting magnitude will have on the surface density of selected targets, assuming either a $[3.6\mu m] < 18.9$ or $[3.6\mu m] < 19.4$ sample. We find that a limiting magnitude of $r \sim 22.5$ or $i \sim 21.5$ should produce a volume density sufficient for the BigBOSS LRG sample. Given the photometric survey magnitude limits of PTF and PS1, we expect that the optical spectral flux will be highly accurate at these limits ($\sim 0.05\%$ magnitude error), and therefore the overall target selection efficiency will be dominated by the LRG redshift window efficiency of 90%. Based on our experience with BOSS, we expect to obtain redshifts for 95% of all LRGs down to our chosen magnitude limit (which is designed to achieve BOSS-like levels of signal-to-noise).

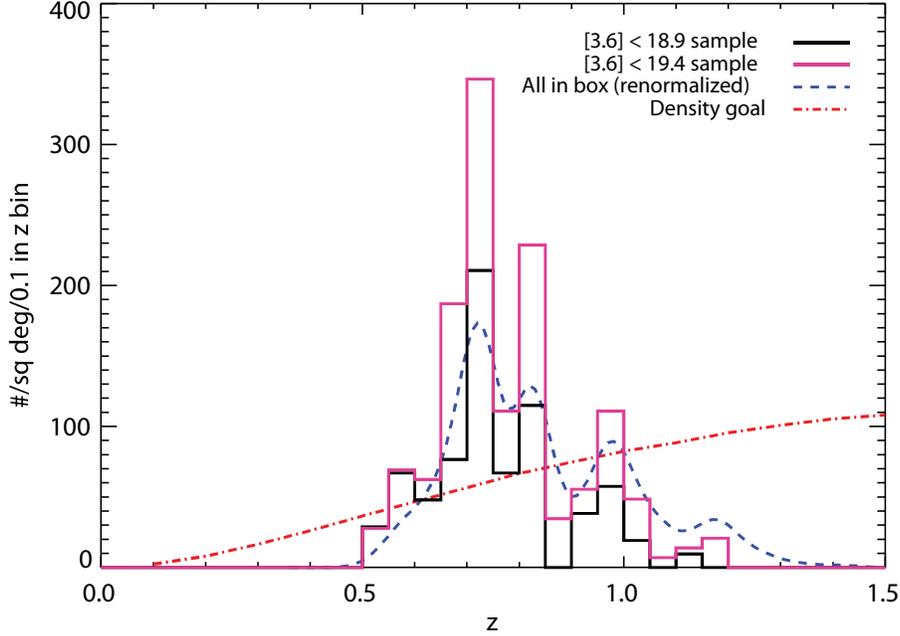

Figure 4.6: Redshift distributions for $z > 0.55$ LRG samples, estimated using data from DEEP2, CFHT LS, and *Spitzer* IRAC. Due to the small area covered and LRG sample sizes, both sample/cosmic variance and Poisson variance are large within small $0.1\Delta z$ bins. Distributions for two possible scenarios are plotted: a shallow sample selected to have $[3.6\mu m]$ magnitude $< 18.9$ (a conservative estimate of the *WISE* depth) and $i < 21.5$ (solid black histogram); and a deeper sample with $[3.6\mu m] < 19.4$ (a more optimistic depth estimate) and $i < 22$ (dot-dashed purple histogram). We also plot the redshift distribution of all galaxies in the LRG selection box (blue dashed line), renormalized to match the average number of galaxies per square degree from the other two samples. The dot-dashed red curve corresponds to the LRG number density goal of $3 \times 10^{-4}$ objects per $h^{-3}$ Mpc$^3$; this goal is easily achievable to $z = 0.8$, and we are within 30% of the goal to $z = 1$.

In Table 2.2, the total LRG target selection efficiency is the product of the LRG spectral feature efficiency ($\sim 100\%$) and the fraction of selected objects that lie in our detection window (90%). The final rate of redshift completeness – the product of fiber completeness, target selection efficiency, and redshift measurement efficiency – for the BigBOSS LRG target sample presented here is estimated to be $\sim 68\%$; this is the fraction of potential LRG targets which will be actually targeted by a fiber, will turn out to be in the desired redshift range, and will yield a redshift. This is significantly higher than for ELG samples, largely because redshift success for a given LRG may be predicted from its magnitude much more easily.

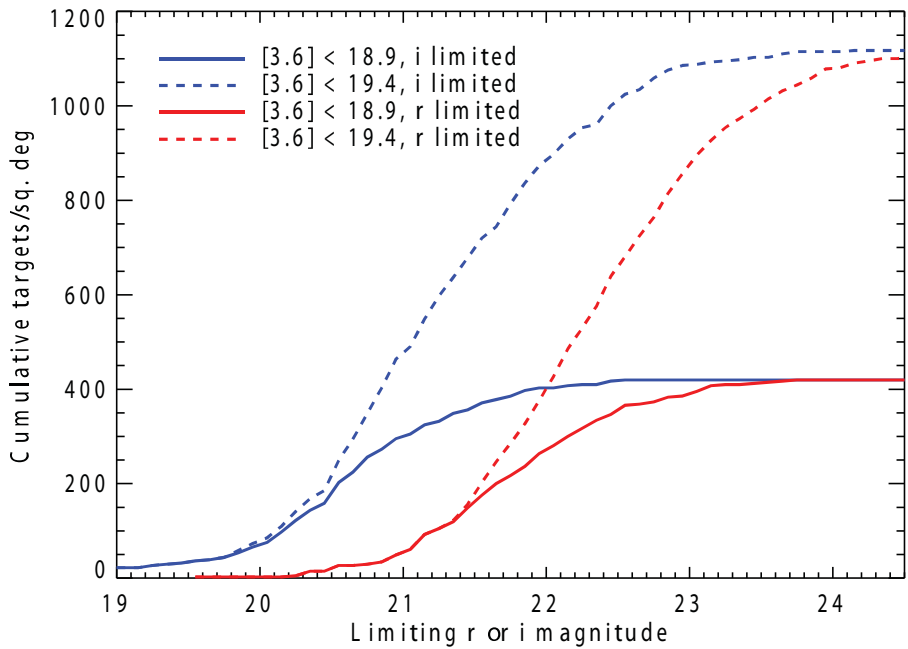

Figure 4.7: Surface densities of $z > 0.55$ LRGs as a function of limiting $r$ and $i$-band magnitude. We consider samples to two possible *WISE* depths, a conservative depth of $[3.6\mu m] < 18.9$ and the extended mission depth $[3.6\mu m] < 19.4$. Target LRG sample sizes are readily achievable so long as satisfactory signal-to-noise is obtained from spectroscopy down to $r \sim 22.5$ or $i \sim 21.5$.

## 4.4   Emission Line Galaxies

### 4.4.1   Target Properties

The largest sample of galaxies that will be selected by BigBOSS are emission line galaxies, typically composed of the brightest late-type spirals. The composite rest-frame colors of these galaxies are typically bluer than those of evolved galaxies such as LRGs due to their active star formation in the recent past; however, as they can exhibit a wide range of internal dust properties, their colors can be significantly dependent on inclination effects. In the local universe, ELGs of a constant emission line luminosity threshold are much less numerous than at high redshifts ($z > 1$). This predominantly reflects the fact that the overall star formation rate of the Universe was $\sim 10\times$ higher at that time [Hopkins & Beacom, 2006]. The correlation of emission lines to star-formation is established well enough to measure the star-formation rate (SFR) to $z \sim 2$, around the peak of the cosmic SFR [Kennicutt, 1998; Moustakas, Kennicut & Tremonti, 2006; Hopkins & Beacom, 2006].

In regions where star formation has recently occurred, short-lived, blue massive stars will provide large numbers of energetic photons into the local interstellar medium, resulting in ionized HII regions. As ions and electrons in these regions recombine, a variety of emission lines will result; the most luminous lines in the optical are members of the Hydrogen Balmer series or are emitted by oxygen ions. The total rate of ionizations and recombinations from a galaxy will be proportional to the total number of massive stars; hence, emission line fluxes provide a useful diagnostic of a galaxy's star formation rate.

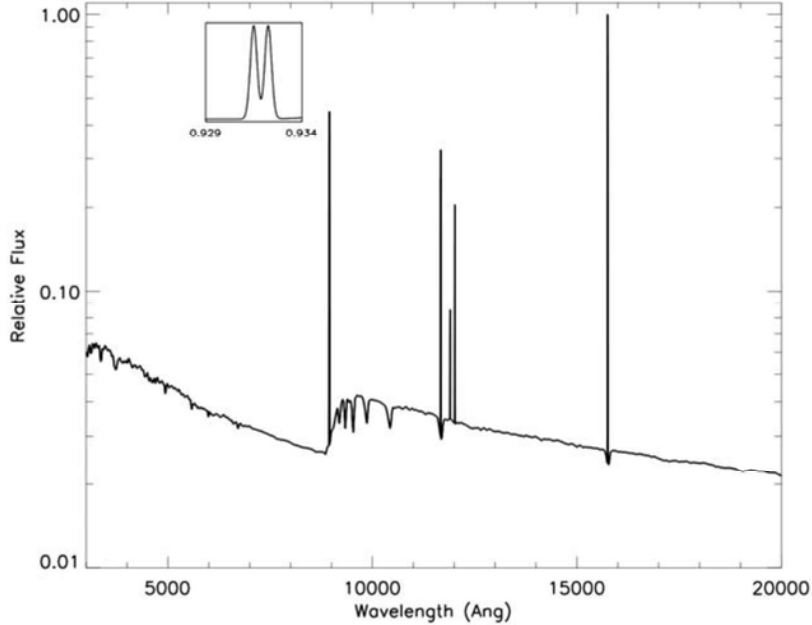

Figure 4.8: A template star-forming, emission line galaxy spectrum at $z = 1.4$ sampled at a constant 0.76Å per pixel interval, similar to the resolution provided by the BigBOSS visible and red spectrographs. The inset figure shows that the [OII] doublet is resolved at this sampling frequency and is split almost evenly between the line components, as is generally observed.

Figure 4.8 shows an example synthetic $z = 1.4$ emission line galaxy spectrum constructed from a star-forming template SED [Bruzual & Charlot, 2003] with emission line fluxes calibrated to match zCOSMOS observations at lower redshifts (see Section 4.4.2 for details). The strongest of the emission lines are typically the H$\alpha$ line at 6563Å rest-frame and the forbidden [OII] doublet transitions at 3727Å. Additional strong lines include H$\beta$ $\lambda4861$ and the [OIII] $\lambda4959 + 5007$ doublet. Of all the emission lines, the [OII] doublet will be most useful for probing the redshifts required by BigBOSS ($z < 2$) without requiring observations beyond 11000Åwhere the near-IR sky background increases rapidly. An additional benefit to [OII] is that it is a doublet closely spaced in wavelength (220 km/s FWHM). Each component line contributes roughly one half of the total line flux, since electron densities typically range from 100-1,000 e$^{-\parallel}$cm$^{-3}$ for star-forming galaxies [Pradhan et al., 2006; Kewley & Ellison, 2008].

The doublet nature of [OII] 3727Å emission provides a unique signature for line identification in observations of sufficiently high resolution. If both components are robustly identified, a secure redshift results, in contrast to single-line redshifts which can correspond to a number of possibilities in the absence of other detected features. The DEEP2 survey [Davis et al., 2003] recognized the unique features of the [OII] emission line and used it (as well as the 4000Å break prominent in older stellar populations) to conduct a redshift survey focusing on the regime $0.7 < z < 1.4$. To date, the survey has resulted in 33,000 confirmed redshifts, most of them obtained via the [OII] doublet, measured in four different survey fields totaling $\sim$3 deg$^2$. Experience from DEEP2 shows that the resolution required to nominally split the [OII] doublet is sufficient to produce two recognizably separate line features for the bulk of emission line galaxies, providing high confidence in the line identification [Weiner et al., 2005]. Only a small fraction of galaxies contain sufficiently high rotational velocities to blend the doublet, and those massive galaxies typically exhibit continuum absorption features from the Balmer series or Ca H & K. Further, [OIII] and H$\beta$ emission lines will be detectable at wavelengths below 11000Å to $z \sim 1$, providing additional certainty to redshifts when the lines have sufficient flux to be detected. The success of the DEEP2 survey in identifying and measuring emission-line redshifts serves an excellent test of strategies for BigBOSS.

### 4.4.2 [OII] Luminosity Function

With large [OII] datasets as DEEP2, it is possible to measure the number density of objects as a function of both [OII] luminosity and redshift. Since surveys of line luminosities are generally limited in completeness at either the faintest or brightest luminosities due to choice of survey characteristics, it is important to include multiple samples that cover a wide range of luminosities. Figure 4.9 shows a compilation of the [OII] luminosity functions produced from multiple emission line datasets at a mean redshift of $z \sim 1.2$, including the DEEP2 Galaxy Redshift Survey [Zhu, Moustakas & Blanton, 2009] and narrow-band filter observations of the Subaru Deep Field and the COSMOS field [Ly et al., 2007; Takahashi et al., 2007].

We find that the composite [OII] luminosity function is best represented by an Abell function (rather than a Schecter function) to match the power law behavior measured by DEEP2 at the bright end of the luminosity function. We parameterize the luminosity function according to

$$\frac{dN}{dLog(10)} = \frac{ln(10)}{ln(\sqrt{2})} \frac{N_b^2 L_b^2}{L_b^2 + L^2},$$

(4.2)

where $N_b$ and $L_b$ characterize the luminosity function behavior as a function of redshift with

$$Log(N_b) = -3.5 + 2.0(x - x^2),$$

(4.3)

$$Log(L_b) = 40.95 + 3.0(x - x^2),$$

(4.4)

$$x = Log(1 + z).$$

(4.5)

The redshift dependence of this LF model is derived from observations in multiple redshift bins available from SDF and DEEP2. The result of the model function is displayed in Figure 4.9 for $z \sim 1.2$. Another interesting feature is that for a fixed space density, the [OII] luminosity is greater at higher redshifts; this is a result of the $\sim 10\times$ larger mean star formation rates in blue galaxies of all types at $z \sim 1$ compared to today. To project line fluxes for redshifts at $z > 1.4$, we adopt a conservative scenario in which the star-formation rate remains constant from $1.4 < z < 2$ (roughly 1Gyr of cosmic time) and no more evolution occurs in the [OII] line luminosity (J. Moustakas, priv. comm.).

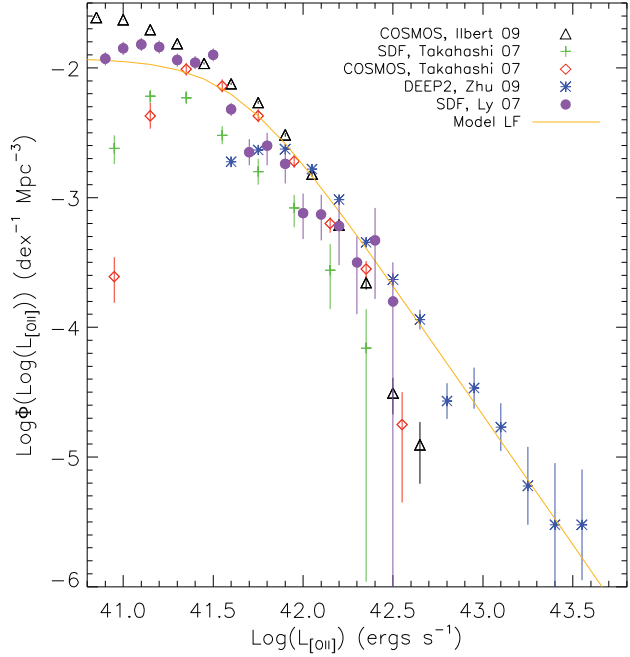

Figure 4.9: The [OII] luminosity function measured from photometric and spectroscopic surveys near $z \sim 1.2$. The luminosity function from DEEP2 spectroscopic measurements behaves as a power law on the bright end and shows good agreement with previous work in the Subaru Deep Field (SDF) and COSMOS field [Ly et al., 2007; Takahashi et al., 2007; Ilbert et al., 2009]. Note that each survey has incompleteness at both the bright and faint ends but the model luminosity function tracks the best sampled data in a given regime.

The black triangles in Figure 4.9 correspond to a catalog of [OII] line luminosities available from the COSMOS survey [Ilbert et al., 2009]. This catalog is composed of photometric redshifts (measured over 30 photometric bands), best-fit galaxy templates, and synthetic magnitudes generated from the *Le Phare* photometric redshift software [Arnouts et al., 2002; Ilbert et al., 2006]. The COSMOS templates also incorporate emission lines calibrated based on [OII] fluxes from VVDS spectroscopic measurements [Le Févre et al., 2005]. For $z > 1.4$, the [OII] fluxes are calibrated by the M(UV)-[OII] relation [Kennicutt, 1998]. This calibration of the [OII] flux with redshift is accurate to 0.2 dex, and this scatter is maintained in the catalog for those objects where the calibration is implemented.

As a check of the COSMOS [OII] flux calibration, we compare the LF measured from the catalog to other [OII] LFs in Figure 4.9. We find that the COSMOS luminosity function is in good agreement with our model Abell function. The COSMOS LF includes more objects than DEEP2 at low luminosities largely because it is based on a photometric-redshift sample (with [OII] emission assigned according to the heuristics described above), and hence includes objects fainter than the DEEP2 limit of $R = 24.1$. However, the DEEP2 LF, which has been based upon spectroscopic redshifts, appears to better track the observed LF from the deepest narrow-band imaging (SDF) at higher line luminosities. It appears that the methods used to assign [OII] fluxes to objects with photometric redshifts from COSMOS may break down at the highest luminosities, possibly due to the VVDS calibrators consisting predominantly of redder galaxies lying at $z < 1$ and the applied $i = 22.5$ magnitude limit.

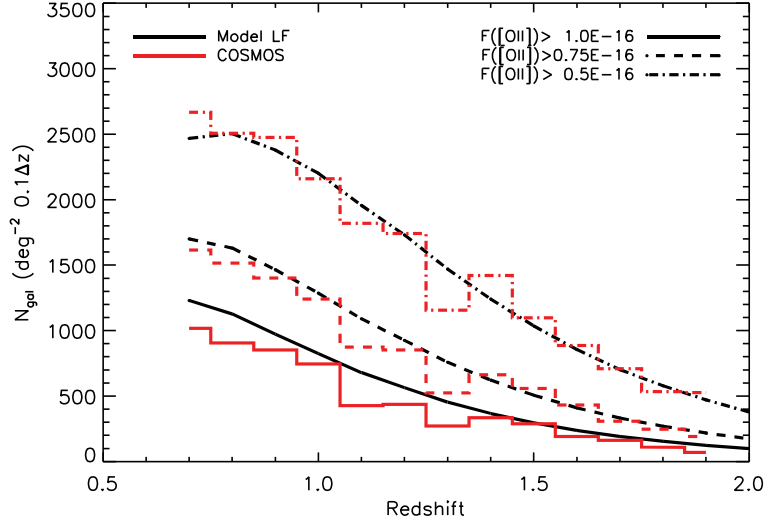

Figure 4.10: Comparison of redshift distributions at three limiting [OII] line fluxes as predicted from COSMOS photometric redshift and restframe spectrum fits calibrated with VVDS [OII] line flux data [Ilbert et al., 2009] and from the model [OII] luminosity function depicted in Fig. 4.9. The agreement is extremely good save at the highest fluxes, for which we would expect the COSMOS estimates to be low based on the previous figure.

By integrating the model luminosity function above a given flux limit, we can construct redshift distributions representative of the ELGs available for targeting by BigBOSS (Figure 4.10). As an additional check on the COSMOS catalog and our model LF, we also plot the redshift distribution resulting from applying the same cuts to the COSMOS sample. We find that the two methods predict similar redshift distributions over a range of [OII] flux limits near the minimum detectable line flux for BigBOSS. This result is reasonable since the bulk of the sample comes from objects in the range where the LFs are in good agreement, having $L_{[OII]} \sim 10^{41.5} - 10^{42.5}$ (ergs s$^{-1}$). The agreement between the model luminosity function and COSMOS predictions increases our confidence that the number density of bright [OII] emitters is well measured up to $z < 1.4$ and conservatively estimated for $1.4 < z < 2$.

**Linear Bias**  The linear clustering bias of bright emission line galaxies relative to their dark matter halos is a matter of current study, but several sources have made relevant measurements. DEEP2 looked at the bias as a function of restframe color at a median redshift of z=0.9 ([Coil et al., 2008], hereafter C08). They found that the blue galaxies, those with the strongest star-formation and [OII] emission line measurement, had an absolute linear bias of $b = 1.28 \pm 0.04$ over a scale length of $1 - 10h^{-1}$ Mpc at $z = 0.9$. C08 also found that this clustering strength is consistent with similar ELG bias measurements from other samples [Marinoni et al., 2005] and that the absolute linear bias at $z \sim 1$ is greater than that in the nearby universe.

Other studies have looked at the clustering as a function of [OII] luminosity to investigate whether there is any correlation between halo mass and line brightness. Using Subaru X-ray Deep Field and semi-analytic models of the relationship between baryonic gas mass and dark matter halos, Sumiyoshi et. al. (2009) estimated the linear bias for various UV-calibrated [OII] flux limits over three redshift bins between $0.5 < z < 1.7$. They found that the bias was largely insensitive to their [OII] flux estimates except for the very brightest objects ($F_{[OII]} > 1 \times 10^{-15}$ ergs s$^{-1}$ cm$^{-2}$) but the overall bias increased with redshift. For our initial BigBOSS projections, we assume that the bias increases with redshift to preserve a constant clustering amplitude; this assumption provides a rough fit to measured galaxy correlation function amplitudes for blue star-forming objects at redshifts from $z \sim 0$ [Zehavi et al., 2010] to $z \sim 3$ [Steidel et al., 2010; Adelberger et al., 2005]. Based on the clustering of $z \sim 1$ samples, we adopt the model $b = 0.76/g(z)$ where $g(z)$ is the cosmological growth function normalized by a factor of $1/(1 + z)$.

### 4.4.3   Selection Technique

Because the vast majority of spectroscopic targets for BigBOSS will be ELGs, the overall survey efficiency will largely depend on the efficient selection of ELG targets from photometric data. Given that DEEP2 efficiently selects ELGs with $z > 0.7$ using broadband optical photometry, we expect that BigBOSS can use similar methods to select objects in a similar redshift range with a high confidence in success. We will therefore first describe the methods applied for DEEP2 and then discuss how they may be adapted for BigBOSS.

Figure 4.11 shows the expected CWW and Kinney-Calzetti tracks [Coleman, Wu & Weedman, 1980; Calzetti et al., 1994] over the redshift range $0 < z < 2$ in CFH12K Mould $BRI$ photometry for a range of galaxy spectral energy distributions (SEDs). As can be seen from this figure, galaxies of all types have $BRI$ colors that rapidly become redder in $R - I$ as the 4000Å break transitions into the $R$-band at $z \sim 0.7$. This effect is strongest for red galaxies and weakest for starbursts. This allows an efficient division between $z < 0.7$ and $z > 0.7$ objects; the dot-dashed line in the figure shows the color selection actually used by DEEP2, which was optimized for completeness at $z > 0.75$ using redshifts of objects in the Extended Groth Strip (where no color preselection is applied). Star-forming galaxies with $z > 0.75$ – roughly equivalent to our emission-line sample – occupy a region in color space below and to the right of the dashed line. The DEEP2 selection had a selection completeness of 97% for galaxies at $z > 0.75$ (i.e., 97% of $z > 0.75$ galaxies pass the color cut) and a redshift window efficiency (i.e., fraction of the selected sample which has $z > 0.75$) of 85%.

Due to the depth limits of available photometric surveys (see Section 4.2) and differing survey goals, BigBOSS will likely use a shallower imaging dataset than DEEP2 with a smaller color selection box to maximize the probability of obtaining [OII] detections. To simulate the expected photometry, we have generated synthetic magnitudes from the COS-MOS fit galaxy templates described in 4.4.2 in both the Sloan $gi$ bands (PS1) and the Mould $R$ band (PTF). Here we choose PTF over PS1 because we expect PTF to have deeper co-added photometry than PS1 in $R$. We also add random magnitude errors onto the synthetic magnitudes based upon the models described

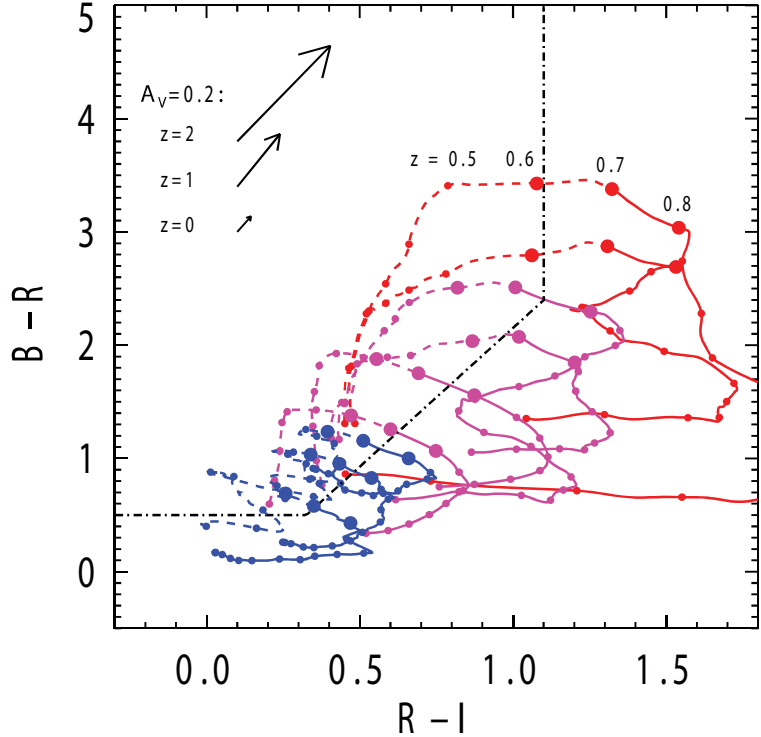

Figure 4.11: BRI color-magnitude diagram illustrating the target selection techniques applied for the DEEP2 Galaxy Redshift Survey utilizing CFH12K photometry. The colored tracks are the trajectories of objects with CWW and Kinney-Calzetti [Coleman, Wu & Weedman, 1980; Calzetti et al., 1994] template spectra through this color space over the redshift range $0 < z < 2$. Red lines correspond to early-type elliptical galaxies, magenta lines to intermediate spiral galaxies, and blue to late-type starbursting galaxies; dots indicate intervals of 0.1 in $z$. The black line (dot-dashed) shows the DEEP2 color selection applied, which has been optimized using observed redshifts in the Extended Groth Strip (where no color cut is applied) to select $z > 0.75$ objects below and to the right of this line.

in Section 4.2.4.

Figure 4.12 shows the location of objects in the $gRi$ color plane using the COSMOS synthetic photometry and the expected PTF and PS1 photometry to $R_{AB} < 23.4$. The figure also color codes galaxies which have [OII] flux above $9 \times 10^{-17}$ ergs s$^{-1}$ cm$^{-2}$ in three redshift bins: $0.7 < z < 1.2$, $1.2 < z < 1.6$, and $1.6 < z < 2.0$ (refer to Appendix A for a calculation of the expected BigBOSS [OII] line flux limit). As was also seen for the DEEP2 BRI selection, low-redshift star-forming galaxies have bluer $(R-i)$ colors than $z \sim 1$ objects, but the SEDs migrate towards bluer colors as redshift increases. We also show an illustrative color selection box which we will use to predict BigBOSS sample properties in the next section. This selection is not unique; one can choose a variety of other selections that will generally modify the target densities at $z \sim 1$ as opposed to higher or lower redshifts.

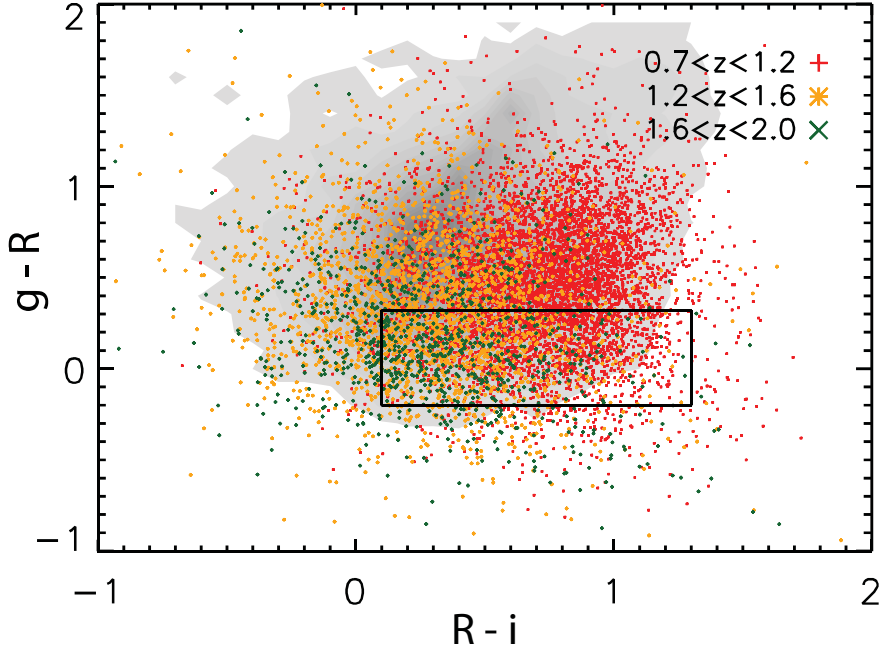

Figure 4.12: Emission line galaxy color selection using synthetic photometry for the 1.3 deg² COSMOS sample described in §4.4.2, applying PTF $R$ and PS1 $gi$ magnitude errors estimated as described in §4.2. The gray contours indicate all galaxies with $R < 23.4$ and the data points indicate those galaxies which have an [OII] flux greater than $9 \times 10^{-17}$ ergs s$^{-1}$ cm$^{-2}$. The black box indicates a simple color cut that would select the brightest [OII] emission line galaxies with $z > 0.7$ with high efficiency.

### 4.4.4  Sample Properties

Figure 4.13 shows the redshift distribution (based on COSMOS photometric redshifts) for objects located in the simple selection box shown in Figure 4.12. The selection produces a distribution of ELGs with a redshift range of $z > 0.7$ where the number density of targets exceeds the BigBOSS requirements (Section 4.1) to a redshift of $z = 1.7$. Our initial optimization studies have shown that the FoM is optimized best when the greatest volume of the Universe can be sampled with the greatest efficiency and number density in the allotted survey time, in line with previously FoM studies [Parkinson et. al., 2010]. The particular shape of the redshift distribution is a second order effect in optimizing the dark energy FoM.

Based on the redshift distribution shown in Figure 4.13, we estimate a redshift window efficiency of 70% for selecting [OII] ELGs in the BigBOSS target range of $0.7 < z < 1.7$. A full 92% of the objects reside at $z < 1.7$, where BigBOSS will have sensitivity to [OII] and other prominent emission lines (e.g., H$\alpha$, H$\beta$, and [OIII]). The redshift window efficiency is therefore significantly affected by the high redshift tail of the distribution and may be improved by reducing the magnitude limit of the selection (at the expense of overall target density), by including additional color information, or by pushing the selection box redder in $R - i$.

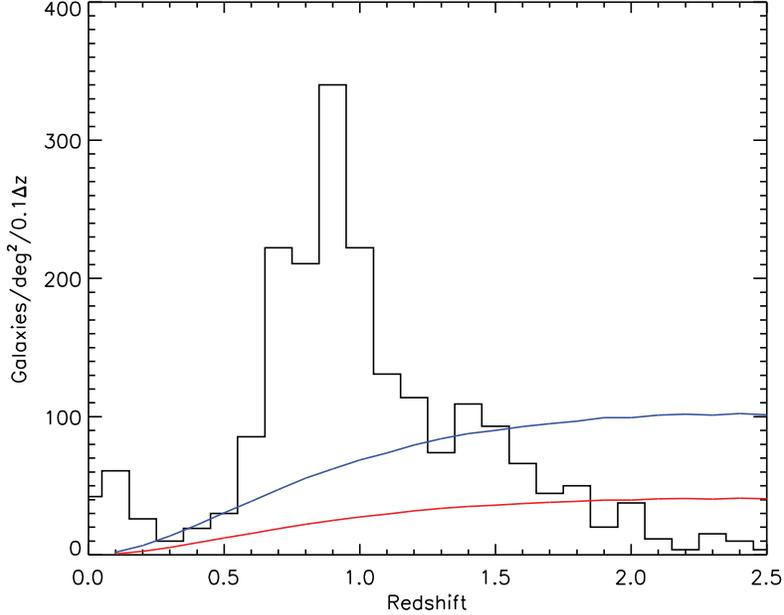

Figure 4.13: The predicted redshift distribution of objects found in the ELG color selection box in Figure 4.12, based on COSMOS photometric redshifts and synthetic photometry. The red line represents a constant volume number density corresponding to the minimum target goal of $n = 1 \times 10^{-4} \ h^{-3}$ Mpc$^3$ and the blue line corresponds to $n = 2.5 \times 10^{-4} \ h^{-3}$ Mpc$^3$. Our target selection meets the minimum density goal at $z < 1.7$ and the higher volume density at $z < 1.5$.

In Figure 4.14, we plot the total surface density of ELGs that have $F_{[OII]} > 9 \times 10^{-17}$ ergs s$^{-1}$ cm$^{-2}$ within the $gRi$ color selection box as a function of $R$-band magnitude limit. In order to use the focal plane fibers with $> 80\%$ efficiency, we project that the ELG target density should be $\sim 20\%$ higher than the fiber density (cf. §6), or about 2300 ELGs deg$^{-2}$. We see that the ELG selection provides this target density for an $R$-band limit of $R < 23.4$. Figure 4.14 also shows the color-selected fraction of objects that will have [OII] line fluxes above various limits as a function of the limiting $R$ magnitude.

We find that the fraction of objects lying in the selection box that have $F_{[OII]} > 9 \times 10^{-17}$ ergs s$^{-1}$ cm$^{-2}$ is roughly 70%; this will be our expected spectral feature efficiency for ELGs. It should be noted that objects with [OII] fluxes below this limit may well yield redshift measurements, but they will have a lower signal-to-noise than the require S/N=8 per line. Higher values of spectral feature efficiency could be obtained by lowering the magnitude limit of the selection, resulting in a loss in the total number density of selected targets unless the color selection box is revised. However, even if an object intrinsically possesses a line flux of $F_{[OII]} > 9 \times 10^{-17}$ ergs s$^{-1}$ cm$^{-2}$, we will not always successfully obtain a redshift for it from a BigBOSS observation due to bright sky emission lines. Based on tests with the spectral simulator (Appendix A), we estimate that we will fail to obtain redshifts 10% of the time, yielding a redshift measurement efficiency of 90%.

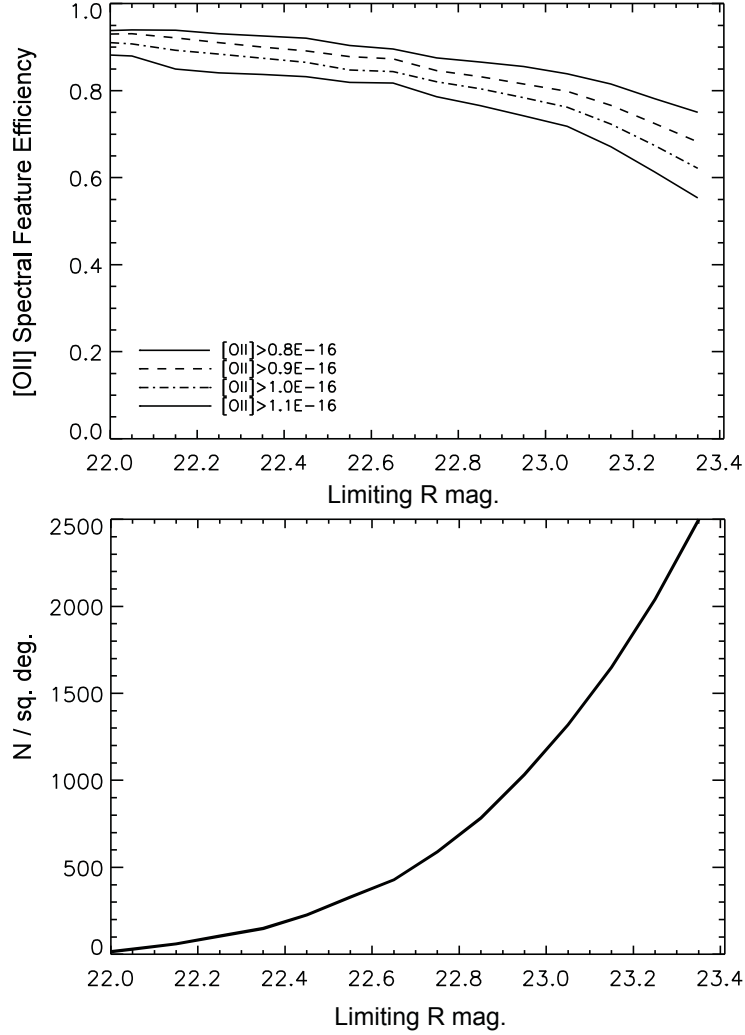

Figure 4.14: *top:* The fraction of color-selected objects in Figure 4.12 having [OII] flux above a certain value, as a function of limiting $R$ magnitude. The overall spectral feature efficiency for our expected $F_{[OII]} > 9 \times 10^{-17}$ ergs s$^{-1}$ cm$^{-2}$ flux limit (see Appendix A), assuming an $R$-band limit of $R < 23.4$, is therefore $\sim 70\%$. *bottom:* Cumulative ELG target density as a function of $R$-band magnitude limit after applying the [OII] flux limit of $9 \times 10^{-17}$ ergs s$^{-1}$ cm$^{-2}$. The BigBOSS survey requires $\sim 2300$ total ELG targets per square degree to efficiently use the focal plane fibers.

In Table 2.2, we record the total ELG target selection efficiency as the product of the ELG spectral feature efficiency (70%) and the fraction of selected objects that lie in our detection window of $z < 1.7$ (92%). The total ELG target selection efficiency is therefore 65%. The final rate of redshift completeness – the product of fiber completeness, target selection efficiency, and redshift measurement efficiency – for the BigBOSS ELG target sample presented here is estimated to be $\sim 47\%$; this is the fraction of potential ELG targets which will be actually targeted by a fiber, will turn out to be in the desired redshift range, and will yield a redshift. Although this may seem low at first glance, it significantly exceeds the redshift completeness of the DEEP2 survey (as DEEP2 had a redshift window efficiency of 82%, was able to observe 70% of potential targets, and yielded redshifts for 70% of objects observed, yielding 40%) or of zCOSMOS-bright (which has a redshift window efficiency of $\sim$ 100%, obtains spectra of $\sim$67% of potential targets, and gets secure redshifts for 61% of objects with spectra, yielding 41%). The BigBOSS rate of redshift completeness is therefore not unusual for a color-selected survey within the $z \sim 1$ universe.

## 4.5  Quasi-Stellar Objects

### 4.5.1  Target Properties

Quasi-stellar objects (QSOs) are extremely luminous extragalactic sources associated with active galactic nuclei (AGN). QSOs are fueled by gravitational accreation onto supermassive black holes (SMBHs) at the centers of these galaxies; almost all continuum and broad-line emission originates within a few parsecs of the SMBH. Even in the nearest QSOs, the emitting regions are too small to be resolved, so QSOs will appear as point sources in images, in contrast to the extended, more easily resolved emission from a galaxy's stars and gas. QSOs commonly exhibit hard spectra in the X-ray wavelength regime, bright Ly-$\alpha$ emission in the rest-frame UV, and a power-law spectrum behaving as $F_\nu \propto \nu^{-\alpha}$ in the mid-infrared bands [Stern et al., 2005]. The specific physical processes that trigger high-luminosity QSO-mode emission in galactic nuclei is a ongoing topic of study, though the basic requirements (a high-mass black hole to have nonnegligible Eddington luminosity, with an ample fuel supply to reach a high Eddington ratio) are well-understood; a variety of scenarios can duplicate this [Hopkins et al., 2006; Croton, 2009] and match observed properties of the QSO population [Croom et al., 2005; Ross et al., 2009]. Similar to the very blue star-forming galaxy population, the number density of quasars was much greater in the distant past, peaking at $z \approx 2-2.5$ [Richards et al., 2009].

Although broad-line (Type 1/unobscured) quasar spectra exhibit characteristic features that separate them from typical star-forming galaxy SEDs, their point-like morphologies, relatively bright apparent magnitudes, and exponential frequency dependence give them *photometric* characteristics that mimic faint blue stars in optical wavelengths. Figure 4.15 shows how QSOs overlap the stellar locus for several Sloan *ugriz* color-color planes. The greatest separation from the stellar locus comes from *ugr* colors where the "UV excess" in $u - g$ produces bluer colors than that of most stars. However, the UV excess is less strong for $z > 2$ quasars, for which the Ly-$\alpha$ forest dampens the hard QSO spectrum. While sophisticated neural-network algorithms have been developed to utilize all available SDSS color information to produce quasar photometric redshifts [Yeche et al., 2010], the simpler photometric selection used by BOSS to target Ly-$\alpha$ QSOs from $2.2 < z < 3.5$ already reaches a 50% targeting efficiency. The BOSS selection produces 20 measured Ly-$\alpha$ QSOs deg$^{-2}$ down to the SDSS photometric limit of $g < 22.1$.

To increase the number of Ly-$\alpha$ forest sightlines over those measured in BOSS, the BigBOSS target selection goal is to deliver a highly-complete Ly-$\alpha$ QSO sample to a fainter magnitude limit. This selection goal presents multiple photometric targeting challenges. First, there is a larger uncertainty in the form of the faint end of the underlying QSO luminosity function. Figure 4.16 shows the integrated surface density for $2.2 < z < 3.5$ QSOs from the Jiang et al. (2006, hereafter J06) luminosity function. At $g = 23$, it is predicted that a *complete* sample would give 45 QSOs / deg$^2$ in this redshift range, whereas the luminosity function used to make predictions for LSST [Hopkins et al., 2007b; Abell et al., 2009] predicts 85 QSOs / deg$^2$, almost a factor of two more. To reach the same number of targets with an incomplete QSO sample, one must go to even fainter magnitude limits. In that case, the multiband photometric data used in the selection must be deeper than that of SDSS but cover a similar area on the sky. While the PTF and PS1 co-added survey data will fulfill this requirement, neither of the surveys will acquire deep $u$-band photometry, which is vital for specifically selecting $z > 2$ QSOs.

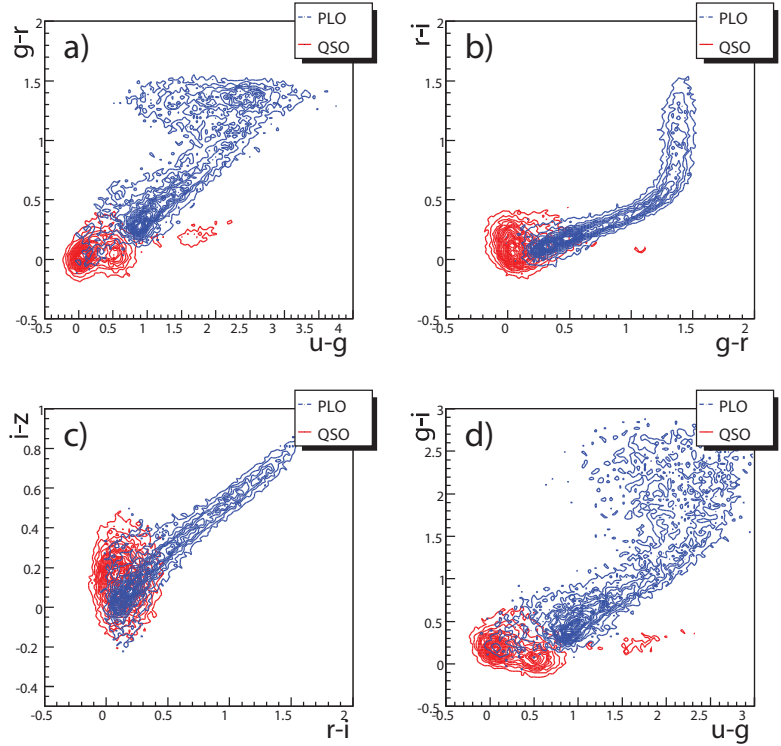

Figure 4.15: The *ugriz* colors of SDSS objects photometrically classified as stellar point-like objects (PLO) and those spectroscopically classified as QSOs (see Yeche et al. [2010]). For the BOSS Ly-$\alpha$ QSO selection, a neural-network algorithm uses available SDSS colors and spectroscopic templates to select the objects most likely to be QSOs with $z > 2.2$.

### 4.5.2   Selection Technique

Efficient selection of QSOs based on integrated photometry will be difficult without deeper $u$-band imaging than SDSS obtained. The availability of such deep imaging is possible but uncertain (cf. §4.2); in its absence, BigBOSS will exploit the intrinsic variability of QSOs to target them. Because the accretion region around a quasar is highly compact, its luminosity can vary on timescales ranging from days to years. The time-variability of astronomical sources can be described using a structure function, a measure of the amplitude of the observed variability as a function of the time delay between two observations; the structure functions of quasars and variable stars differ strongly from each other. Selecting quasars by their structure function has been successfully tested in the QUEST survey [Rengstorf et al., 2004; Bauer et al., 2009].

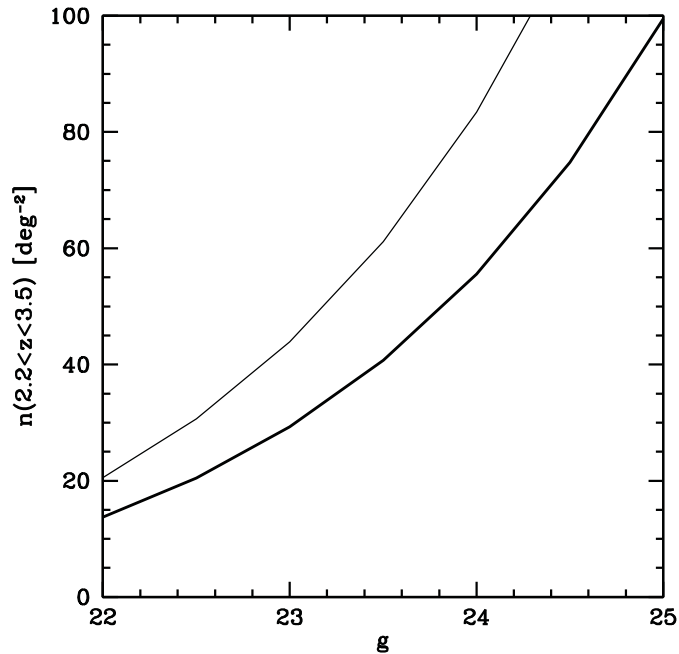

Figure 4.16: Surface density of quasars (objects per sq. deg.) in the redshift range $2.2 < z < 3.5$, derived from the Jiang et al. [2006] luminosity function (thin line). We also plot a second line with surface density 33% smaller, illustrating the source counts expected for a 67% complete sample (thick line) .

QSO-variability selection techniques have recently been refined by incorporating a model of the structure function which is a power law in observed time lag (as opposed to the more commonly-used rest-frame lag, which requires a known redshift to compute; [Schmidt et al., 2010], hereafter S10). This model can be parameterized in terms of $A$, the mean amplitude of the variation on a one year time scale (in the observer's reference frame) and $\gamma$, the logarithmic slope of the variation amplitude with respect to time. Figure 4.17 shows the structure function selection cuts for PanSTARRS-like data defined by S10. This selection was determined using SDSS Stripe 82 data down-sampled to match PS1 observations for objects with known spectroscopic classifications. The S10 selection is effective in separating QSOs from typical stellar contaminants such as F/G stars and RR Lyrae variables, while selecting 75% of all known QSOs in the field.

While structure function information can be drawn from multi-epoch data a single deep band (such as PTF $R$), a similar variability selection can be be obtained by measuring the structure function separately in multiband ($gri$) temporal data. Figure 4.18 shows the structure function derived from SDSS Stripe 82 $gri$ co-added data for $z > 2.2$ QSOs;

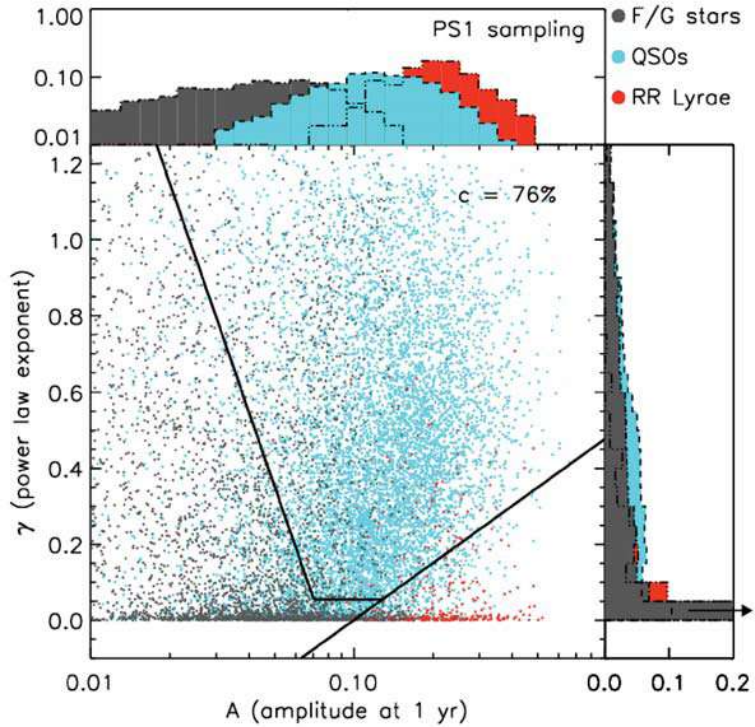

Figure 4.17: An illustration of the QSO variability-based selection from Schmidt et al. [2010]. The plotted parameters $A$ and $\gamma$ describe the amplitude and logarithmic slope of the time variability structure function of each object. The points indicate the positions of 15,000 spectroscopically-classified objects in SDSS Stripe 82 in this plane, determined using structure functions measured from $griz$ SDSS photometry which has been down-sampled to 6 epochs, matching expectations for PS1 $3\pi$ survey. The gray points are for F/G stars, the red points are RR Lyrae stars, and the aqua points are confirmed QSOs. The solid lines illustrate a selection in the amplitude-slope ($A - \gamma$) plane which efficiently targets quasars rather than stars.

this data has photometric limits similar to those expected for the PS1 survey. In the overplotted fit curves, the amplitude $A$ of the structure function in each band is allowed to vary independently, while the amplitude variation $\gamma$ is simultaneously fit from all bands. An efficient selection algorithm has been developed which begins with a loose selection of all blue point sources with $(g - r) < 0.9$ and $i_{AB} < 23.5$. Three-band structure function information is then fed to a Neural Network (NN) to separate QSOs from stars, employing the global structure function fits in the process [Palanque-Delabrouille et al., 2010]. This method is currently being tested in BOSS in SDSS Stripe 82 and will produce results soon.

By employing a broad color cut followed by rejection of stars using variability information, we can select a QSO sample with a high degree of completeness. For the strawman selection algorithm described above, Figure 4.19 shows the fraction out of all $z > 2.2$, $g < 23$ QSOs that are included in the BigBOSS QSO sample ("QSO completeness") as a function of the restrictiveness of the QSO targeting algorithm. If only the highest-confidence objects are selected (corresponding to a low surface density), a significant fraction of the useful targets would be rejected. A high completeness (80–90% out of all optimal QSOs) may be reached by targeting 180–250 targets deg$^{-2}$. As is illustrated in Figure 4.20, the estimated QSO completeness as a function of redshift is relatively flat; this selection technique is effective throughout the desirable redshift range.

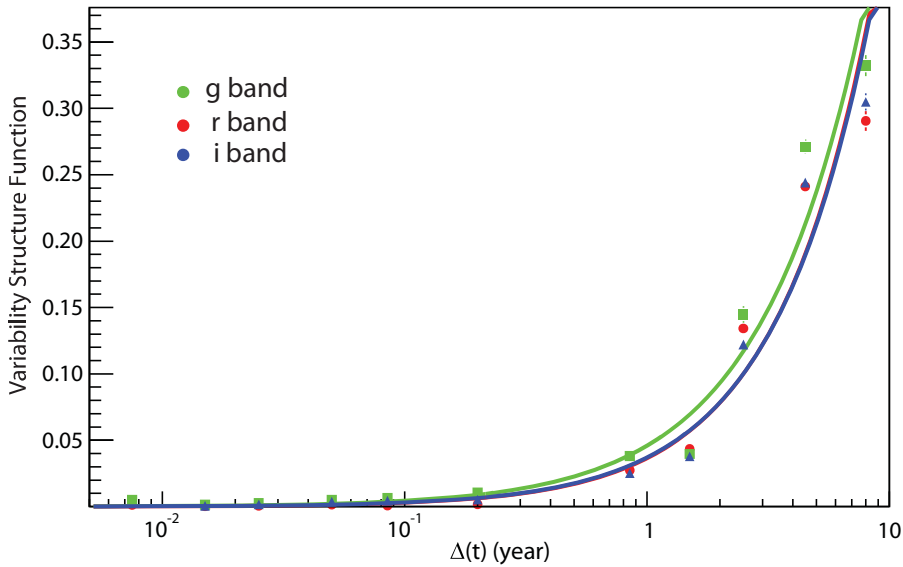

Figure 4.18: Structure functions measured in SDSS Stripe 82 data for $z > 2.2$ QSOs with $(g - r) < 0.9$ and $i_{AB} < 23.5$, determined for each of the $g$, $r$, and $i$ bands separately. The solid curves have been simultaneously fit to all three bands, with the amplitude in each band varying but the power-law slope ($\gamma$) required to be the same. These models are then used in a neural network-based algorithm to separate QSOs from stars, with results shown in Figs. 4.19 and 4.20.

However, roughly 75-80% of the $\sim 250$ objects deg$^{-2}$ selected as candidate Ly-$\alpha$ QSOs will either not lie in the desired redshift range or are interloping variable stars. Since each Ly-$\alpha$ QSO must be observed multiple times to achieve adequate signal-to-noise, targeting such a high surface density of objects with a $\sim 25\%$ success rate would have a significant adverse impact on the LRG and ELG samples. We can increase the effective redshift window efficiency by using the first tiling pass of the survey to determine redshifts to all QSOs (which requires much less signal) and then rejecting low-$z$ objects (see §6 for details). In this way, we can minimize the impact of Ly-$\alpha$ QSO targeting on other BigBOSS science programs and still achieve 90% completeness in $z > 2.2$ QSO targeting sample.

### 4.5.3   Sample Properties

Assuming an average of the J06 and LSST QSO luminosity functions and 80-90% selection completeness, we expect there to be $\sim 65$ Ly-$\alpha$ QSO targets deg$^{-2}$ to $g < 23$ suitable for repeated BigBOSS observations. We may estimate the redshift distribution of this sample by using the confirmed QSO redshifts from the BOSS survey and rescaling the distribution to the expected total areal density. The resulting BigBOSS Ly-$\alpha$ QSO redshift distribution is shown in Figure 4.21. While the redshift distribution for BOSS may differ from that for BigBOSS because the selection criteria are different, we expect that the BOSS sample should only be *more* weighted towards low redshifts than BigBOSS, given that the QSO selection for the former depends heavily on relatively shallow $u$-band photometry. Figure 4.21 should therefore be considered a conservative estimate for the redshift distribution of BigBOSS Ly-$\alpha$ quasars, particularly at higher redshifts extending to $z > 3.5$. The final redshift distribution, after accounting for losses to fiber completeness and redshift measurement failures, is recorded in Table 2.4.

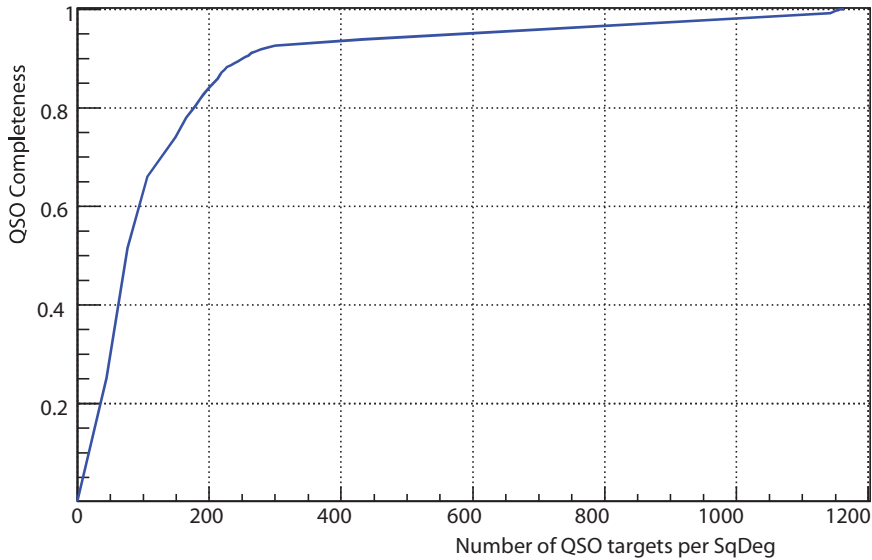

Figure 4.19: The fraction of all $z > 2.2$, $g < 23$ QSOs targeted ("QSO completeness") as a function of the looseness of the selection criteria (parameterized by the number of objects selected per square degree). This plot results from tests of a new color-variability algorithm applied to SDSS Stripe 82 data. BigBOSS will target roughly 250 QSOs per square degree in its first visit to each field, and use the resulting redshifts to reject low-$z$ objects for all return visits. In this way, a large fraction of all desired targets will be observed, while spending relatively little time on variable stars and low-redshift QSOs.

The proposed selection scheme will initially target any objects that have similar colors and intrinsic variability as Ly-$\alpha$ QSOs. We expect that the largest contaminant population will be faint horizontal branch stars. It is likely that there will also be a significant fraction of $z < 2$ QSOs targeted in the first pass. The extent to which we will sample Ly-$\alpha$ QSOs, as opposed to lower-redshift objects, using the color-variability technique is currently being tested via a BOSS ancillary targets program; we expect to have a complete sample in Winter of 2010 (early results are quite promising). Based on such tests, we will be able to optimize our target selection algorithms, potentially reducing the number of candidate QSOs that must be tested in the initial targeting pass compared to the estimates above. An example of one possible optimization method, incorporating near-infrared data from the

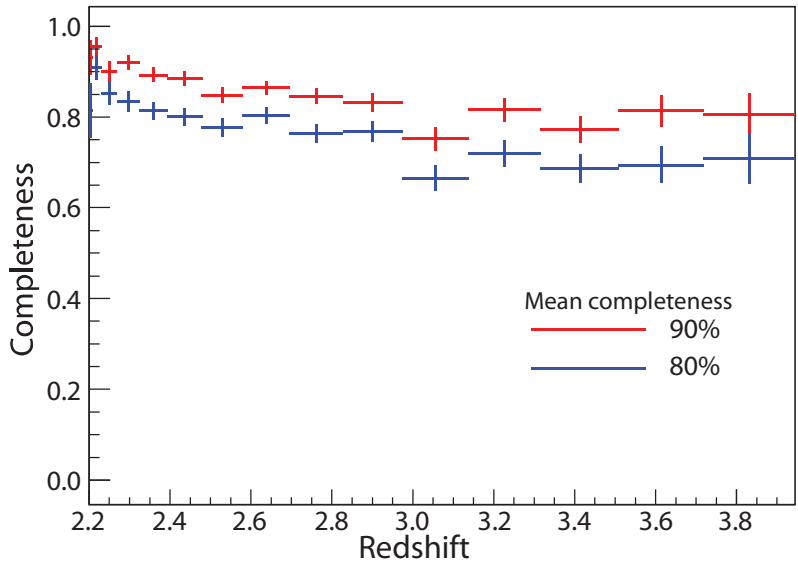

Figure 4.20: The QSO completeness as a function of redshift for samples with two different overall QSO target density levels, based on the same data as Figure 4.19. For target sample sizes similar to those which BigBOSS will use, the QSO selection completeness is relatively flat with redshift.

*WISE* satellite, is presented at the end of this section.

The $g < 23$ limit chosen for the BigBOSS QSO sample should ensure sufficient signal-to-noise to detect Ly-$\alpha$ forest correlations from co-adding 5 observations; hence, the sample's spectral feature efficiency should be $\sim 100\%$. We conservatively estimate that we will fail to obtain redshifts for QSO targets in the first visit 10% of the time, yielding a redshift measurement efficiency of 90%. Since $\sim 75\%$ of objects selected by our strawman algorithm are at $z < 2.2$, our redshift window efficiency will be $\sim 25\%$ in the first pass over a region. Hence, in this initial visit, the rate of redshift completeness – the product of fiber completeness, target selection efficiency, and redshift measurement efficiency – will yield $\sim 18\%$. However, in the remaining four visits, the target selection efficiency and redshift measurement efficiencies will be $\sim 100\%$ as objects which are not known to be at $z > 2.2$ will not be re-observed. This scheme will yield an 80% redshift completeness for the targeted $z > 2.2$ QSOs in the subsequent observed tiles. The average redshift completeness over all visits will be 67%, comparable to the LRG sample and significantly greater than the completeness for ELGs.

### 4.5.4   Identifying Low-redshift QSOs

Our proposed target selection techniques can achieve a highly complete QSO sample without the use of $u$-band dropout information. However, because the color cut used is relatively broad $((g - r) < 0.9)$, it will also select QSOs over a wide redshift range. Additional information from other bands could be used to sculpt the redshift distribution as desired.

In particular, many $z < 2$ QSOs can be identified using mid-IR photometry from the

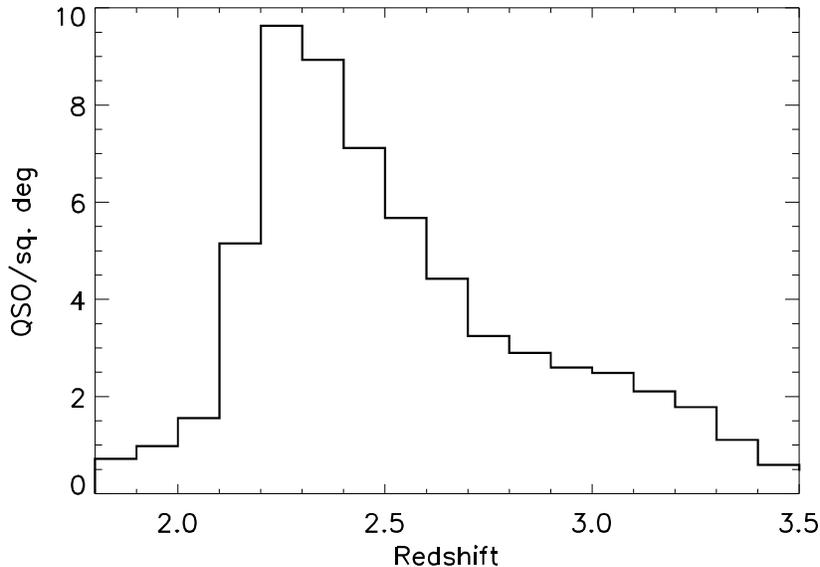

Figure 4.21: Estimated BigBOSS redshift distribution for Ly-α QSOs. This sample is taken by rescaling the confirmed BOSS quasar redshift distribution and rescaling to the surface density of targeted BigBOSS Ly-α QSOS with $2.2 < z < 3.5$. The targeted surface density is $\sim 65$ deg$^{-2}$ (cf. Table 2.2).

*WISE* satellite. The constraints from *WISE* can be used either to veto these QSOs from the BigBOSS survey sample or to help select a wider range out of the overall QSO population to benefit ancillary science. To determine the utility of *WISE* photometry for selecting $z < 2$ QSOs, we used data from the Spitzer observations of the Böotes Field of the NOAO Deep Wide Field Survey, the so-called *Spitzer* Deep Wide-Field Survey (hereafter SDWFS; [Ashby et al., 2009]). SDWFS reaches 80% completeness limits of 18.2, 18.1, 16.8 and 16.1 Vega mag in the 3.6μm, 4.5μm, 5.8μm and 8.0μm bands respectively over an area of over 9 deg$^2$, and is essentially complete at depths corresponding to the *WISE* 5σ point source limits.

The IRAC four-band color-color diagram for those SDWFS sources with flux densities brighter than the *WISE* 5σ point source limits in the 3.6μm and 4.5μm bands (hereafter, the SDWFS/*WISE* sample) is shown in Figure 4.22. The dashed lines show the "AGN wedge" as defined by [Stern et al., 2005], which is highly effective at discriminating AGN from IR-bright galaxies. While *WISE* will not provide photometry near 5.8 or 8.0μm, one could construct a similar four-color diagram using all of the *WISE* bands (3.4, 4.6, 12 and 22μm). However, this would restrict any *WISE*-selected sample to only sources that are detected in all four bands. Instead, a simple two-color cut of [3.6]-[4.5]≥0.6 results in selecting the bulk of the sources in the "AGN wedge", and relies on the bands for which *WISE* photometry goes deepest.

Applying this selection to the sample in Figure 4.22 (i.e., SDWFS/*WISE*) results in 407 sources, which corresponds to a surface density of $\sim 50$ deg$^{-2}$. Of these sources, 91% lie within the AGN wedge and 98% have $I_{Vega} < 22$. The main contaminants are likely to be very low-redshift star-forming galaxies with strong PAH emission and a few high-redshift obscured galaxies. The former could be easily excluded using a star-galaxy separation based on ground-based optical survey imaging, while the latter are rare at bright optical magnitudes.

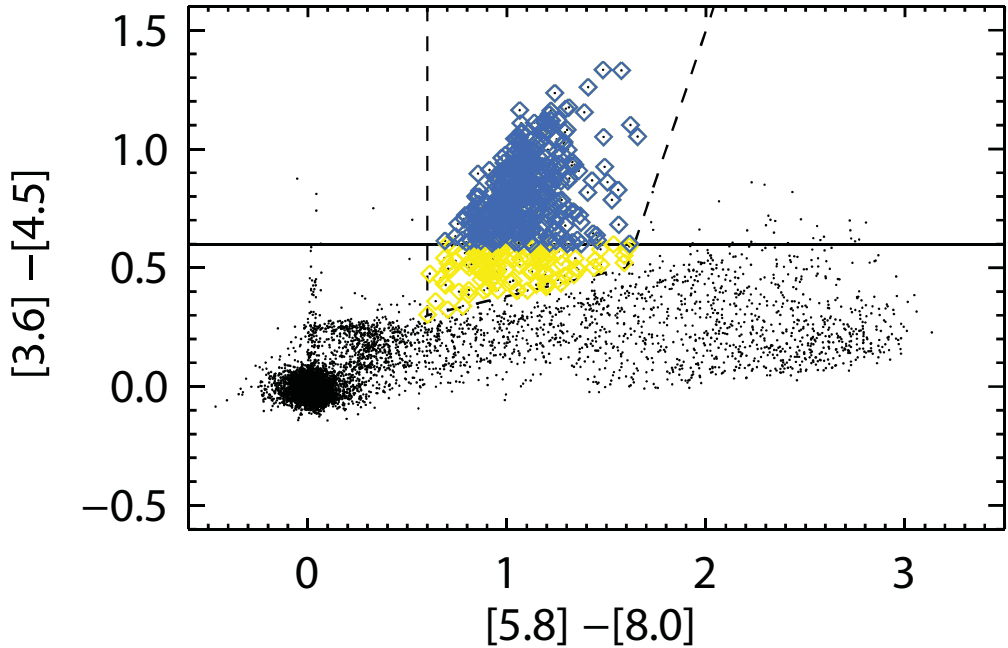

Figure 4.22: The *Spitzer*/IRAC color-color diagram for those SDWFS sources [Ashby et al., 2009] which would be detectable by *WISE* at both 3.6μm and 4.5μm (i.e., the SD-WFS/*WISE* sample). The dashed lines show the "AGN wedge" as dened by [Stern et al., 2005], which is highly effective for selecting bright AGN. Note that the bulk of the wedge AGN may be selected using a simple color cut of [3.6]-[4.5]≥0.6.

Approximately 62% of the objects in the SDWFS/*WISE* sample have spectroscopic redshifts from the AGN and Galaxy Evolution Survey (AGES; Kochanek et al. (2010), in preparation; see also [Kochanek et al., 2004]) and other spectroscopic campaigns using the W. M. Keck Observatory telescopes. The redshift histogram in Figure 4.23 shows that 46% of the objects with redshift information have $z \geq 1$ and only ∼3% have $z \geq 2.2$. We find that the magnitude distributions of sources with and without spectroscopic redshifts is roughly similar, and so we can expect that the redshift distribution of the full sample would be comparable. Thus, by employing *WISE* photometry to reject low-redshift AGN, we can reduce the number of objects that must be sifted through in the first-pass search for Ly-α QSOs by ∼ 20% (from ∼ 250 objects deg$^{-2}$ to ∼ 200), at the cost of rejecting ∼ 1.5 Ly-α QSOs per square degree. More sophisticated techniques may be able to sculpt the redshift distribution with smaller loss; we will explore these with actual *WISE* photometry when it becomes publicly available.

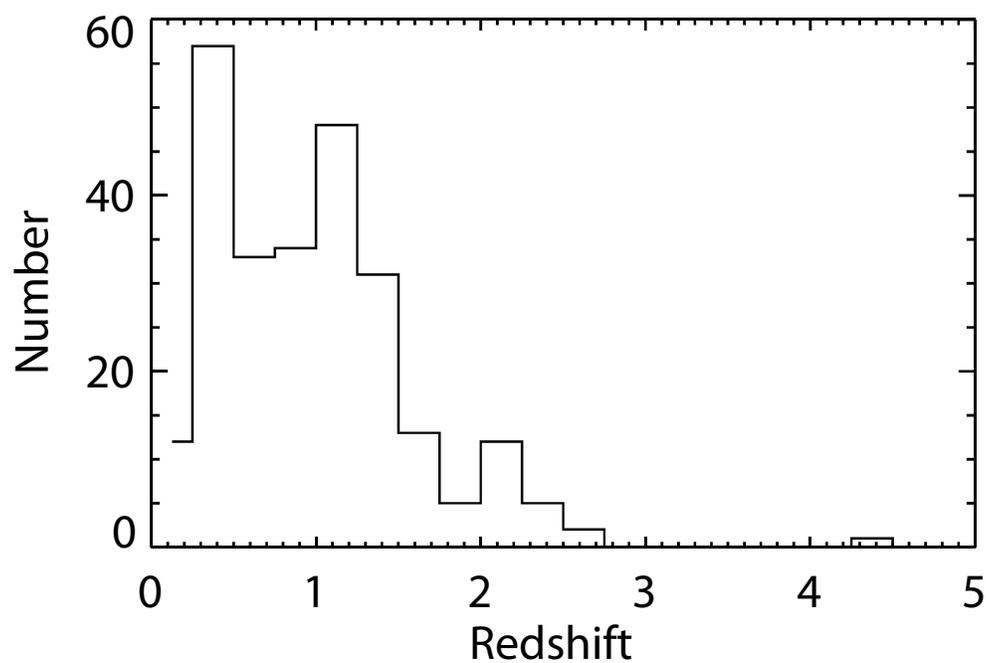

Figure 4.23: Redshift distribution of the SDWFS/*WISE* sample of sources, based on spectroscopy from the AGES [Kochanek et al., 2004] survey and the Keck telescopes. 46% of the sources lie at redshifts $z \geq 1$, but only $\sim 3\%$ are at $z > 2.2$.

# 5   The BigBOSS Instrument

## 5.1   Overview

The BigBOSS instrument is composed of a set of telescope prime focus corrector optics, a massively multiplexed, roboticized optical fiber focal plane, and a suite of fiber-fed medium resolution spectrographs, all coordinated by a real-time control and data acquisition system. The conceptual design achieves a wide-field, broad-band mulit-object spectrograph on the Mayall 4-m telescope at KPNO.

Table 5.1 summarizes the key instrument parameters such as field of view, number of fibers, fiber size and positioning accuracy, spectrograph partitioning, and integration time, were derived from a blend of science requirements and technical boundaries. These were derived from a confronting the science requirements for the Key Science Project with realistic technical boundaries.

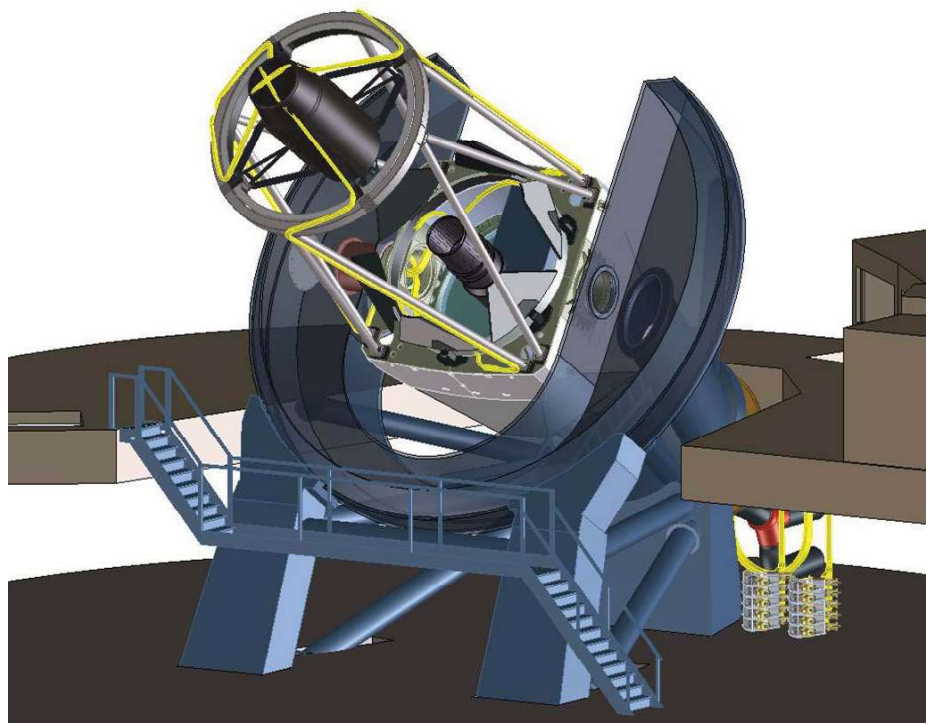

Figure 5.1: BigBOSS instrument installed at the Mayall 4-m telescope. A new corrector lens assembly and robotic positioner fiber optic focal plane are at mounted at the prime focus. The yellow trace is a fiber routing path from the focal plane to the spectrograph room incorporating fiber spooling locations to accommodate the inclination and declinaton motions of the telescope. The two stack-of-five spectrograph arrays are adjacent to the telescope base at the end of the fiber runs.

The instrument wavelength span requirement of 340–1060 nm was determined by the need to use galaxy [O II] doublet(3727Å and 3729Å) emission lines to measure the redshift of Luminous Red Galaxies ($0.2 < z < 1$) and Emission Line Galaxies ($0.7 < z < 1.7$) and the Ly-$\alpha$ (1215Å) forest for Quasi-Stellar Objects($2 < z < 3.5$) as described in Section 2.

Table 5.1: Instrument Parameters.

| | |
|---|---|
| **Telescope** | |
| 3° linear FOV | |
| 3.8 m diameter aperture, f/4.5 | |
| 1.8 m linear obscuration | |
| Focal length 17.1 m | |
| Wavelength response 340–1060 nm | |
| Blur <28 $\mu$m RMS (0.35 arcsec) | |
| Focal surface | |
|     4000 mm convex sphere | |
|     950 mm diameter | |
| **Fiber System** | |
| 5000 robotic fibers | |
| Fiber diameter 1.45 arcsec (120 $\mu$m) | |
| Fiber actuator spacing 145 arcsec (12 mm) | |
| **Spectrographs** | |
| Bandpasses | |
|     Blue: | 340–540 nm |
|     Visible: | 500–800 nm |
|     Red: | 760–1060 nm |
| Resolution | |
|     Blue: | 3000 |
|     Visible: | 2960 |
|     Red: | 4140 |
| **Cameras** | |
| 4k×4k pixels per channel | |
| 3 pixel minimum sampling | |
| Pixel size | |
|     Spatial: | 0.75 arcsec |
|     Blue: | 0.488 |
|     Visible: | 0.732 |
|     Red: | 0.732 |
| QE (400–1000 nm) | >80% |
| Read noise | <2.5 e |
| Dark current | <0.03 e/s/pixel |
| Pixel rate | 100 kpixel/sec/port |
| **Instrument cycle time (parallelizable)** | |
| CCD readout | 40 s |
| Fiber positioning | 60 s |
| Telescope slew and guide lock | <60 s |

Our large, $3°^{\parallel}$ linear FOV was set by a requirement to accomplish a 14,000 deg$^2$ survey area in 500 nights at the required object sensitivity. The field was selected following feasible designs that were demonstrated in earlier NOAO work and expanded upon with BigBOSS studies. In our implementation, the existing Mayall prime focus is replaced with a six element corrector illuminating the focal plane with a f/4.5 telecentric beam that is well matched to the optical fibers acceptance angle. In this way, the large FOV can be accomplished within a total optical blur budget of 28 $\mu$m RMS.

Given the FOV and required number of objects to observe, we design the focal plane to accommodate 5000 fibers by using demonstrated 12 mm pitch fiber actuators. A 120 $\mu$m fiber core size is chosen to fit a 105 $\mu$m FWHM image of a galaxy (after telescope blur and site seeing of 1 arcsec RMS) while minimizing inclusion of extraneous sky background. The fiber size choice allows for fiber tip placement of 5 $\mu$m RMS accuracy.

In order to achieve spectral resolutions of 3000–4000 for resolving the [O II] doublet lines while keeping the optical element small and optimized for high throughput, we have chosen to divide the system into ten identical spectrographs each with three bandpass-optimized arms. with each spectrograph recording 500 fibers and each arm instrumented with 4k×4k CCD.

The exposure time of 16.6 minutes is based on the requirement that at least one of the lines of the [OII] doublet lines 3727Å and 3729Å from an Emission Line Galaxy with a line flux of $0.9 \times 10^{-16}$ ergs/cm$^2$/s is detected with a signal-to-noise ratio of 8. The time was derived using the BigBOSS exposure time calculator described in Appendix A and including known detector characteristics (readnoise, dark current, and quantum efficiency), effective telescope aperture, mirror refection, fiber coupling and transmission losses, and spectrograph throughput, A one minute deadtime between exposures was set to maintain the needed observing efficiency while allowing for spectrograph detector reads, fiber positioning, and telescope pointing.

Figures 5.2 and 5.3 schematically show the content and interplay of instrumental systems.

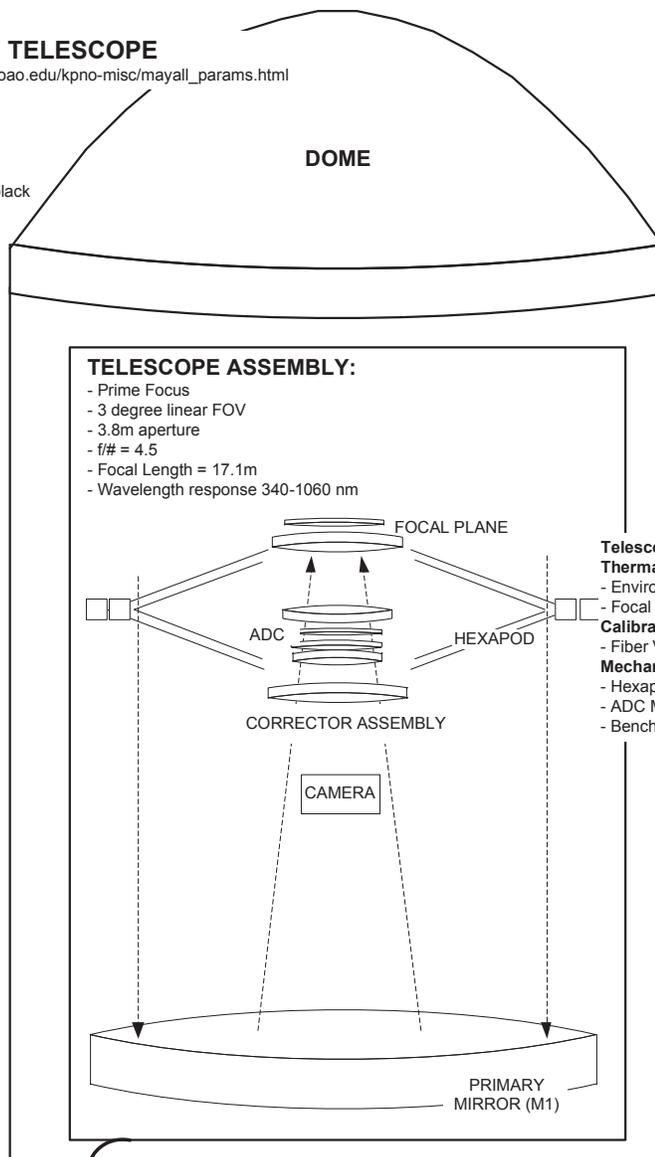

**KITT PEAK 4-m (Mayall) TELESCOPE**
- Parameters: http://www-kpno.kpno.noao.edu/kpno-misc/mayall_params.html

**FACILITY MODIFICATIONS:**
- Primary mirror edge sensing
- Thermal control to 3 deg w/ glycol
- Cooling pipes
- Paint M1 support structure diffuse black

DOME

**FOCAL PLANE and FIBERS:**
- 1-m Curved Surface
- 5000 Broad Spectrum Fibers
- Provides 3 deg dia on sky

**CORRECTOR ASSEMBLY:**
- 4 lens corrector
- 4 plate ADC
- Hexapod Motor Controlled

**ADC:**
- 2 opposite-rotating Risley prisms
- LLF1 or N-PSK3 material
- 1 degree tolerance on rotation

**Hexapod:**
- 5 to 10 um step size
- +/-2mm despace
- +/-1mm lateral
- 1 deg tilt

**FIBER VIEW CAMERA:**
- Measures fiber location
- Views illuminated fiber tips
- Kodak 50Mpix CCD
- Canon f/2.8 Lens
- 71mm Aperture
- 25 demagnification
- Uses existing M2 mounting IF

**OBSCURATION (not shown):**
- 1.8m obscuration
- Required for stray light

**PRIMARY MIRROR (M1):**
- Existing 4m mirror
- F/number = f/4.5
- Concave, hyperbolic

**TELESCOPE ASSEMBLY:**
- Prime Focus
- 3 degree linear FOV
- 3.8m aperture
- f/# = 4.5
- Focal Length = 17.1m
- Wavelength response 340-1060 nm

FOCAL PLANE

ADC      HEXAPOD

CORRECTOR ASSEMBLY

CAMERA

**Telescope
Thermal Control:**
- Environmental Monitors
- Focal Plane Thermal Control
**Calibration Control**
- Fiber View Lamps
**Mechanism Control:**
- Hexapod Motor Control
- ADC Motor Control
- Bench Environment Monitors

PRIMARY
MIRROR (M1)

**FIBER RUN (5000 Fibers, Sub-bundle units):**
- 30-40m fiber run from Focal Plane to Spectrographs
- Low OH fused silica (340-1060nm)
- Core 120 um diameter
- Provides maximum attenuation of 30% @ 340nm
- Fiber testing (Korea)

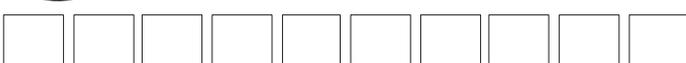

**FTS CONTROL ROOM**
- 10x Spectrograph Bench-mounted in FTS control room
- Floor has cooling pipes
- Vibration-isolated
- Thermally controlled to +/-1K
- (Remote control room in Tucson)

Figure 5.2: BigBOSS telescope system block diagram.

Figure 5.3: BigBOSS focal plane and spectrograph systems block diagram.

## 5.2 Telescope Optics

### 5.2.1 Design

BigBOSS employs a prime focus corrector to provide a telecentric, seeing-limited field to an array of automated fiber positioners. Basic design requirements are listed in Table 5.2. The optics team received considerable guidance and assistance from Ming Liang in the areas of corrector, atmospheric dispersion compensator, and stray light design. The final corrector and ADC design was developed from a concept presented by on the NOAO web site ([Liang, 2009]).

Table 5.2: Telescope Requirements.

| Requirement | Value | Description |
|---|---|---|
| Compatibility | N/A | Use existing telescope mount and M1 of Mayall and Blanco 4-m telescopes. Include mount for existing f/8 M2. |
| f/# | 4.5 | 3.8 m aperture, 17.1 m focal length |
| Geometric blur | < 0.8 arcsec FWHM | RMS across field |
| Zenith Angle | 0–60°‖ | Will require atmospheric dispersion corrector (ADC) to meet blur requirements |
| Field of View | 3°‖ | Full field of view |
| Wavelength Range | 340–1060 nm | Simultaneous correction required across entire band |

Cassegrain and prime focus options were explored. Prime focus was selected for its superior stray light performance, increased throughput due to simplified baffling and smaller central obscuration, and lower cost. The corrector includes four corrector elements, and a pair of ADC elements (each consisting of two powered prisms). Materials and design of the corrector and ADC were selected for manufacturing feasibility. All elements of the corrector are long-lead items, and initial contacts have been made with raw material suppliers and lens manufacturers. Corning can supply the large fused silica pieces, and N-BK7 and LLF1 are current production glasses at Schott.

Figure 5.4 shows the optical layout of the BigBOSS prime focus corrector and ADC. The four singlet corrector elements are fused silica, each with one aspheric and one spherical surface. Element C1 is the largest lens, 1.25 m fused silica. Lens elements were sized to have more 15 mm of radius beyond the clear aperture to allow for polishing fixturing and mounting. The ADC consists of two wedged doublets, with spherical external surfaces, and a flat, cemented wedge interface. ADC elements are made of LLF1 and N-BK7, and all are within the current production capability of Schott. A minimum 300 mm gap exists between Element 4 and the central fiber positioner (focal surface).

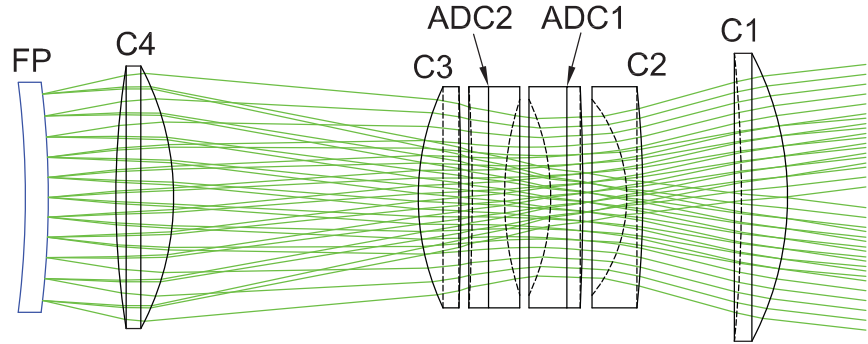

Figure 5.4: BigBOSS prime focus corrector consists of four corrector elements and two ADC prism doublets.

Figure 5.5 shows the ideal rms geometric blur performance (no manufacturing, alignment or seeing errors) of the BigBOSS corrector mounted on the Mayall telescope. For reference, the required FWHM geometric blur of 0.8 arcsec corresponds to a blur RMS of 28 $\mu$m, so realistic manufacturing margins exist.

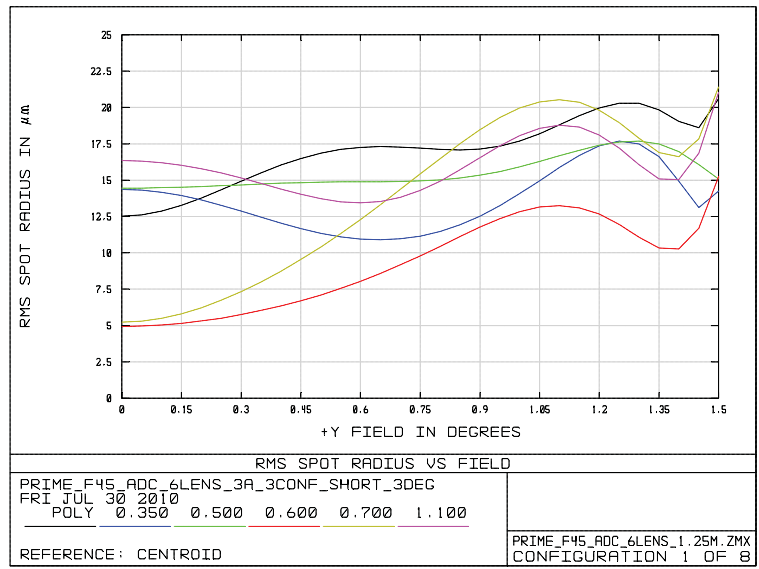

Figure 5.5: Ideal geometric blur performance of BigBOSS corrector on Mayall 4-m telescope. The required 0.8 as FWHM corresponds to a geometric blur radius of 28 $\mu$m.

### 5.2.2   Focal Surface

The focal surface is a convex sphere of 4000 mm radius of curvature and has a diameter of 950 mm. This is the surface that the optical fiber tips must placed on to 10 $\mu$m accuracy.

### 5.2.3   Tolerancing

The Mayall telescope is seeing-limited with an atmospheric FWHM of 0.9 arcsec, or 72 $\mu$m FWHM. For the 17.1 m focal length of BigBOSS, this corresponds to an RMS radius of 32 $\mu$m. Peak geometric blur (multispectral) of the perfect telescope across the 3°‖FOV is 18.3 $\mu$m, or 43 $\mu$m FWHM. With manufacturing, alignment and thermal drift, the telescope geometric blur is 28 $\mu$m RMS, or 66 $\mu$m FWHM. The overall peak budget for seeing, residual phase error, manufacturing, alignment error and thermal drift is 100 $\mu$m FWHM, or 1.2 arcsec. This is a worst-case number, and performance of the telescope is better over the majority of the field.

Tolerances for the telescope are broken down into three major categories: compensated manufacturing errors, compensated misalignment, and uncompensated errors. Manufacturing errors such as lens radius of curvature and thickness may be compensated to a certain degree by varying the spacing of the lens elements during assembly and alignment. Table 5.3 shows compensated manufacturing tolerances on the individual optical elements.

Table 5.3: Compensated Manufacturing Tolerances. $V$ is vertex lateral error ($\mu$m), $T$ is thickness error ($\mu$m), $W$ is wedge ($\mu$m at edge), $R_1$ is surface one radius error ($\mu$m sag), $K_1$ is surface one conic (%), $R_2$ is surface two radius ($\mu$m sag), $K_2$ is surface two conic (%), and $n_h$ is the homogeneity of the index of refraction (ppm).

|             | $V$  | $T$  | $W$ | $R_1$ | $K_1$ | $R_2$ | $K_2$ | $n_h$ |
|-------------|------|------|-----|-------|-------|-------|-------|-------|
| C1          | 250  | 1000 | 50  | 1000  |       | 300   |       | 5     |
| C2          | 200  | 1000 | 50  | 200   |       | 1000  | 0.01  | 5     |
| ADC1-1      | 1000 | 1000 | 150 | 10    |       | Flat  |       | 5     |
| ADC1-2      | N/A  | 1000 | N/A | Flat  |       | 150   |       | 5     |
| ADC2-1      | 200  | 1000 | 150 | 1000  |       | Flat  |       | 5     |
| ADC2-2      | N/A  | 1000 | N/A | Flat  |       | 1000  |       | 5     |
| C3          | 1000 | 1000 | 100 | 200   |       | 1000  | 0.1   | 5     |
| C4          | 1000 | 1000 | 100 | 150   | 1     | 100   |       |       |
| Focal Plane |      |      |     | 10000 |       |       |       |       |

Residual alignment errors and thermal drift in the assembled corrector are compensated by a motion of the entire corrector barrel and focal plane via motorized hexapod. Residual errors after these compensations are primarily higher-order aberrations, and are budgeted as compensated tolerances in Table 5.4.

The current operations plan involves characterization of the telescope for gravity sag as a function of elevation, and thermal drift of telescope focus. These are compensated continuously by motion of the hexapod. Other manufacturing errors may not be compensated (between observations) by motion of the hexapod, for example, corrector glass inhomogeneities. Such effects are currently being quantified, but the optical performance of the corrector (geometric blur) has more allowance than other existing and planned (e.g., DES) designs.

Table 5.4: Compensated Alignment Tolerances.

| | Lateral Error ($\mu$m) | Despace Error from previous surface ($\mu$m) | Tilt Errors ($\mu$m at edge) |
|---|---|---|---|
| C1 | 150 | Compensator | 200 |
| C2 | 150 | 200 | 150 |
| ADC1 | 150 | 500 | 250 |
| ADC2 | 500 | 500 | 250 |
| C3 | 200 | 500 | 175 |
| C4 | 250 | 500 | 300 |
| Focal Plane | 1000 | 100 | 100 |

### 5.2.4 Optical Mounts

Corrector and ADC elements have coefficients of thermal expansion between 0.5 and 8.1 ppm/°C. The largest element is corrector element C1 (1.25 m in diameter). Operational temperatures range from -10°∥to 30°∥ Although larger transmissive elements have been built, detailed design and careful attention will be necessary during the design, fabrication and test phases in order to achieve the science goals of BigBOSS.

Overall responsibility for mounting and aligning the large glass elements of the corrector lies with University College London (UCL), who is also responsible for the similar corrector barrel assembly for DES. Requirements and goals for the optical mounts are listed in Table 5.5.

Table 5.5: Glass mount design guidelines

| Item | Tolerance |
|---|---|
| No metal on glass interfaces | Reduce surface contact stress on substrate |
| RTV athermalized glass mounts | Near zero-stress at glass/metal interface, metal ring slightly higher CTE than glass. |
| Flexure link from metal lens mount (low expansion) to barrel | Maintain alignment while allowing compliance between lens barrel and lens mount. |
| Modular, pinned construction, barrel assembled in sections | Ability to disassemble corrector to access individual lenses. |

The glass elements of the corrector will be mounted in rigid lens cells with a compliant layer interface. A nickel/iron alloy will be used as the cell material with RTV rubber pads as the interface between cell and lens. With a suitable choice of the Ni/Fe alloy mix, a particular CTE can be chosen which in combination with precise thickness RTV pads around the perimeter of the silica lens allow an athermal design that allows the lens to expand and contract with minimal stress. Once mounted in the cell, standard fastener construction can be used to mount the lens cell via a metal ring to the barrel. Titanium cell rings ($9.2 \times 10^{-6}$/°C) are used with RTV to similarly athermalize the higher expansion N-BK7 and LLF1 lenses ($6.2 \times 10^{-6}$/°C and $7.1 \times 10^{-6}$/°C respectively). A circular array of flexure blades allows for thermal expansion between the lens cell and the corrector barrel. This heritage design is currently being implemented by UCL on the DES project (see Figure 5.6b).

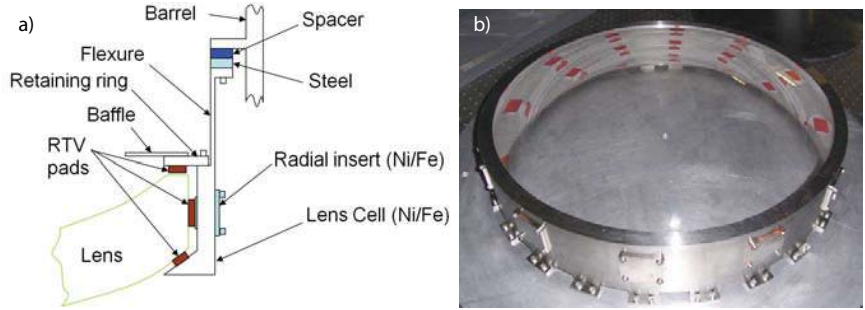

Figure 5.6: Schematic of athermalized, flexure-mounted lens cell design (left) and 550 mm prototype lens and lens cell at UCL (right).

### 5.2.5   Coatings

The preferred coating technology for the BigBOSS lenses is a hard (and durable) coating of $MgF_2$, with a tuned Sol-Gel coating. While Sol-Gel can be tuned to some degree for the bandpass of the telescope, it is not as durable as $MgF_2$. Cost, risk, performance and alignment constraints on the various coating technologies will be investigated during the fabrication of the corrector lenses, and a final decision is not necessary at this time. The likely configuration for the coatings is a hybrid $MgF_2$ undercoat with a tuned Sol-Gel overcoat (demonstrated performance of <0.5% loss over the visible band). At least two vendors (REOSC and SESO) are capable of coating the optics, including the 1.25 m diameter C1 element. It is expected that improved capability will be available subsequent to lens polishing.

### 5.2.6   Stray Light and Ghosting

A major benefit, and reason for selecting the prime focus option over Cassegrain, is the simplified stray light baffling. A wide-field Cassegrain design appropriate for BigBOSS would require a 50% linear obscuration, with carefully designed M1 and M2 baffles in order to block direct sneak paths to the detector. With a prime focus design, out-of-field rays miss the focal plane entirely. The main sources of stray light (first order stray light paths) are surfaces illuminated by sky light, and directly visible to the focal plane. Chief among these surfaces are the structure surrounding M1, which will be painted with durable diffuse and specular black stray light coatings (Aeroglaze Z302, Z306 and Ebanol). Other first order stray light paths include particulate contamination on M1 and the surfaces of the correctors.

The BigBOSS corrector was designed to ensure internal reflections within the corrector do not contribute significantly to stray light at the focal surface. The main causes of reflections are typically reflections off concave surfaces (facing the focal surface), and are most significant for elements in close proximity to the focal surface. Figure 5.7 shows a ghost stray light path from the C4 corrector element, which has been reduced by ensuring the radius of curvature of the first optical surface is smaller than its separation from the focal plane. As shown, the focus of the ghost is located off the focal surface, and only a diffuse reflected return, off two surfaces with >0.98% transmission surfaces contributes to the stray light at the focal plane. Additional point source transmittance analysis with realistic contamination and surface roughness is currently underway with the existing stray light model of the telescope and corrector.

Reflections between the focal plane array and nearby corrector surfaces are a typical source of stray light in an imaging wide-field corrector system. Because the fiber positioners can be made rough, and painted black, this source of stray light may be virtually eliminated on a robotic fiber array.

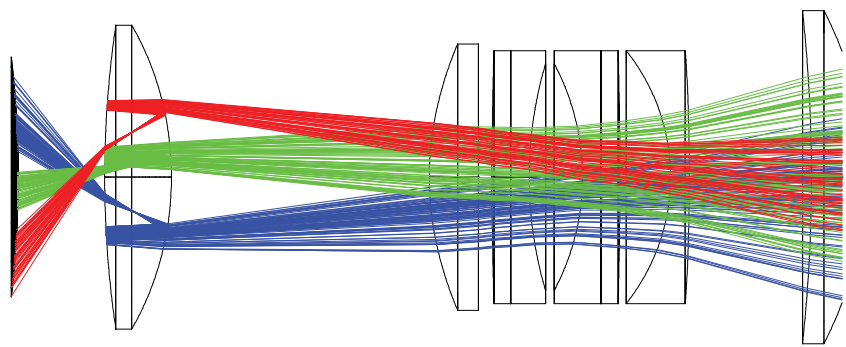

Figure 5.7: Reflections off corrector lens surfaces could contribute ghost background noise. Elements are designed to reduce bright ghost irradiance on the focal plane to acceptable levels.

### 5.2.7   Atmospheric Dispersion Corrector

Chromatic aberration must be sufficiently small to place incoming light between 0.34–1.060 nm within the geometric blur allocation. Because observations will be between $0-60°^{\parallel}$ from zenith, an atmospheric dispersion corrector will be necessary. The ADC elements are 0.9 m in diameter and made of Schott LLF1 and N-BK7. Wedge angles within the two elements are roughly $0.3°^{\parallel}$. Figure 5.8a shows the PSF across a $3°^{\parallel}$FOV at an angle $60°^{\parallel}$from zenith with the ADC rotated to correct for atmospheric dispersion. Figure 5.8b is for the uncorrected case. Rotational tolerance requirements for the ADC are greater than $1°$, and the ADC rotator is consequently not a high-precision mechanism.

### 5.2.8   Hexapod Adjustment Mechanism

Compensations for gravity sag, temperature change and composites dryout will be provided by a six-degree-of-freedom hexapod mechanism, Figure 5.9. The focal plane and corrector elements are positioned relative to one-another during alignment, and moved as a unit by the hexapod. Requirements for the hexapod are as listed in Table 5.6.

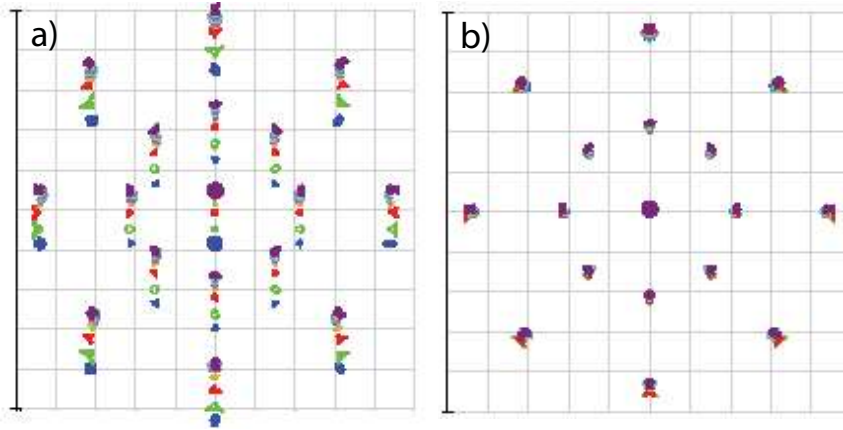

Figure 5.8: Geometric raytrace shows effects of atmospheric dispersion on telescope point spread function. a) A heavily chromatically aberrated view of the sky 60°∥ from zenith. Overall scale is 1m square, PSF exaggerated by factor of $10^6$. This chromatic aberration is removed by rotating the ADC prisms 85°∥ as shown in b). The dispersion being compensated here is 3 arcsec.

Table 5.6: Hexapod Mechanism Requirements

| Motion Requirement | Value | Comments |
| --- | --- | --- |
| Despace | ±2 mm | Focus direction |
| Lateral | ±1 mm | Lateral translation |
| Tilt | TBD | Tilt of corrector, pitch and yaw directions |
| Step size | 5 $\mu$m | Also called actuator granularity |

### 5.2.9   Telescope Simulator

The corrector integration and testing will be performed at FNAL using a telescope simulator developed for the DECam Project at the Blanco telescope, the twin of the Mayall. The telescope simulator (see Figure 5.10) will allow us to prove that the corrector passes the technical specifications independent of the expected orientations of the telescope. This platform will also allow us to develop the procedures that will be used to install the focal plane on the telescope. Performing this work in the lab, rather than in the field, we reduce the risk of extended telescope down-time when we install the instrument on the telescope for the first time, and we minimize the amount time required for integration and commissioning at KPNO.

The telescope simulator base is 4.3 m tall and 7.6 m wide. The four rings weigh 14,500 kg. The outer one has a 7.3 m diameter. Two motors from SEW Eurodrive can orient the camera to any angle within the pitch and roll degrees of freedom. The pitch motor is 1/3 HP, 1800 RPM, directly-coupled, torque-limited and geared down 35,009:1 for a maximum speed of 20 minutes per revolution. The roll motor is 1/2 HP, 1800 RPM, geared down 709:1 for a maximum speed of 11.5 minutes per revolution. The coupling is by means of a 18.9 m chain attached to the inner race (3rd ring out). Of course, the motor controls allow the assembly to be moved more slowly. When the assembly is not moving, the motors automatically engage brakes. The motors are controlled from a panel located on the exterior of the base. These controls are simple power on/off, with forward/backward and speed control for each ring. Four limit switches prevent the rings from being oriented in any undesired location.

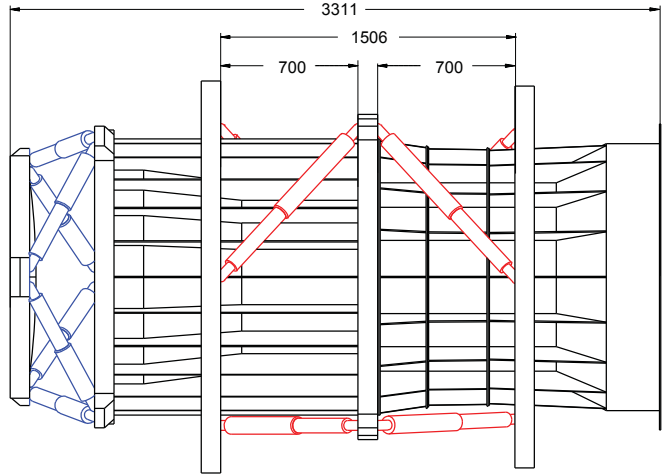

Figure 5.9: The corrector is supported on a six-degree-of-freedom hexapod mechanism that attached to the telescope structure through a series of rings and struts. Four of the hexapod actuators are the cylinders in the center of the figure (red). The cylinders at the left (blue) form the passive kinematic mount of the focal plane to the corrector.

### 5.2.10 Fiber View Camera

During the course of the survey, before any given exposure, after the mechanism to arrange the position of the 5000 fibers has completed its task the Fiber View Camera will take a picture of the fibers on the fiber plane to check the accuracy of all of the fiber positions, and if needed allow the correction of any misplaced fibers. The camera will be located on the axis of the telescope at a distance of 1 m below corrector element C1, as shown on Figure 5.11. The camera will be supported in this position by thin spider legs from the ring supporting the first element of the corrector optics. In this position the lens of the fiber view camera will be 5 m from the fiber plane. To get an image of the 950 mm diameter fiber plane on a 40 mm CCD will require the camera to have a demagnification of about 25. This can be accomplished with a 200 mm focal length lens. A detailed ray tracing through the camera lens and corrector optics finds that the image sizes are too small so we will have to defocus the camera to get the images to spread out over enough pixels to allow good interpolation. The fibers will be back-illuminated at the spectrograph end by a 10 mW LED.

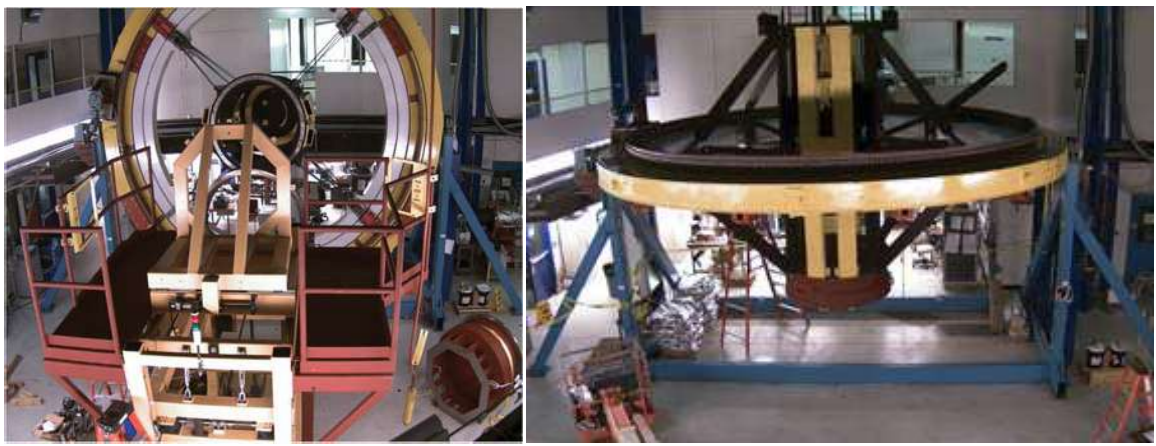

Figure 5.10: The telescope simulator at Fermilab with the DECam Prime Focus Cage. The structure in the background supports a copy of the two rings (white) at the top of the Serrurier Truss. The yellow rings are connected to motor drives that enable the Prime Focus Cage to be pitched and rolled, for the purpose of testing and pre-commissioning of the instrument prior to delivery to the telescope. The foreground structure (left) is the secondary (f/8) mirror handling fixture, enabling the installation of the mirror over the end of the optical corrector for the use of instruments at the Cassegrain Focus.

The view camera images the fiber tip focal plane through the corrector optics. This introduces some distortions in the images. A detailed study of the image shapes, using the BEAM4 ray tracing program, shows that these distortions are at an acceptable level. A set of fixed and surveyed reference fibers will be mounted in the focal plane and imaged simultaneously with the movable fibers. These can be used to deconvolve any distortion and any motion of the camera with respect to the focal plane due to gravity sag.

**5.2.10.1 Design Considerations.** The performance requirements for the fiber view camera are summarized in Table 5.7. We note further that since we plan to illuminate the fibers with a monochromatic LED, the CCD of choice should be monochromatic. For this reason we will use a monochromatic CCD, the Kodak KAF-50100. The plan is then to build a custom camera body (see Figure 5.12) using a commercially available lens, the Canon EF 200 mm f/2.8 L II USM. There also exists commercially available clocking and readout electronics for the Kodak CCD that we plan to use.

**5.2.10.2 Fiber Illumination.** We plan to illuminate each fiber at their end in the spectrometer with a monochromatic 10 mW LED. We estimate that each fiber will emit $2 \times 10^9$ photons/sec into a $30°^{\parallel}$ cone, full angle, at the focal surface. The solid angle of the fiber view camera lens will capture $3 \times 10^5$ photons/sec/fiber image. With 25% quantum efficiency this gives 75,000 electrons/sec/fiber image on the CCD.

**5.2.10.3 Dark Current and Read Noise.** It is desirable to run the camera at room temperature. The dark current in this Kodak CCD is advertised as 15 e/pixel/sec at 25°C and the read noise is 12.5 e at a 10 MHz read out rate. Both of these are quite negligible compared to the high fluxes expected from the fibers.

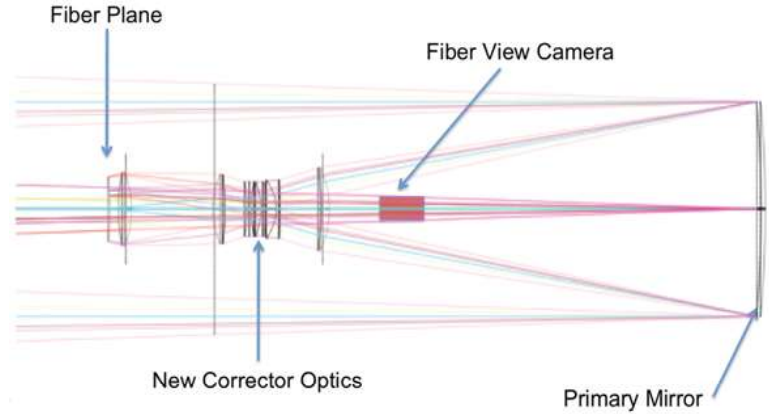

Figure 5.11: The Mayall Telescope showing the placement of the corrector optics and the fiber view camera

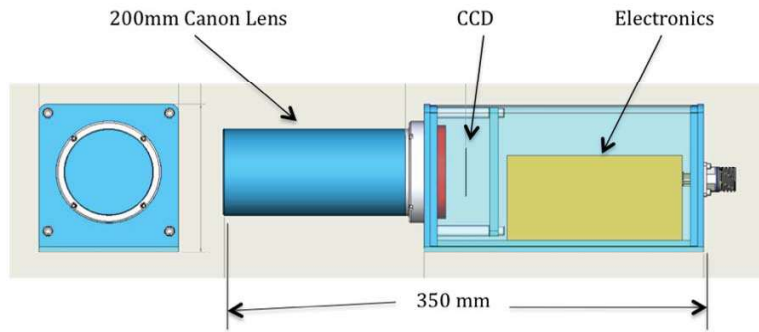

Figure 5.12: Schematic of the Fiber View Camera

**5.2.10.4  Fiber Position Precision.**  With a demagnification of 25 the 120 $\mu$m diameter fiber will have a 5 $\mu$m diameter image on the CCD. Including the optical distortions we still expect image sizes well under 10 $\mu$m, not a match to the 6 $\mu$m CCD pixels. We plan to defocus the lens slightly to produce large enough images to allow interpolations to much better than the pixel size. A Monte Carlo calculation was performed to determine the optimal image size. For the discussion here we will assume a 25 $\mu$m diameter image with significant flux spread over 16 pixels. With a one second exposure we expect 75,000 electrons per image. With such large signal to noise, we expect to centroid the fiber position to ∼0.1 $\mu$m. Systematic effects can double this to 0.2 $\mu$m. With the factor of 25 demagnification this translates to a 5 $\mu$m measurement error on the fiber plane.

**5.2.10.5  Occupancy and Ability to Resolve Close by Fibers.**  With 16 pixels in a fiber image, the 5000 fibers will occupy 80,000 pixels. Compared to the $50 \times 10^6$ pixels on the CCD, this gives an acceptable occupancy of ∼$2 \times 10^{-3}$. The fiber positioning mechanism sets the closest separation between any two fibers to be 3 mm. With the factor of 25 demagnification this means 120 $\mu$m or a 20 pixel separation on the CCD, so that overlap of the images will not be a problem. We have developed code to carry out a simultaneous fit to two images when two images are close together so that the tails of one image under the other and vice versa are correctly taken into account.

**5.2.10.6   Modelling and Scene Calibration.** The fiber position measurement precision quoted in Table 5.7 is based on the assumption that given the high statistics we can measure the position of the centroid of the image on the CCD to 3% of the pixel size. A Monte Carlo simulation will be useful to determine the optimum image size on the CCD for the best precision. In addition to the statistical error there will be systematic effects that limit the precision, such as variations in pixel size and response, lens distortions, etc. Before installation of the camera on the telescope a measurement and calibration of the precision in a test set up is anticipated.

Table 5.7: Fiber View Camera Requirements

| Feature | Req. | Goal |
|---|---|---|
| Centroid Precision on Focal Plane | $\leq$15 $\mu$m | $\sim$5 $\mu$m |
| Absolute Position Calibration | $\leq$15 $\mu$m | $\sim$5 $\mu$m |
| Nearest Neighbor Distance | 3 mm | 1.5 mm |
| Exposure Time | $\leq$2 sec | $\sim$1 sec |
| Readout Time | $\leq$2 sec | $\sim$1 sec |
| Thermal Stability | $\leq\pm1°$C | 1°C |
| Mounting (Vibration) | $\leq$15 $\mu$m/sec | 5 $\mu$m/sec |
| Scattered Light | None | None |

## 5.3   The Focal Plane Assembly

The BigBOSS focal plane assembly includes three main items: the support structure (adapter), the focal plate (which supports the $\sim$5000 actuators) and the set of actuators. The focal plane system is being studied by the Instituto de Astrofísica de Andalucía (IAA-CSIC, Granada, Spain). The IAA-CSIC, in collaboration with the company AVS, is working on its conceptual design. The focal plane parameters depend heavily on the support structure (corrector barrel) and on the final design of the fiber positioners (actuators). The focal surface is as a convex spherical cap with 4000 mm radius of curvature and 950 mm in diameter. The focal plate is foreseen to be an aluminum alloy plate $\sim$100 mm thick. Its primary purpose is to support the fiber positioners such that the fibers patrol area form tangents to the focal surface. The focal plate will be attached to the corrector barrel through a support structure, that we call "adapter". Figure 5.13 shows a view of the whole system.

### 5.3.1   Interfaces

The focal plate must be held at the back of the corrector barrel, facing the last lens of the corrector. Due to the distance to the corrector (about 200 mm), the focal plate cannot be directly attached to the corrector barrel and some structure in between (adapter) will be necessary. This will need to provide manual adjustment for initial focusing.

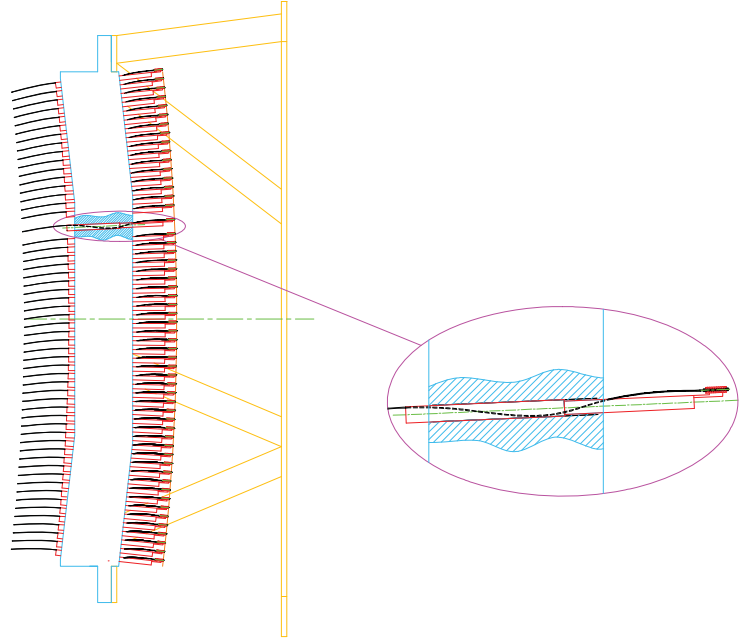

Figure 5.13: A cross section of the focal plane assembly with a possible shape of the focal plate and the adapter to attach the focal plate to the corrector barrel (orange color). The detail of a single actuator supporting a fiber is shown enlarged.

The focal plane supports several systems, most importantly being the 5000 fiber positioners. These will be inserted from the back of the focal plane to facilitate replacement. Insertion depth, tilt and rotation angle are precisely controlled, tolerances allocated from an overall focus depth budget. An array of back illuminated fixed fibers will serve as fiducials for the fiber view camera. Guiding and focus sensors also reside on the focal plane.

The focal plane is electrically connected to the power supplies for the fibers positioners, positioners wireless control system, electronics for guiding and focusing sensors, fiber view camera lamps, and environment monitors. Electromagnetic interference, both received and transmitted, will need careful study.

The large number of fibers and cables coming from the prime focus necessitate careful placement. They will be routed from the focal plane to the telescope support cage while minimizing obscuration of the primary mirror. Careful packing within the footprint of the primary optics support vanes coming from the telescope Serrurier truss will be designed.

### 5.3.2   Focal Plate Adapter

A structure is needed in order to attach the focal plate to the corrector barrel. A simple structure made of two circular flanges linked by a number of trusses should be able to cope with the flexures and sag. A few reference pins will be used to obtain mounting repeatability.

Interface of the adapter will be the corrector barrel on one side, and the focal plate edge on the other side. The adapter requirements are shown in Table 5.8.

Table 5.8: Focal Plate Adapter Requirements.

| | |
|---|---|
| Outer diameter | ∼950 mm |
| Positioning error (XY) | < 100 $\mu$m (absolute) |
| Wedge error (between interfaces) | < 0.25 mrad |

### 5.3.3   Focal Plate

The focal plate will be a solid piece of metal, probably an aluminum alloy, with multiple drills for actuator housing. The plate does not need to have a spherical shape, but the holes hosting the actuators must have their axes converging to the focal surface center and the plate must support the actuators so that their tips lie on the spherical focal surface. An example of suitable shape is shown in Figure 5.13. The edge of the plate must match the adapter flange that attaches to the corrector barrel. A few reference pins will be used to obtain a repeatable positioning onto the flange. Because the holes hosting the actuators do not follow any regular pattern (see Fiber Positioner Topology section), they will have to be machined from the model coordinates via a 5-axes machine tool. Care must be taken with the thermal expansion of such a large metal plate (aluminum alloys might not be ideal in this respect), which could easily overcome the actuators required positioning precision. A detailed materials study will be conducted in order to constrain the thermal effects as much as possible. Eventually, it could be necessary to set up a thermally controlled environment around the plate and the actuators, which could be obtained by enclosing the back of the focal plane with a vacuumed box, the other side of the focal plane being enclosed by the corrector last lens. Analysis is ongoing for studying the behavior of the honeycomb focal plate structure in terms of stiffness, deformations, sag etc. Such parameters vary with the inter-actuator wall thickness and plate thickness, but also depend on the actuator characteristics (once mounted, the actuators will contribute importantly to the stiffness and weight of the focal plane). In general, during the focal plane integration and test, it will be necessary to characterize all the reference positions of the actuators via the fiber view camera (with fiber back-illumination), which makes it much easier to fulfill the positioning precision requirement over such a large array of actuators. The requirements for the focal plate are shown in Table 5.9.

Table 5.9: Focal Plate Requirements.

| | |
|---|---|
| Number of actuators | ∼ 5,000 |
| Outer diameter | 950 mm |
| Positioning error | < 100 $\mu$m (absolute) |
| Perpendicularity error | < 0.4 mrad |
| Actuator housing tilt error | < 2 mrad |
| Focal plate thickness | 100 mm |
| Tot max allowed weight | 700 kg |

### 5.3.4 Fiber Positioners

A key enabling element for an efficient survey is a robotically manipulated fiber positioning array. The ability to reposition the fiber array on a timescale of $\sim 1$ min greatly improves on-sky operational efficiency when compared to manual fiber placement methods. Requirements for the fiber positioner system are shown in Table 5.10.

Table 5.10: Fiber Positioner System Requirements.

| | |
|---|---|
| Number of actuators | $\sim 5{,}000$ |
| Actuator pitch | 12 mm center to center |
| Patrol radius | 12 mm/$\sqrt{(3)}$=6.93 mm (filled survey) |
| Defocus over patrol disk | $< 10$ $\mu$m including spherical departure |
| Patrol disk tilt error | $< 2.4$ mrad |
| Positioning accuracy (w/o fiber view camera feedback) | $< 100$ $\mu$m (absolute) |
| Local positioning accuracy (over 200 $\mu$m distance) | $< 5$ $\mu$m (absolute) |
| Power (per actuator) | $< 0.4$ W peak during actuation, $< 5$ mW while waiting for command |
| Fiber termination | 1.25$\times$10 mm ferrule, replaceable without disassembly of actuator |
| Stray light treatment | Diffuse black paint on upper surfaces of fiber positioner. |

The fiber positioners selected for BigBOSS will be assembled by the China USTC group, who has experience designing and manufacturing the actuators for the LAMOST project. Several variants of the LAMOST actuators, specifically designed for BigBOSS, are currently under test at USTC. Current BigBOSS designs include a 10 mm diameter (12 mm pitch) actuator, as well as a 12 mm and 15 mm diameter variant. The 12 mm diameter version (see Figure 5.14) has a measured positioning repeatability (precision) of better than 5 $\mu$m. Key changes to the LAMOST design include installation of smaller diameter motors with co-linear axes, and a redesign of the gear system. The LBNL/USTC team is currently working to achieve a 10 mm diameter (12 mm pitch) actuator. The Instituto de Astrofísica de Andalucía (IAA-CSIC, Granada, Spain) in collaboration with the company AVS is in parallel working on a separate design for a BigBOSS 10 mm diameter actuator. The two designs will be reviewed and, finally, the best of each will be used for the BigBOSS actuator. IAA-CSIC/AVS has extensive experience with the design, construction and testing of a high precision fiber positioner prototype for the 10 m Gran Telescopio Canarias.

The choice of power and command signaling architecture for the robotic actuators is driven by packaging constraints. LAMOST experience shows that fiber, power and command line routing space is at a premium on a high-density robotic focal plane array. LAMOST opted to implement a hybrid wire/wireless scheme, in which only power lines and fibers were connected to each actuator and commanding was implemented by a ZigBee® 2.4 GHz wireless link. Based on this experience, BigBOSS has baselined ZigBee wireless communication. We have discussed with UC Berkeley and LBNL people working with the ZigBee standards committee and they indicated that this is good application and will be more so with an upcoming revision to the standard. Five transmitters will communicate with 1,000 actuators. The thermal cover on the aft end of the corrector will serve as a faraday cage to contain RF transmission from the ZigBee array. Although ZigBee commanding is currently baselined, power-line commanding is also under consideration.

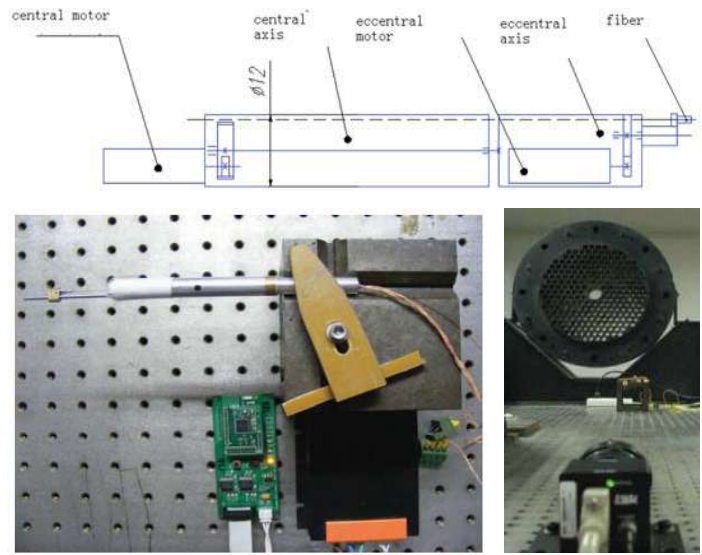

Figure 5.14: 12 mm diameter actuator under test at USTC

Figure 5.15 shows the baseline actuator control board as implemented by USTC, and the overall architecture. In order to reduce the size of the board compared to that of LAMOST, a smaller microcontroller (without integral ZigBee) was selected. The power converter of the LAMOST board was made unnecessary by selecting a motor driver, microcontroller and ZigBee IC that operate at the same voltage. Each group of 250 actuators will be powered by one dedicated 250 W power supply.

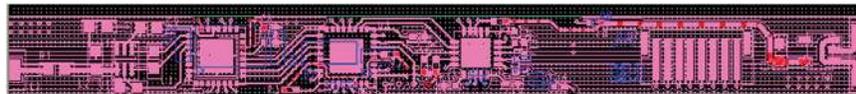

Figure 5.15: BigBOSS wireless actuator control board is 7 mm wide, with 4-layers. This USTC design is simplified compared to that of LAMOST.

### 5.3.5   Fiber Positioners Placement

Each fiber tip must be positionable within a disc (patrol disc) in order to gather the light of a targeted galaxy. In order to cover all the focal plane, the patrol discs must overlap as shown in Figure 5.17, top-left panel. This is possible thanks to the fiber being mounted on an arm which can protrude from the actuator chassis envelope. The positioning software must be clever enough so as to avoid collisions between arms of different actuators (simulations performed at IAA-CSIC show that this is possible). The patrol disc is necessarily flat because of the positioner characteristics, but the fiber tips should lie on a convex spherical focal surface, with a radius of curvature of 4000 mm and a diameter of 950 mm. Two questions arise:

**5.3.5.1  What is the best position of the disc with respect to the spherical focal surface?**  Placing the disc tangent to the sphere is not ideal because the borders of the disc would be affected by defocusing. The same is true if the circumference of the disc is embedded in the spherical surface; in this case the center of the disc would suffer the defocusing. The best position must be somewhere between these two extreme positions, and we assume it to be that position for which the defocus is the same at the center and at the border of the patrol disc (other positions could be used, with little practical difference). Figure 5.16) illustrates the trades. For a 4000 mm radius of curvature, and 6.93 mm patrol radius, and imposing $W = S$ (see Figure 5.16), the amount of defocus is identical at the center and at the border of the patrol disc. The best position is found with the patrol disc 2.5 $\mu$m away from being tangent to the focal surface. It also tells us that the defocus at the center and border of the patrol disc is also 2.5 $\mu$m.

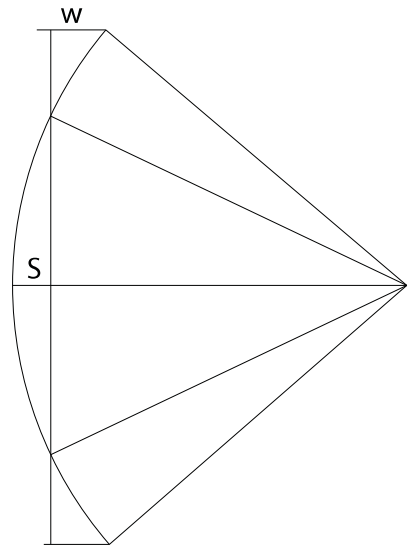

Figure 5.16: Cross section of a patrol disc (the vertical line) when intersecting the focal surface (curved line). $S$ and $W$ are the deviations in and out of the focal plane.

**5.3.5.2  Is it possible to distribute the actuators uniformly over the spherical surface?**  Here, "uniformly" means that the distances between the centers of one patrol disc and its six neighbors (a hexagonal pattern is assumed) is the same over the whole focal plane. A sphere can not be tessellated with uniform size hexagons, thus a compromise must be accepted and the final pitch between actuators will necessarily vary across the focal plane. The task is then to find a distribution as uniform as possible over a sphere, and, ideally, a distribution which can be easily transferred to a drilling machine for fabrication. The process adopted here is to stretch a flat, uniform distribution of hexagons onto a sphere. Figure 5.17 shows two possible types of deformations that can be used. The bipolar mapping follows the opposite process to that of mapping a portion of the Earth onto a plane map— the regions close to the poles (at the top of the figure) have a greater density than those close to the center. The multipolar mapping yields a different distribution, which gives a rotation-invariant pattern about the center, thus a slightly more uniform distribution. Other distributions could be used (for example an orthogonal projection, or a central projection centered at the curvature center), but it is found that the multipolar mapping gives the best results, which means that the pitch varies *less* across the focal plane.

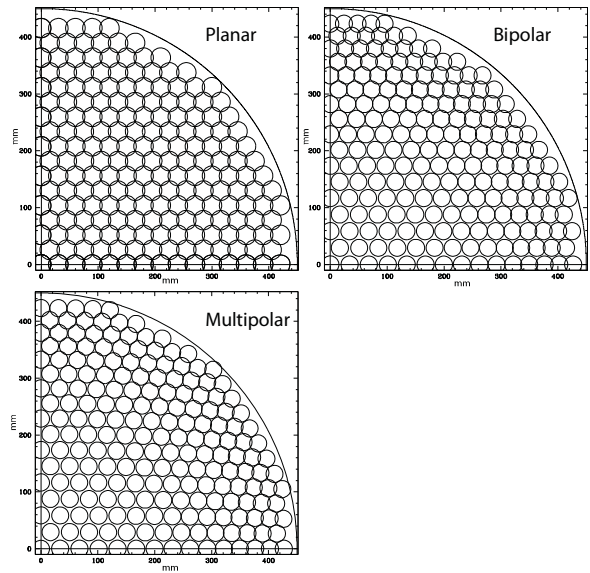

Figure 5.17: Distribution of patrol discs over a flat disc (top left), over a spherical cap with the bipolar mapping (top right) and with the multipolar mapping (bottom left). The radius of curvature of the spherical cap is greatly reduced here (600 mm) in order to exaggerate the deformations. The patrol discs cannot be nested properly in the case of the spherical cap, thus their size is arbitrary. The picture is meant to just give the idea so only one quarter of the focal plane is shown, the rest being symmetric.

For the spherical focal surface of 950 mm diameter and 4000 mm radius of curvature with a 12 mm actuator pitch (5549 possible locations for actuators and other specialized fibers), the multipolar mapping has a peak center-to-center difference of 26 $\mu$m; the orthogonal projection yields 78 $\mu$m, the bipolar mapping yields 79 $\mu$m and the central projection 153 $\mu$m. Thanks to the large radius of curvature and relatively small diameter of the focal surface, the differences can be mitigated to the order of tens of microns, but they cannot be neglected and will make the machining of the focal plane challenging because no regular pattern can be followed. We note that the anti-collision software algorithm for moving the actuators also must take into account the varying safe distances across the focal plane.

### 5.3.6 Guide Sensors

The Mayall telescope control is expected to point the telescope to within ∼3 arcseconds of the desired observation field. The BigBOSS star guidance system (SGS) is required to assist in telescope pointing at levels below 10 mas and ensure each of the optical fibers is located to within 15 $\mu$m of the desired target on the sky. Trade studies between two star guider designs are in progress. Regardless of the final design, the system must contain at least two viewing fields with radius of 30 arcsec. This will allow the SGS to determine the current pointing of the telescope once the Mayall control has finished slewing to a new location. By comparing an observed star field to a star catalog (NOMAD, for example) the current telescope pointing can be determined.

The system must also be large enough to ensure that several guide stars are available for tracking. The resolution of the star centroids must be better than several microns. The difference between observed and desired telescope pointing are then sent back to the telescope control system for adjustment. Finally, the fibers can be arranged relative to the observed pointing direction. The telescope is then updated periodically with correction requests for telescope pointing from the tracked star locations.

The two designs being considered have both been used in other systems. The first and more common design incorporates imaging sensors within the focal plane. A baseline design would be four optical CCDs located in each focal plane quadrant. Despite the added complexity of optical sensors on the focal plane, it provides a relatively stable location between fiber positioners centers and the guider. Figure 5.18 shows an example of this layout on the LAMOST telescope.

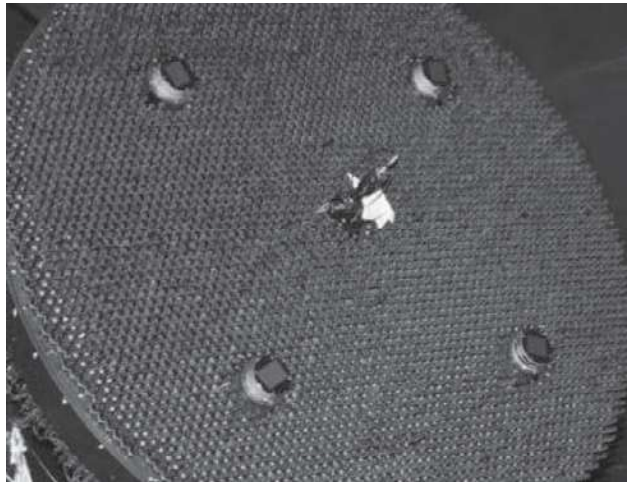

Figure 5.18: Photograph of the LAMOST focal plane with star guiders.

A second design is similar to that deployed in SDSS-III (Figure 5.19). All star guidance would be obtained via imaging optical fibers that are fixed in the focal plane. The fibers would transport star field images to remote cameras. At least two of the imaging fibers would need to be at least 30 arcseconds in diameter in order to acquire the current pointing direction. This design offers focal plane simplicity as it simply replaces approximately 20 science fibers with coherent fiber bundles and transmits the star images to a remote location.

This design is limited by light losses in the fibers and cost consideration.

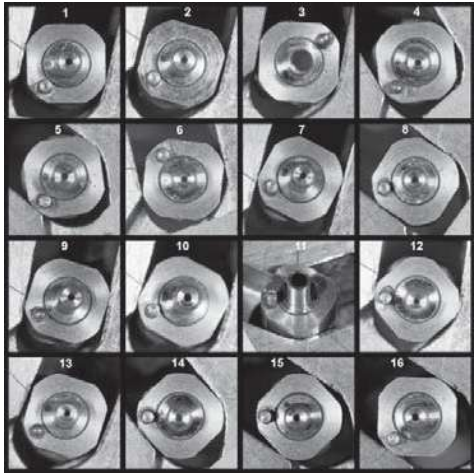

Figure 5.19: Photograph of the BOSS guider fibers. Two are large field fibers for star acquisition.

In either design, at least 240 arcmin$^2$ of sky would need to be covered by the fixed imaging fibers or the guider sensors. This area ensures that enough guide stars would always be available in a magnitude range both sufficiently bright for detection and within the dynamic range of the sensor. Using the average star density over a 10 degree radius at NGP, there are 0.14 stars per arcmin$^2$ in the magnitude range $15 < g < 17$ or 0.07 stars per arcmin$^2$ in the magnitude range $14 < g < 16$. The star density can be as much as a factor of two lower than this average. In these case, sufficient guide stars should still be available. Additionally, in acquisition mode, the guidance system can integrate for a much longer period of time. This deeper observation will provide significantly more stars in the acquisition field. Saturated stars in the field still provide useful information in the pattern detection algorithm. In the worst case scenario when there are not at least two quality guide stars visible, the guidance system can request a small pointing adjustment and all fibers will be repositioned. The guider can also run in a mode of longer integration allowing it to track fainter stars at a slightly reduced update frequency. In order to maximize live time, the system will be designed to keep the operator informed of the quality of the guidance signal. If the quality deteriorates or is lost, new fields in another part of the sky will be recommended.

### 5.3.7   Focus Sensors

Two separate instruments will be deployed to monitor the telescope focus. First, a Shack Hartman sensor will be installed in the center of the field of view. This is a well known technology and will provide wavefront errors.

Second, a focus sensor comprised ~11 steps of viewing above and below focus in the focal plane. The defocus steps are provided by varying thicknesses of glass above the imaging sensor. Nominal steps are 0, ±50, ±100, ±250, ±500 and ±1000 $\mu$m. Stars imaged above

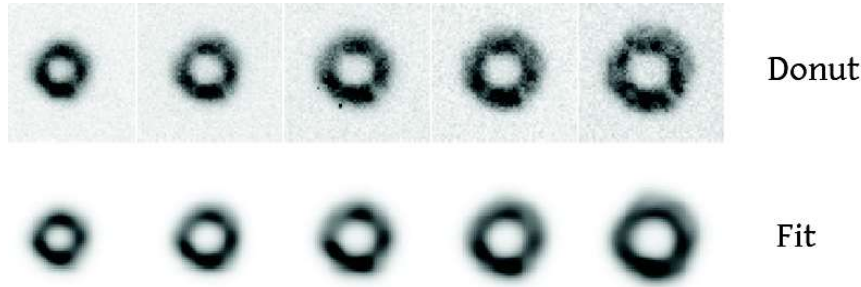

Figure 5.20: Star images taken at the Blanco telescope and the associated Zernike expansions.

<div align="center">Table 5.11</div>

| | |
|---|---|
| $Z_4$ | Defocus |
| $Z_5$ | Astigmatism x |
| $Z_6$ | Astigmatism y |
| $Z_7$ | Coma x |
| $Z_8$ | Coma y |
| $Z_9$ | Trefoil x |
| $Z_{10}$ | Trefoil y |
| $Z_{11}$ | Spherical Abberation |

and below focus will form an annular shaped pattern. Analysis of these many donut shapes will provide corrections needed in focus.

The sensor technology used in this focus sensor will mirror that of the fine guidance star sensors. The first option is a single imaging sensor (CCD) in the center of the field of view. The other option is several imaging fibers each positioned at varying positions above or below focus. Focus information derived from the sensors will drive the six-axis corrector barrel hexapod to perform a focus adjustment at an update period yet to be determined.

The focus sensors will image stars above, below and in focus. The current focus and alignment of the telescopes can be determined from the coeffficients of a Zernike expansion of these images (Eq. 5.1). Table 5.11 shows the optical meaning of several Zernike terms. The in focus star images provide seeing information that assists in the above and below focus image calculations. Figure 5.20 shows an example taken with the Mosaic 2 camera at the Blanco telescope.

$$W(u,v) = \sum_{i=4}^{i=37} c_i Z_i(\rho, \theta) \tag{5.1}$$

In BigBOSS, donut images will be captured and processed with a frequency of around one minute. After an ∼15 minute data integration, many measurements of focus and alignment (changes) will be available. The hexapod can then apply any needed corrections to the optical system during the data readout period.

### 5.3.8   Fiber View Camera Fiducials.

As described in the fiber view camera section, a set of fixed fibers in the focal plane are used as fiducials. The number and deployment await detailed studies from the development of the view camera fiber position reconstruction code. At the moment, it is thought that these fibers would be illuminated by lamps in the focal plane region, saving routing them off the telescope structure.

### 5.3.9   Thermal Control

Source light is collected at the prime focus by 5,000 robotically controlled actuators. Each actuator has a peak power of 0.4 W while actuating, and an idle (waiting for ZigBee command) power of roughy 2 mW. On average, each repositioning of the array is estimated to dissipate 150,000 joules, which could raise the temperature of the focal plane assembly by roughly 1°C. This temperature increase is not negligible, and would be expected to degrade telescope seeing unacceptably. We are trying to better estimate these numbers.

Other potential heat sources are guider and focus sensor electronics, lamps for fiber view camera ZigBee base stations.

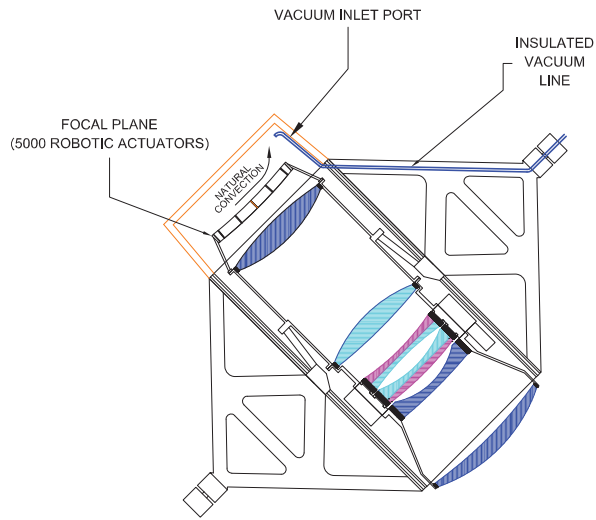

Figure 5.21: Heat generated by the fiber positioners is capped, and exhausted by an insulated vacuum line.

The cooling approach adopted on BigBOSS employs an insulated cap behind the focal plane, and an insulated suction line to draw away warm air from the focal plane. Figure 5.21 shows the nested corrector mount on the radial spider vanes. The corrector moves within a barrel assembly capped at the top by an insulated cover. Heat generated in the focal plane region is removed by natural convection (gravitational pumping) and contained by the cover. An insulated vacuum line located at the top corner of the cover removes warm air directly from the top of the corrector, and removes it from the dome (where it is released into the atmosphere downwind of the telescope). Ambient air is drawn into the corrector barrel through the gap between the corrector and outer barrel. Flow of air into a vacuum port is essentially irrotational (potential) flow, and does not generate vorticity and turbulence, as would an air outlet line.

Forced air cooling was rejected due to due to its consequences for seeing, and glycol loops were rejected due to risk to the primary mirror.

## 5.4  Fibers

### 5.4.1  Overview

The fiber system consists of 5000 fibers, ∼30 m in length. The input ends are each mounted in one of a close-packed, focal-plane array of computer-controlled actuators. The fibers are grouped to reach each of ten spectrographs at their output ends. The output ends are held in precision-machined ("V-groove") holder blocks of 100 fibers each, five blocks per spectrograph, to align the output fiber ends with the spectrograph slits. The planar-faced fiber input ends are each placed by an actuator with 10 $\mu$m accuracy. Each fiber input end can be non-destructively removed and replaced from its actuator assembly. The fiber run uses guides, trays, and spools to reach from the focal plane to the spectrograph room. To facilitate installation and maintenance, the fiber system concept includes intermediate breakout/strain relief units and fiber-to-fiber connectors within the fiber run. The fiber termination at the spectrograph input is a linear arc slit array containing 500 fibers. The arcs are modularized into sub-slit blocks of 100 fibers. The fiber system and its requirements are summarized in Table 5.12. Key performance and technology issues are discussed in detail hereafter.

### 5.4.2  Technology and Performance

Fiber throughput can be affected by transmission losses in the glass of the fiber, and losses at the fiber ends due to polishing imperfections and surface reflection. Losses cause increased exposure time and have the effect of limiting the total sky coverage during the survey life. Therefore fiber losses are a critical issue in the BigBOSS design. Low-OH silica fibers such as Polymicro FBP or CeramOptec Optran (Figures 5.22 and 5.23) are well matched to the desired pass band and have a minimum of absorption features that are inherent in high-OH, UV enhanced fibers. The fiber ends will be treated with AR coatings so that light loss at each fiber end can be reduced from ∼5% to <1.5% each (see Figure 5.24). The performance of every fiber will be tested by a group independent of the manufacturer.

Light incident on a fiber at a single angle will exit the fiber with a distribution of angles. Consequently, a cone of radiation entering the fiber at a certain focal ratio will exit the fiber spread into a smaller focal ratio, i.e., suffer from focal ratio degradation (FRD). The FRD is caused in part by imperfections in the fiber manufacturing process and by the quality of the fiber-end mechanical treatment, e.g., bonding and polishing stresses induced on the fiber's terminus. Actual measured FRD for a selection of fibers made for BOSS are shown in Figure 5.25. We use values from this experience in establishing our performance parameters. FRD is exacerbated by stresses, bends and micro-cracks caused by fiber handling and routing. Demonstrated control of FRD is important in order to achieve the desired throughput because light distributed beyond the acceptance of the spectrograph may be lost or scattered. Quality control inspection will be used to verify the net FRD of each fiber so that an accepted fraction of f/4.5 input flux will be projected within the f/4.0 acceptance of the spectrograph including an allowance for the fiber angular output tolerance. critical for this relatively low-resolution spectral application where the spectral resolution requires only a modest sub-aperture of the grating.

Table 5.12: Fiber Specifications.

| **Fibers** |
|---|
| Low OH fused silica (340–1060 nm) |
| Core 120 $\mu$m diameter |
| Cladding and jacketing combined outside diameter < 240 $\mu$m |
| Fiber performance robust to repeated actuator flexing (30k cycles) |
| Focal ratio degradation: > 90% energy within f/4.0 output for f/4.5 input |
| **Input end at focal plane** |
| Output face, flat polished, AR coated (340–1060 nm) <1.5% total loss |
| Ferrule terminated, removable from actuator |
| Ferrule 5 $\mu$m axial position accuracy in actuator |
| Fiber end angle $\pm0.25°$ |
| **Fiber run** |
| Length <30 m |
| Bulk transmission |
|     340–450 nm > 50 − 70% |
|     450–550 nm > 85% |
|     550–1060 nm > 90% |
| Sub-bundle maximum cross-dimension for focal plane routing 50 mm |
| Bundle performance robust to repeated flexing (30k cycles) |
| Coupling connector |
|     >100 fibers per connector |
|     <2% loss |
|     Verified performance life 100 mates |
| **Output end at spectrograph** |
| 500 fibers per slit assembly |
| Slit height 120 mm |
| Slit radius of curvature 330 mm |
| Fiber ends within $\pm45$ $\mu$m of spectrometer slit radius of curvature |
| Slit made of 5 blocks each with 100 fiber units |
| AR coated (340–1060 nm) <1.5% loss |
| Provision for diffuse back illumination of fiber ends |
| **Environmental** |
| Input end and fiber run operational temp range -10 to +20°C |
| Output end operational temp range 15 to 25°C |

We also consider the potential for FRD over the course of the thousands of random motions of the fiber positioner that represent the observation lifetime. Propagation of *ab initio* microcracks as the fiber is flexed during actuator motion may lead to a time dependent degradation of transmission efficiency. Various fiber types differ in their cladding overcoats, which according to vendors can affect flex performance. The Polymicro fibers used for BOSS, a hard clad silica with a single hard polyimide overcoat, have proven FRD robust to hand insertion flexing cycles. CeramOptec makes a fiber construction to minimize internal fiber stresses by using a two-layer clad (hard then soft glass) and a two-layer coat (hard then soft plastic). We will conduct degradation tests for the different fiber constructions to determine their life FRD properties given the mechanical requirements of the actuator rotation cycles.

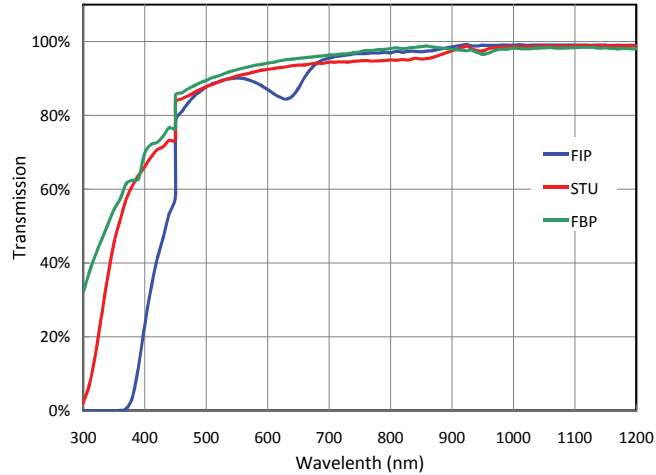

Figure 5.22: Polymicro 30 m length fiber transmission comparison for three types of fiber, FIP, STU, and FPB. BigBOSS expects to use the low OH content FPB fiber.

The fibers will also be flexed in their bundled run assemblies as the telescope slews over the sky. These repeated motions may also induce worsened FRD. Bundle and sub-bundle bending will be constrained to the rated long-term life radii by using guides belts, rails and soft clamps. Stress propagation to the fiber ends cause by friction induced wind-up over many motion cycles will be mitigated by using spiral wind construction and low friction sleeves over the sub-bundles. A fiber bundle assembly mock-up will be exercised over the designed routing system to verify its life performance.

### 5.4.3  Positioner Fiber End

At the focal plane, each fiber end is terminated individually to a positioner. The termination will be made by bonding the fiber into a ferrule and then finishing the optical surface (Figure 5.26) with flat polishing. To maximize throughput, AR coating will be applied to the polished fiber ends. A low temperature ion-assisted-deposition coating process will be used to avoid compromising the fiber/ferrule bond since the coating must be applied after fiber/ferrule assemblies are polished. The ferrule will be coupled to the metal actuator arm using a removable interface that provides the required 5 $\mu$m axial precision for matching the focal surface. The ferrule-actuator interface must not induce thermal stress on the fiber tip over the broad thermal range found at prime focus. A low-stress semi-kinematic fitment will be used that accounts for the variable amount of material removed during optical polishing. The lateral fiber-end positioning accuracy is less critical because the fiber tracking camera will calibrate the fiber's lateral position. Nonetheless, the position needs to be repeatable and stable between camera calibrations.

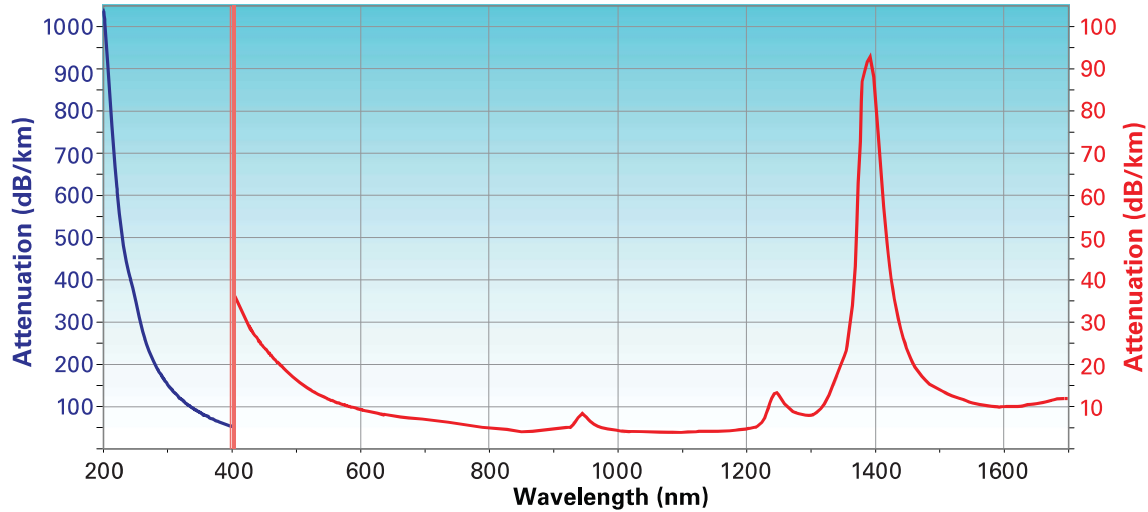

Figure 5.23: Polymicro FPB Low OH fiber attenuation from 200 to 1700 nm. The red line corresponds to an attenuation scale from 0 to 100 dB/km. The blue line corresponds to an attenuation scale from 0 to 1000 dB/km.

Fiber performance is affected by surface stress and damage both during installation of the fiber into the ferrule and during polishing. Low stress fiber assemblies have been achieved using combinations of specific fiber glasses/buffers, ferrule ceramics/steels and specialized epoxies. Critical factors include the balance of material's coefficient of thermal expansion (CTE) and adhesive shrinkage, strength and modulus. We anticipate using a polyimide buffered fiber as polyimide survives the AR coating process temperature. Polyimide also has high diametrical and concentric tolerances and high stiffness, which allows for bonding the un-stripped, buffered fiber within the ferrule and for good optical quality end-polishing, respectively. Depending on the final design of the positioner ferrule, either the fiber/ferrule CTE's will be matched and used with a relatively brittle epoxy (e.g. Epotek 353-ND) or the fiber/ferrule CTE's will differ and be used with a relatively elastic epoxy (e.g. Epotek 301-2). The brittle epoxy method allows for a thermally accelerated cure which can yield manufacturing efficiencies. We plan to verify the fiber ferrule fabrication process as well as its performance over lifetime temperature cycles.

Protective sleeving will terminate at each ferrule assembly and will be bonded in place, serving as reinforcement at the high stress region where the fiber enters the ferrule assembly. The sleeve should be sufficiently flexible to allow unimpeded movement of the actuator and must allow repeated movement within the actuator's guide channel while having adequate wear resistance. Woven polyimide sleeve (Microlumen Inc.) or close-wound PEEK polymer helical tubing are our chosen candidate sleeving types, The jacketed fibers from a localized region of actuators will be collected into sub-bundles of 100 fibers. The collection ports of the sub-bundles will be suspended from a fiber-harnessing support grid located near the aft of the focal plane's back surface. At the support grid, each sub-bundle will enter a protective sheath to commence the fiber run.

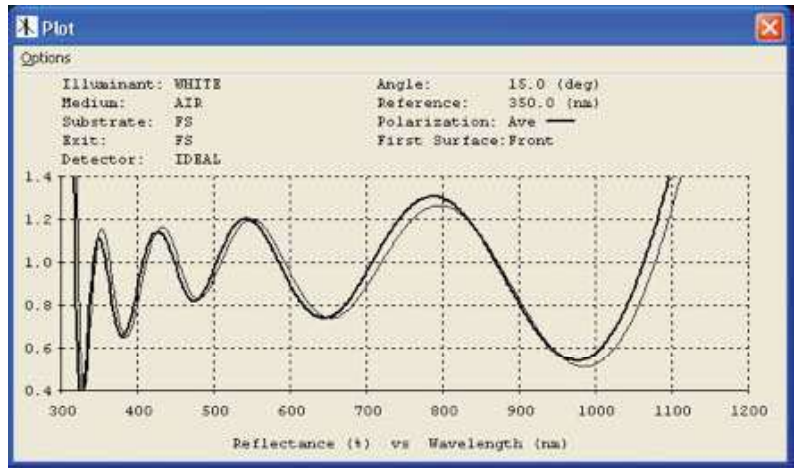

Figure 5.24: Modeled AR coating at 0°‖and 8°‖ incidence (by Polymicro on FBP).

### 5.4.4   Fiber Run

Fibers and actuator electrical power lines will run in a cable bundle that starts at a harnessing support grid located behind the focal plane and then runs across the secondary support vanes, down the telescope structure toward the primary cell, through a new access port in the primary mirror shutter's base, and into the telescope elevation bearing. The fiber run feeds within an existing large air-conditioning conduit, exiting about the polar bearing with a spool loop, and then entering the spectrograph room where the bundle branches to feed each spectrograph assembly. Guides, spools, and link-belts will constrain bundle motion to limit twist and enforce minimum bend radii. We plan to use standard outer cabling products, such as PVC clad steel spiral wrap (e,g, ADAPTAFLEX) and furcated sub-bundles in a segmented polymer tube (e.g. MINIFLEX) that exhibits a desirable trade of mechanical properties such as flexibility, toughness, crush and extension resistance, and minimum bend radius. Figure 5.27 shows a cross section through the cable.

The full cable uses an Aramid yarn tensile element to limit length extension. The Aramid yarn is built up with a polymer coating to a diameter around which the loose-fiber carrying furcation tubes can be sprial wound in a uniform radial packing. The spiral avoids cumulative tension in the end terminations sue to differential length strain on the furcation tubes when bending the conduit. The helical cable core is wrapped with a protective ribbon of polymer tape and a hygroscopic gel layer that maintains a dry environment within the cable volume. We will evaluate a range of cable size options while considering routing constraints and ease of fabrication, assembly, and maintenance. An appealing approach is to use five primary cables consisting of industry standard 33 mm conduits, each supporting ten 5 mm diameter furcation tubes carrying 1000 loose-packed (<80%) fibers.

### 5.4.5   Breakout-Relief Boxes and Fiber Connectors

Cable breakout and strain-relief boxes will be located at the primary focus support ring. The boxes contain free loops of fiber that equalize differential tension within the main cable and isolate longitudinal fiber movement from causing stress at the fiber terminations. The fiber run is also rearranged at the boxes into a non-obscuring profile for the run across the light-path on the spider structure arms. Cable breakout-relief boxes will also be located in the spectrograph room to divide fiber runs with each conduit to their respective instruments and to provide a fiber length reservoir for stress free routing to the instrument slit assemblies.

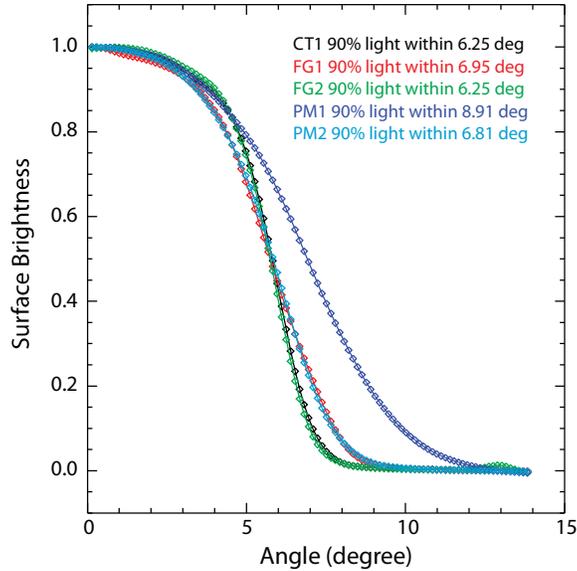

Figure 5.25: The surface brightness profiles for 5 prototype BOSS fibers with identical material requirements and core size (120 μm) as BigBOSS. The fiber vendor and manufacturer designated "CT1" was chosen for construction, and 3 other prototypes nearly met specifications on throughput, focal ratio degradation (FRD) and physical characteristics. As for BOSS, each delivered fiber will be mechanically inspected and tested for throughput.

We anticipate that including fiber to fiber connector(s) in the fiber run will ease fabrication, integration, installation, and schedule demands. A fiber connector will allow both the focal plane and the spectrograph to be fully and independently assembled and tested off-site. For example, the exit fiber slits can be aligned and tested with their spectrographs in the laboratory, and the fiber input ends can similarly be installed and tested in their actuators at the focal plane. However, the use of connectors will incur some optical loss. Bare fiber, index-matching gel filled connectors can exhibit losses < 2% and lenslet arrays or individual GRIN lens connectors can be limited to similar loss.

We are conducting trade studies on adapting commercial devices or constructing custom connectors. Commercial modules under consideration include: US Conec MTP connectors, presently in use on the BOSS project, that is available in standard sizes up to 72 fibers and 2) Diamond S.A. MT series connectors, in standard sizes up to 24 fibers, that have been ganged into larger multiples by LBNL for the ATLAS project. Alternates include custom bare-fiber or lensed connector designs based on integral field unit (IFU) schemes.

A key parameter for the coupler is the number of fibers per mating. Considerations include the fabrication cost, coupler size impact on the fiber run routing, and integration, test and service modularity. One logical unit would use 100 fibers each on 50 connectors where each unit corresponds to a spectrograph slit-blocks of 100 fibers, as described below. An alternative unit would use 500 fibers each on 10 connectors, where each unit supplies one spectrograph's slit. The number of fiber connector couplings over the project life is limited as the coupling will be made for testing the fiber run, the focal plane, and the spectrograph, and for telescope installation or maintenance. We anticipate that a proven lifetime of 100 couplings will suffice for the project - a value that is factors of several within the rated life of commercial connectors under consideration.

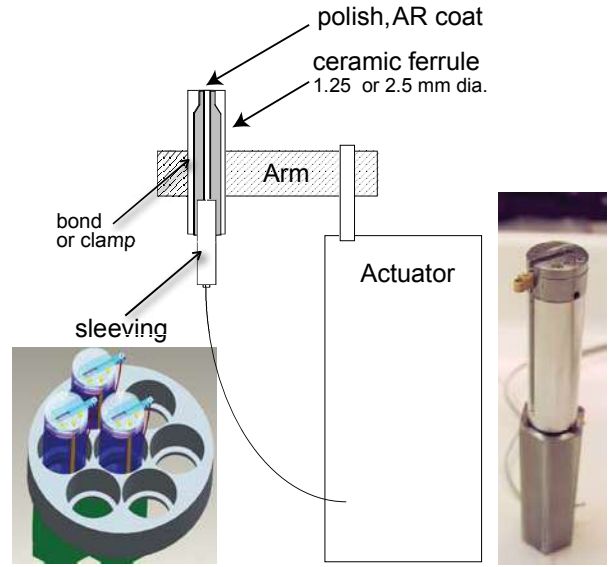

Figure 5.26: An $r$–$\theta$ actuator is shown at the bottom right and a simple array is at the bottom left. The illustration shows a fiber attached to the position arm. The fiber is first glued into a ferule (such as the zirconia ferrule shown in the image ,upper left) and the tip is then polished and anti-reflective coated.

Commercial vendors customarily deliver performance-verified fiber connector pairs after their assembly into an optical fiber bundle. We expect to obtain verified connectors for the long cable run and join these to connectors that 'pig-tail' to the focal plane and to the spectrograph slits, where the pig-tails are obtained as a verified pair on a fiber bundle and then cut and end finished for the focal plane and slit assembly. The location of the connectors will be determined following further study of the fiber routing scheme and connector methodology. The connectors are best used in clean and controlled environments for reliable coupling. We include an environmental enclosure at the junction to limit foreign debris or other environmental intrusions about the connectors.

### 5.4.6 Slit Array

The output end of the fibers terminate in 10 slit arrays, one per spectrograph assembly. Each slit array consists of a group of 500 fibers arranged in a planar arc specified by the spectrograph optical prescription. Fiber ends are directed toward the spectrograph entrance pupil and represent the illumination input, i.e., the spectrograph entrance slit (Figure 5.28).

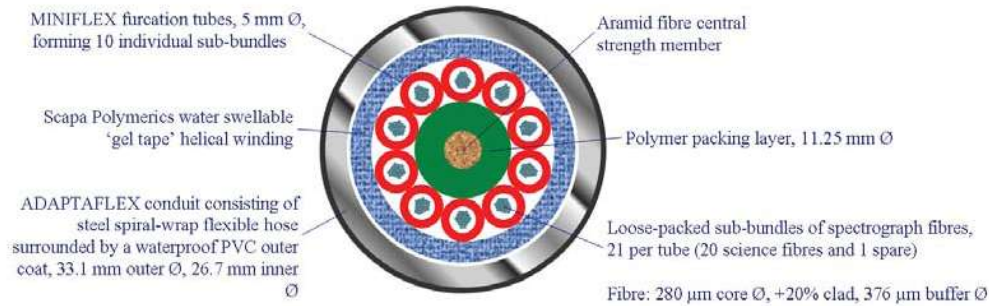

Figure 5.27: Proposed cable cross section. Five cables would be made with the above cross section, each carrying 1000 fibers. Ten furcation tubes are spiral wrapped circumferentially about the central strength member and then surrounded with protective layers.

The slit arc is concave toward the spectrograph with a radius of 330 mm to match the pupil. The fiber's center spacing of 240 $\mu$m is established by the spectrograph field size together with the desired dark regions between each fiber's spectral trace on the sensor. Optical tolerances demand a precise location for the fiber tips with respect to focal distance, i.e., the fiber tips must lie within 10 $\mu$m of the desired 330 mm radius input surface. Lateral and fiber center spacings are not demanding.

The slit array is a mechanical assembly that includes five blocks of 100 fibers each which are precisely arranged to a strong-back metal assembly plate. The plate provides the mechanical interface to the spectrograph and is installed using registration pins for accurate location. The assembly plate also supports and constrains each block's fiber bundle and terminates the bundles' protective sheaths. The subset 100-fiber blocks are the basic fabrication unit for the fiber system. The ends of the individual fibers are bonded into V-shaped grooves. The fiber ends are cleaved and then co-polished with the block surface. The V-grooves are EDM machined into a metal planar surface at radial angles that point each fiber toward the radius of curvature. Fiber jacketing is removed prior to bonding and terminated into a larger V-grooves and the jacketed fiber is supported by adhesive on a free bonding ledge to enforce minimum curvature radii and strain relief of the fibers before their entry into bundle sleeving. Following finish polishing, fiber support, and tested for throughput, FRD and alignment, the face of the fiber block will be AR coated. The method, materials and process for the block production follow the same considerations discussed for the fiber input end and will be verified through pilot development and test, including the impact of fiber bonding and finish schemes on throughput and FRD and the robustness of jacket termination, free fiber support and bundle termination.

Each slit array assembly also includes a provision to flood the spectrograph focal plane with continuum flux so that a spectral flat field can be obtained. We intend to install an illuminated 'leaky' optical fiber on the slit assembly plate that runs parallel to and nearby the slit. Lamp illumination of the leaky fiber will flood the spectrograph to provide a diffuse continuum field for spectral flat-fielding across local detector regions. An internal shutter in each spectrograph camera will back-reflect flux into the fiber slit ends so that the fiber tracking camera can calibrate the position of the fiber input ends on the focal-plane.

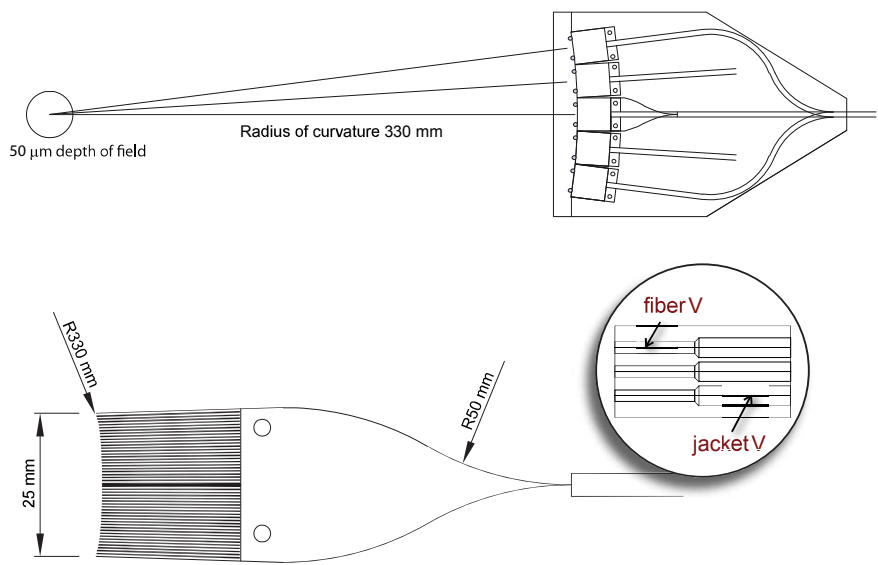

Figure 5.28: At the top is an illustration of 500 fibers focusing on the input of a spectrograph, forming the input slits. Below, a 100 fiber subset is glued in a plane and the fiber tips machined to the focal length of the spectrograph input. The inset shows V-shaped groves into which fiber and its jacketing are bonded. The fiiber tips are polished and AR coated after bonding.

### 5.4.7   Additional Fibers

Additional fibers will be included in the fiber system that will terminate at fixed positions on the focal plane. These fibers will be back illuminated by LED flood of the spectrograph shutter and provide fiducial marks for calibration of the focal plane geometry. The fiber system will also include spare fibers to allow for performance or damage mitigation during major maintenance episodes. The options for replacing damaged fibers is limited by the bonded design of the fiber slit array. Either spare fibers can be ganged and replaced as 100-fiber slit unit blocks or individual fibers can be fusion spliced within the fiber unit run. The Breakout-Relief Boxes can allow for a spare fiber reservoir so that fiber can be drawn from the boxes to the appropriate replacement location. We will establish a spare-fiber maintenance plan and method following further development of the fiber connector scheme.

## 5.5   Spectrographs

The spectrograph performance specifications are given in Table 5.13. These values lead to the optical design. The system is divided into three channels to enhance the throughput and decrease the complexity of each individual one. The overall efficiency is enhanced despite the addition of dichroics by selection of detectors, glasses, AR coatings and gratings optimized for each band. The moderate complexity of each channel allow compact packaging. This optimization will impact dramatically integration, test and maintenance procedures.

The Figure 5.29 shows the proposed architecture. The full bandpass is divided in three channels: blue (340–540 nm), visible (500–800 nm) and red (760–1060 nm). This separation is accomplished with dichroics each reflecting the shorter bandwidth and transmitting the longer one. Each channel consists of a two lens collimator, a grism and a six lens camera. A cooled CCD in a dedicated cryostat terminates the optical path. The pupil size is about 85 mm and the lens diameters vary from 80 mm to 120 mm. The lens thicknesses are constrained to be less than 25 mm, resulting in small lens volumes, which helps to keep the mechanic support simple and light.

Table 5.13: Spectrograph Optical Specifications.

| Geometry Specifications | | | |
|---|---|---:|---|
| Fiber diameter | | 120 | $\mu$m |
| Detector pixel pitch | | 15 | $\mu$m |
| Spectral detector elements | | 4096 | pixels |
| Spatial detector elements | | 4096 | pixels |
| Minimum resolution elements | | 3 | pixels |
| Demagnification | | 2 | |
| Fiber pitch (detector) | | 8 | pixels |
| Fiber spacing (slit plane) | | 120 | $\mu$m |
| Number of fibers (spatial) | | 500 | |
| Fiber f/# | | 4.5 | |
| Collimator f/# | | 4 | |
| Spectrograph channels | | 3 | |
| **Spectral Specifications** | | | |
| Bandpasses | Blue: | 340–540 | nm |
| | Vis: | 500–800 | nm |
| | Red: | 760–1060 | nm |
| Dispersions | Blue: | 0.488 | Å/pix |
| | Vis: | 0.732 | Å/pix |
| | Red: | 0.732 | Å/pix |
| Resolutions | Blue: | 3004 | |
| | Vis: | 2958 | |
| | Red: | 4142 | |
| **Optical Performance** | | | |
| End-to-end throughput | All | > 40% | |
| Throughputs w/ dichroics | Blue: | > 50% | |
| | Vis: | > 70% | |
| | Red: | > 70% | |
| Grating throughput | Blue: | > 80% | |
| at maximum | Vis: | > 80% | |
| | Red: | > 80% | |
| Encircled energy | All | > 85% | In 8 pixels |
| | All | 50% | In < 3.5 pixels |
| Scattered light | All | < 2% | w/o grating |
| Shutter | All | < 0.1% | Closed |
| | All | > 99% | Opened |

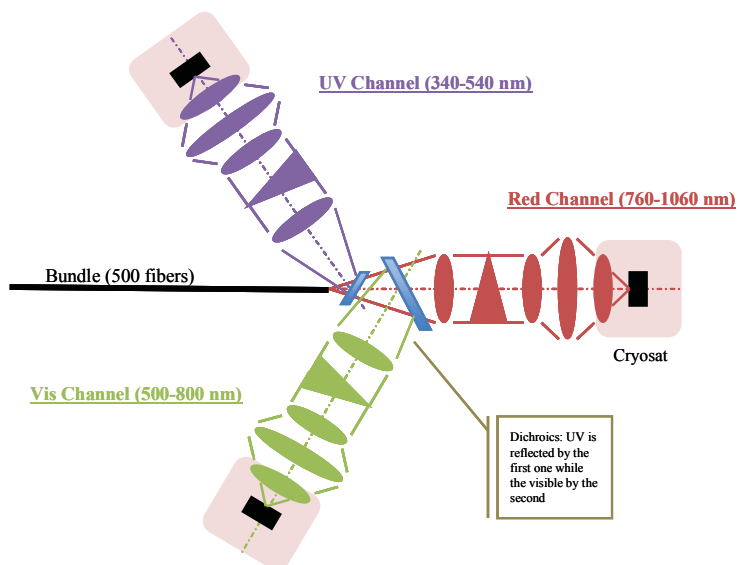

Figure 5.29: Schematic view of the spectrograph channel division.

### 5.5.1 Entrance Slit

The entrance slit of the spectrograph is made by the 500 fibers. They are aligned along a 330 mm radius circle creating a curved slit (Figure 5.30). The pitch of the fibers is 240 $\mu$m while its fiber core diameter is 120 $\mu$m. This configuration delivers a 120 mm long slit at the entrance of the spectrograph. The collimator accepts light within a cone whose axis that passes through the center of curvature of the slit to mimic the entrance pupil. The cone of each fiber is an f/4 beam, implying a 82.5 mm pupil diameter. Each fiber end will be located within $\pm 45$ $\mu$m to the 330 mm circle in the light beam direction.

### 5.5.2 Dichroics

The dichroics split the light beams from the fibers into the three bands. The transition between reflection to transmission permits the two parts of the spectrum to be matched by cross-correlation. Table 5.14 summarizes the specifications of the dichroics, and Figure 5.31 shows their configuration.

### 5.5.3 Optical Elements

The collimator is based on a doublet and, as mentioned above, the lenses all have reasonable diameters. The grating is within the prism's body. Each face of this prism is perpendicular to the local optical axis, which reduces aberration. The exit face of this prism has a spherical surface. Three doublets compose the camera. The last one is the entrance window of the detector cryostat. The f/2 beam at the detector favors a short distance between the last lens and the image plane. A flat entrance window for the cryostat would lead to longer distance, a less than optimal design. The current capabilities of the optical manufacturers allows us to use a multiple number of aspherical surfaces. In the current process of optimization, we decided to have one aspherical surface per lens. This is not seen as a risk, or even as cost driver, by several vendors. The proposed solution is very compact and elegant. As described further in the description of the structure, the entire spectrograph array will have a volume of about 2 m$^3$, impressive for 30 detectors and 5000 fibers.

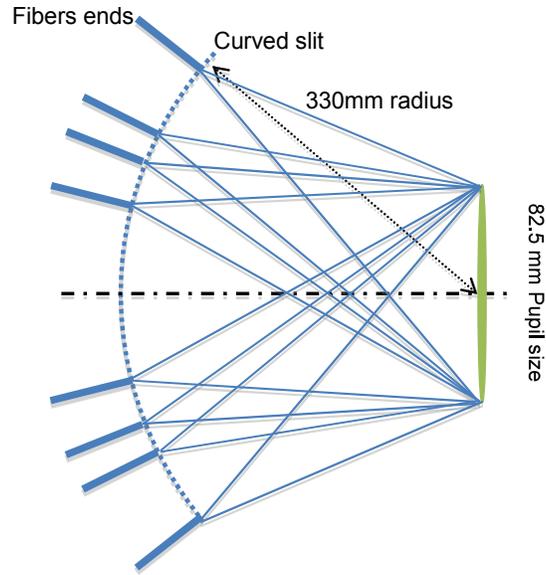

Figure 5.30: Schematic view of the optical interface to the fibers (only 8 fibers are represented).

### 5.5.4 Gratings and Grisms

The likely grating technology will be the volume phase hologram grating (VPHG) to ensure a high throughput. The lines density (900 to 1200 lines/mm), the beam diameter (80 mm) and the groove angle (12 to 18°) are fully compatible with the standard use of VPHG.

### 5.5.5 Optical Layout

Figures 5.32 to 5.34 present the layouts of the three channels of a spectrograph. Notice that all optical elements (at the right) are within a very small volume. The dichroics are at the left of Figures 5.33 and 5.34. This is a favorable mechanical implementation and is similar to the concept used for the VLT/MUSE instrument. Manufacturing and integrating the ten copies of each channel in a short period of time has been demonstrated by the MUSE project and the WINLIGHT Company.

### 5.5.6 Optical Performance

The first performance evaluation is the spot diagram. For BigBOSS, diffraction limited performance is not required. The fiber core is to be imaged onto four 15 $\mu$m pixels while the diffraction limit varies from 1 to 3 $\mu$m. Figure 5.35 shows wavelength versus field position spot diagrams for the three channels. The specification is to have 50% of the encircle energy (EE) within 3.5 pixels and more than 85% in 8 pixels. Figure 5.36 shows the 50% and 95% encircled energies for the three channels as a function of wavelength and field of view. The results for both performance metrics are summarized in Table 5.15. We note that we are lower than the specifications with less than 3.5 pixels for 50% EE and that 8 pixels contain 95% EE.

Table 5.14: Dichroics Specifications.

| Item | Value | Range | Comment |
|------|-------|-------|---------|
| **UV Dichroic** | | | |
| Transition | 520 nm | | |
| Reflection | $> 98\%$ | 340–500 nm | TBC |
| Transmission | $> 98\%$ | 540–1100 nm | TBC |
| Absorption | $< 1\%$ | 340–1100 nm | TBC |
| Length | 120 mm | | |
| Width | 30 mm | | |
| Flatness | $\lambda/4$ | | pk-pk over 30 mm patch |
| Working angle | f/4 | 0–13°$^{\parallel}$ | |
| Substrate | Silica | | |
| **Visible Dichroic** | | | |
| Transition | 780 nm | | |
| Reflection | $> 98\%$ | 500–760 nm | TBC |
| Transmission | $> 98\%$ | 800–1100 nm | TBC |
| Absorption | $< 1\%$ | 500–1100 nm | TBC |
| Length | 100 mm | | |
| Width | 80 mm | | |
| Flatness | $\lambda/4$ | | pk-pk over 30 mm patch |
| Working angle | f/4 | 0–13°$^{\parallel}$ | |
| Substrate | Silica | | |

Table 5.15: Number of 15 $\mu$m CCD pixels containing 50% and 95% encircled energy and the RMS variation on the detector.

| EE | UV | Visible | Red |
|-----|------|---------|------|
| 50% | 3.2±0.14 | 3.2±0.16 | 3.1±0.13 |
| 95% | 5.1±0.52 | 5.3±0.70 | 5.1±0.51 |

### 5.5.7 Shutter

A shutter will be placed between each spectrograph channel body and its cryostat. A clear aperture of 100 mm diameter is sufficient and a commercial shutter can be used. A candidate shutter can be found at http://www.packardshutter.com/

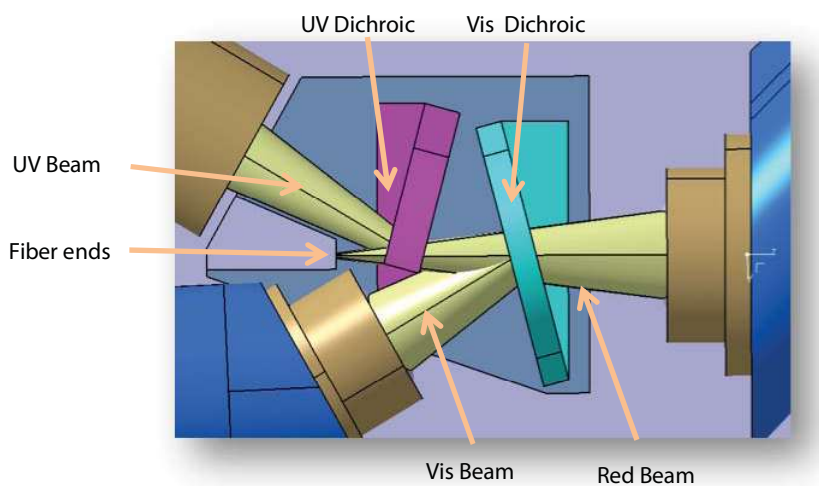

Figure 5.31: Dichroics configuration view.

### 5.5.8 Mechanical

**5.5.8.1 Optical elements support.** The optical elements are grouped in doublets. Each lens will be glued to one side of a doublet barrel. Each doublet will be integrated in the spectrograph channel body (see Figure 5.37). Mechanical alignment and positioning will be enough to insure the image quality. Since the entire system will be thermalized in the instrument room, the criterion on the differential thermal expansion is not be the driver in terms of image quality. The only time thermal stress of the glass is consideration is for transport and storage.

**5.5.8.2 Light baffling.** The spectrograph body will completely block external light. In the same way, the dichroic support will be a good place to block stray light. The only places were the light could leak into the path are the interface between the fiber and the dichroic body and the dichroic body and the spectrograph body. Interfaces with light traps will be designed to eliminate stray light contamination.

### 5.6 Cryostats and Sensors

The 30 BigBOSS cameras (10 for each channel of the instrument) contain a single CCD housed in a small cryostat.

### 5.6.1 Cryostats and Sensors

The preliminary requirements for the design of the cryostats are as follows. CCDs are to be cooled down to 160–170K and their temperature must be regulated within 1K. Cryostats include the last two lenses of the spectrographs and must allow CCDs to be adjusted within $\pm 15$ $\mu$m along the optical axis to maintain image quality. The design must be simple to give easy access to the instrument, it must require low maintenance and make fast replacements possible (typically, one cryostat to be replaced in less than 24 hours by 2 persons). Finally, the system once in operation must be insensitive to electromagnetic discharges.

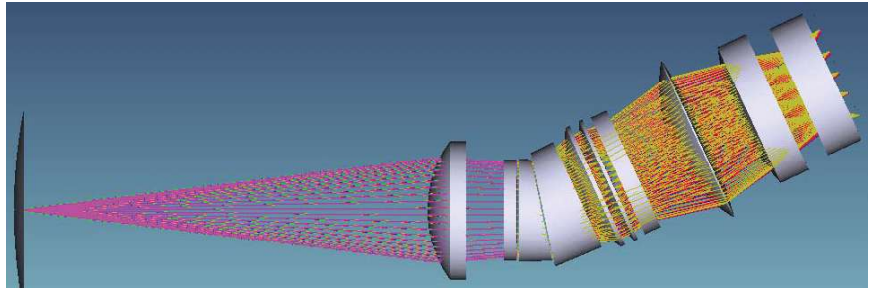

Figure 5.32: Optical layout of the blue spectrograph channel. This channel is fed with the reflected beam from the first dichroic, not shown.

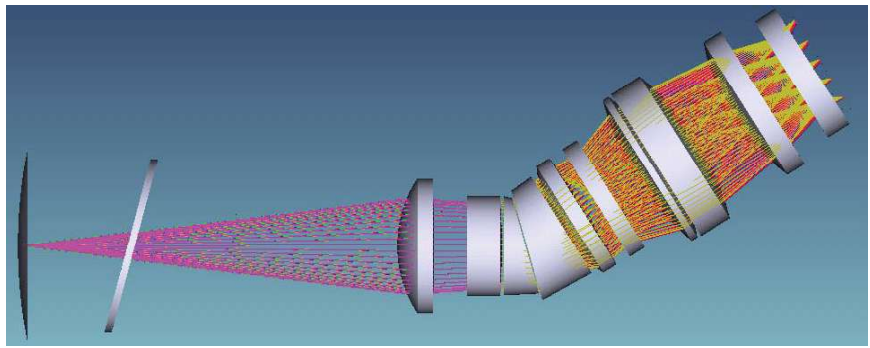

Figure 5.33: Optical layout of the visible spectrograph channel. This channel is fed with the reflected beam from the second dichroic. The transmitting first dichroic is shown.

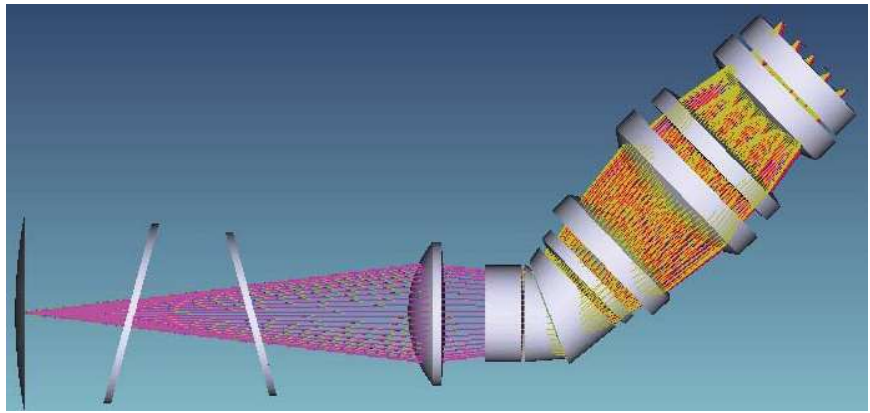

Figure 5.34: Optical layout of the red spectrograph. This channel is fed with the transmitted beam through both dichroics, which are shown at the left.

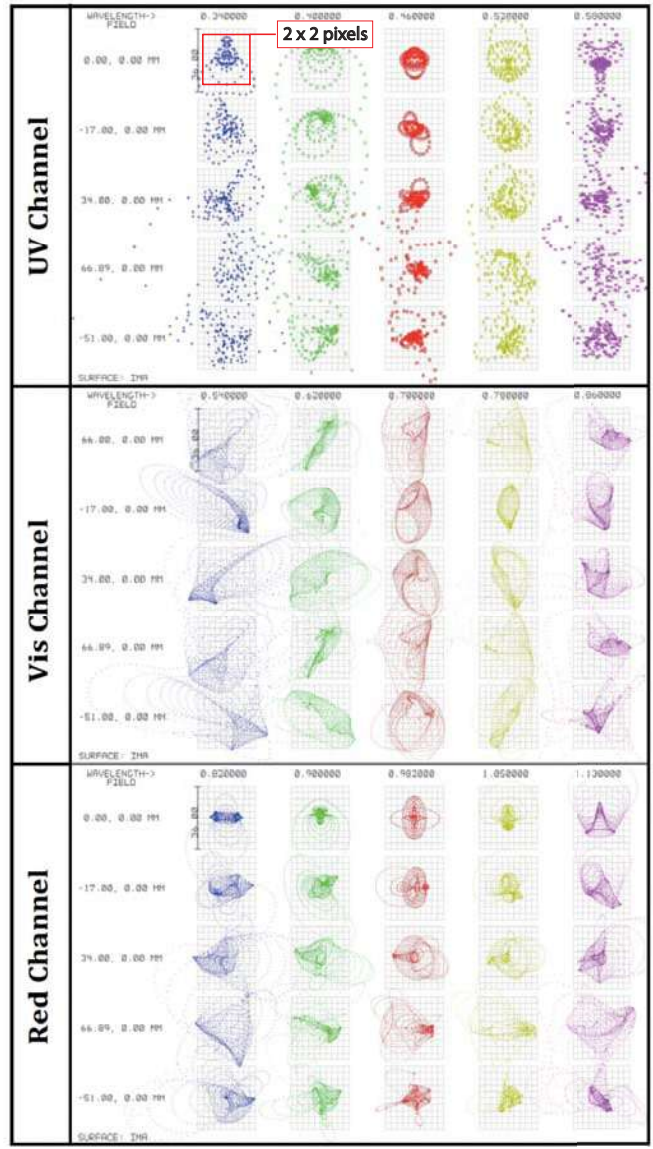

Figure 5.35: Point source spot diagrams for the three spectrograph channels for five wavelengths and five field positions.

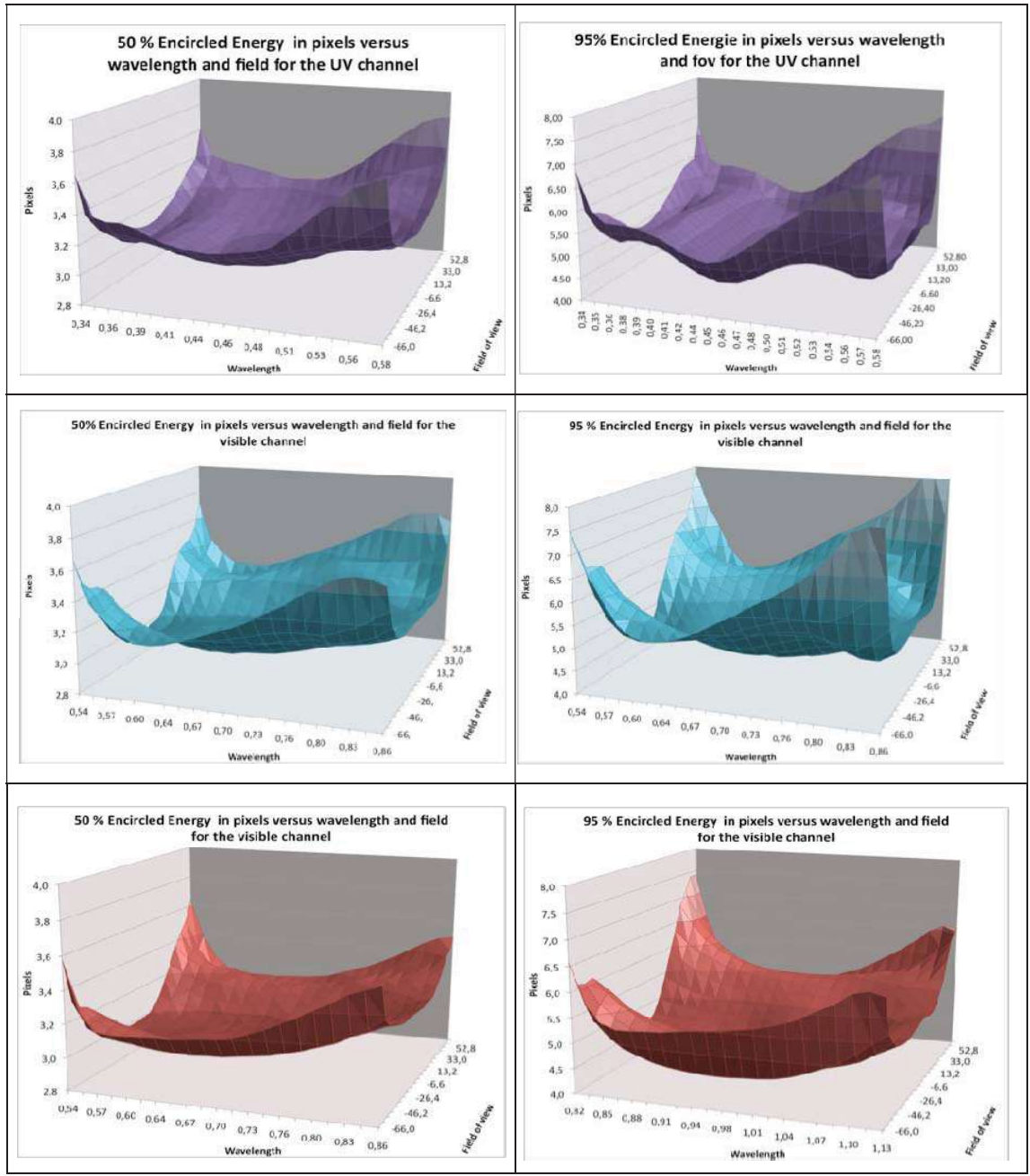

Figure 5.36: Encircled energy contours, 50% on the left and 95% on the right, for the three spectrograph channels as function of wavelength and field of view.

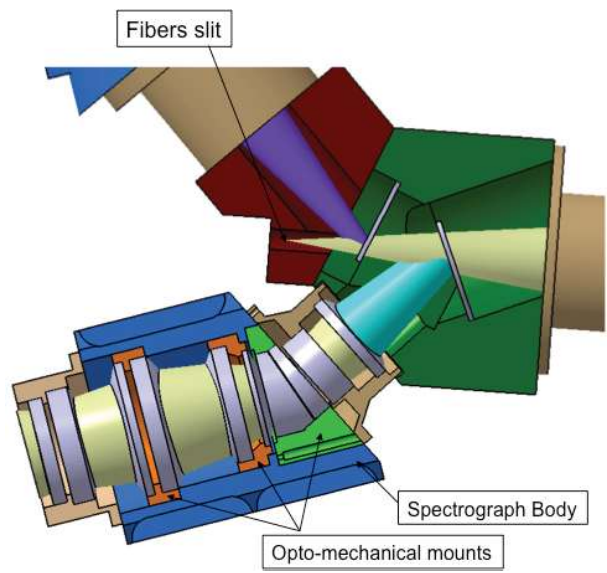

Figure 5.37: Spectrograph mechanics showing the dichroics box (green), visible channel structure housing two lens doublets and the grism (lower blue),and the cryostat with the final lens doublet.

One of the most important requirements is to have independent units in order to be able to react quickly in case of changes or failures. To produce cooling power for the 30 cryostats, we will thus use one closed cycle cryocooler per cryostat, each with its own CCD temperature monitoring. The above requirement led us to adopt the same mechanical design for all cryostats except for the support of the front optics.

### 5.6.1.1   Focal plane

The focal plane is determined by the optical configuration of the spectrographs and will be slightly different in each channel. The last two lenses of each spectrograph arm have to be integrated in the cryostat due to their short distance to the CCD plane. They will act as the window of the cryostat vessel. These lenses will be aligned (at room temperature) by mechanical construction. Each cryostat has to provide a mechanism to align its CCD under cold conditions. As a reference for the alignment, we use the interface plane between the last mechanical surface of the spectrograph housing and the front surface of the cryostat (see Figure 5.38).

The first lens, CL1, will support the pressure difference between ambient conditions and the internal cryostat vacuum, whereas the second one will be in vacuum. The lenses will be assembled in the cryostat front flange and fixed to the spectrograph. The assembly will use specific high precision parts to meet the alignment requirements given in Table 5.16.

The alignment of the cryostat part which supports the CCD will rely on the roll-pitch system developed for MegaCam at CFHT. The system is composed of a pair of outer flanges with 3 micrometric screws positioned at $120°^{\parallel}$(see Figure 5.39), inserted between the front flange and the moving part of the cryostat. In order to prevent any lateral displacement, locking will be provided by balls in V-grooves located inside the flanges. Once in position, the balls will be locked by a screw.

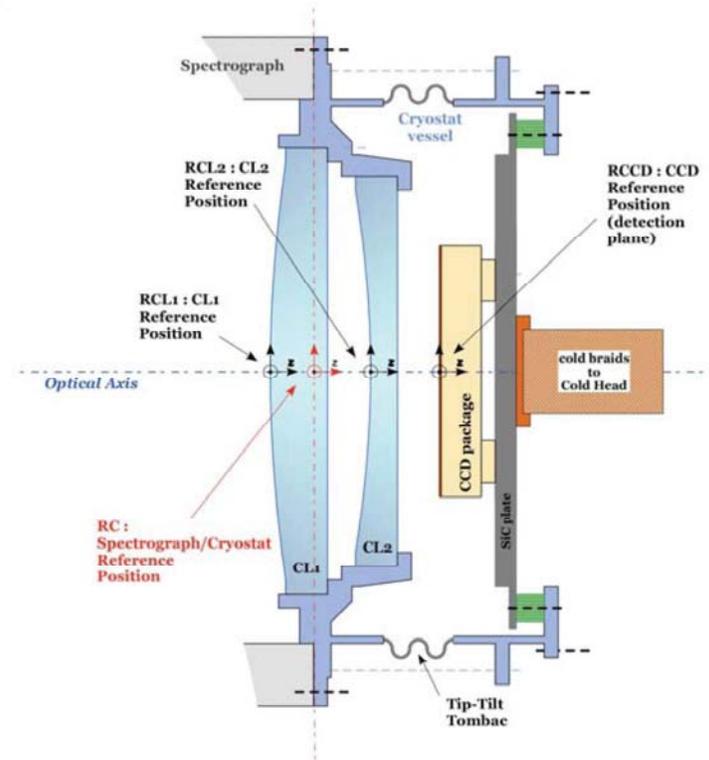

Figure 5.38: Positions and reference (RC) of the last pair of lenses and CCD plane of one spectrograph arm.

This system should allow us to align the CCD plane within 15 $\mu$m along the optical axis and within 1.5 arcmin in Rx and Ry. A design study of the mechanical assembly of the lenses and tip-tilt system has been performed with simulations at Irfu. The final validation of the design will require a cryostat prototype to be mounted and tested at Irfu. Final values of the lens and focal plane positions will be given by the spectrograph design studies.

### 5.6.1.2 Cryostat vessels

The cryostat vessel ensures the mechanical connection with the spectrograph, the thermal and vacuum conditions for the CCD and the interface with the control system and the CCD electronics.

The cryostat is a metal cylinder that will receive a front flange that integrates the last pair of lenses and the tip-tilt system, and a rear flange to support the cold head. The cylinder sides will be equipped with several connection pipes: one for the vacuum, one for the CCD flex connector and one or two for the electrical connection to the control system.

Figure 5.40 shows cryostats assembled on the three arms of a spectrograph, with the CCD electronics (black boxes), the cold heads (dark green), their compressors (small light

Table 5.16: Lenses Positioning Requirements.

| Errors | Cryostat vs Spectrograph | CL1 | CL2 |
|---|---|---|---|
| Along X or Y | $\pm 35\ \mu m$ | $\pm 50\ \mu m$ | $\pm 50\ \mu m$ |
| Along Z | $\pm 15\ \mu m$ | $\pm 40\ \mu m$ | $\pm 35\ \mu m$ |
| X or Y rotation in arcmin | $\pm 1.5$ | $\pm 3$ | $\pm 1.5$ |

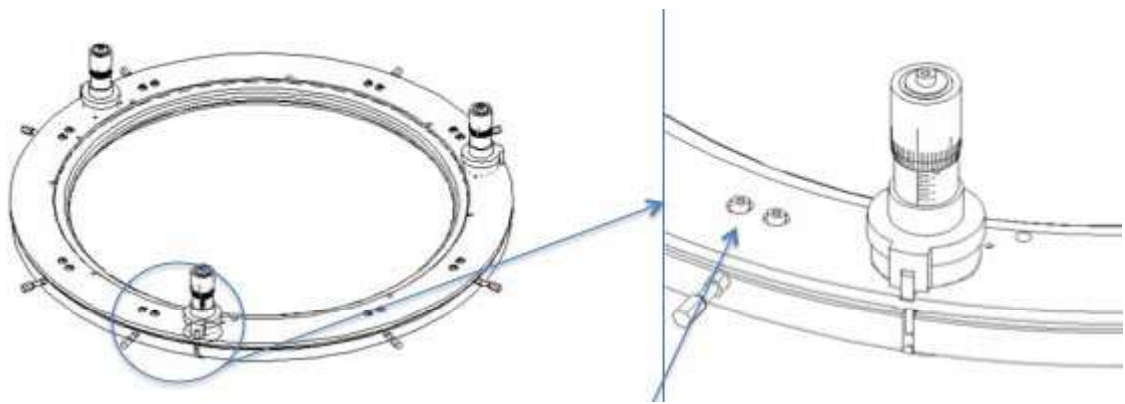

Figure 5.39: Sketch of the tip-tilt system mechanism for fine alignment of the CCD by micrometric screws based on the roll-pitch system developed for MegaCam. The locking system is implemented as stainless steel balls in a grove.

green cylinders) and the tip-tilt system (screws in white).

Cooling power is supplied from the cold machine to the CCD through a set of mechanical parts. As shown in Figure 5.41, the CCD, mounted on its SiC package, is followed by a SiC cold plate connected to the Cu cold tip of the cold machine through flexible cryo-braids. The SiC cold plate ensures the mounting of the CCD and supplies cold power with minimal thermal losses. The CCD package and cold plate will be made of the same material to reduce stresses from thermal contraction. The cold plate will be equipped with a Pt100 resistor as a temperature sensor. Braids will be dimensioned to have a thermal capacitance suitable for the CCD temperature regulation, which will be achieved by tuning the electrical power of a resistive heater glued on one side of the tip of the cold machine.

Thermal shielding of the cryostat will be provided in three pieces, one for the vessel sides, one for the rear flange and one for the front lens. The latter will differ for the three arms of the spectrographs, which have lenses of different diameters. The shielding will be provided by polished Al plates or MLI foils. The final choice will be based on the results of the tests with the cryostat prototype.

Finally, the design of the vacuum system takes into account the mechanical assembly of the spectrographs which will be mounted in two towers of five spectrographs each. To allow easy access, each tower will be equipped with three vacuum units. A vacuum unit will be composed of a primary/secondary pumping machine and a distribution line to five vertically aligned cryostats (see Figure 5.42). Each pipe to a cryostat will be equipped with an isolation valve. One full-range vacuum sensor will allow pressure to be measured. This sensor will be isolated by a manual valve in case of maintenance operation.

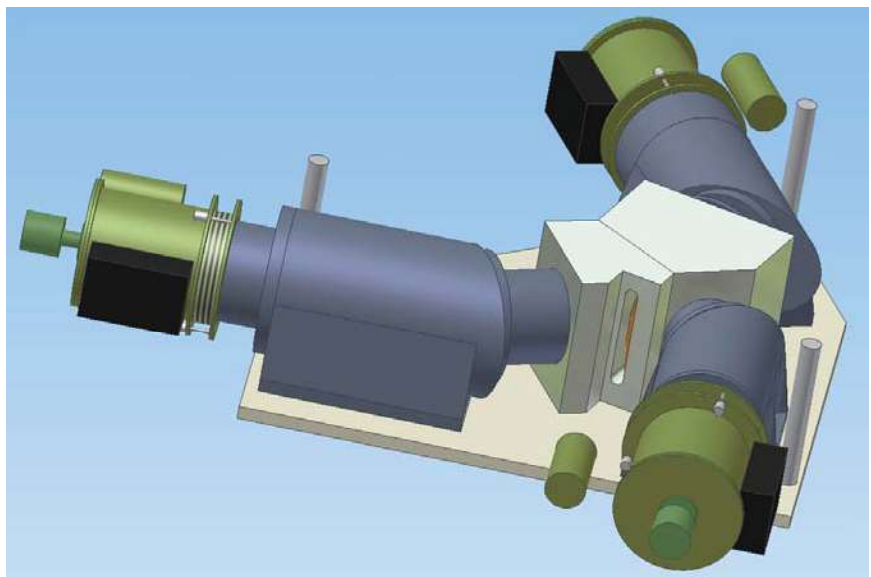

Figure 5.40: 3D model of a complete spectrograph with its 3 cryostats.

We plan to run with static vacuum during the observation periods, cryo-pumping keeping vacuum conditions inside the cryostats. The procedure of pumping between these periods has to be discussed and defined.

### 5.6.2  Cryogenic System

The cryogenic system uses independent and autonomous cooling machines, based on pulse tube technology, in order to have a simple and robust system for the control of the 30 cryostats that also allows easy integration, assembly and maintenance operations.

Linear Pulse Tubes (LPT) were developed by the Service des Basses Temperatures

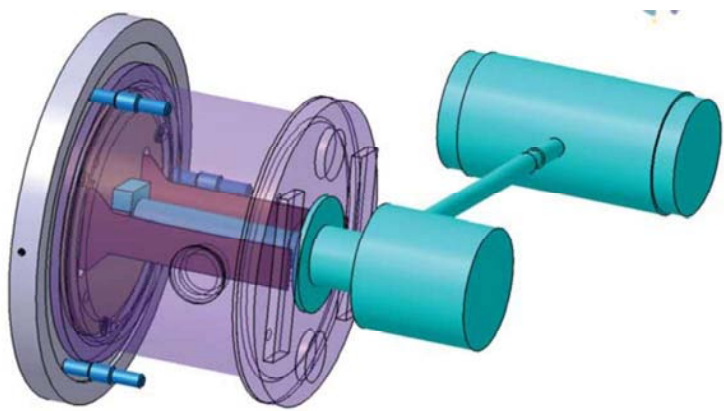

Figure 5.41: Sketch of a cryostat.

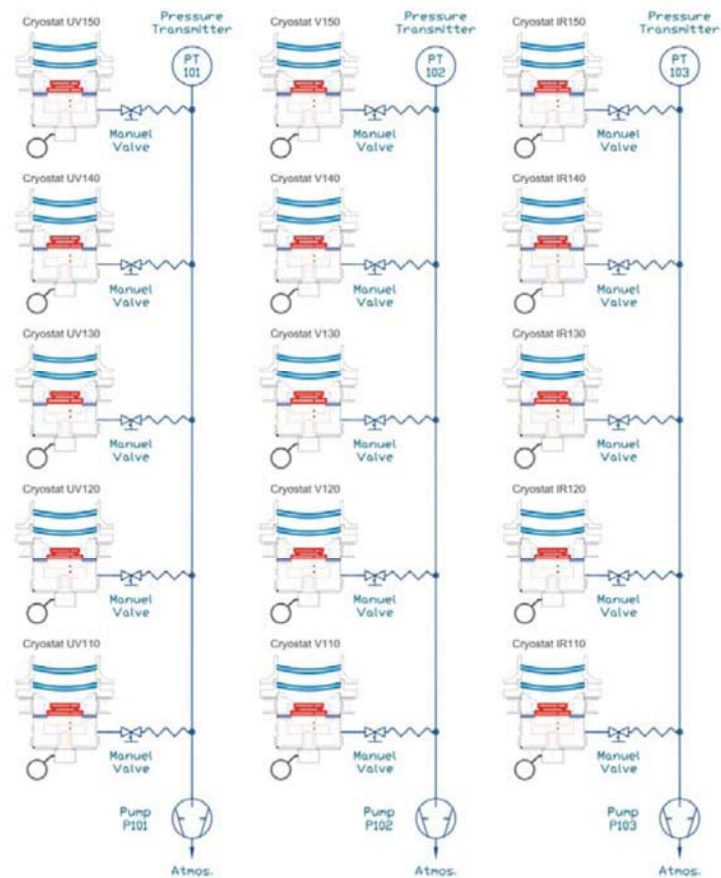

Figure 5.42: Vacuum system for a tower of spectrographs.

(SBT) from CEA in Grenoble (France). The technology was transferred by CEA/SBT to Thales Cryogenics BV Company which provides several models of LPTs with different power and temperature ranges. To define an appropriate LPT model for BigBOSS cryostats, a preliminary estimate of the power and temperature budget of the different elements of the cryostat was done, as shown in Table 5.17. The values are meant for a CCD temperature of 170K and a maximum difference of -20K with respect to the cold finger of the cold head. A 3 W, 150K cold machine appears adequate.

#### 5.6.2.1   Linear pulse tubes (LPT)

The Linear Pulse Tube (LPT) is a miniature closed-cycle pulse tube cooler, made of a compressor module connected by a metal tube to a pulse tube cold finger (see Figure 5.43). The compressor pistons are driven by integral linear electric motors and are gas-coupled to the pulse tube cold finger. The pulse tube has no mechanical moving parts. This technology, combined with the proven design of the ultra reliable flexure bearing compressors, results in extremely reliable and miniature cryocoolers with a minimum of vibrations. In addition, the compact magnetic circuit is optimized for motor efficiency and reduction of electromagnetic interference.

Table 5.17: Radiative and conductive thermal losses.

| Element | Loss (W) | % of total loss |
|---|---|---|
| Lens (radiative) | 2.0 | 69 |
| CCD dissipation | 0.1 | 3.5 |
| CCD electronic cables | 0.1 | 3.5 |
| Cold plate / vessel (radiative) | 0.2 | 7 |
| Cold plate supports (conductive) | 0.2 | 7 |
| Cold base regulation capacity | 0.3 | 10 |
| TOTAL | 2.9 | 100 |

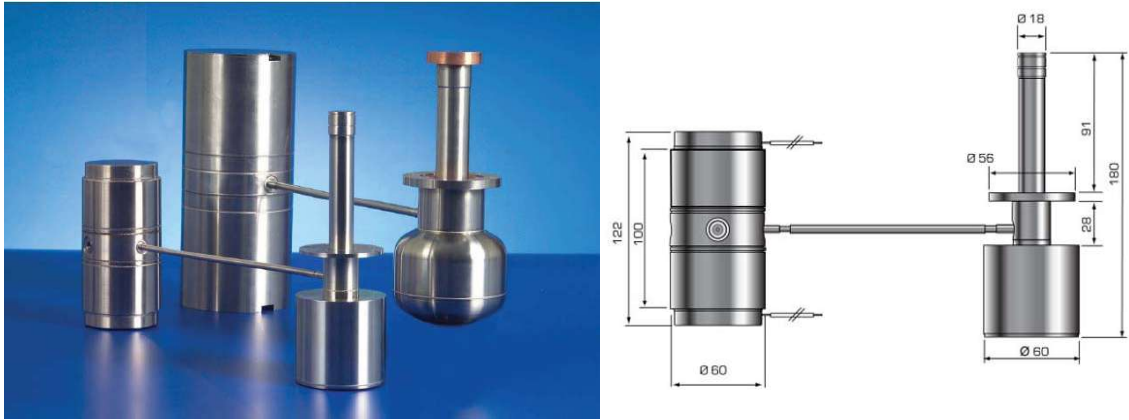

Figure 5.43: Left: two models of LPT, LPT9510 (in the foreground) and LPT9310, with powers of 1 W and 4 W at 80K, respectively Right: dimensions of the LPT9510 model.

### 5.6.2.2 Device monitoring and temperature regulation

The LPT compressor is powered with an AC voltage signal which sets the cold finger operating point in power and temperature. Changing this voltage allows the thermal performance to be tuned in a given range (see Figure 5.44).

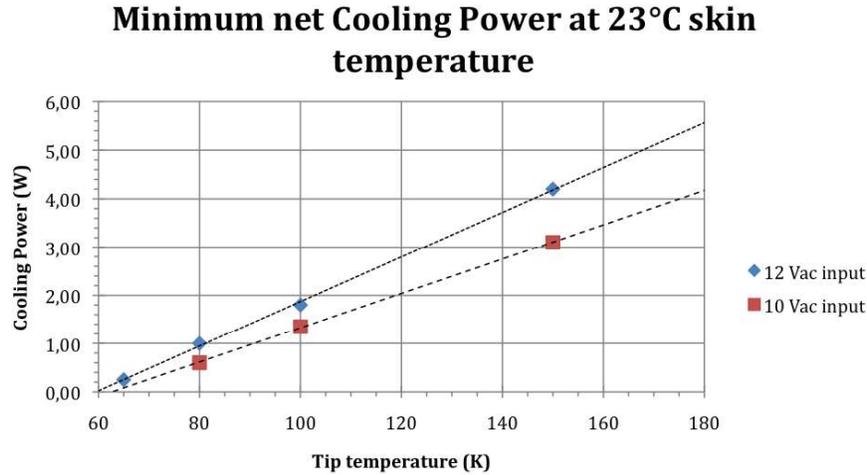

Figure 5.44: Power vs. temperature diagram for the LPT9510.

The LPT machine is provided with an electrical interface called CDE (Cooler Drive Electronics) powered by an input DC signal. The CDE converts the input signal from DC to AC and adjusts the output voltage. A pre-tuning is usually done by the manufacturer to meet specific customer requirements.

A CDE with higher functionality is also available. It can be used to drive the LPT in order to achieve extreme temperature stability and provides internal feedback about the thermal control process itself (see Figure 5.45). Combined with the thermal capacitance provided by the cold base and its heater (see Sec. 5.6.1.2), the CDE could offer a second solution to set and regulate the CCD temperatures. The final configuration of the regulation system will be discussed with the LPT manufacturer and will depend on the results of cryogenic tests.

### 5.6.3 Cryostat Control System

We have adopted a well-tested control system for the 30 CCDs and cryostats that has been working reliably on many projects for several years (MegaCam, Visir/VLT, LHC Atlas and CMS experiments at CERN). The three main components are a programmable logical controller (PLC), measurement sensor modules and a user interface on a PC. The general architecture of the system is presented in Figure 5.46.

The PLC is a Simatic S7-300 unit type from Siemens with a system core based on a UC319 mainframe. The program implemented in the PLC will acquire in real time all variables corresponding to the monitoring and control of the instrument: vacuum and temperature monitoring, control of the cold production unit, CCD cooling down and warming up, safety procedures on cryogenics, vacuum and electrical power. Safe operations all systems will be insured. A local network (based on an industrial bus, e.g., ProfiBus or ProfiNet) ensures communication with the remote plug-in I/O modules and with the PLC.

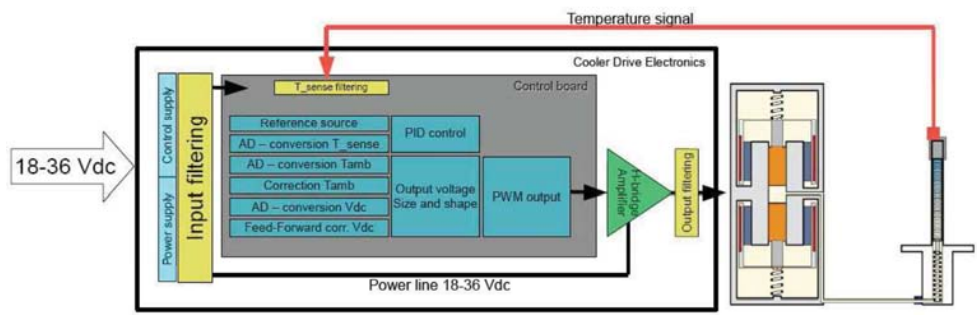

Figure 5.45: Block diagram of a Cooler Drive Electronics.

Temperature measurements are provided by Pt100 temperature probes directly connected to the PLC. The other analog sensors (heaters, vacuum gauges) are connected to a 4–20 mA or 0–10 V module. All measurement sensors will be located in two cabinets, each dedicated to one spectrograph tower (see Figure 5.47).

Supervision software (with user interface) is implemented in the industrial PC connected to the PLC via a dedicated Ethernet link. It will ensure the monitoring and control of all variables, with possibly different levels of user access rights. This system will also allow the set-up to be remotely controlled via the Ethernet network that will be accessible from Internet through a secured interface.

### 5.6.4  Detectors

Each of the three arms of a spectrograph will use a 4k×4k CCD with 15 $\mu$m pixels. For the blue arm we are baselining the e2v CCD231-84 with its good quantum efficiency down to 340 nm. For the visible arm we are baselining the LBNL 4k×4k CCD as used by BOSS. The red arm will also use the BOSS format CCD except that it will be a thick CCD ($\sim$ 650 $\mu$m to achieve usable QE out to 1060 nm. Figure 5.48 show the two types of CCDs. CCD performance characteristics and cosmetics will be the same as established by BOSS. Typical achieved value are shown in Table 5.18.

The quantum efficiency performance of the BOSS e2v and LBNL CCDs is well established and is shown in the two left curves in Figure 5.49. The high-side cutoff of a CCD is determined by its thickness as the absorption length increases rapidly above 900 nm. The absorption is also a function of temperature, decreasing with increasing temperature. To maximize the near infrared reach we propose to use a very thick CCD, 650 $\mu$m compared to 250 $\mu$m used in BOSS and the visible arm. Such a CCD can achieve a QE of around 25% at 1050 nm (at 175K). Measurement of dark current of CCDs of this thickness combined with signal-to-noise simulations for BigBOSS indicate that this temperature can be tolerated. QE simulations to date have been done for a 500 $\mu$m thick CCD and are shown in Figure 5.49.

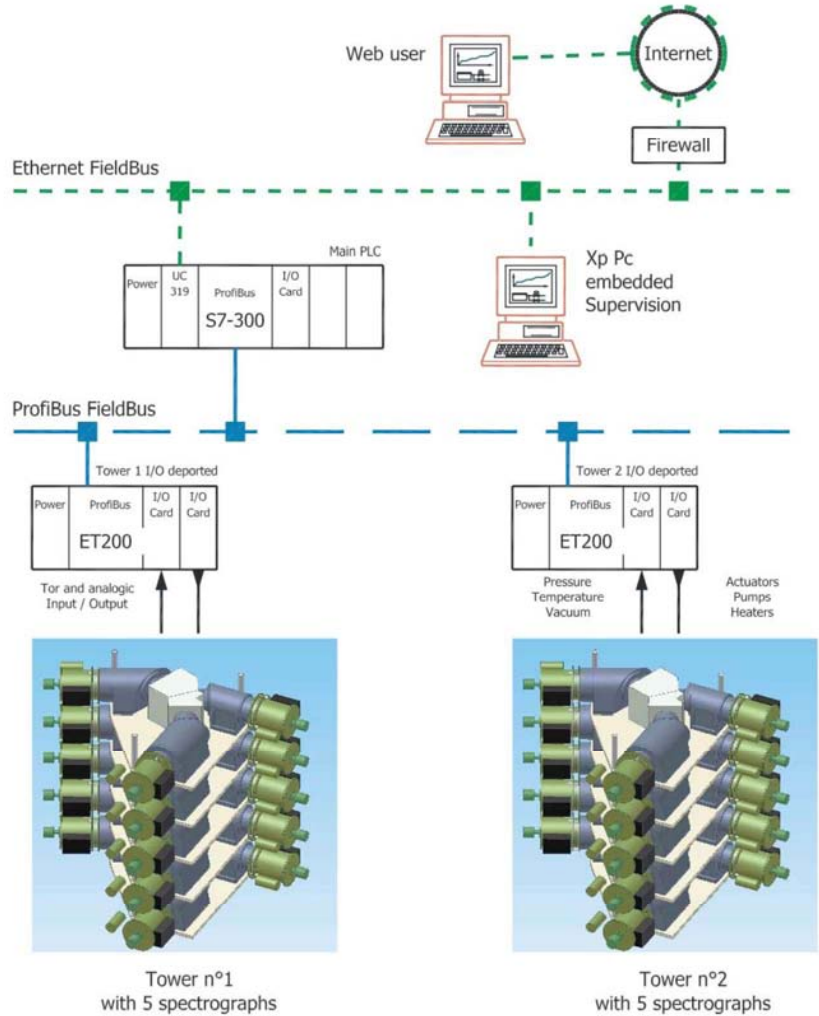

Figure 5.46: Architecture of the BigBOSS cryostat control system.

A concern with the thick CCD is the depth of focus variation that is rapidly changing between 900 nm and 1060 nm. We have simulated this for an f/2 beam focused at the optical surface of a 650 $\mu$m thick CCD. We include the measured effects of lateral charge diffusion.

Figure 5.50 shows the projected conversion charge distributions at the pixel plane for several wavelengths. The 950 nm photons mostly convert at the surface of the CCD and the distribution is essentially gaussian, determined by lateral charge diffusion during the 650 $\mu$m charge drift to the pixel plane. For increasing wavelengths, there is less lateral charge diffusion on average but this is offset by the spread in the conversion area as the f/2 beam diverges in the CCD thickness. We note that the relative areas under the curves in the figure scale like the relative quantum efficiencies. Also shown in Figure 5.50 is the PSF of the convolved fiber and spectrograph optics response. Simulations indicate that the contribution from the CCD blurring is not important.

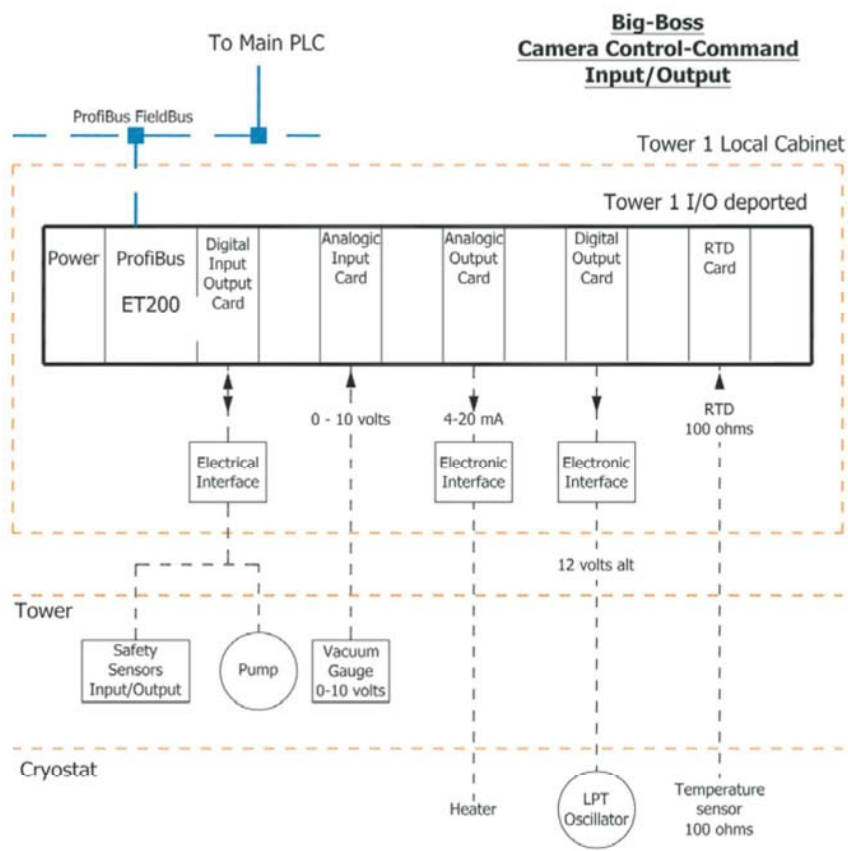

Figure 5.47: Configuration of the control system for one spectrograph tower.

### 5.6.5   Detector Readout Electronics

The electronics for each CCD will be mounted on the warm side of the cryostat wall. This provides easy access for replacement without disturbing the detector. This will include local power generation from an isolated single input voltage, CCD bias voltages generation, programmable clock levels and pattern, CCD signal processing and digitization, and set voltage readback. Configuration and control of the electronics and delivery of science data will be over Ethernet links, possibly optically isolated. A block diagram is shown in Figure 5.51.

There is a level of complexity introduced into this electronics because the mixture of n-channel (e2v) and p-channel (LBNL) CCDs. The CCD output structures required opposite sign DC biasing voltages and the electron-to-voltage gains are of opposite sign. Common clocking circuitry can work for both, but the e2v devices require four-phase parallel clocking while the LBNL devices require three. In addition, the LBNL devices require a HV depletion supply.

The analog signal processing and digitization can be accomplished with the CRIC ASIC that can accommodate either n- or p-channel devices. The n-channel device exists; the version that supports both types of CCDs is in fabrication. CRIC contains a programmable gain input stage, a single- to double-ended current source followed by a differential dual-slope integrator correlated double sampler. The voltage output of the integrator is converted by a 14-bit pipeline ADC. Two integrator range bits plus the ADC bits provide 14-bit resolution over a 16-bit dynamic range to encode the pixel charge. The CRIC chip contains four channels of the above. The data is transmitted off-chip with a single LVDS wire pair. A differential serial LVDS configuration bus is used to configure, command and clock the device.

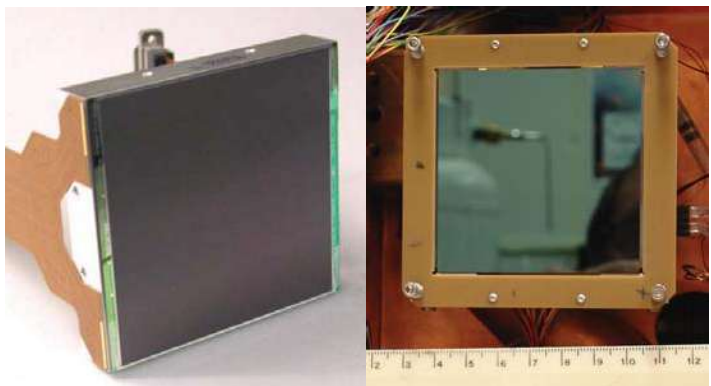

Figure 5.48: 4k×4k, 15 $\mu$m CCDs: left, e2V and right, LBNL. A four-side abuttable package similar to that shown for the e2v device is underdevelopment for the LBNL CCD.

Table 5.18: BOSS achieved CCD performance of detectors proposed for use in BigBOSS. Readnoise is for 70 kpixel/s.

| | | LBNL | | e2v | |
|---|---|---|---|---|---|
| Parameter | Req. | Red 1 | Red 2 | Blue 1 | Blue 2 |
| Read Noise | Blue: < 3 | 2.1 | 2.5 | 1.8 | 1.8 |
| (e-) | Red: < 5 | 2.2 | 2.5 | 1.6 | 2.2 |
| | | 2.5 | 2.4 | 1.7 | 1.9 |
| | | 2.1 | 2.4 | 1.8 | 1.8 |
| Dark Current | Blue: < 4 | ∼1 | ∼1 | 1.5 | |
| (e-/pix/hr) | Red: < 8 | | | | |
| Cosmetics | < 15 | 2 | 0 | 0 | 4 |
| (bad columns) | | | | | |

We belive that a single configurable board design can service the two types of CCD technologies.

## 5.7 Calibration System

### 5.7.1 Dome Flat Illuminations

Continuum and emission line lamps illuminating the dome flat exercise the entire instrument light path and generate spectra placed on the CCDs as galaxies do. The line lamps are useful for verifying the corrector focus and alignment. Whether these can be intense enough for dome illumination needs to be investigated. Laser comb lamps may be an alternative, but typically the wavelength spacing is finer than the spectrograph resolution. Again, further study is required. The lamps will be mounted at the top of the prime focus cage. The dome flat screen is already in place.

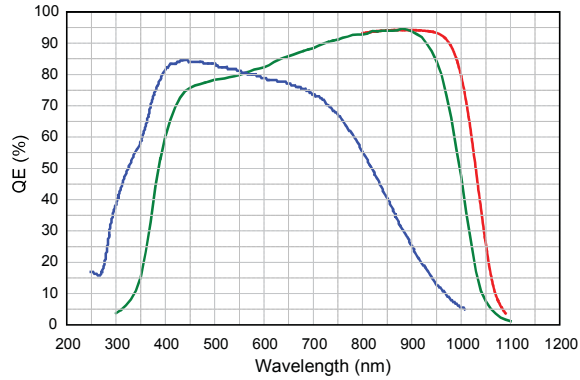

Figure 5.49: Quantum efficiency for the three types of BigBOSS CCDs. Left curve is for e2v CCD231-84, center curve is for LBNL BOSS 250 $\mu$m thick CCD, and the right curve is the simulation for an LBNL 500 $\mu$m BOSS-like CCD.

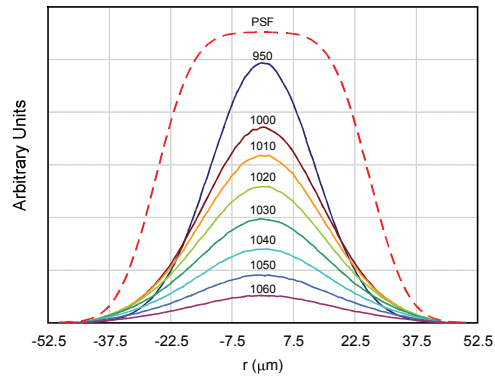

Figure 5.50: Thick CCD PSF. An f/2 beam for wavelengths near cutoff is focused at $r = 0$ on the surface of a 650 $\mu$m thick CCD with 100 V bias voltage. The curves show the radial charge distribution collected in the pixel plane. The horizontal bins correspond to 15 $\mu$m pixels. The dashed curve is the optical PSF from the fibers and spectrograph optics.

### 5.7.2 Spectrograph Slit Illumination

As described earlier, the fiber slit array assemblies will have a lossy fiber that can illuminate the entire spectrograph acceptance angle with white light or line lamps. This allows the entire CCD area to be illuminated with arc and line lamps. By this means, the four dark pixel rows between spectra can be illuminated.

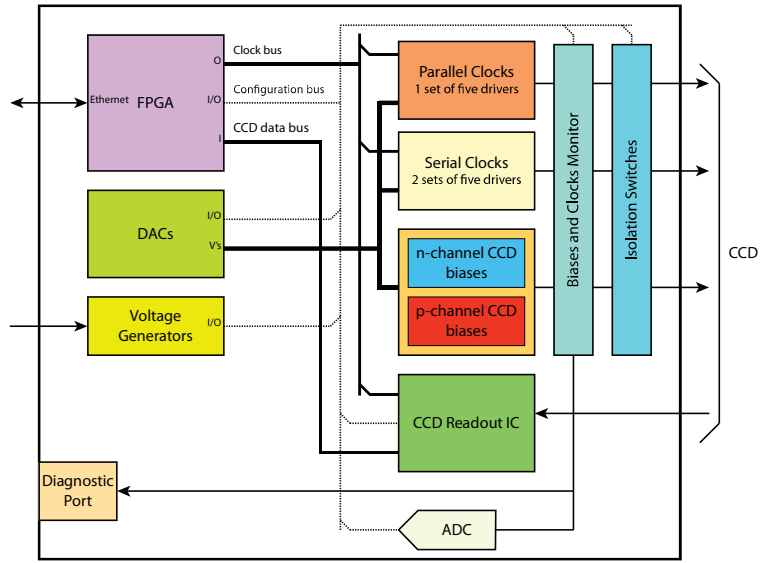

Figure 5.51: CCD frontend electronics module block diagram supporting both n-channel and p-channel CCDs.

## 5.8 Instrument Readout and Control System

The BigBOSS data acquisition system (DAQ) is responsible for the transfer of image data from the frontend electronics to a storage device. It has to coordinate the exposure sequence, configure the fiber positioners and it provides the interface between BigBOSS and the Mayall telescope control system. The instrument control system (ICS) is designed to aid in this effort. Every component of the instrument will be monitored and detailed information about instrument status, operating conditions and performance will be archived in the facilities database. In the following sections we first discuss a typical exposure sequence to introduce some of the requirements for the DAQ and ICS systems. This is followed by a description of the exposure control system which includes the fiber positioners and a section on readout and dataflow. Later sections cover the instrument control system and the interface to the Mayall telescope. We conclude with a discussion of the online software we envision for BigBOSS.

### 5.8.1 Exposure Sequence

A typical BigBOSS exposure sequence is shown in Figure 5.52. The observation control system (OCS) is responsible for coordinating the different activities. In order to maximize survey throughput we will set up for the next exposure while the previous image is being digitized and read out.

At the end of the accumulation period of an exposure after the shutters are closed, the OCS instructs the frontend electronics to read out the CCDs. At the same time the guider and focus control loops are paused. Information about the next pointing has already been loaded to the OCS during the previous accumulation phase. Once the shutter is closed the OCS transmits the new coordinates to the telescope. The focal plane systems are switched to positioner mode and the fiber positioners moved to a new configuration. The first snapshot of actual fiber locations is then acquired by the fiber view camera. It will require a second cycle to complete the positioner setup. After the telescope reaches the new target position, the OCS activates the guider to close the tracking feedback loop with the telescope control system. Guider correction signals are sent at a rate of about 1 Hz. Once the telescope is tracking, the OCS re-enables the focus control loop. At the end of the second fiber positioning cycle, the focal plane systems switch back to low power mode. The OCS waits for the CCD readout to complete and for the fiber view camera to signal that fibers are in position before it commands the shutters to open. While the spectra are being acquired information about the next exposure, including telescope coordinates and target positions, is loaded into the OCS.

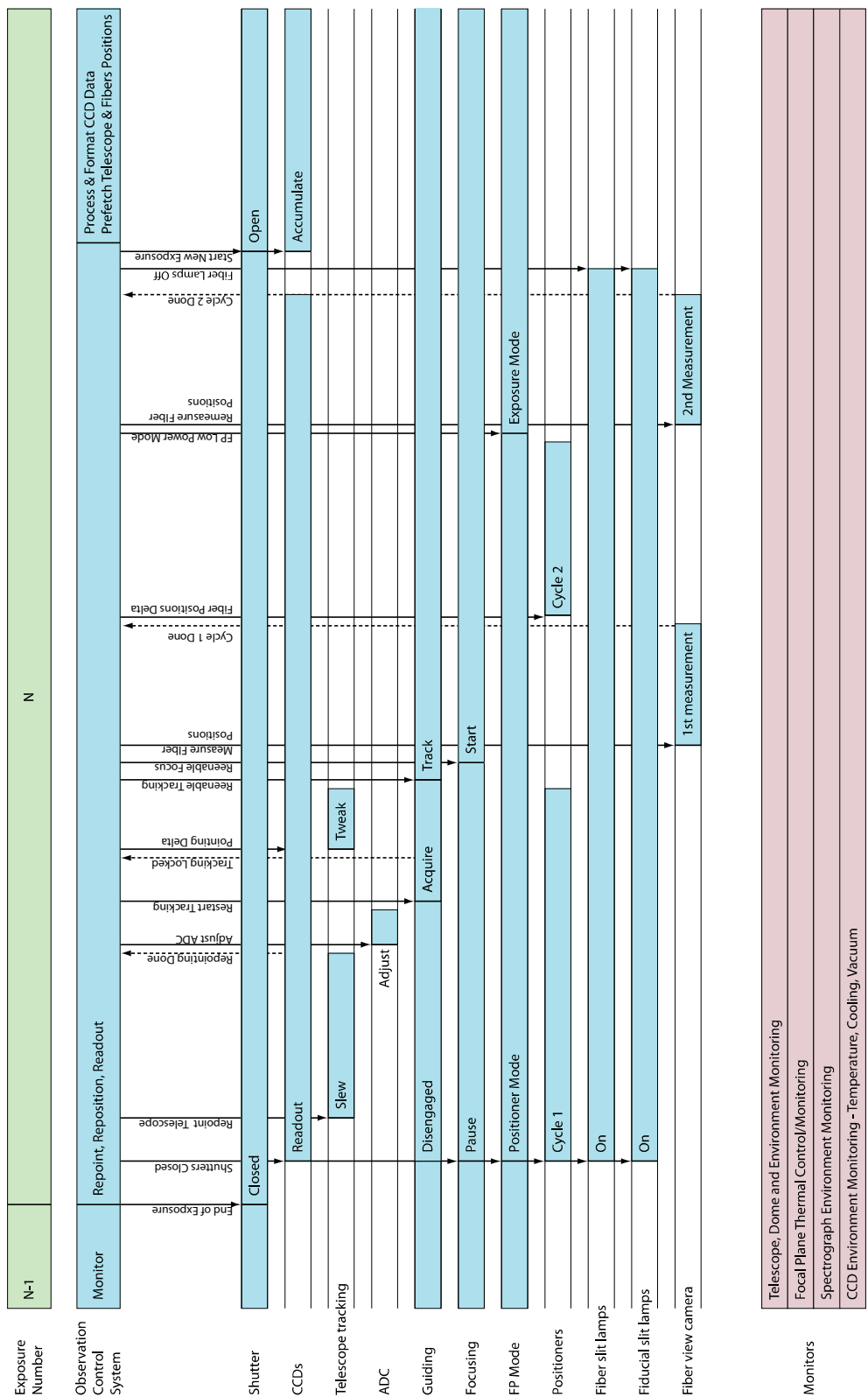

Figure 5.52: An example of a BigBOSS exposure sequence.

At a typical pixel clock rate of 100 kHz CCD readout will take approximately 42 seconds. The BigBOSS DAQ system is designed to complete the entire sequence outlined above in a similar amount of time so that the time between exposures will be no longer than 60 seconds.

### 5.8.2  Readout and Dataflow

The BigBOSS instrument consists of ten identical spectrographs each with three cameras covering different wavelength regions. Each camera uses a single 4k×4k CCD with four readout amplifiers that operate in parallel. A default pixel clock of 100 kpixels/s results in a readout time of approximately 42 seconds. The charge contained in each pixel is converted with 16-bit ADCs yielding a data volume of 34 MBytes per camera or about 1 GByte per exposure for the entire instrument. A schematic view of the BigBOSS DAQ system is shown in Figure 5.53. While we are still evaluating different options we are considering a system consisting of 30 identical slices, one for each camera.

In the block diagram (Figure 5.53) data flows from left to right starting with the CCDs and ending with the images stored as FITS files on disk arrays in the computer room. Each CCD is connected to a camera frontend electronics module that will be located directly on the spectrographs. Optical data and control links connect each camera to its data acquisition module which includes a full frame buffer and a microcontroller with a high speed network interface to the online computer system in the control room. Several architecture and technology options are still being investigated at this time. This includes the placement of the Camera DAQ modules. The best location might be close to the frontend electronics near the cameras but because of the data/control link we could also choose a more convenient location in the Mayall dome.

We need to determine that data and control links can be combined and establish the package form factor for the Camera DAQ modules. For the BOSS/SDSS-III data acquisition system we combined the functionality provided by the DAQ module with the backend of the frontend electronics. We intend to explore this option for BigBOSS as well.

Our baseline for the network link on the DAQ module is (optical) Gigabit Ethernet with the assumption that the Camera Controller supports the TCP/IP software protocol. This feature combined with the modular design allows us to operate individual cameras with only a laptop computer, a network cable and of course the online software suite. We expect this to become a very valuable tool during construction, commissioning, and maintenance.

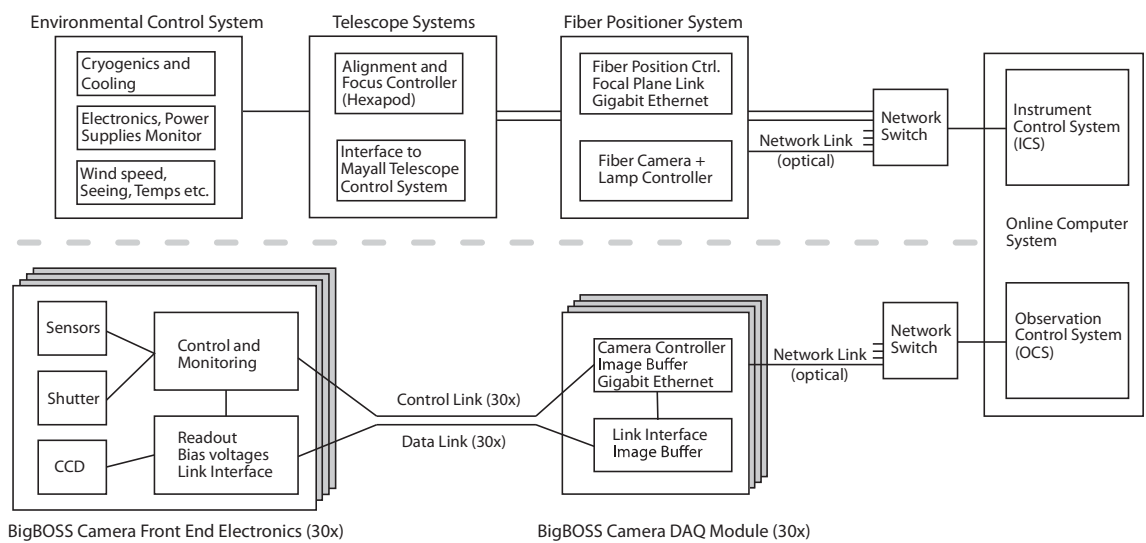

Figure 5.53: Block diagram of the BigBOSS data acquisition system.

Data transfer from the frontend electronics to the Camera DAQ modules will begin shortly after the start of digitization and will proceed concurrently with CCD readout. System throughput will be designed to match the CCD readout time of 42 seconds to avoid additional dead time between exposures. The required bandwidth of approximately 10 Mbits/s is easily achievable with current technology. A small buffer memory on the frontend electronics module provides a certain level of decoupling between the synchronous CCD readout and the transfer over the data link. The Camera DAQ module will have a full frame buffer. The Camera Controller assembles the pixel data in FITS format and transfers the image over a standard network link to the online computer system in the control room. The BigBOSS online software performs the necessary book keeping to ensure that data from all 30 cameras have been received. Initial quality assurance tests are performed at this stage and additional information received from the telescope control system and other sources is added to the image files. The need for an image builder stage to create a combined multi-extension FITS file is currently not foreseen. The final image files will be transferred to the BigBOSS processing facility at LBNL NERSC.

### 5.8.3 Instrument Control and Monitoring

Hardware monitoring and control of the BigBOSS instrument is the responsibility of the instrument control system (ICS). Shown schematically in Figure 5.53 we distinguish two sets of ICS applications. Critical systems such as cooling for the CCDs and the monitor system for the frontend electronics have to operate at all times. Fail-safe systems and interlocks for critical and/or sensitive components will be implemented in hardware and are the responsibility of the device designer. Control loops and monitor functions for these applications will use PLCs or other programmable automation controllers that can operate stand-alone without requiring the rest of the BigBOSS ICS to be online. Measured quantities, alarms, and error messages produced by these components will be archived in the BigBOSS facility database where they can be accessed for viewing and data mining purposes.

The second set of instrument control applications consists of components that participate more actively in the image acquisition process such as the shutters, the fiber positioning mechanism and the focus and alignment system. The control interface for these devices typically consists of a network enabled microcontroller with firmware written in C. The online system interacts with the hardware controller via a TCP/IP socket connection although other interfaces will be supported if required. We envision that the DAQ group provides the higher level software in the instrument control system while the microcontroller firmware will be developed by the groups responsible for the respective components. Similar to the first set of ICS devices, this group of applications will also use the facility database to archive the instrument status.

Each of the BigBOSS spectrographs will include three shutters, one per CCD camera. Each shutter will be individually controlled by the camera frontend electronics module, or a dedicated system that will control all 30 shutters (TBD). Commercial shutters typically use an optoisolated TTL signal. The length of the control signal determines how long the shutter is open. We will control exposure times to better than 10 ms precision and keep the jitter in open and close times among the 30 shutters to less than 10 ms. Details of the interface to the shutter will depend on the actual shutter system selected for the BigBOSS cameras.

BigBOSS controls applications can be categorized by location into spectrograph-based systems, telescope-based system and external systems. Spectrograph-based systems include the fiber slit array lamps, the shutters, electronics monitoring, cryostat thermal and vacuum control and some environmental monitors. The group of telescope-based systems consists of the fiber view camera and fiber view lamps, the hexapod and corrector controllers, the fiber positioner, the focal plane thermal control system as well as additional environmental monitors. Components in both these groups will be integrated with the ICS using the architecture discussed in the previous paragraph. The third category consists of external instruments such as a seeing monitor, an all sky cloud camera and the dome environmental systems. The interface to these devices will be discussed in the next section.

### 5.8.4   Telescope Operations Interface

The BigBOSS online system has to interface with the existing Mayall telescope control system (TCS) to send new pointing coordinates and correction signals derived from the guider. In return BigBOSS will receive telescope position and status information from the Mayall TCS. Since the dome environmental system and most of the observatory instrumentation for weather and seeing conditions is already connected to the TCS we will not access these devices directly but control and monitor them through the TCS. Similar to the design developed by BigBOSS collaborators for the Dark Energy Camera and the Blanco telescope the BigBOSS online system will include a TCS interface process that acts as conduit and protocol translator between the instrument and the telescope control systems.

During an exposure, the BigBOSS guider and the telescope servo systems form a closed feedback loop to allow the telescope to track a fixed position on the sky. For an imaging survey it is sufficient to have a stable position. BigBOSS, however, requires a precise absolute position so that the fibers are correctly positioned on their targets. Given a pointing request, the Mayall slews into position with a typical accuracy of 3 arcsec. Using the guide CCDs in the focal plane we will then locate the current position to 0.03 arcsec accuracy. If the offset between requested and actual position is larger than a certain fraction of the fiber positioner motion we will send a pointing correction to the TCS to adjust the telescope position. Details of this procedure need to be worked out and depend on the pointing precision of the Mayall control system.

### 5.8.5 Observation Control and Online Software

The BigBOSS online software will consist of a set of application processes built upon a layer of infrastructure software that facilitates message passing and information sharing in a distributed environment. The application layer can be divided into several functional units: the image pipeline, the instrument control system including the connection to the Mayall TCS, data quality monitoring and the user interfaces with the observer console. The Observation Control System (OCS) is the central component of the BigBOSS image pipeline coordinating all aspects of the observation sequence. Connected to the OCS is an application that proposes an optimized sequence of pointings for the telescope based on a number of inputs including survey history, current time and date and the current observing conditions. At the end of an exposure the OCS will initiate readout and digitization and the DAQ system transfers the image data to a disk cache. The OCS notifies the data transfer system developed by NOAO that image data is available to be transferred to the NOAO archive and the BigBOSS image processing center. Continuous monitoring of both the instrument and the image quality is required to control systematic uncertainties to achieve the BigBOSS science goals. Quality assurance processes will analyze every spectrum recorded by the instrument and provide immediate feedback to the observer. Feedback on the performance of BigBOSS is also provided by the instrument control system (ICS) which monitors and archives a large number of environmental and operating parameters such as voltages and temperatures. In addition, the ICS provides the interfaces to the BigBOSS hardware components and the telescope control system as outlined in the previous sections. The BigBOSS user interface architecture will follow the Model-View-Controller (MWC) pattern now commonly in use for large applications. We intend to evaluate different technologies including those developed for SDSS-III/BOSS and the Dark Energy Survey.

The infrastructure layer of the BigBOSS online software provides common services such as configuration, access to the archive database, alarm handling and processing as well as a standard framework for application development. Due to the distributed architecture of the BigBOSS online software, inter-process communication takes a central place in the design of the infrastructure software. We will evaluate several options including openDDS, an open source implementation of the Data Distribution Service standard used by LSST and the Python-based architecture developed for DES.

## 5.9 Assembly, Integration and Test

### 5.9.1 Integration and Test

Several large subsystems of the BigBOSS will be integrated and tested before delivery to the Mayall. These are the telescope corrector barrel, the focal plane with fiber positioners, fiber slit arrays, the spectrographs and cameras, and the instrument control system. Figure 5.54 pictorially shows the integration flow. Below is a broad brush description of the integration process, which will require much greater elaboration during the conceptual design phase.

Prior to shipment, the corrector barrel lens elements are aligned and demonstrated to image to specifications. Actuators for the hexapod and the ADC are installed and operational. The fiber view camera mount attachment is verified. A focal plane mock-up is test fitted. When delivered to the Mayall, the secondary mirror mount will be verified.

For the systems that contain fibers, we assume that intermediate fiber optic connector blocks will be used between the positioners and the spectrographs. This enables more comprehensive integration and testing before delivery to the Mayall, and makes installation easier.

Prior to delivery to the Mayall, the focal plane will be integrated with the fiber positioners, guider sensors, focus sensors, fiber view camera fiducial fibers, and cable/fiber support trays. The positioners will be installed with their fibers in place, which be terminated in connector blocks. A myriad of tests can be performed by individually stimulating fibers in the connectors. Positioner operation will tested and positioner control address, location and fiber slit array position will be mapped. There is no requirement that any one fiber be placed in a specific focal plane position, only that, in the end, a map from positioner position to spectrograph spectral position be determined. A fiber view camera emulator can verify the performance of all the positioners.

This focal plane assembly is delivered to the Mayall and fitted to the corrector barrel. An acceptance testing plan will need to be developed that defines when the Mayall top can be disassembled and the BigBOSS prime focus structure installed.

The fiber slit array assembly precision can be measured by stimulating individual fibers in the connector blocks. This will also generate map for slit array position to connector location. This can be repeated with the actual spectrographs after their installation at the Mayall site. The fiber bundles can then be routed to and through the telescope to mate up with the fiber positioner connectors. Support of the fibers will require attachment of several structures to the telescope. The details are yet to be determined.

Spectrographs will be fully assembled and tested prior to shipment. This includes the cameras, cooling and vacuum systems, and the control system. Prior to their delivery, the Mayall FTS room will be reconfigured. The spectrographs and support equipment can be installed during day shifts and tested with the online software system, including acquiring spectra from internal lamps.

The instrument control system will have been developed in parallel with the other systems and will have been used in the commissioning and testing of other assemblies.

In summary, installation activities at the Mayall will entail replacing the existing prime focus structure including the mount ring with the BigBOSS equivalent. The focal plane will then be mounted and the fiber strung to and from the spectrograph room. In parallel, the spectrographs will installed, plumbed and the fiber slit arrays inserted. Interfacing to the instrument control system and its interface to telescope operations also occurs. Commissioning will then commence.

### 5.9.2 Commissioning

A goal for commissioning is to have equipment delivered to the Mayall and run through preliminary shakedown tests so they are ready for the annual August shutdown. The major disruption to Mayall, the disassembly of the top end occurs then. If we take the Dark Energy Survey model, four to six weeks comprise the shakedown period, requiring that the corrector and focal plane arrive in June. DES allocates six weeks for installing and testing the new cage and the f/8 support, a similar activity to that for the BigBOSS corrector and focal plane.

DES uses time over the following 11 weeks to complete on-sky commissioning. For BigBOSS, activities during this time will be demonstrating combined fiber positioning and telescope pointing, achieving and maintaining focus, end-to-end wavelength calibration using dome arc lamps or sky lines, and focusing the f/8 secondary using the corrector internal adjusters.

As described above, the major instrument subsystems will be fully integrated and tested before delivery to the Mayall. The hoped-for outcome is that commissioning time will only go into the first-time co-operation of these subsystems.

We note that once the f/8 support and positioning are verified in the telescope, Cassegrain instruments can be once again operated. This, of course, precludes BigBOSS commissioning when in operation.

## 5.10 Facility Modifications and Improvements

Improvements to the Mayall telescope and its dome are speculative at this time. We describe below potential issues and fixes that have been identified by NOAO and others.

### 5.10.1 Dome Seeing Improvements

There are dome and telescope improvements that can or might improve seeing. These need further study.

#### 5.10.1.1 Stray light

The Dark Energy Survey did a stray light study of the Blanco telescope. They identified the outer support ring of the primary mirror as the dominant stray light source. This flat annular ring is already painted black at the Mayall, but a conical shape may be more effective. The Serrurier truss is presently white and there may be a benefit to change this to a matte black. These will be studies with our stray light codes.

#### 5.10.1.2 Primary mirror

The Mayall primary mirror support system is current and no improvements are required. A wavefront mapping prior to BigBOSS operation should be performed to confirm that it is positioned correctly.

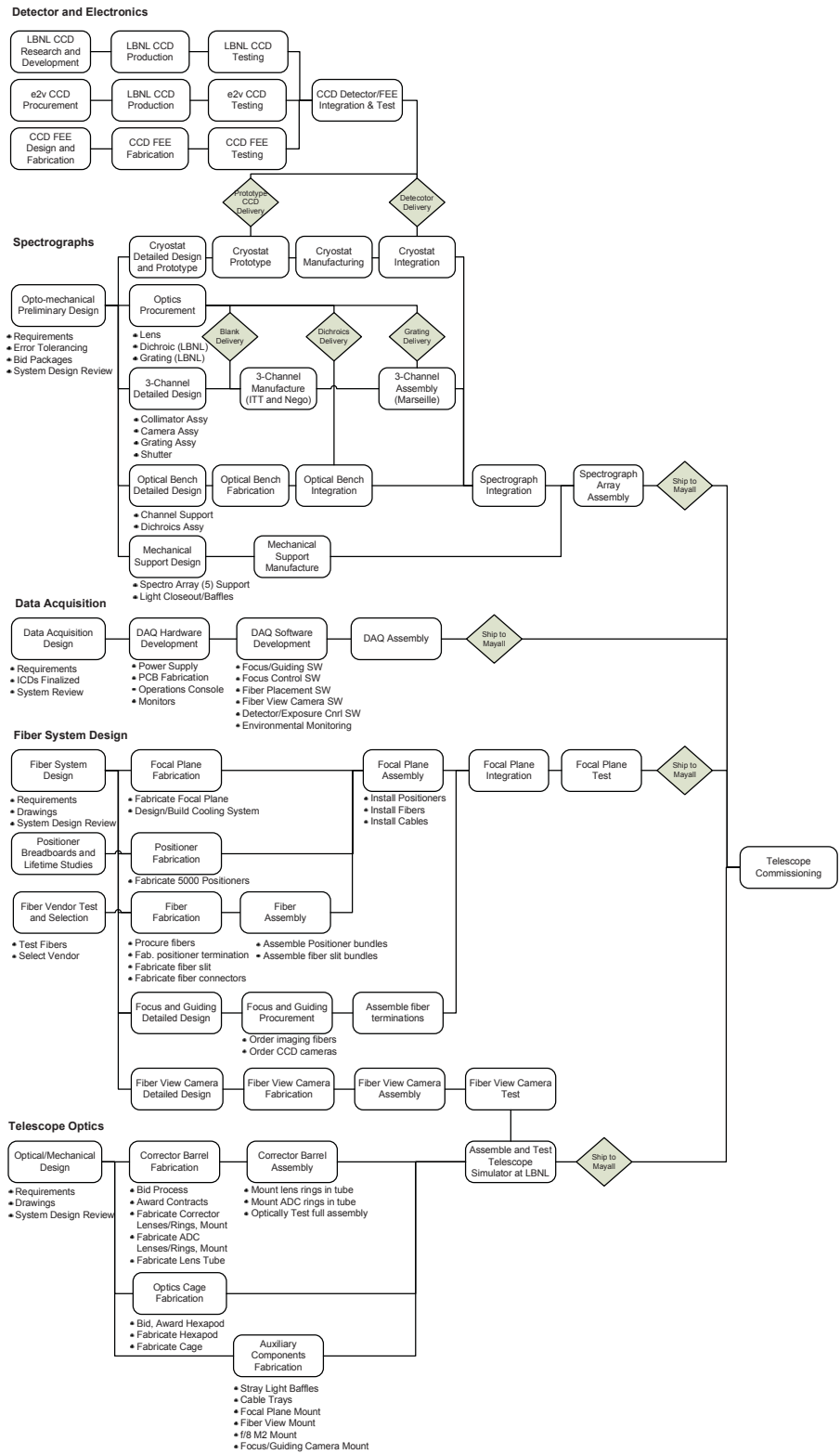

Figure 5.54: Assembly, integration and test flow.

### 5.10.1.3   Thermal sources

Air currents and heat sources in the dome impact seeing. The telescope control room is presently located on the telescope floor. The room will be relocated to a lower level at the Blanco and a similar solution is being considered for the Mayall in support of BigBOSS. It may be possible to study the impact of the control room in its present position under heated and unheated conditions.

The mass of the primary mirror central baffle impacts its thermalization to ambient temperature. Reconstructing this with a lighter design may be desirable.

One difference between the Blanco and the Mayall is that the former has a two-sheet protective cover for the primary mirror that does not trap air when open. The Mayall has a multi-petal system that partially traps a 1 m column of air above the primary. Again, it is speculative that a redesign of this can improve dome seeing.

### 5.10.2   Telescope Pointing

Historically the Mayall has shown absolute point accuracy of 3 arcsec in both declination and right ascension. More recently, right ascension accuracy is of order 15 arcsec. This will be corrected.

Telescope slew times have been recently measured, $<20$ sec for moves $<5°^{\parallel}$ Unexpectedly, the primary mirror was observed to take 40–50 sec to settle. This impacts the 60 sec deadtime between exposures that we have established as a goal. The cause may be a software issue in the drive of mirror supports. Further study and corrective action, hopefully, should be supported.

### 5.10.3   Remote Control Room

A long term goal of NOAO is to remote the telescope operations to Tucson. A remote instrument control room for BigBOSS is also desirable. The practicality of this and the cost are not yet understood.

### 5.10.4   Secondary Mirror Installation

It is required that BigBOSS provide a mounting mechanism for the existing secondary mirror to support Cassegrain focus instruments. This will require procedures and fixtures to remove the fiber view camera and support and to rig in the secondary. These will have to be jointly developed with NOAO.

### 5.10.5   Spectrograph Environment

The preferred location for the spectrograph system is in the FTS room adjacent to the telescope. A large part of this room in on the telescope support pier. The general area is presently partitioned into multiple rooms with removable walls and will need to be reconfigured for BigBOSS use. There appears to be an air handling in place already, but it may require rework to provide a temperature controlled environment at the appropriate level.

# 6    Survey Operations Plan

## 6.1    Survey Strategy

### 6.1.1    Baseline Survey Plan

The BigBOSS survey will target emission line galaxies, luminous red galaxies, and QSO candidates over a field of view of $\approx$14,000 deg$^2$. These three sets of primary candidates will be selected using color criteria from broad-band imaging from either (or both) the PanSTARRS and Palomar Transit Factory surveys (see Chapter 4 for further details). We anticipate having the imaging data and candidate catalogs in hand one year prior to the start of regular survey operations, in order to enable preparatory studies of the sample definition strategy and to understand the sample selection function(s). In addition, we intend to undertake a short (13 night) Pilot Survey in order to fully characterize the sample selection and refine the color selection criteria. During the course of the regular survey, some of the fields targeted by the Pilot Survey will be repeatedly targeted as calibration fields to track the survey performance.

The survey fields will be selected to include the 10,000 deg$^2$ region covered by the BOSS SDSS-III survey, plus an additional 4,000 deg$^2$ which, in our current baseline, covers a strip $\sim$10$-$20 deg wide that extends the northern Galactic BOSS region to lower Galactic latitudes. Figure 6.1 shows the footprint of the survey. The desire for $\approx$80% completeness results in a requirement for 5 pointings per sky position. Given the BigBOSS field of view of 7.07 deg$^2$, this results in 9824 pointings (or "tiles") on the sky. We are exploring other footprints which distribute the additional 4000 deg$^2$ between both spring and fall fields (i.e., northern and southern Galactic regions) and which will be better optimized to cover areas of low Galactic extinction. All of the fields currently chosen are observable at Kitt Peak at airmass less than 2.0 at some point during the year, while the majority have foreground extinction of $E(B-V) < 0.15$ mags (see figure 6.2).

In each 3$^\circ$‖diameter pointing, the BigBOSS fibers will target (on average[6]) approximately 2550 ELG candidates, 800 LRG candidates, and 640 QSO candidates. The remaining (more than 1000) fibers per pointing will be available for sky, calibration stars, and ancillary targets. Each position on the sky will be covered by approximately 5 tiles. The first tile will be observed at least one night prior to the remaining tiles, and will include a large number of QSO candidates. The time delay between the first tile and the remaining ones is to ensure that the QSOs relevant for the Ly$\alpha$ BAO experiment (i.e., those with redshifts $2.2 \leq z \leq 3.5$) can be identified by their spectra, and then targeted more efficiently by the remaining pointings. By the end, the baseline survey will have successfully obtained redshifts for roughly 15.3 million ELGs, 3.4 million LRGs, and 0.63 million QSOs (e.g., see Table 2.2 for details).

### 6.1.2    Calibration Fields

In order to properly characterize the BigBOSS survey performance and accurately measure the sampling and completeness functions, we will define 4 to 6 calibration fields, at least two of which can be targeted in any given observing run. These fields will be targeted at least once during each BigBOSS run, and over the period of the survey will build up a total area of $\approx$30-40 deg$^2$ which is densely sampled with deep spectroscopy. By carefully choosing the calibration fields to lie in regions which have wide-area multi-wavelength and archival spectroscopic coverage (e.g.: selected PS1 calibration fields; overlapping with M31; the best studied portions of the Sagittarius stream; the 9 deg$^2$ NDWFS Boötes field; the 2 deg$^2$ COSMOS field; the SXDF and UDS fields; the LSST deep pointings; etc.), these fields will be invaluable for many ancillary science programs (e.g., galaxy evolution, Galactic structure, etc.) and have high legacy value.

---

[6]That is, averaged over all five tiles (see Table 6.2).

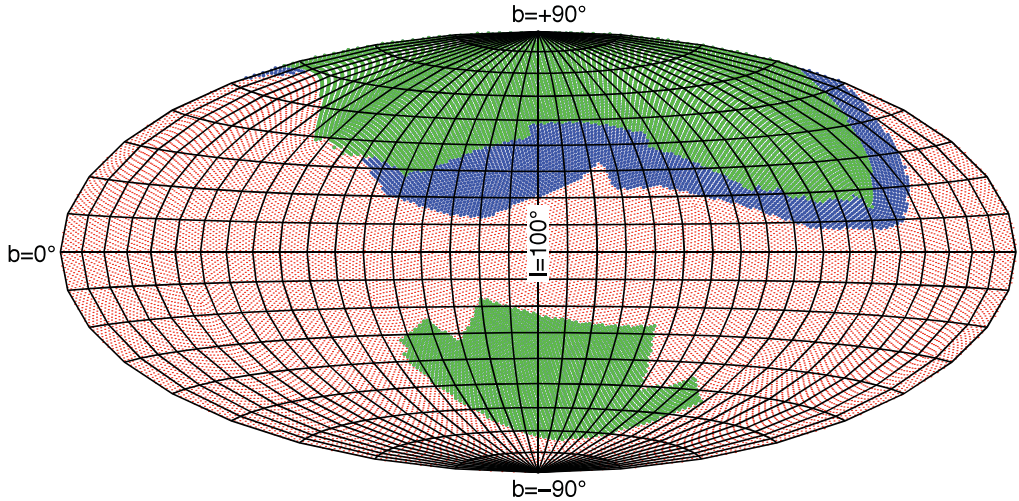

Figure 6.1: Footprint of the survey (in Galactic coordinates), which includes the BOSS LRG Survey currently underway (green) and a 4,000 deg$^2$ strip, chosen here to be at northern Galactic latitudes (blue).

### 6.1.3  Optimizing the Survey Observing Strategy

Kitt Peak observing conditions are strongly affected by the Southwestern monsoon season, which primarily affects the months of July and August (see Figure 6.3). The seeing conditions at the Mayall have not been systematically characterized, but Figure 6.4 shows the measured distribution of $I$-band seeing FWHM measurements from the KPNO 4m Mosaic prime focus camera. In the $I$-band, the median seeing is $\approx$1.0 arcsec, while the average is $\approx$1.1 arcsec. The seeing is likely to be largely due to the turbulence around and within the dome, since the mountain seeing is known to be much better (e.g., see http://www.wiyn.org/DIQ.pdf). It is possible that modifications to the telescope environment (such as moving the location of the Control Room) can improve the seeing even further; this would translate directly into improved survey performance.

In order to optimize the survey, we modeled the entire BigBOSS Key Project in the following manner. Given the 9,824 field positions defined previously, our software calculates arrays of target airmass, moon position and distance from the target field for all times during the year, in one hour intervals. The software then uses these arrays to decide the order in which these fields should be observed, calculating the exposure time needed (given the airmass, sky brightness, the desired output signal-to-noise ratio per emission line of 8 for the redshifted [OII] doublet at the line flux limit of the survey (see Chapter 4.4.1 and Appendix A), and an efficiency factor that accounts for the typical clear fraction at the time of the observation) and adding in the relevant overheads (estimated to be a total of 60 seconds per field). For the exposure time calculation, we assumed an average seeing of 1.1 arcsec and an intrinsic target half-light radius of 0.3 arcsec, and calculated fiber losses based on these spatial profiles. The exposure time calculation included all of the efficiency losses identified by the instrument design team in the telescope/instrument system, as well as all of the readily identifiable sources of noise (object photon shot noise, location-dependent sky noise, and detector read noise). We set a minimum exposure time of 1,000 sec, which achieves a S/N>8 for one component of the redshifted [OII] doublet under clear dark-sky conditions at zenith. We constrained the schedule such that no observations were permitted in the three summer months of June, July and August (traditionally summer shutdown for KPNO due to the monsoon) nor during brightest Moon conditions. The night sky brightness was based on a combination of the dark night sky as observed with the VLT/UVES (Hanschik 2003), normalized to a surface brightness of 18.8 mag arcsec$^{-2}$ (in the $z$-band) and with a dependence on airmass consistent with the airglow arising in a thin layer at 86 km altitude, and a model for the reflected solar spectrum of the moon with a normalization set by the formula of [Krisciunas & Schaefer, 1991], which includes dependences on the lunar phase, the Moon position, and the distance between the target field and the moon.

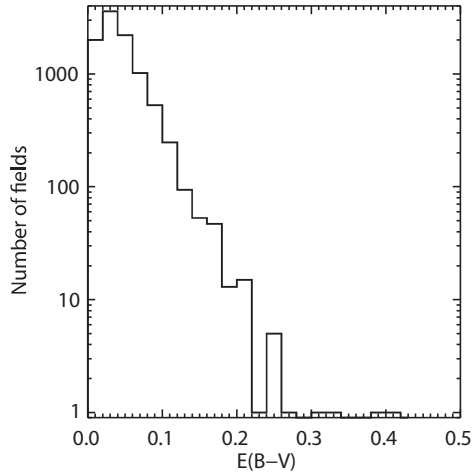

Figure 6.2: Distribution of Galactic extinction for the 9,824 survey fields in the baseline survey. The mean and median of the distribution are 0.043 and 0.035 respectively. Further optimization of the field footprint may result in a smaller fraction of high extinction fields.

Figure 6.6 and Table 6.1 show the distribution of nights for the optimal survey as a function of year. Figure 6.7 shows the distribution of time as a function of moon phase for this optimal baseline survey. Figure 6.8 shows the equatorial projection of the BigBOSS survey footprint, this time color-coded according to the year of observation. In the current plan, the survey strategy is to aggressively schedule time to complete the survey in the minimum number of years. As a result, the bulk of the observations are completed within four years, with the last year being used primarily for "clean up". During years 3 and 4, many of the nights are only scheduled as partial nights (i.e., the green and blue bars in figure 6.6). We continue to experiment with different strategies for covering the survey footprint to determine which approach is optimal, and the current strategy will be further optimized.

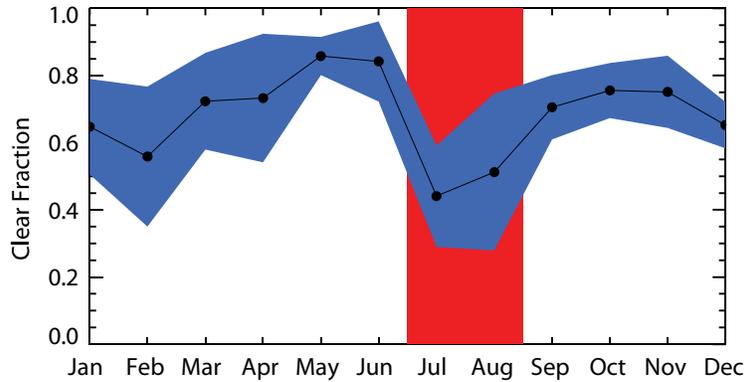

Figure 6.3: Average fraction (determined for the period 2001 to 2009) of clear conditions at Kitt Peak as a function of month. The black dots and line show the average, and the blue range defines the rms from the mean. The red bar represents the months of July and August, during which the Mayall is shut down for maintenance.

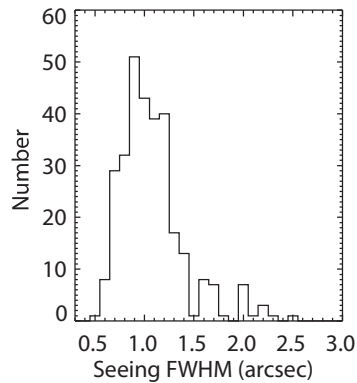

Figure 6.4: Distribution of *I*-band seeing FWHM measurements from the Mayall prime focus instruments. While we assumed this distribution for our modeling, the seeing is dominated by dome effects and improvements may be possible.

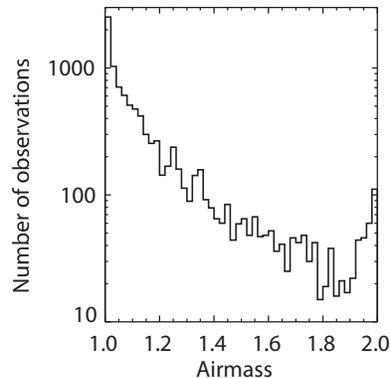

Figure 6.5: Distribution of airmass for the fields observed by the baseline strategy. Observations are restricted to airmass ≤ 2.0; the median and mean of the distribution are 1.08 and 1.18 respectively. Further optimization of the field footprint and survey strategy are likely to result in fewer fields observed at high airmass.

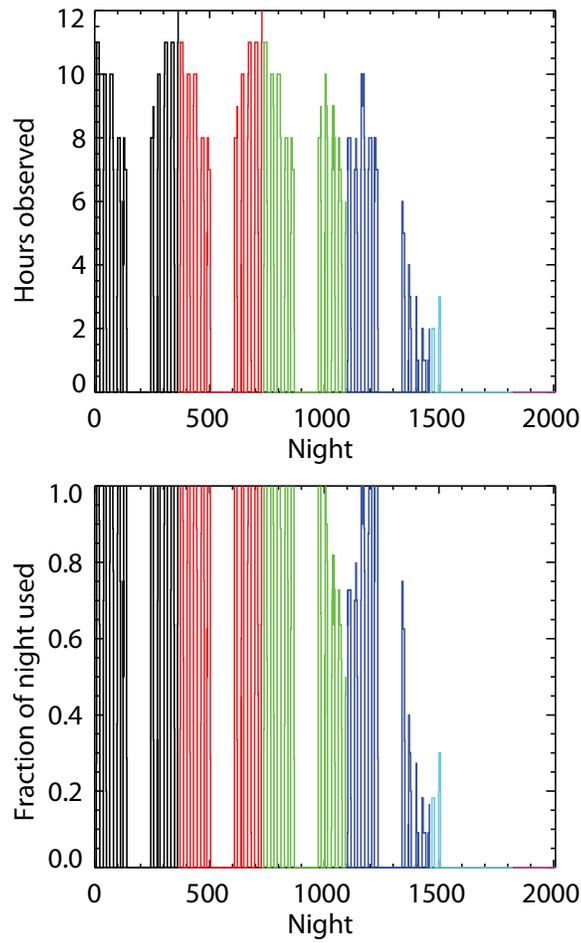

Figure 6.6: Distribution of nights used for the entire BigBOSS Key Project, as computed by our optimization routine. Different colors denote different years. The total time shown here includes the effects of the site seeing and weather conditions. The large gaps are due to the summer shutdowns at KPNO, and the narrow gaps are times near the full moon.

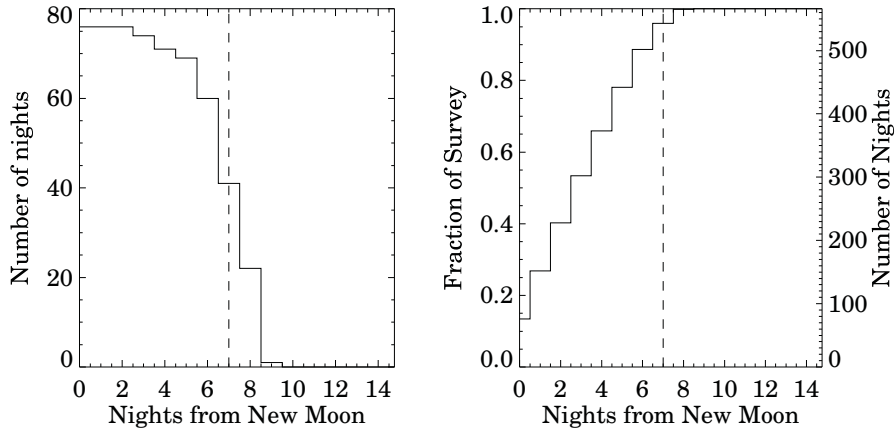

Figure 6.7: *Left:* The distribution of nights on which some data are obtained (i.e., including partial nights) as a function of nights from new moon for the current baseline BigBOSS survey. *Right:* The cumulative fraction of nights of observation as a function of days from new moon. The right axis shows the total number of calendar nights used by the proposed survey (includes partial nights).

The total number of scheduled hours used by the survey is 4,265, or $\approx$450 equivalent 9.5-hour nights. These hours are distributed over 565 calendar nights, a higher number than 450 because it includes partial nights shared with NOAO users. The baseline plan is to schedule $\approx$130-140 partial or full nights per year for the BigBOSS survey during the first four years, and the remaining 20 nights during the last year. A more evenly paced program spread over 5 years can also be accommodated; the exact distribution of nights per year can be negotiated by NOAO. The total survey time presented here does not include the extra overhead that may be required to accommodate any unforeseen issues. Doubling the overhead (from 1 min to 2 min) to account for the current telescope settling time will add an additional 17 equivalent 9.5-hour nights of time to the program; this remains within the 500 night limit. Despite the fact that our current optimization results in only 450 equivalent nights, our total request remains 500 nights in order to accommodate contingency due to possible weather and instrumentation issues.

As shown in Figure 6.7, the current BigBOSS baseline survey mainly uses nights during dark and grey time, no more than $\pm$ 7 to 8 nights from the new moon. This requirement is driven mainly by the Ly$\alpha$ forest BAO program, which necessitates observing a large number of faint QSOs at blue wavelengths. However, this requirement does <u>not</u> mean that access to dark time will not be available to NOAO users. During the regular survey operations, 20% of the fibers remain unassigned to primary targets and a significant fraction of these could be made available for synchronous observing programs by the NOAO community (see Chapter 3). Over the course of the proposed baseline survey, this corresponds to between 5 and 10 million potential targets that would be available to the NOAO user community for scheduling. In addition, the survey makes use of partial nights, especially in the latter years, which can be shared with the community. Finally, the Key Project may be able to use any available unused fibers during time assigned by NOAO to community-led programs. We anticipate that the distribution of nights, scheduling of the survey, and sharing of fibers on the instrument will be coordinated by NOAO in order to ensure community access to dark/grey time every year.

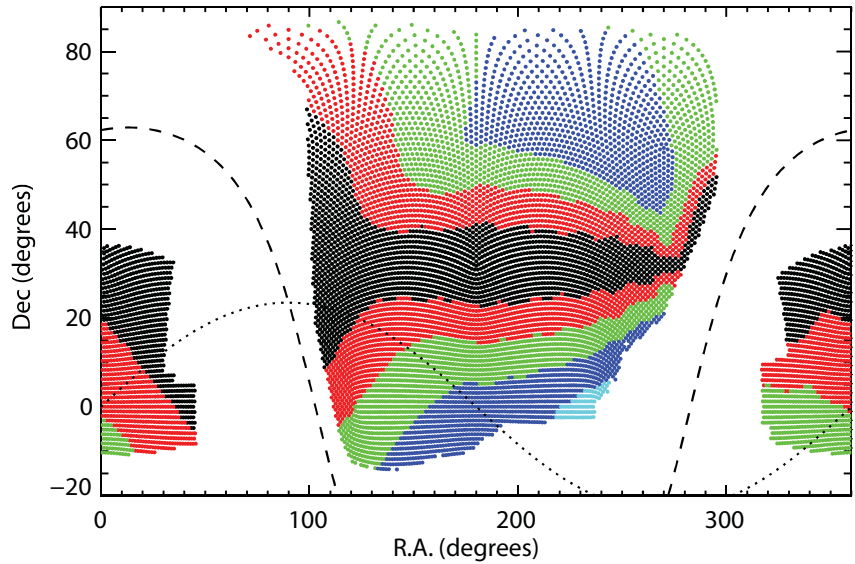

Figure 6.8: Equatorial projection of the field distribution for the BigBOSS baseline survey, color-coded according to the year of completion. The color-coding corresponds to the years shown in figure 6.6. The dashed line represents the Galactic plane, and the dotted line the Ecliptic plane. Note that while this footprint depicts the current baseline, it may be reoptimized.

## 6.2   Fiber assignment

In Table 2.2, we quoted a "fiber completeness" for the efficiency to observe our targets of 80%. That is, of all available targets, we can expect to reach the desired exposure times for 80% of them, given the physical constraints of the minimum fiber spacing. In this subsection, we justify this number.

The fiber completeness depends on the density of available targets (discussed and justified in §4) as well as the density of observations in the survey area (discussed and justified in §6.1.1). Given these numbers, we can evaluate the resulting fiber completeness using fiber assignment simulations.

In the simulations used in this model, fibers are assigned randomly to available targets that have not been previously observed. Given a list of targets, we track whether each target has achieved enough effective exposure time. At the time of each observation, each available actuator is assigned randomly to a target in its patrol area (see §5). We step through the actuators in order, and do not allow two actuators to observe the same target in the same exposure. After each exposure the effective exposure associated with each target is evaluated, and, if reached, the target is removed from consideration for further observation.

Table 6.1: Total Time for the Baseline BigBOSS Survey[1]

|         | Calendar Nights | Hours | Effective Nights |
|---------|-----------------|-------|------------------|
| Year 1  | 138             | 1211  | 127.5            |
| Year 2  | 139             | 1210  | 127.4            |
| Year 3  | 138             | 1089  | 114.6            |
| Year 4  | 131             | 710   | 74.7             |
| Year 5  | 19              | 45    | 4.7              |
| Total[2]| 565             | 4265  | 449              |

[1] Based on median weather conditions.
[2] Does not include contingency or Pilot Survey.

For the default instrument configuration and the baseline survey plan, we find that the fiber completeness is 80%. Meanwhile, the efficiency (or utilization) of the fibers is also about 80% — in each observation 80% of the fibers are assigned to primary science targets. A greater density of fibers or a greater number of observations would increase the completeness but decrease the efficiency.

In the real observations, a more optimized approach would yield slightly higher completeness than this simple "greedy" algorithm. Initial work has indicated a gain of about 2% in efficiency by using a more optimal algorithm, or about 600,000 more targets total.

### 6.2.1   Fiber Allocation Fractions

To execute the survey in a manner that achieves the key science goals, BigBOSS must assign each target type (ELG, LRG, or QSO) a fraction of the available fibers. Further, because the LRG and QSO targets require continuum measurements, multiple fiber exposures must be allocated to these objects. We therefore quantify the allocated exposure times per square degree by the *effective* target density, which is the surface density for each target type multiplied by the number of required exposures. Table 6.2 shows the effective target density for the BigBOSS targets and the expected number of exposures per target. These values are also shown in the survey overview Table 2.2.

An additional complication to the survey strategy is the manner in which the QSO targets are confirmed. As discussed in Chapter 4, achieving a highly-complete Ly$\alpha$ quasar sample using only *gri* photometry requires targeting across a large portion of the stellar locus where contamination from objects with similar colors can be high [e.g., Richards et al., 2002]. To remove the contaminants from the initial target distribution of $\sim 250$ QSO targets deg$^{-2}$, BigBOSS will allocate more fibers to the QSO targets in the first tile of a given sky field position. This strategy allows for a confirmation of the Ly$\alpha$ QSOs within the sample before allocating any further exposure time to measure the Ly$\alpha$ forest. While the fraction of ELG targets must decrease in the first tiling to accommodate this strategy, the overall decrease in the number of ELG redshifts is $< 10\%$ while the QSO target completeness increases to $> 80\%$.

Since the primary targets for the Key Project have an 80% utilization fraction of fibers, 20% remain unallocated for the average tile (this number is much smaller for the first tile of a given pointing, but increases to 24% for tiles 2 through 5; see Table 6.2). These fibers will be randomly distributed over the 7 deg$^2$ BigBOSS field of view. The BigBOSS Key Project will allocate a fraction of these unallocated fibers in each tile for calibration (i.e., sky and standard stars); the remainder will be available for ancillary targets. The exact number of calibration targets needed is not known at present (it will be the subject of tests during the construction and Pilot Survey phases), but we anticipate that on average between 10% and 20% of fibers will be available to the NOAO community for ancillary targets. Community science projects that could use these unallocated fibers are discussed in Chapter 3.

Table 6.2: Target and Fiber Allocation for the BigBOSS Survey

|  | ELGs | LRGs | QSO Candidates | Ly-$\alpha$ BAO QSOs[2] | Other[1] |
|---|---|---|---|---|---|
| *Tile 1:* | | | | | |
| Potential targets deg$^{-2}$ | 281 | 135 | 250 | 0 | 49 |
| Assigned fibers[3] deg$^{-2}$ | 281 | 135 | 250 | 0 | 49 |
| Target split[4] | 0.39 | 0.19 | 0.35 | 0 | 0.07 |
| Time split[5] | 1 | 0.5 | 0.2 | 0 | 1 |
| *Average of Tiles 2-5:* | | | | | |
| Potential targets deg$^{-2}$ | 510 | 143 | 0 | 65 | >173 |
| Assigned fibers deg$^{-2}$ | 381 | 107 | 0 | 52 | 173 |
| Target split | 0.54 | 0.15 | 0 | 0.07 | 0.24 |
| Time split | 4 | 2 | 0 | 0.8 | 4 |

[1]Includes fibers allocated to sky, calibration and ancillary (i.e., including community) targets, based on a total BigBOSS fiber density of 713 deg$^{-2}$
[2]QSOs selected (from tile 1 observations) to be at $2.2 < z < 3.5$
[3]Allocated fiber density including fiber completeness
[4]Division of targets within each tile
[5]Fraction of total exposure time this set represents

## 6.3   Use by External Community

As described in Section 3, we envision that BigBOSS will benefit the NOAO user community in three ways.

First, astronomers will be able to propose for telescope time with the BigBOSS instrument through either the regular or survey Time Allocation Committees (TACs) administered by NOAO. In this instance, the BigBOSS instrument will be considered a regular facility instrument: full (or partial) nights will be assigned to successful proposals by the NOAO TAC and supported primarily by the NOAO staff. The BigBOSS team will provide the tools required to plan and schedule the observations with the instrument and for basic pipeline reductions. These will be the same tools that will be used for the BigBOSS Key Project, and the BigBOSS collaboration will collaborate with NOAO on a plan on how best to implement and support the use of the instrument.

Second, we envision that a number of fibers will be made available for use by external users even during time committed to the proposed BigBOSS Key Project. The expectation is that on average 20% of the fibers will not be utilized for primary targets. While some of these will be needed for calibration (i.e., to target sky and standard stars), a significant fraction could be made available through the NOAO TAC process for "synchronous" science programs.[7] This enables other large survey programs for targets sparsely distributed across the 14,000 deg$^2$ Key Project footprint. Those programs would be subject to the magnitude limits and cadence of the Key Project. This mode has proven very successful in SDSS-I, SDSS-II and SDSS-III/BOSS, although in those cases limited to the SDSS community.

In order to enable such synchronous observations, we envision collaborating with NOAO to create a mechanism whereby proposals for use of BigBOSS during its regular survey operations are first ranked by the regular NOAO TAC; these targets are provided to the BigBOSS collaboration to be included in the targeting lists used by BigBOSS for its survey operations. Observations of these targets will be obtained as part of the regular survey strategy, reduced as part of the standard survey pipeline and made available to the proposers as the reduced data become available. Note that proposals for use of fibers during normal BigBOSS survey operations will be subject to various constraints: time critical requests cannot be considered in this mode; exposure times will be limited to typical BigBOSS single-pointing integration times; and non-standard reduction methods cannot be supported.

Finally, the reduced data from the BigBOSS Key Project will be publicly released through NOAO and made available for archival research. This is discussed in more detail in Section 3. We note again the large impact of the archives resulting from the SDSS spectroscopic surveys on astrophysics, and expect that the science yield of the BigBOSS archive will have similarly high legacy value.

As mentioned in section 6.4.3, BigBOSS Key Project observations will be carried out with the help of an Observing Scientist, who will be located in the Mayall control room with the telescope operator. The Observing Scientist will be responsible for the afternoon checks, nightly implementation of the observing plan, monitoring of the instrument and observing, and quality assessment. One possible operational mode would be for this Observing Scientist to be responsible for *all* the observations with the BigBOSS instrument, i.e., by the NOAO community and the BigBOSS team. In this mode of operations, all observations with the BigBOSS instrument would be uniformly obtained by an expert observer (i.e., essentially in a queue mode), and all the data would be reduced by the same pipeline used for the BigBOSS Key Project survey data. In this model, the cost of the Observing Scientists would be proportionately shared between the BigBOSS project and NOAO. The advantages of this operational approach are that NOAO would not have to "start" new observers with this complex instrument, would not have to maintain a separate (and different) pipeline, and would be able to ensure a smooth transition (through the Observing Scientists) when, at the conclusion of the BigBOSS Survey, responsibility for the maintenance and operation of the instrument is transferred from the BigBOSS team to NOAO.

---

[7]The number of calibration targets needed per field is yet to be determined, since it will be based on the final instrument performance caharacteristics (e.g., stability, throughput uniformity, fiber properties, etc.) and observational characteristics (e.g., spatial and temporal variations in the sky spectrum, extinction variations, etc.). We are planning tests during the construction phase and will experiment with different strategies during the Pilot Survey in order to determine the number and distribution of calibration fibers. At this stage, we estimate that on average between 10% and 20% of the 5000 fibers will be available for "synchronous" science observations during the Key Project survey.

## 6.4 Day-time Support, Instrument Support and Maintenance Plan

In order to minimize the overheads of switching between instruments and the failure / breakage of any component, we propose that BigBOSS remain permanently mounted at the Prime Focus of the Mayall telescope for the entire duration of its operation at Kitt Peak. This will render the Mayall prime focus unavailable for other instruments during this period. (We note that the only existing prime focus instrument is the $36'$ field-of-view MOSAIC imager, which will be superceded by the WIYN One Degree Imager and the Blanco Dark Energy Camera, and that no other prime focus instruments are currently being planned.) However, as described in Chapter 5, the BigBOSS prime focus assembly will enable the use of the Ritchey-Chretien secondary focus by providing mounting points for the existing Mayall F/8 secondary mirror. We propose to leave the entire fiber train intact and the end-to-end system operational even during periods when the F/8 is being used, so that we can continue to monitor the health of the system and minimize the time used for vetting the instrument during the switch back from F/8 to Prime Focus operations.

BigBOSS is a complex instrument undertaking a complex survey, and proper maintenance and monitoring of the various hardware and software modules will be critical to the success of the survey. In the following subsections, we outline the various tasks that we envision will be required during routine survey operations.

The basic requirement is that there be a sophisticated automated daemon that monitors the instrument status and produces a regular report for the use by the Survey Team. This daemon should schedule tests, analyze the results, and provide regular updates on the status of the different parts of the instrument (e.g., the dewars, vacuum and cooling systems, fiber positioning system, software and hardware supporting the acquisition system, ADC positioning, focus, etc., fiber throughput, etc.). The schedules for testing / monitoring of different systems are likely to be different (for instance, dewar temperatures need to be monitored hourly, whereas CCD gain and noise measurements could be scheduled once every month), and the required response times for problems are also likely to vary.

### 6.4.1 Routine Maintenance

- Yearly test of Prime Focus optics (throughput, focus, etc.)
- Yearly test of Fiber Positioning and calibration system
- Yearly test of Spectrograph and CCD system

- Yearly throughput tests of end-to-end system
- Monthly test of vacuum on each of 30 dewars (i.e., two per week-day)
- Monthly test of UPS systems
- Monthly test of calibration sources
- Monthly test of temperature control of spectrograph room

### 6.4.2   Maintenance During BigBOSS Runs

*Tasks Prior to each BigBOSS Run:*

- If an F/8 to Prime Focus swap is required prior to each run, a few-hour on-sky commissioning period may be required at the start of each run in order to, e.g., exercise the primary instrument modes, determine any zero point offset in the fiber alignment camera calibration, test the pointing, fiber alignment, and guide/acquisition systems on sky, perform system throughput verification tests, etc.

- Test end-to-end installation of instrument (end-to-end system functionality check; elaborate version of the afternoon checkout listed below, but with added rigorous optical quality checks)

- Review survey status and targeting plan for each run (from BigBOSS Team + NOAO PI targets), with weather options

*Day-time Tests and Calibrations:*

- Daily generation of instrument monitoring report and quality assessment of previous night's data / survey status

- Daily test of CCD health (i.e., test exposures to measure read noise, dark current, sensitivity, gain, and exercise readout software and real-time reduction software)

- Daily test of Instrument control software (i.e., test of fibers, alignment camera imaging, measurement of fiducial fiber positions, actuator response, ADC fiducial / motors test, shutter test, etc.)

- Daily spectrograph calibrations (i.e., arc lamps, flat fields, focus check, optical quality check, connectivity to data transfer system, disk space availability, etc.)

- Daily review of targeting and scheduling plan (with options or modifications as needed based on weather predictions and NOAO PI usage)

*Night-time Tasks and Calibrations:*

- Telescope initialization and checkout (verify telescope health and check pointing, focus, tracking, etc.)

- Beginning of night calibrations (twilight observations, standard star / cluster fields to vet end-to-end system and build on-sky throughput history)

- Execution of BigBOSS Key Project observing plan

- Real-time evaluation of observations and strategy (i.e., monitor pointing/airmass/weather/seeing/through issues, and reassess plan for night in real time; monitor instrument health; check real time reductions and quality assessment; monitor proper operation of data archiving, etc.)

- End-of-night calibrations (twilight observations, standard star / cluster fields, etc.)

- Telescope shutdown

- Instrument shutdown

### 6.4.3   Staffing During Survey Operations

We assume that NOAO will provide a telescope operator who will be responsible for operating and maintaining the functionality of the telescope during nights assigned to the BigBOSS survey. In addition, the BigBOSS team will provide an Observing Scientist who will be responsible on each night for implementing the schedule, running the instrument, checking / vetting the pointing, acquisition, and fiber alignment, vetting the data and the proper operation of the real-time pipelines. The Observing Scientist should (ideally) be co-located with the telescope operator in the control room. In addition to the on-site Observing Scientist, other scientists from the BigBOSS Team (located at the collaboration institutions) will provide remote assistance with the day-time checkout tasks (described above). Scientists from the LBNL group will be responsible for running (and maintaining) the pipeline reductions and data quality assessment routines and for preparing a daily report for consideration by the BigBOSS Observing Scientist (and the relevant BigBOSS Team personnel) prior to each night. We anticipate a pool of no more than 3 Observing Scientists employed by the project for the duration of the survey, with typically only one on duty on any given observing night. These Observing Scientists will likely have other duties when they are not covering observing runs, and support for their positions could be shared with NOAO.

### 6.4.4   Plan for Instrument Part Replacement

Detailed maintenance and spares plans will be highly dependent on the components selected during the preliminary design phase and experience gained during development of BigBOSS specific components. We anticipate that special (i.e., not off-the-shelf) spares of critical systems will be constructed and provided as part of the BigBOSS construction process. Here we describe what we anticipate will be encompassed in the plans. Most failures in the items below will require a day shift to repair.

### 6.4.5   Telescope Systems

**Corrector Motors and Drives.**   The corrector barrel is mounted on a hexapod actuator to provide focus and axial alignment with the primary mirror. There are six motors involved and an electronics module to control the motors. When a design is complete and a vendor is selected, we will establish the inspection and maintenance schedule and the parts inventory that needs to be stored at the Mayall. The design phase needs to consider *in situ* repair of the hexapod, if possible.

The ADC also has a pair of motors, a mechanical drive system and control electronics. The maintenance and repair issues are the same as those for the hexapod.

**Fiber View Camera.**   The fiber view camera is a critical component in the operation of BigBOSS. It is comprised of a CCD and control/readout electronics, probably packaged as a unit. A working spare needs to be available.

**Lamps.**   Two types of lamp systems exist in the corrector region. One type provides dome flat illumination, both broadband and line lamps. The other type illuminates the fiber view camera fiducial fibers and is probably LED based and built in redundancy is possible. Both systems contain a small amount of electronics. Easily accessible locations for these items should make replacement fairly easy with locally stored spare parts.

### 6.4.6   Focal Plane Systems

**Guider and Focus Sensors.**   These are critical items to the operation of BigBOSS. Spares of the detectors and their electronics must be maintained at the telescope. If implemented as CCDs on the focal plane, opposed to being remoted via fiber bundles, repair times may be lengthy. In either case, redundancy using multiple sensors can reduce the demand for immediate repair capabilities.

**Fiber Positioners.**   The impact of failed fiber positioners is a complex issue. An individual positioner is not easily accessed, but individual failures are not a serious threat to successful operation. What has not yet been established is the threshold for the total number of failed positioners that would trigger maintenance intervention. This limit is probably in the low single-digit percent range. Since partial disassembly of focal plane infrastructure will be required for servicing, positioner servicing can only practically be done during summer shutdowns.

### 6.4.7   Spectrographs

**Cryostats and CCDs.**   The only components in the cryostats that have any likelihood to fail are the CCDs, temperature monitors and heaters. The cryostats are designed for replacement in 24 hours with on-site spares. Since there are three CCD types in three cryostat types, a spare of each must be maintained at the telescope.

**CCD Frontend Electronics**   The CCD frontend electronics is mounted on the outside of the cryostat. There are two configurations of one board design reflecting the difference between the e2v and LBNL CCDs. Spares must be maintained at the telescope for daytime replacement.

**Shutters.**   Shutter replacement requires removal of a cryostat. Spares will be kept locally.

**Lamps.** The fiber back-illumination lamps are probably LEDs mounted on the dichroic box. They should be inherently reliable and redundancy is easily implemented. The slit array illumination is probably a lossy fiber driven remotely by a fiber bundle itself illuminated by arc lamps. The dome flat lamps are a mixture of arc lamps and halogen lamps. The latter two systems are only used occasionally and are not required for routine operations; repairs can be scheduled with longish lead time.

**Cryocoolers.** The LPTs are given for a MTTF of 40,000 hours and do not require maintenance. The monitoring of their performance during lifetime could be implemented in the cryostat control system, especially if we use the higher-functionality version of the LPT. In case of one unit failure, the cold machine will be replaced by a spare one belonging to the same series and the faulty unit will be returned to the manufacturer for examination and repair.

**Vacuum System.** Six vacuum pumps are envisioned and are normally valved off. Most likely, one spare will be kept on site.

### 6.4.8 Controls Systems.

The control system and computers are commercial using industry standard interfaces. During the design phase, a list of critical spares will be identified. During the course of BigBOSS operations, we will need to monitor parts becoming obsolete.

While it is desirable to maintain software versions, both our own and operating systems, control system parts replacement over the lifetime of the instrument will probably require updates to the software.

## 6.5 Pilot Program

In order to verify the fiber assignment and scheduling algorithms, mitigate the risks associated with the target selection, test the efficiency of various tiling strategies, optimize observing strategies with the aim of minimizing overheads (i.e., resulting from pointing, slewing, fiber positioning, readout, etc.), understand the calibration requirements, and commission the data reduction and archival pipelines, we intend to undertake a short ($\approx 13$ night) Pilot Survey at the end of the instrument's on-sky commissioning phase. This Pilot Survey will provide an important ground truth which will allow us to finely optimize and finalize our plans for executing the BigBOSS Key Project.

We envision this Pilot Survey as targeting 5 fields, three in the north Galactic cap accessible during the spring semester and two in the southern strip accessible during the fall semester. These fields will be targeted to a depth roughly 1 magnitude fainter than the selection depth for the BigBOSS Key Project, and with much higher completeness. In order to reach 95% completeness in our targeting, we expect each field to be targeted by 6 different fiber configurations. Each fiber configuration will be targeted for a total exposure time of 7 times our nominal survey exposure time (i.e., $7 \times 15\text{min} = 105$ min), which will result (with overheads) in roughly 2 hours per fiber configuration per pointing. This portion of the Pilot Survey requires 6.7 clear nights, or 10 assigned nights (accounting for 65% clear fraction). In addition, we will require an additional 3 nights to experiment with different tiling and calibration strategies and to interface the calibration, reduction and analysis pipeline seamlessly with the survey operations. The Pilot Survey will therefore require a total of 13 nights.

We anticipate requesting this time at the end of the on-sky commissioning phase of the instrument, and note that this time is not considered part of the Key Project, since it will be used to verify many aspects of the instrument, pipelines, and survey prior to the start of actual survey operations.

# 7  Data Management Plan

## 7.1  Data-Taking System

### 7.1.1  Operations Database

At the heart of the BigBOSS data-taking system will be an operations database, which will log information obtained from the Instrumental Control System, the Observing Control System, the online software (see Section 5), as well as the survey strategy software (see Section 6). Currently the SDSS-III data-taking system uses a similar model. A central database logs telescope and instrument status and meta-data. It tracks current observations, priorities, weather, airmass etc. so that observers can make informed decisions about upcoming observations.

A web-based interface to this database will provide observers a complete picture of the observation status. Our team has experience building such tools for the SDSS 2.5m telescope and for PTF. This tool will allow members of the BigBOSS team to remotely monitor observations, which is expected to be especially valuable during commissioning and early operations. This could also allow PIs to monitor their observations during NOAO PI programs.

### 7.1.2  Fiber Location Specification

The Observing Control System needs to be fed fiber locations, and thus we will develop a standardized system for specifying fiber location control, and store the specifications for each exposure in the operations database. The details of the fiber location specification will depend on the engineering details of the final system. We propose that the fiber positions will be specified in terms of a radial and angular offset from a central position, that the central positions will be mapped relative to a fixed point on the focal plane and that this fixed point can be mapped to a definite celestial coordinate (Right Ascencion and Declination) during exposures. Excursions outside of the known mechanical range of motion of the fibers will be flagged. Requests to move fibers to specific positions will be logged in the operations database, as well as the response of the instrument, allowing the request to be verified against fiber camera images.

### 7.1.3  Observation Planning Specification

For BigBOSS observations, targets derived from the target database will be consolidated into a pointing by the survey strategy software. The constraints of coordinates (airmass), moon, and priority provide a number of days and range of hour angle where each pointing is observable. Scheduling is a difficult, non-linear problem, but our experience with scheduling observations on the SDSS Telescope and planning tools for the WFMOS survey will form the basis of our system. Observation plans and results will be stored in the operations database.

For non-BigBOSS observations, pointings defined by the PIs will be inserted into the operations database, and accessible for observation through the online software using the same tools available for BigBOSS observations. We anticipate that PI users will be observing in a broader range of conditions and with a more variable use of the instrument (different exposure times, for example, or weaker lunation constraints), and the software will have the freedom to handle such conditions.

The BigBOSS project will work with the NOAO Science Data Management group of the NOAO System Science Center to determine the appropriate data formats, header protocols, and meta-data formats. Numerous standards have already been established by NOAO, the Virtual Astronomical Observatory, and the requirements of the broader astronomical community. Many of these will pertain to the BigBOSS surveys, and we will address how this data set fits into National Observatory and OIR System operations as a whole.

### 7.1.4 Telescope and Instrument Status

As described in the Section 5, telescope and instrument status are monitored by the Telescope Control System on the Mayall and the ICS and OCS systems. All requests and responses for these systems will be logged in the operations database.

### 7.1.5 Raw and Meta-data Structure

Raw spectroscopic data will take the form of FITS images. Meta-data will be stored in FITS headers (as needed) and in a database. It will also be possible for the operations database to generate flat files for the meta-data, which might be needed for downstream processing.

Fiber camera images will be stored as FITS images. Fiber camera meta-data will also be placed in the operations database.

### 7.1.6 Archive: Permission Locks, Backups, Checksums

Raw data will be transferred daily to a central data repository. There it will be backed up to an tape storage system and copied to one or more mirror facilities. Before transfer, directories containing nightly data will be permission-locked so that no further data can be written. Checksum of files will be computed before transfer to insure data integrity at every stage of transfer and backup. As part of this process, the operations database will also be backed up, both in a flat file form and on a remote clone database.

Failures of the checksums at any point in this chain of steps will trigger human investigation of the problem, and recovery from the original files at KPNO.

## 7.2  Pre-Survey and Target Selection Data Management

During the pre-survey phase of BigBOSS we will assemble a target catalog based on photometric data from PanSTARRS-1, Palomar Transient Factory, and WISE. These will be tied to the astrometric and (where appropriate) photometric system of the final SDSS-III/BOSS imaging data, to be publicly released in December 2010. We will store and curate the photometric data files used to construct the targeting catalogs. Furthermore, any spectroscopic data taken prior to BigBOSS for the purposes of target selection testing or verification we will also curate. As described below, all target catalogs and preparatory spectroscopic data sets will be stored in the BigBOSS science database.

   Connected to the data management effort will be the development and testing of the target selection software. A major part of this effort is building the software system for target selection, including its interaction with the target database. We will develop modules for each target selection category which attach to the science database in order to flag objects according to each set of criteria.

   The initial target selection will be available approximately 1 year before observations commence. The redshift success rates and redshift distributions in the Pilot Survey inform any improvements to the targeting algorithms before the Survey proper begins.

   A final important presurvey data management activity is the development of survey mock catalogs, for testing of target selection and tiling effects on scientific analyses. Based on $N$-body simulations and other techniques, these simulations will be stored in the same formats and database structures as the real data, allowing the development and testing of science analysis pipelines in a realistic environment. These mock catalogs and analyses will allow "science grade" testing of the survey planning decisions.

## 7.3   Quality Assessment System

We will build on the experience of the BOSS team which has a very good quick reduction pipeline in operation on the Sloan Telescope. The pipeline is a stripped-down version of a full reduction pipeline, replacing the most expensive computational steps with simpler (and in some cases more robust) algorithms. In addition, the stability of the BigBOSS spectrographs will mitigate the need to obtain and process quick-look calibrations (arcs and flats) in real time; using one or two calibrations per night will be sufficient. This reduces the resources necessary for the real-time reductions.

   This system gives an estimate of S/N per exposure as a function of wavelength and object magnitude. This allows a robust, near-real-time decision of whether a tiling on the sky has been observed to completion. The system also flags problems with the telescope or instrument, which has proved valuable at Sloan for quickly identifying problems such as failed shutters or electronics glitches.

   The quick reduction pipeline will log its results in the operations database. The web front-end will allow real-time access to visualizations of the results (and the results themselves) to its users.

   The computational resources needed for quick extractions will thus be fairly modest at 1 CPU for each of the 30 CCDs. If the more expensive row-by-row optimal extractions used by full BOSS reductions were used, the computing requirements would increase by a factor of several but still be very manageable.

## 7.4   Data Processing and Analysis Strategy

### 7.4.1   Extraction Strategy

Extraction is the problem of inferring one-dimensional input astronomical spectra from two-dimensional digital spectrographic images. Since BigBOSS will operate in a very low signal-to-noise regime, it is imperative that our extraction strategy be *statistically* optimal, so that every bit of significant information recorded by the spectrograph CCDs is faithfully propagated into the 1D spectra. Furthermore, we must keep *systematic* extraction errors to an absolute minimum: even small systematic mis-estimates of the night sky spectrum will lead to large non-Gaussian residual errors in the extracted spectra, when considered relative to the flux levels of BigBOSS core science targets.

The "optimal extraction" algorithm described in detail by [Horne, 1986] represents the current standard of quality, and it has many mature implementations including the `idlspec2d` software used for the analysis of SDSS and (currently) BOSS spectroscopic data. However, this algorithm has a key shortcoming when applied to fiber spectroscopy, in that it treats the spectrograph PSF (i.e., the convolution of the optical fiber image with the spectrograph camera aberrations) as a separable function of $x$ and $y$ coordinates on the CCD detector. Residual coma, astigmatism, and core/wing effects in real spectrographs conspire to falsify this assumption of separability, and thus traditional optimal extraction does not generate a mathematically correct model for the two-dimensional spectrograph data. While this shortcoming can be safely ignored at higher signal-to-noise levels, it must be tackled head-on for BigBOSS.

The BigBOSS extractions will therefore be carried out following the algorithm described by [Bolton & Schlegel, 2010]. This algorithm extracts spectra using a fully correct two-dimensional model to the CCD data. The method accounts for optical heterogeneity among the fibers, and propagates all information and resolution forward to the final extracted spectra. The resolution and statistical covariance of the extracted spectra are accurately characterized, and the extracted samples have (by construction) no covariance from one pixel to the next. This permits straightforward and correct $\chi^2$ comparisons of models against the extracted spectra. The implementation of this method will be carried out initially as part of the ongoing BOSS survey, and we expect to have substantial practical experience and usable code in place by the time that the BigBOSS instrument is commissioned.

One of the greatest challenges to the implementation of the Bolton & Schlegel algorithm is the need for a detailed and accurate representation of the "calibration matrix" that relates input flux as a function of wavelength and fiber number to the response of all CCD pixels. In this regard, the anticipated stability of the bench-mounted and thermally controlled BigBOSS unit spectrographs affords a great advantage. Calibration libraries will be assembled on a monthly or yearly basis using standard arc-lamp and flat-lamp illumination systems, with alternating sparse masking of the input fibers to allow measurement of the fiber PSF wing profiles in the absence of fiber-to-fiber cross-talk. Calibrations may also be obtained using narrow-band tunable-laser illumination; several BigBOSS collaborators are actively exploring this method in collaboration with researchers at the National Institute for Standards and Technology, in the context of the BOSS instrument.

### 7.4.2   Sky Subtraction

Sky subtraction is an important problem for fiber-based spectrographs. In the background-limited faint-galaxy regime in which BigBOSS will operate, it is perhaps *the most* important problem. The most significant challenge is posed by the many OH rotational emission lines that become extremely prominent redward of 7000Å. The resolution of BigBOSS, while not *high resolution*, will be approximately twice as high as that of SDSS-I and BOSS, and thus the wavelength regions strongly affected by these emission features will be cut roughly in half (since the strongest OH lines are already resolved from one another at SDSS-I and BOSS resolution). Nevertheless, optimal handling of the problem of sky subtraction will be crucial to the full scientific success of the BigBOSS instrument and survey program.

Several key strategies for effective sky subtraction have been proven in the SDSS and other surveys already, and we will adopt these strategies in our approach to BigBOSS sky subtraction. Most importantly, the sky must be modeled and subtracted *before any rebinning or combination of the spectra* [Kelson, 2003], so as not to degrade native resolution and introduce ill-characterized correlations. We will also decompose our wavelength solution into *relative* and *absolute* components [Bolton & Burles, 2007]. Relative wavelength calibration is a crucial ingredient to the success of sky subtraction, and it can be determined with much greater accuracy than absolute wavelength calibration. We will also factor our flat-fielding images into pixel-flat and fiber-flat components so as to ensure the most accurate relative calibration between fibers, which will be crucial for the accurate transfer of model sky spectra between fibers. Finally, we will map the large-scale spatial illumination pattern differences between sky and calibration frames using periodically acquired twilight flat frames.

Traditional shortcomings of sky subtraction in multi-fiber spectrographs can be traced to three principal causes: (1) variation of the spectrograph PSF between sky and object fibers (due to fiber non-uniformity and spatially varying camera aberrations); (2) systematic errors due to the use of mathematically inaccurate models in the extraction of 1D spectra from 2D CCD pixel data; and (3) insufficient spatial sampling of the sky by dedicated background fibers. The extraction algorithm of [Bolton & Schlegel, 2010] will directly remedy the first two problems. By modeling the input night sky spectrum "upstream" from the optical system, and convolving with the varying PSF over the fiber array and camera field of view before computing $\chi^2$ against the data, the algorithm avoids the problem of subtracting spectra with varying extracted line-spread functions between sky and object fibers. In addition, by extracting the raw CCD data using an image-modeling basis composed of *two-dimensional* PSF profiles, the algorithm avoids the systematic shortcomings of the traditional "row-by-row" optimal extraction algorithm [Horne, 1986] that implicitly assumes a separable form for the 2D spectrograph PSF – an assumption that is violated most strongly in the case of narrow emission features such as OH sky lines.

The problem of sufficient spatial sampling will be addressed in two ways. First, a substantial number of blank sky fibers will be allocated in each BigBOSS pointing, to allow a first-pass modeling of the sky spectrum and its variation across the telescope field of view. In general, the blank sky assignments will vary among the fibers from pointing to pointing, providing a greater constraint on measuring the fiber responses. Second, all faint galaxy targets, once extracted and modeled with a sufficient basis of eigenspectra, will be subtracted from the data to permit a second-pass modeling of the sky with finer spatial sampling. At approximately one fiber per 5 arcmin$^2$, the full set of fibers will provide a dense sampling of these features on the sky.

Our science goals require the subtraction of night-sky flux to better than 2%. The algorithm of [Bolton & Schlegel, 2010] formally permits "perfect" Poisson-limited sky subtraction, *provided sufficiently accurate system calibration.* We anticipate that our calibration plan, described in the previous subsection, will be sufficient in this regard. Once again, we note that the stability of the bench-mounted BigBOSS spectrographs will make this accurate calibration problem more tractable than for the telescope-mounted SDSS-I and BOSS spectrographs, which experience significant flexure and routine fiber-cartridge changes. To increase the accuracy of BigBOSS sky subtraction, we will investigate the benefits of "tweaking" our high-precision calibration libraries against daily calibrations and individual science frames themselves.

The implementation of all of the above strategies for accurate sky modeling and subtraction (with the exception of tunable laser applications) are included within the Project Execution Plan for the ongoing SDSS-III BOSS project, to be incorporated in the next-generation extraction pipeline for the survey that will be developed and tested over the coming 1 to 3 years. This software will be written within a modular, object-oriented framework so as to allow for maximum generalizability and re-use for future instruments and surveys such as BigBOSS. Hence, we expect to have substantial experience and code base at the ready for accurate extraction and sky subtraction of BigBOSS first-light data.

### 7.4.3  Redshift Measurement

Redshift measurements from extracted BigBOSS spectra will be made using forward-modeling techniques similar to those that have proven successful in the SDSS and BOSS projects. We will use deep BigBOSS data and (where necessary) spectral models to construct "Eigenspectrum" basis sets for each of a number of object classes: LRGs, ELGs, QSOs, and stars of all spectral types. For each spectrum and each object class, we will: (1) redshift the Eigenspectrum basis to a trial redshift; (2) fit the data with the best error-weighted least-squares linear combination of Eigenspectra at that trial redshift; (3) record the resulting value of $\chi^2$ for that trial redshift; (4) increment the trial redshift value differentially; and (5) repeat from step 1 until the entire plausible range of redshifts for that object class is covered. The classification and measured redshift for the spectrum will then be established by the global minimum reduced $\chi^2$ from this process. We may furthermore place photometric priors on the allowable classes and redshifts of targets, if this strategy is found to objectively improve redshift success metrics. Automated flags for redshift confidence will be set based upon the difference in reduced $\chi^2$ values between the best and next-best classification/redshift for each spectrum, the presence of excessive negative flux in the best-fit template model, reduced $\chi^2$ values that are too large even when minimized, and absence of a sufficient number of good data pixels in the extracted spectra.

The two main target categories for the BigBOSS galaxy BAO survey—LRGs and ELGs—each have characteristic narrow-band features that will make these redshift measurements robust. In the case of LRGs, the strong 4000Å continuum break and prominent Fraunhofer metal absorption lines will provide a clear and unambiguous redshift signal. ELG spectra will be characterized by [O II] 3727 doublet emission, which will be split at BigBOSS resolution and will therefore provide secure emission-line redshifts. For both target categories, the attention to statistical and systematic accuracy in extraction and sky subtraction described in Sections 7.4.1 and 7.4.2 will be of crucial importance to minimize the presence of sky-subtraction residuals that could lead to spurious redshift measurements.

### 7.4.4  Development Plan

Many of the research and development elements for BigBOSS are planned as part of the BOSS software development plan. However, there are significant differences between both the details of the instrument and the details of the target categories. Thus, a dedicated effort on BigBOSS reduction software will be necessary.

We are developing a detailed development plan for BigBOSS, centered on a set of data challenges applied to simulated spectra (modeled on those of similar projects like DES and LSST). The first step will be to create realistic data simulation software. Then, a series of challenges will test performance on: spectral extraction, sky-subtraction, wavelength calibration, flux calibration, and redshift finding. Once real data is in hand, the tools developed for these challenges will be essential in estimating completeness in the actual BigBOSS data set. The Eigenspectra used for the redshift-fitting will initially be based upon SDSS and other data sets, then improved with BigBOSS spectra as the Key Project progresses.

Indeed, some of this work has already begun. In preparation for this proposal, simulations of redshift-finding have demonstrated the viability of the BigBOSS instrument design for finding [OII] emitting objects.

## 7.5   Community Deliverables

### 7.5.1   Database Structure

BigBOSS will produce a science database that will contain the relevant imaging and targeting catalogs as well as the spectroscopic data, including the spectra themselves. Outputs from the database will be provided in common formats. Our techniques and tools will be based on the models from the SDSS, and NOAO surveys. The final database structure and interface will be developed in the period prior to BigBOSS commissioning. This will allow us to interface with real data via the database immediately. Mirror sites will provide data security. The NOAO facility, including its disk and tape farms, is designed for long-time data curation. We anticipate ongoing support from NOAO for access and interfaces to the database and other data products.

### 7.5.2   Target Lists and Window Function

The imaging catalogs used for target selection will be stored in the science database, along with their astrometric or photometric recalibrations. The results of the target selection pipeline will also be stored, including the target selection version, the type(s) of target assigned for each catalog object, and the relative priorities. Spectroscopic data will be matched to the photometric and targeting data in the database.

The system will allow for spectroscopic data of objects not in the BigBOSS targeting database, to incorporate ancillary targets and PI-driven observations. PI-driven observations will be tagged as such in the database, so that appropriate protection of the data can be implemented.

Results from the survey strategy software and pointing information will be included in the science database. This allows construction of the window function, necessary for computations of the 2-point statistics of the Key Project for both the real data set and mock catalogs. SDSS has considerable experience in constructing database tables and functions to satisfy this need, and we will build on tools like Mangle ([Swanson et al., 2008]) and the SDSS CAS Regions format.

### 7.5.3   Spectra

The curated form of the spectroscopic data set will be the FITS images for the raw data, the FITS files for the extracted and calibrated spectra, and the FITS files with the measured parameters. Direct access to these files will be available to the collaboration.

The calibrated spectra will be available as vectors of uncorrelated fluxes with errors. These include the fluxes, flux errors, the line spread function (LSF; 1D PSF in the wavelength direction), and bad-pixel masks. $\chi^2$ tests of template spectra against these data will procede as projections of those templates using the LSF for each spectrum. The spectroscopic catalog will consist of the parameters of the best-fit templates to each spectrum, the confidence, redshift error, and object classification. Any photometric information used in the extraction and redshift analysis will be included in the spectroscopic output files.

While the FITS files form the curated archival format for the data, we will enable other "views" of the reduced data products through the science database. As in SDSS, these will include tables of reduced parameters, plots of the spectra, and pointers (in the form of URLs) to the spectra in the FITS files. However, we will also include in the tables the spectra themselves, for ease of searching and analysis. Photometric information will be available by connecting and matching the spectroscopic database to the science database.

### 7.5.4   Documentation and Web Site

We will provide documentation for the BigBOSS science database, the interface tools, the targeting selection algorithms, and the reduction algorithms, as well as all other aspects of the analysis. This documentation will be tailored for developers, collaborators, and for the general research community.

The base-level documentation will be the data model, which will track the directory structure and contents of each survey file. We will develop a system to store, display, and search this data model, as well as to verify specific files against it. Pipeline developers will be required to document all of their outputs in this system. Similarly, any database tables will be required to be documented in a standard format. This base-level documentation will be most useful for developers and collaborators.

All software will be "self-documenting" in industry-standard formats, meaning that it will contain documentation formats that are machine-translatable into HTML or other formats. This policy will allow the developer's documentation notes inside of software to be easily viewable by users.

Web user interfaces, like the online interface, the operations database interface, and the science database interface, will have user guides including FAQs and cookbooks for common tasks.

The algorithms and techniques of target selection and spectroscopic reduction will be documented in full in technical papers published in the literature. We will use archived technical notes to track details of analysis performed for design and development purposes. Our web documentation will feature the "highlights" of the technical papers and notes to help make the database contents more usable.

Through our collaborators on the SDSS, PTF and DES, we have extensive experience in building web portals for accessing and analyzing data, with thoroughly documented results. We will draw on the experiences and lessons of these previous surveys.

### 7.5.5 Public Outreach

The BigBOSS team will bring a great deal of experience from SDSS to our public outreach. The first line of public outreach will of course be a web site, explaining the science goals and technology of the project. In addition to being an impressive piece of hardware, the overall mission is simple to express: BigBOSS will create the largest-volume three-dimensional map of the Universe to date. We will work with the NERSC Visualization Group to develop large-scale visualizations and fly-throughs based on this data. In addition, we will also provide illustrations of the motivation in physics for performing BigBOSS, and its consequences for understanding the nature of dark matter and dark energy. Finally, since BigBOSS will likely find a number of unusual and perhaps unique objects, the web site will serve to present and explain any strange or unusual objects we find.

For the purposes of secondary and higher education, we will provide "example science projects" for the data sets, which could be used for homework or labs in physics or astronomy courses. These science projects can be designed from the most simple (eyeball redshift measurement) up to the rather complex (finding unusual spectra). These science projects would make of the actual research tools provided by the BigBOSS science database.

Our team has experience building all of these sorts of public outreach tools. For example, the SDSS SkyServer[8] provides access to the SDSS data to a wide variety of audiences, from elementary school teachers to research professionals.

### 7.5.6 Software Distribution

Analysis software will be available along with public data. We will release tagged versions of the software used to generate the data in the database. We will provide a complete dependency tree for all software, down to kernel version, so that other researchers can replicate our results. As much as possible, we will provide test cases with the distributed software. However, to simplify software operations, we will provide a central computing platform with stable, proven software versions to ensure that collaborators are all using the same software.

## 7.6 Computing Requirements

We have summarized the computing needs for BigBOSS in Table 7.1. The data storage requirements are relatively modest and can be accommodated by the NERSC Global Filesystem[9]. The on-mountain storage will allow us to store one month of data as a contingency against data transfer failures. The data reduction requirements are based upon scaling the existing cluster that is used for BOSS data reductions. Data reductions will also take place at NERSC, which already has systems available to us that easily meet our CPU requirements. Currently NERSC already provides direct support for five experimental astrophysics programs based at LBNL and is laying the groundwork for the BigBOSS program through an allocation of 50 TB of space on their global filesystem for tests related to using data from the Palomar Transient Factory for BigBOSS target selection studies.

---

[8] http://cas.sdss.org/dr7/en/
[9] http://www.nersc.gov/nusers/systems/NGF/

Table 7.1: Summary of BigBOSS computing requirements.

| | |
|---|---:|
| Targeting Data Storage (flat files) | 150 TB |
| Targeting Database | 2 TB |
| Telescope and Instrument Control | 10 CPUs |
| Quality Assessment Processing | 30 CPUs |
| On-mountain Storage | 1 TB |
| Raw Data Storage | 4 TB/year |
| Data Reduction | 250 (ca 2010, quadcore) CPUs |
| Processed Data Storage | 20 TB/year |

## 7.7   Risk Assessment

### 7.7.1   Personnel Hiring

The risk profile associated with the data management of BigBOSS has three essential components: personnel, software development, and hardware.

By far the most important of these three risks is personnel. For the proposed schedule, it is important to move the data management team into place early enough such that survey planning and execution can move forward. This need motivates the hiring of 3 FTEs to assemble the target selection database, increasing to 4 to finalize the target selection, with an additional hire to help with commissioning. Our team is well-placed to address this risk: most notably LBNL, NERSC, JHU, and NYU have a proven ability to attract, or already have on staff, personnel with the expertise appropriate to handling massive astronomical data sets. In particular, all these groups have members with expertise in distributing the Sloan Digital Sky Survey data. Our plan is transition some of the personnel working on SDSS-III to BigBOSS in order to take full advantage of this experience.

### 7.7.2   Software Development and Performance

Software development is a second major risk, and breaks down into two parts: sources of delay, and software failures. Delay in target selection software can impede and/or complicate survey execution. Putting the data handling structure and personnel in place early is critical to mitigating this risk. The planned Pilot Survey to acquire spectra of large numbers of our targets is another important way to mitigate this risk (§7.2).

Delay in developing quality assessment software to run at the telescope can complicate commissioning; further, delay on the development of the final pipeline can complicate survey execution. In this case we have the world's experts on major spectroscopic pipelines on our team. They developed real-time quality assessment software and the final spectroscopic pipelines used by the SDSS for the past 10 years. In addition, we will have 2 FTEs, increasing to 3 during commissioning, to focus on this effort. We expect to build on their tools and experience to help address this risk.

Software failures can also put the project at risk. QA software failures can lead to underexposure or overexposure of spectra (the latter putting the overall schedule in jeopardy).

Addressing this risk requires significant personnel resources, particularly during commis-
sioning, to be in place to check results visually. In addition, the software tools must be
developed to allow easy access to the nightly data to rapidly address any problems that
develop. As we outline above, our data management plan includes such tools.

The final pipeline can also fail to sufficiently recover redshifts or calibrate spectra at the
level allowed by the data. We mitigate this risk rather simply by saving all metadata and
raw data used in reductions, to allow us to improve the software over time. Our planned
computing facilities are sufficient for numerous reprocessings of the data set over time.

### 7.7.3  Data Hardware and Connectivity

The last source of risk is hardware inadequacy or failure, which can be either in the data
storage or the data transfer. Disk farms are highly reliable systems; however, disk and
server failure is a reality in any data facility. This risk motivates the full spinning-disk
mirror and tape backups described above.

With about 40 GB of data produced each night, connectivity to the observatory is
important. Periodically, as with any network, we will experience a lack of connectivity. The
on-mountain computing system described above will be sufficient to store about 20 nights
of observing in order to bridge any periods of network failure.

# A   Exposure Time Calculator

In order to estimate the exposure time requirements for BigBOSS galaxy targets and overall survey length, we have developed two exposure time calculators (ETCs). The first of these calculators is in the form of a simple spreadsheet where all calculations are open to scrutiny with minimal effort. The second calculator is written in IDL and is based on the spreadsheet but extended to incorporate an entire 2D simulation of the measured spectrum. The full wavelength-dependent simulation, called *bbspecsim*, uses existing measurements of sky brightness, sky transparency, and instrumental throughputs in all possible cases. The results of these exposure time calculators provide a transparent way to forecast the BigBOSS survey requirements.

In the following discussion, the ETCs are described in the context of acquiring sufficient S/N on the [OII] emission line of ELG targets to make a detection and redshift measurement. This criterion results in a *minimum* exposure time for each tile of the survey, and therefore generally governs the speed at which the survey can be performed.

**Sky Properties**   The measured signal in the vast majority of the BigBOSS galaxy spectra will be limited by background sky emission. The BigBOSS spectra will be at sufficient resolution to separate most of the sky emission lines, and therefore, the majority of the ELG [OII] line detections will occur over the background continuum between the sky lines. Historically, this background sky continuum level has been uncertain due to instrumental limiting conditions (such as scattered light) and site variations. Noxon [1978] found the continuum to be 130 photons $s^{-1}$ $m^{-2}$ $arcsec^{-2}$ $\mu m^{-1}$ at 8500Å during dark time at Fritz Peak Observatory. DEEP2 measurements from the Keck DEIMOS spectrograph give a similar value of 133 photons $s^{-1}$ $m^{-2}$ $arcsec^{-2}$ $\mu m^{-1}$, but the SDSS spectrograph measurement is $\sim 180$ photons $s^{-1}$ $m^{-2}$ $arcsec^{-2}$ $\mu m^{-1}$ at Apache Point Observatory (APO). Since BigBOSS will be closer in design to that of the SDSS spectrograph, we assume the conservative value of 180.

Figure A.1 shows the sky emission spectrum measured from the BOSS spectrographs during dark time near zenith out to 10300Å. The spectrum is a resolution of R$\sim$ 3000, so the separation between sky lines underestimates what would be achieved in the BigBOSS spectrographs (R$\sim$ 4500). Beyond 10300Å, the spectrum is supplemented with higher resolution model emission spectrum from Gemini Observatory[10]. This far-red portion of the spectrum is scaled to match the BOSS continuum and emission line peak level in a small overlap region. Note that wavelengths >10300Å correspond to [OII] at a $z > 1.75$ and therefore are not critical to the survey. The current sky emission reflects that observed at APO (at a lower resolution than BigBOSS) and not KPNO. Because Kitt Peak is generally a better site than APO, our assumed sky spectrum should conservatively estimate the sky conditions at KPNO.

The sky extinction for KPNO is derived from kpnoextinct.dat, a widely used extinction curve for Kitt Peak which extends from 3400Å to 9000Å. Beyond this general extinction of the atmosphere, we extend our model to include bands of heavy water absorption. For the water bands below 9000Å, we used high resolution spectra measured from Kitt Peak

---

[10]http://www.gemini.edu/sciops/telescopes-and-sites/observing-condition-constraints/optical-sky-background

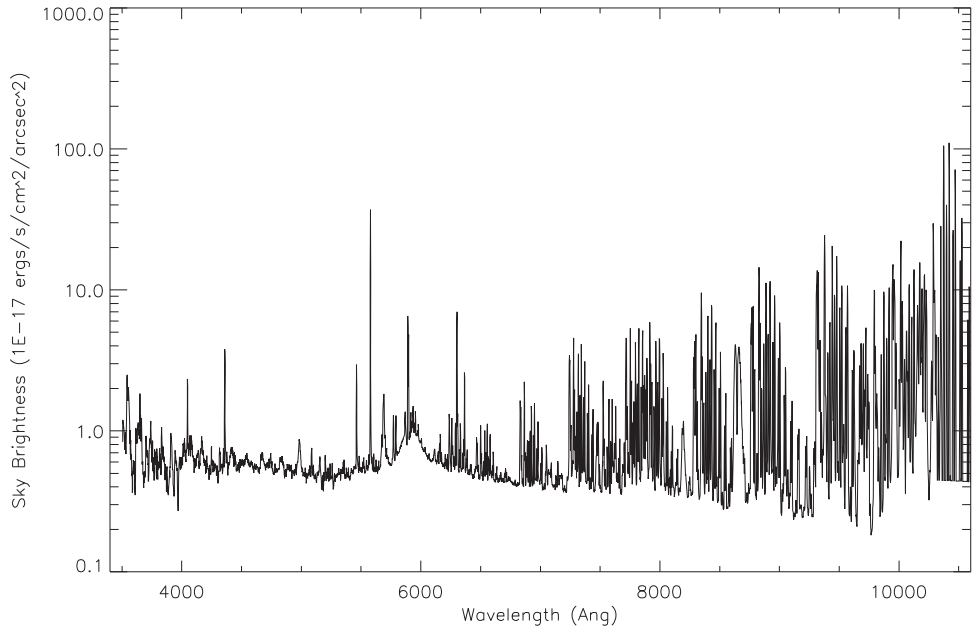

Figure A.1: Sky emission spectrum used in the calculation of exposure times for BigBOSS. The data for $\lambda < 10300$Å is from the BOSS spectrographs and was taken during during dark time near zenith. Data beyond $\lambda > 10300$Å is taken from the Gemini model sky spectrum and scaled to match the BOSS resolution and sky continuum level.

(A. Dey, priv. comm.). Beyond 9000Å, we have supplemented the extinction curve with simulated HITRAN data originally performed for Palomar observatory with 3mm of water vapor at 1700m. The HITRAN data is sampled with a very high resolution (0.1Å) which we downsample to the resolution of BigBOSS. The combined extinction curve is computed for zenith angle and plotted in Figure A.2.

The last adjustable components to our atmosphere mode incorporates the airmass, $X$, and Gaussian RMS variation, $\sigma_s$, of the observation. The current ETC uses default values of 1.2 airmass and 1.1″‖FWHM seeing for the BigBOSS instrument on the Mayall telescope The average airmass term is calculated from the mean of all tiled observations simulated in the survey lifetime (see Section 6). The seeing value is taken from the most recent average seeing measured from the MOSAIC camera mounted at prime focus on the Mayall telescope. The ETC also scales the sky brightness as a function of the airmass, $X^1$, since the column density through the OH sky emission is linear with airmass. Imaging data from the MOSAIC camera indicates that the seeing is not airmass dependent in the $i$-band for $X < 2$, indicating that dome seeing is dominating the delivered point spread function (see Figure 6.4). We therefore currently adopt airmass independent seeing for the Mayall. The ETC does not currently handle effects like differential atmospheric dispersion, although we expect such losses to be small given the use of an atmospheric dispersion corrector for BigBOSS.

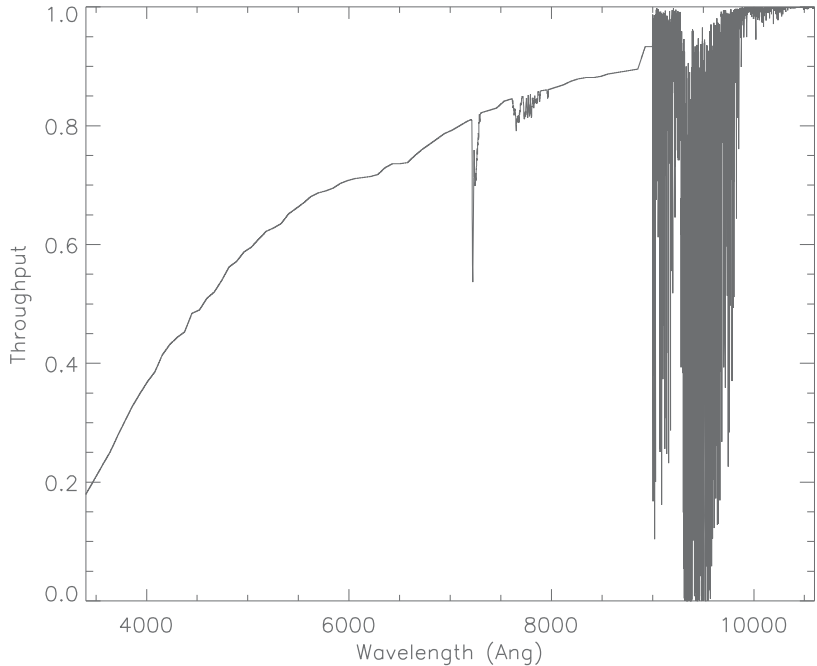

Figure A.2: The transmission of the sky used in the calculation of exposure times for BigBOSS. The data for $\lambda < 9000$Å is from low resolution KPNO observations and has been supplemented with high-resolution water absorption band measurements. Data beyond $\lambda > 9000$Å is simulated from HITRAN for the Palomar observatory with 3mm atmospheric water vapor and rescaled to match the KPNO data at 9000Å. The HITRAN data is sampled every 0.1Å, and therefore the absorptions seen here must be convolved with the object and sky background spectrum before being downsampled to the BigBOSS resolution.

**Telescope and Corrector Throughput**  The telescope collecting area is defined by the primary mirror diameter and reduced by the obscuration of the corrector lens support structure. We compute this area as the geometric throughput of the telescope relative to the full collecting area. For a primary mirror of 3.797m and corrector diameter of 1.8m, BigBOSS will have a geometric throughput of 0.775, and therefore the collecting area of the telescope is 8.72m$^2$. The reflectivity of the primary mirror is measured from witness samples taken during re-aluminization of the Sloan 2.5m primary mirror at Kitt Peak; the average value of reflectance is $\sim 90\%$. The corrector is currently designed to have 12 optical interfaces and glued prisms in the ADC. Detailed optical simulations of the glass component thickness and applied anti-reflection coatings result in an average throughput of 0.78. The full wavelength dependent throughput for both the telescope and corrector is shown in Figure A.4.

**Fiber Throughput**  The BigBOSS baseline design calls for 1.5″$^\parallel$ diameter fibers. The amount of source light that enters the fiber will depend on several factors, including (but not limited to) atmospheric seeing, source size, telescope pointing, and fiber positioning. Assuming that the seeing and source size are the dominant terms for galaxy observations (where telescope pointing and fiber positioning have met their design requirements), we calculate the size of a Gaussian spot convolved with an exponential galaxy profile, which can be expressed analytically as

$$\sigma_{psf}^2 = 2 \left(\frac{h_l}{1.68}\right)^2 + \sigma_s^2 \tag{A.1}$$

where $\sigma_s$ is the seeing RMS and $h_l$ is the galaxy half light radius. We use the measured seeing from the MOSAIC camera (1.1″) as our baseline value and note that the measurement *includes* sub-dominant blur contributions from the camera corrector optics and guiding system (see Jacoby et al. [1998] for details). We anticipate similar, if not better, optical and guiding performance for the BigBOSS focal plane. Using high-resolution imaging from the COSMOS field with HST [Leauthaud et al., 2007], we estimate that the mean $h_l$ size of ELGs between $0.5 < z < 1.5$ will be 0.3″, and therefore the nominal FWHM of the PSF will be 1.25″∥for 1 airmass. We compute a lookup table for the fractional loss of light within a 1.5″∥fiber circle for a given Gaussian spot with $\sigma_{psf}$ and perfectly centered on the fiber. The amount of light captured by the fiber as a function of the spot FWHM is shown in Figure A.3.

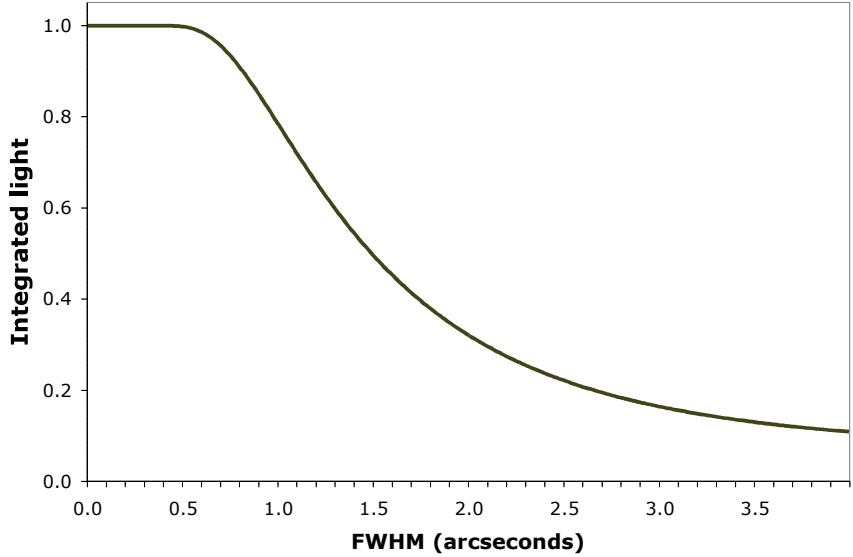

Figure A.3: The integrated fraction of light received by a 1.5″∥diameter fiber for values of Gaussian FWHM. This calculation assumes a perfect centering of the Gaussian profile on the fiber.

For the focal plane plate scale of 82.64$\mu$m arcsec$^{-1}$, BigBOSS will require 120$\mu$m core fibers. We assume that BigBOSS will use the Polymicro FBP fibers which have low water absorption and have broad applications in astronomy. For wavelengths of 8000-10000Å, the attenuation can be as low as 3db/km, translating to 98% transmission for the nominal fiber run length of 30m. However, all silica fibers have heavier attenuation for bluer wavelengths and can have <60% transmission for wavelengths shorter than 4000Å. The full wavelength dependent throughput for the Polymicro fibers is shown in Figure A.4.

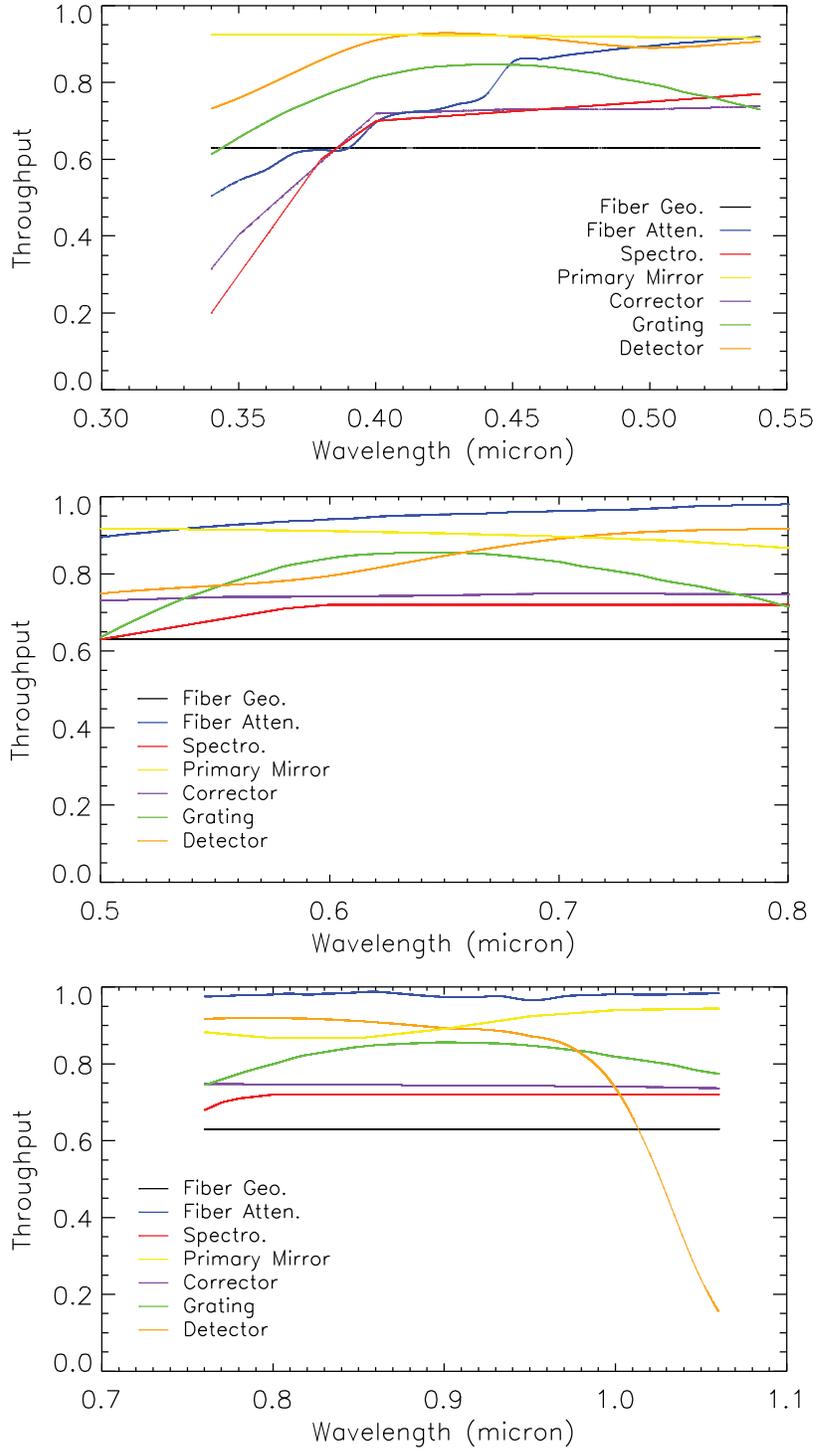

Figure A.4: The assumed throughputs for various instrument components in the blue (top), visible (middle), and red (bottom) BigBOSS optical paths.

**Spectrograph Properties** To first order, the BigBOSS spectrograph will image the $120\mu m$ core fiber end onto the BigBOSS detectors with a 2.67 demagnification, producing a circular image with a $45\mu m$ diameter. The BigBOSS spectrograph design calls for 3 pixel FWHM sampling of the imaged monochromatic spot. In the red and visible arms of the spectrograph, this sampling corresponds to a 2.2Å resolution at a dispersion of 0.732Å $pixel^{-1}$, and 1.13Å at a dispersion of 0.48Å $pixel^{-1}$ in the blue spectrograph arm. In *bbspecsim*, the ETC uses monochromatic, 2D images produced by a photon-level Monte Carlo simulation of the optics. These spots are generated over the entire spatial reach of each spectrograph arm and therefore include more subtle effects such as distortion and coma produced by the spectrograph optics (see Figure A.5). The simulation linearly interpolate between these monochromatic images in the dispersion direction to generate a full spectral image.

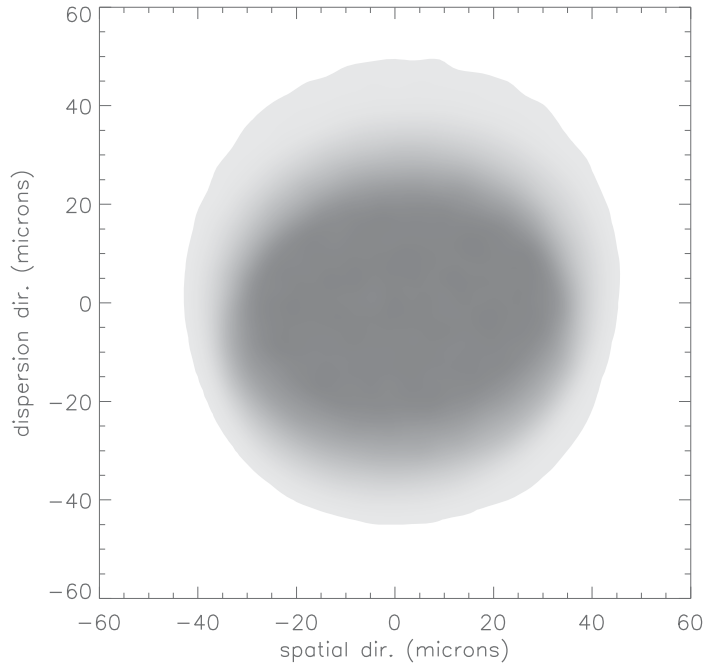

Figure A.5: A 2-dimensional monochromatic spot image generated by a Monte Carlo simulation of photons in the BigBOSS spectrograph optics. This particular image is near the center of the CCD in the red spectrograph and has been stretched to show the behavior of the PSF wings. The pixel sampling is 1 micron.

Along with optical images, the simulated model of each spectrograph arm also generates a wavelength-dependent throughput that takes into account the effects of all the glass materials needed to construct the image. The throughputs are currently > 50% in the blue arm for wavelengths longer than 4000Å and > 70% in both the visible and red arms of the spectrograph. The throughputs of the VPH gratings are considered separately from the spectrograph optics. The grating throughputs are scaled from the measured VPH designs used in the BOSS spectrographs by moving the blaze wavelength to the central wavelength in each BigBOSS arm. This simple scaling represents what is reasonably achievable given current technology and may be improved upon in the final BigBOSS design. The through-puts for both the spectrograph optics and VPH gratings are plotted in Figure A.4.

**Detector Properties**  Similar to the BOSS instrumentation, each arm of the BigBOSS spectrographs will hold a single 4k×4k CCD detector with 15$\mu$m pixel size provided by either e2v (blue arm) or LBNL (visible and red arms). The critical properties of these devices are the read noise and the quantum efficiency (QE) as a function of wavelength. We nominally assume the read noise, $\sigma_{read}$, is 3.0 electrons pixel$^{-1}$, consistent with the experience of BOSS. The QE for these devices are also measured from the BOSS experiment; the wavelength curves are provided along with all other instrumental thoughputs in Figure A.4.

**Sky Subtraction**  As seen in Figure A.1, the emission from the sky background creates bright spectral lines that can both mask the detection of [OII] emission lines and degrade the signal from the astronomical source. The signal from the sky background must therefore be removed from each object spectrum and done so carefully to avoid imprinting residuals which lead to false detections and inaccurate measurements. Current algorithms used in BOSS achieve near Poisson-limited sky subtraction of the sky background emission, but in a few cases, non-Poisson errors remain from minor variations in the the OH emission lines. A new analysis of the multi-object fiber spectral data from BOSS using the modeled PSF of the spectrographs and decomposed templates for the sky spectrum have produced evidence for achieving the Poisson limit in sky foreground subtraction [Bolton & Schlegel, 2010].

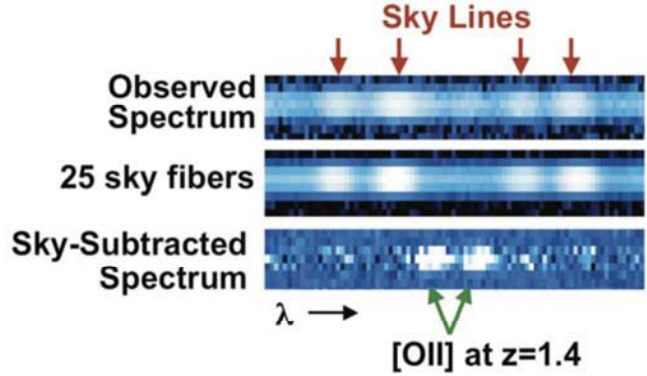

Figure A.6: A graphical representation of the subtraction of sky emission from simulated $z = 1.4$ ELG spectrum in BigBOSS. The top panel is the observed spectral image, the middle panel is the averaged sky spectrum formed from 25 sky fibers, and the bottom panel shows the sky-subtracted spectrum with prominent [OII] doublet signal.

For the purposes of a quick estimation of the noise contributed from the sky subtraction process, we presently skip the details of the "perfect" sky-subtraction process (involving sparse matrix inversion of multi-object spectra, see §7.4.2 for details) and assume that the sky measured from each exposure is the mean value of $n_{sky}$ co-added sky-only fiber spectra. Therefore, the residual error of the sky subtraction, $\sigma_{skysub}$, is proportional to $\sqrt{n_{sky}}$ and is added in quadrature to the noise from the sky-subtracted object spectrum. Given that all ELG object spectra ($\approx 300 - 500$ targets deg$^{-2}$) will contain sky emission with little to no contamination from the galaxy continuum, we should expect to construct a mean sky spectrum with a minimum of $n_{sky} = 25$ and easily resolve field-dependent variations in the sky emission background. Figure A.6 shows a graphical representation of the sky subtraction process for a [OII] emission line doublet at $z = 1.4$, where each line of the doublet has a S/N=8. This process represents a reasonable estimate of Poisson noise contributed from the sky-subtraction process that should be possible with BigBOSS. We note that residual, systematic (non-Poisson) errors are not implemented in *bbspecsim* (although the extraction techniques of Bolton & Schlegel [2010] should avoid such errors), but these systematic effects can be studied in the future.

**Signal to Noise Calculation** The single-pixel signal-to-noise per exposure is calculated using the following equation:

$$(S/N)_{pix} = \frac{S_{obj}}{\sqrt{(S_{obj} + S_{sky}) + \sigma_{skysub}^2 + \sigma_{read}^2}} \tag{A.2}$$

where $S_{obj}$ and $S_{sky}$ are the photon signals from the object and sky. All terms are computed on a per pixel basis as subtended by the monochromatic spot with 1 pixel of dispersion. To compute the S/N for a single line of the [OII] doublet, we simply multiply $(S/N)_{pix}$ by $\sqrt{n_{pix}}$, where $n_{pix}$ is the effective number of pixels used in the optimal extraction of the emission line, or

$$n_{pix} = 4\pi(\sigma_{psf}^2 + \sigma_{line}^2), \tag{A.3}$$

where both $\sigma_{psf}$ and $\sigma_{line}$ are in units of pixels. In our baseline calculations, we assume the emission line has a nominal velocity dispersion of 70km s$^{-1}$.

For the full spectral simulation, we compute $(S/N)_{pix}$ for each spectral pixel (spaxel) in each spectrograph arm. The simulation uses the 2D images of the fiber spot and convolves all throughputs, sky, and object spectra with sub-pixel sampling. The 2D spectrum is then reduced to a 1D spectrum by linearly fitting the signal in each spaxel with the spatial profile generated by the simulated PSF collapsed along the spectral dimension and weighted by the spaxel variance. Figure A.7 shows the S/N of a $z = 1.4$ [OII] emission line doublet at $9 \times 10^{-17}$ ergs s$^{-1}$ cm$^{-2}$ line flux and a 1000 second exposure. The S/N produced by one half of the [OII] doublet is consistent with the simple single-wavelength calculation. Computing the variance per pixel also allows us to apply random variation from a normal distribution and therefore produce realistic monte carlo spectra. These spectral simulations will aid development of software that optimizes [OII] line detection and redshift measurement.

**Exposure Times** Tables A.1 and A.2 shows the S/N and exposure time values based on the above instrumental parameters at 10300Å (where [OII] is at the redshift limit of $z = 1.75$). Specifically, Table A.1 calculates exposure times for a constant S/N=8 of half the [OII] doublet given various values of [OII] line flux and seeing. We find in this simple calculation that the requirement of S/N=8 for a line flux of $9 \times 10^{-17}$ ergs s$^{-1}$ cm$^{-2}$ is achieved in $\sim 1000$ seconds. Table A.2 uses the same values of line flux and seeing but computes the line S/N for a fixed 1000 second exposure. In general, $(S/N) \propto t^2$ and depends linearly on the line flux signal. Therefore a 10% difference in time or even line flux will only have a marginal impact on the line S/N. However, the S/N degrades rapidly with seeing as signal is lost outside the fiber but the sky brightness remains fixed.

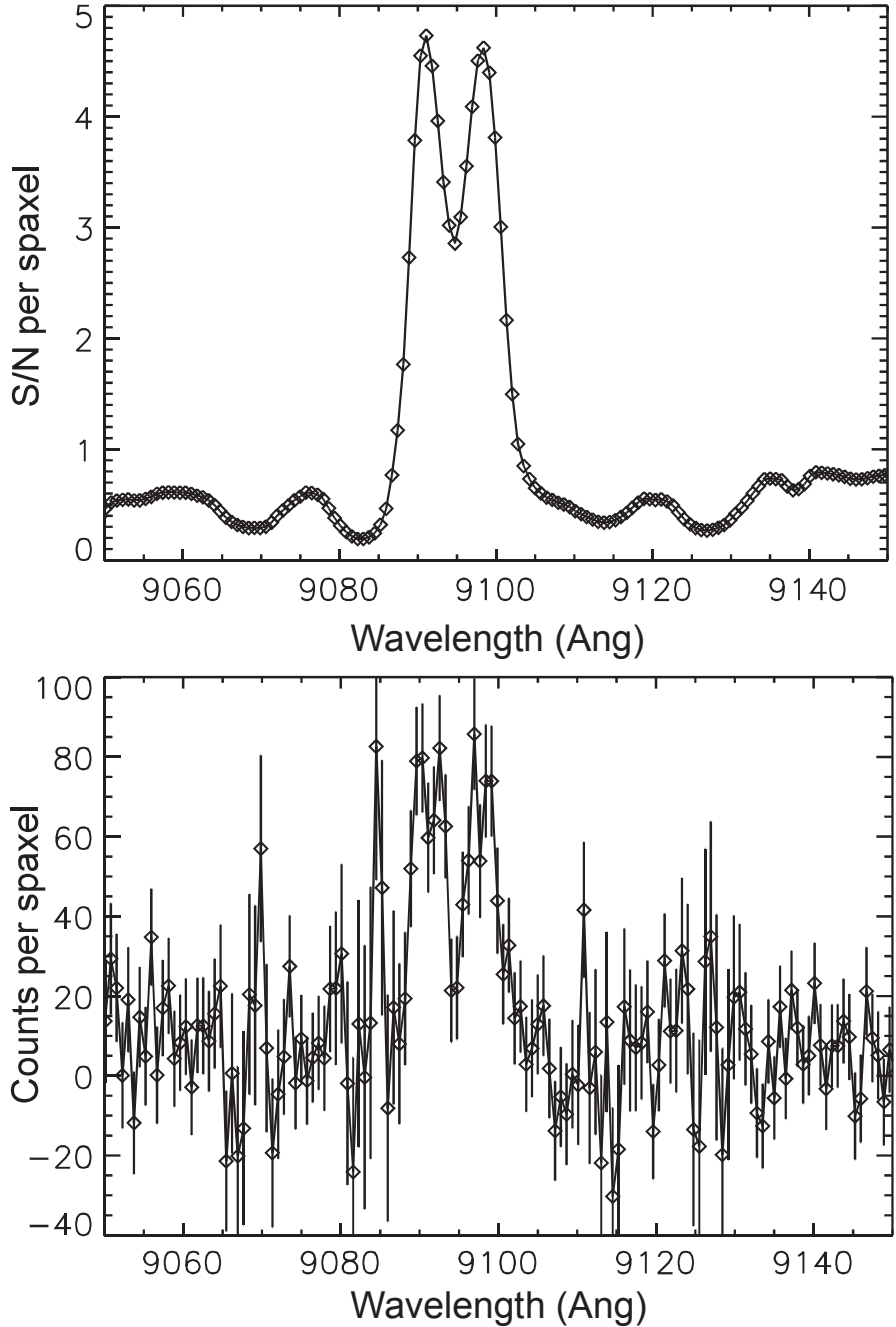

Figure A.7: Extracted 1D spectra for simulated $z = 1.4$ [OII] emission line with $9 \times 10^{-17}$ ergs s$^{-1}$ cm$^{-2}$ in a 1000 second exposure at the BigBOSS red spectrograph resolution. The top figure shows the S/N per spectral pixel (spaxel) for the simulated emission line. The total quadrature sum of the S/N over one line of the doublet achieves S/N=8. The bottom figure is the same simulated spectrum with errors applied from a normal distribution.

Table A.1: Calculated exposure times for a fixed S/N=8 for one line of the [OII] doublet.

| [OII] Flux | Seeing FWHM | | |
|---|---|---|---|
| (ergs s$^{-1}$ cm$^{-2}$) | 0.9$''^{\|}$ | 1.1$''^{\|}$ | 1.3$''^{\|}$ |
| $10 \times 10^{-17}$ | 645 | 874 | 1186 |
| $9 \times 10^{-17}$ | 758 | 1031 | 1402 |
| $8 \times 10^{-17}$ | 911 | 1241 | 1694 |

Table A.2: Calculated S/N values for one line of the [OII] doublet in a fixed 1000 second exposure.

| [OII] Flux | Seeing FWHM | | |
|---|---|---|---|
| (ergs s$^{-1}$ cm$^{-2}$) | 0.9$''^{\|}$ | 1.1$''^{\|}$ | 1.3$''^{\|}$ |
| $10 \times 10^{-17}$ | 10.1 | 8.7 | 7.5 |
| $9 \times 10^{-17}$ | 9.3 | 8.0 | 6.9 |
| $8 \times 10^{-17}$ | 8.5 | 7.3 | 6.2 |

**Comparisons and Cross-Checks**    In order to validate our calculations, we looked at the throughputs of two similar multi-fiber systems: the BOSS spectrograph on the 2.5m Sloan Telescope at APO and the AAOmega spectrograph at the 3.9m Anglo-Australian Telescope (AAT). Both systems image fibers in a slit plane configuration and disperse the spectra between multiple spectral arms at moderate resolutions ($\sim 1000 - 3000$). Additionally, both systems are currently being used for redshift measurements in dark-energy studies.

The top plot in Figure A.8 shows a comparison of the end-to-end throughputs for the BOSS and AAOmega[11] systems along with the throughputs used in *bbspecsim*. The throughputs include all instrumented optical components, slit losses due to point-source seeing, and atmospheric extinction at an airmass$\approx 1.2$. The points used in the plot are chosen to show the general values of end-to-end throughput between instruments and avoid bands of heavy atmospheric absorption. The plot shows that the throughput of BigBOSS is similar to BOSS and AAOmega at visible wavelengths. BigBOSS will continue to have response down to 3400Å where BOSS and AAOmega have little to no throughput below 3700Å. However, the largest gain of BigBOSS will be at wavelengths longer than 8000Å where BOSS and AAOmega has diminishing response. This gain comes primarily from the third arm of the BigBOSS spectrograph where the VPH grating have higher throughput over a relatively smaller wavelength range and the thick LBNL CCDs continue to have response.

While BigBOSS clearly has throughput advantages at the longest optical wavelengths, the full advantages of the BigBOSS instrument comes from the spectral sensitivity as a function of wavelength. The sensitivity includes instrumental differences such as primary mirror size, the sky background within the fiber diameter, and the spectroscopic resolution. To compare the sensitivities, we plot the signal-to-noise ratio (SNR) per angstrom for a flat spectral source with $mag_{AB} = 20$ in one hour of observation in $1.1''$ seeing. The values for AAOmega are calculated from the online SNR calculator[12] and adjusting the five fixed Johnson band magnitudes to AB magnitudes and using $1.3''$ seeing. The BOSS SNR is actual spectral data from a one hour observation of a V=20 white dwarf (approximately flat in AB magnitudes) in $1.3''$ seeing conditions at 1.0 airmass. All data is either generated or rescaled to an airmass of 1.2.

---

[11]http://www.aao.gov.au/local/www/aaomega/#thrumos

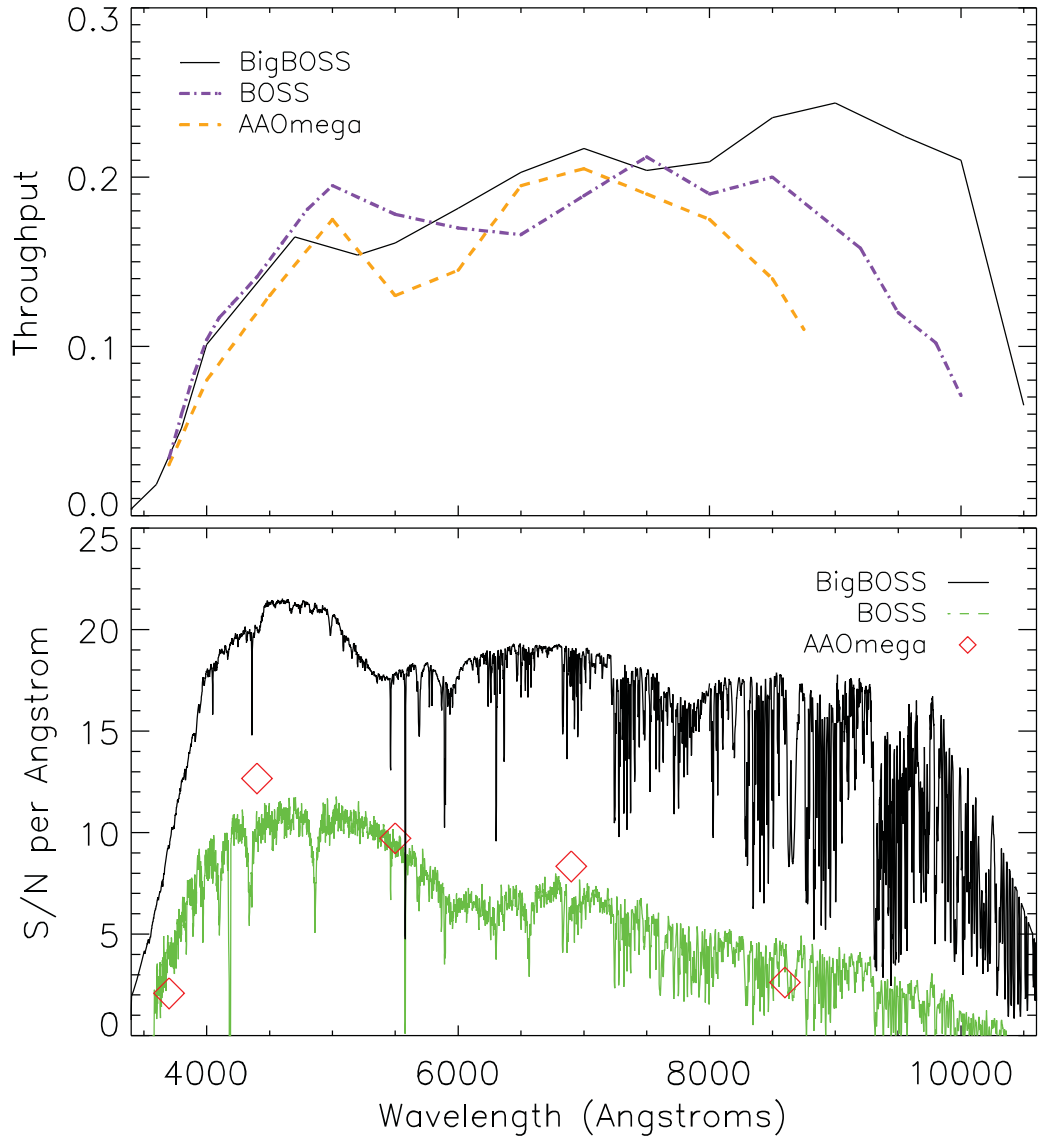

Figure A.8: (*top*) The end-to-end throughputs for BigBOSS, BOSS, and AAOmega. The throughputs assume atmospheric absorption at an airmass of 1.2 at each observing site and fiber losses due to point-source seeing blur (1.5″ median seeing and 2.1″ diameter fibers at AAT, 1.4″ median seeing and 2″ diameter fibers at APO). The plotted points are chosen to show the overall throughput trends and avoid bands of heavy atmospheric absorption. AAOmega and BOSS have red cutoff wavelengths of 8800Å and 10000Å, respectively. (*bottom*) The spectroscopic sensitivity, computed as the signal-to-noise per angstrom, for BigBOSS, BOSS, and AAOmega. The signal for BigBOSS is computed from *bbspecsim* using a one hour exposure on a point source with mag$_{AB}$ = 20 in median seeing conditions. The plotted data for AAOmega comes from the online SNR calculator assuming the same constant AB magnitude, exposure time, and 1.3″ seeing.

The bottom plot of Figure A.8 shows the comparison of spectroscopic sensitivity for BigBOSS, BOSS, and AAOmega. Here we see the real benefit to the BigBOSS spectrograph design. The combination of improved throughputs, better site seeing, smaller fiber diameters, and higher resolution all combine to improve the sensitivity of BigBOSS over similar spectroscopic dark-energy surveys. At the bluest wavelengths, the scaling of sensitivity between BOSS and BigBOSS is about a factor of 1.8 and we expect that the primary mirror size difference alone should contribute a factor of 1.6. For the reddest wavelengths, the $1.5''$ fiber size and $1.1''$ seeing of BigBOSS reduce the sky background by a factor of 2 over $2''$ diameter fibers, and the increased throughput over BOSS and AAOmega improve the overall spectroscopic sensitivity by a factor of $\sim 3$ or more. The improved sensitivity of BigBOSS can be exploited to increase the survey speed, returning more galaxy redshifts over a wider redshift range than is possible from BOSS or AAOmega.

---

[12]http://www.aao.gov.au/cgi-bin/aaomega_sn.cgi

# References

Abazajian, K. N., et al. 2009, Astrophys. J. Supp., 182, 543.

Abbott, T., et al. 2005, arXiv:astro-ph/0510346v1

Abell, A. A., et al., 2009, LSST Science Book, arXiv:0912.0201v1

Adelberger, K. L., Steidel, C. C., Pettini, M., Shapley, A. E., Reddy, N. A., & Erb, D. K. 2005, Astrophys. J., 619, 697

Albrecht, A., et al., 2006, Report of the Dark Energy Task Force, arXiv:astro-ph/0609591.

Albrecht, A., et al., 2009, Findings of the Joint Dark Energy Mission Figure of Merit Science Working Group, arXiv:0901.0721.

Alcock, C. & Paczinsky, B. 1979, Nature, 281, 358.

Allen, J. T., Hewett, P. C., Maddox, N., Richards, G. T., & Belokurov, V. 2010, arXiv:1007.3991 (Mon. Not. R. Astron. Soc.in press)

Amanullah, R., et al. 2010, Astrophys. J., 716, 712.

Arnouts, S., et al. 2002, Mon. Not. R. Astron. Soc., 329, 355

Ashby, M. L. N., et al. 2009, Astrophys. J., 701,428.

Athanassopoulos, C., et al. 1996, Phys. Rev. Lett., 77, 3082

Baldwin, J. A., Phillips, M. M. & Terlevich, R. 1981, Proc. Astron. Soc. Pacific, 93, 5.

Baldauf, T. et al. 2010, PRD, 81, 6.

Barmby, P., et al. 2008, Astrophys. J. Supp., 177, 431.

Battaglia, G., et al. 2005, Mon. Not. R. Astron. Soc., 364, 433.

Bauer, A., Baltay, C., Coppi, P., Ellman, N., Jerke, J., Rabinowitz, D., & Scalzo, R. 2009, Astrophys. J., 696, 1241.

Beers, T. C., Preston, G. W., & Shectman, S. A. 1985, Astron. J., 90, 2089.

Begelman, M. C., Blandford, R. D. & Rees, M. J. 1980, Nature, 287, 307.

Belokurov, V., et al. 2006, Astrophys. J. Lett., 642, L137.

Bernstein, G. M. 2009, Astrophys. J., 695, 652.

Bhattacharya, S. et al. 2010, Phys. Rev. D, submitted.

Bianchi, S., Chiaberge, M., Piconcelli, E., Guainazzi, M. & Matt, G. 2008, Mon. Not. R. Astron. Soc., 386, 105.

Blanton, M. R., Eisenstein, D., Hogg, D. W., Schlegel, D. J. & Brinkmann, J. 2005, Astrophys. J., 629, 143.

Bolton, A. S., Burles, S., Koopmans, L. V. E., Treu, T. & Moustakas, L. A. 2006, Astrophys. J., 638, 703.

Bolton, A. S., & Burles, S. 2007, New Journal of Physics, 9, 443.

Bolton, A. S., Burles, S., Koopmans, L. V. E., Treu, T., Gavazzi, R., Moustakas, L. A., Wayth, R. & Schlegel, D. J. 2008a, Astrophys. J., 682, 964.

Bolton, A. S., Treu, T., Koopmans, L. V. E., Gavazzi, R., Moustakas, L. A., Burles, S., Schlegel, D. J. & Wayth, R. 2008b, Astrophys. J., 684, 24.

Bolton, A. S. & Schlegel, D. J. 2010, Proc. Astron. Soc. Pacific, 122, 248.

Bonoli, S., Marulli, F., Springel, V., White, S. D. M., Branchini, E., & Moscardini, L. 2009, Mon. Not. R. Astron. Soc., 396, 423.

Booth, C. M., & Schaye, J. 2009, Mon. Not. R. Astron. Soc., 398, 53.

Bower, R. G., Benson, A. J., Malbon, R., Helly, J. C., Frenk, C. S., Baugh, C. M., Cole, S., & Lacey, C. G. 2006, Mon. Not. R. Astron. Soc., 370, 645.

Brown, M. J. I., Dey, A., Jannuzi, B. T., Brand, K., Benson, A. J., Brodwin, M., Croton, D. J. & Eisenhardt, P. R. 2007, Astrophys. J., 654, 858.

Bruzual, G. & Charlot, S. 2003, Mon. Not. R. Astron. Soc., 344, 1000.

Calzetti, D., Kinney, A.L., & Storchi-Bergmann, T. 1994, Astrophys. J., 429, 582.

Cannon, R., et al. 2006, Mon. Not. R. Astron. Soc., 372, 425.

Cappellari, M., et al. 2006, Mon. Not. R. Astron. Soc., 366, 1126.

Chamballu, A., Bartlett, J. G., & Melin, J. -. 2010, arXiv:1007.3193.

Chang, T., Pen, U., Peterson, J. B., & McDonald, P. 2008, Phys. Rev. D, 100, 091303

Christlieb, N., Wisotzki, L., Reimers, D., Homeier, D., Koester, D., & Heber, U. 2001, Astron. Astrophys., 366, 898.

Cimatti, A., et al. 2008, Astron. Astrophys., 482, 21.

Clewley, L., Warren, S. J., Hewett, P. C., Norris, J. E., & Evans, N. W. 2004, Mon. Not. R. Astron. Soc., 352, 285.

Coil, A. L., Newman, J. A., Croton, D., et al. 2008, Astrophys. J., 672, 153.

Cole, S., et al. 2005, Mon. Not. R. Astron. Soc., 362, 505.

Coleman, G. D., Wu, C. C. & Weedman, D. W. 1980, Astrophys. J. Supp., 43, 393.

Comerford, J. M., et al. 2009, Astrophys. J., 698, 956.

Comerford, J. M., et al. 2010, Astrophys. J., in preparation.

Cooper, M. C., et al. 2006, Mon. Not. R. Astron. Soc., 370, 198.

Cooper, M. C., et al. 2008, Mon. Not. R. Astron. Soc., 383, 1058.

Cooray, A., Sarkar, D. & Serra, P. 2008, Phys. Rev. D, 77, 123006.

Cresswell, J. G. & Percival, W. 2009, Mon. Not. R. Astron. Soc., 392, 682.

Croom, S. M., et al. 2004, Mon. Not. R. Astron. Soc., 349, 1397.

Croom, S. M., et al. 2005, Mon. Not. R. Astron. Soc., 356, 415.

Croom, S. M., et al. 2009, Mon. Not. R. Astron. Soc., 392, 19.

Croom, S. M., et al. 2009, Mon. Not. R. Astron. Soc., 399, 1755.

Croton, D. J., et al. 2006, Mon. Not. R. Astron. Soc., 365, 11.

Croton, D. J. 2009, Mon. Not. R. Astron. Soc., 394, 1109.

Daddi, E., et al. 2005, Astrophys. J., 626, 680.

Dalal, N., Doré, O., Huterer, D. & Shirokov, A. 2008, Phys. Rev. D, 77, 123514.

Daniel, S. F., et al. 2010, Phys. Rev. D, 81, 123508.

Daniel, S. F. & Linder, E. V. 2010, arXiv:1008.0397.

Davis, M., et al. 2003, Proc. of SPIE, 4834, 161.

Davis, M., et al. 2007, Astrophys. J. Lett., 660, 1.

De Lucia, G., & Blaizot, J. 2007, Mon. Not. R. Astron. Soc., 375, 2.

Desjacques, V., Seljak, U. & Iliev, I. 2009, Mon. Not. R. Astron. Soc., 396, 85.

Dressler, A. 1980, Astrophys. J., 236, 351.

Drexlin, G. et al 2005, Nucl. Phys. B, 145, 263.

Eisenhardt, P. R. M., et al. 2008, Astrophys. J., 684, 905.

Eisenstein, D. J. & Hu, W. 1998, Astrophys. J., 496, 605.

Eisenstein, D. J., et al. 2001, Astron. J., 122, 2267.

Eisenstein, D. J., et al. 2003, Astrophys. J., 585, 694.

Eisenstein, D. J., et al. 2005, Astrophys. J., 633, 560.

Eisenstein, D. J., et al. 2006, Astrophys. J. Supp., 167, 40.

Eisenstein, D. J., Seo, H.-J. & White, M. 2007, Astrophys. J., 664, 660.

Eisenstein, D. J., Seo, H.-J., Sirko, E. & Spergel, D. 2007, Astrophys. J., 664, 675.

Emsellem, E., et al. 2004, Mon. Not. R. Astron. Soc., 352, 721.

Eracleous, M., Halpern, J. P., Gilbert, A. M., Newman, J. A. & Filippenko, A. V. 1997, Astrophys. J., 490, 216.

Faber, S. M., et al. 2007, Astrophys. J., 665, 265.

Feldman, H., Kaiser, N. & Peacock, J. 1994, Astrophys. J., 426, 23.

Ferguson, A. M. N., Irwin, M. J., Ibata, R. A., Lewis, G. F., & Tanvir, N. R. 2002, Astron. J., 124, 1452.

Franx, M., et al. 2003, Astrophys. J. Lett., 587, L79.

Frieman, J. A., Turner, M. S., & Huterer, D. 2008, Annu. Rev. Astron. Astrophys., 46, 385.

Garnavich, P. M., et al. 1998, Astrophys. J., 509, 74.

Gates, E., et al. 2004, Astrophys. J. Lett., 612, L129.

Gavazzi, R., Treu, T., Koopmans, L. V. E., Bolton, A. S., Moustakas, L. A., Burles, S. & Marshall, P. J. 2008, Astrophys. J., 677, 1046.

Gerke, B. F., et al. 2007, Astrophys. J. Lett., 660, L23.

Gibson, R. R., et al. 2009, Astrophys. J., 692, 758

Gilks, W. R., Richardson, S. R. & Spiegelhalter, D. J., "Markov chain Monte Carlo in practice", Chapman & Hall (Florida, 1996).

Glazebrook, K., et al. 2007, Proceedings of the Durham "Cosmic Frontiers" ASP conference eds. Metcalfe & Shanks.

Glazebrook, K., et al. 2005, White paper submitted to the Dark Energy Task Force, arXiv:astro-ph/0507457v2.

Guhathakurta, P., Ostheimer, J. C., Gilbert, K. M., Rich, R. M., Majewski, S. R., Kalirai, J. S., Reitzel, D. B., & Patterson, R. J. 2005, arXiv:astro-ph/0502366.

Guzzo, L., et al. 2008, Nature, 451, 541.

Gwyn, S. D. J. 2008, Proc. Astron. Soc. Pacific, 120, 212.

Haiman, Z., & Hui, L. 2001, Astrophys. J., 547, 27.

Hamilton, A. J. S. 1998, in "The Evolving Universe. Selected Topics on Large-Scale Structure and on the Properties of Galaxies" Kluwer Academic Publishers, v. 231, p. 185, ISBN: 079235074X.

Hamuy, M., Phillips, M. M., Maza, J., Suntzeff, N. B., Schommer, R. A., & Aviles, R. 1995, Astron. J., 109, 1.

Hamuy, M., Phillips, M. M., Suntzeff, N. B., Schommer, R. A., Maza, J., & Aviles, R. 1996, Astron. J., 112, 2391.

Harris, H. C., et al. 2006, Astron. J., 131, 571.

Heavens, A. F., Matarrese, S., & Verde, L. 1998, Mon. Not. R. Astron. Soc., 301, 797

Hildebrandt, H., van Waerbeke, L. & Erben, T. 2009, Astron. Astrophys., 507, 683.

Hill, G. J., Gebhardt, K., Komatsu, E. & MacQueen, P. J. 2004, Mitchell Symposium, AIP Conf. Proc. 743, 224.

Hill, G. J., et al. 2008, Proc. SPIE, 7014,

Ho, S., Lin, Y., Spergel, D. & Hirata, C. M. 2009, Astrophys. J., 697, 1358.

Holland, S. E., Bebek, C. J., Dawson, K. S., Emes, J. H., Fabricius, M. H., Fairfield, J. A., Groom, D. E., Karcher, A., Kolbe, W. F., Palaio, N. P., Roe, N. A. & Wang, G. 2006, "High-voltage compatable, fully depleted CCDs," SPIE 6068-12.

Hopkins, A. M. & Beacom, J. F. 2006, Astrophys. J., 651, 142.

Hopkins, P. F., et al. 2006, Astrophys. J. Supp., 163, 1.

Hopkins, P. F., Richards, G. T., & Hernquist, L., 2007b, Astrophys. J., 654, 731.

Horne, K. 1986, Proc. Astron. Soc. Pacific, 98, 609.

Howell, D. A., et al. 2009, arXiv:0903.1086 (Astro2010 White Paper)

Hu, W. 1999, Astrophys. J., 522, 21.

Ilbert, O., et al. 2006, Astron. Astrophys., 457, 841

Ilbert, O., et al. 2009, Astrophys. J., 690, 1236.

Ivezic, Z., et al., for the LSST collaboration, arXiv:0805.2366v1.

Jacoby, G. H., Liang, M., Vaughnn, D., Reed, R., & Armandroff, T., 1998, Proc. of SPIE, 3355, 721

Jeong, D. & Komatsu, E. 2009, Astrophys. J., 691, 569

Jiang, L., et al., 2006, Astron. J., 131, 2788.

John, T. L. 1988, å, 193, 189.

Kaiser, N. 1987, Mon. Not. R. Astron. Soc., 227, 1.

KamLAND Collaboration 2008, Phys. Rev. Lett., 100, 221803

Kazin, E., et al. 2010, Astrophys. J., 710, 1444.

Kelson, D. D. 2003, Proc. Astron. Soc. Pacific, 115, 688.

Kennicutt, Jr., R. C. 1998, Annu. Rev. Astron. Astrophys., 36, 189.

Kewley, L. J., Geller, M. J. & Jansen, R. A. 2004, Astron. J., 127, 2002.

Kewley, L. J., Groves, B., Kauffmann, G. & Heckman, T. 2006, Mon. Not. R. Astron. Soc., 372, 961.

Kewley, L. J. & Ellison, S. L. 2008, Astrophys. J., 681, 1183.

Kilic, M., et al. 2006, Astron. J., 131, 582.

Kochanek, C., et al. 2004, Bulletin of the AAS, 36, 1495.

Komatsu, E., et al. arXiv:1001.4538.

Komossa, S., Burwitz, V., Hasinger, G., Predehl, P., Kaastra, J. S. & Ikebe, Y. 2003, Astrophys. J. Lett., 582, L15.

Koopmans, L. V. E., Treu, T., Bolton, A. S., Burles, S. & Moustakas, L. A. 2006, Astrophys. J., 649, 599.

Kowalski, M., et al. 2008, Astrophys. J., 686, 749.

Krisciunas, K., & Schaefer, B. E. 1991, Proc. Astron. Soc. Pacific, 103, 1033.

Kubo, J. M., Allam, S. S., Annis, J., Buckley-Geer, E. J., Diehl, H. T., Kubik, D., Lin, H., & Tucker, D. 2009, Astrophys. J. Lett., 696, L61.

Lauer, T. R., et al. 2007, Astrophys. J., 662, 808.

Law, N. M., et al. 2009, Proc. Astron. Soc. Pacific, 121, 1395.

Le Févre, O., Vettolani, G., Paltani, S., et al. 2005, å, 439, 845.

Leauthaud, A., et al. 2007, Astrophys. J. Supp., 172, 219.

Liang, M. 2009, http://www-kpno.kpno.noao.edu/kpno-misc/mayall_widefield.html.

Linder, E. V. 2005, Phys. Rev. D, 72, 043529.

Linder, E. V. & Cahn, R. N. 2007, Astropart. Phys., 28, 481.

Linder, E. V. 2008, Astropart. Phys., 29, 336.

Liu, X., Shen, Y., Strauss, M. A. & Greene, J. E. 2010, Astrophys. J., 708, 427.

López-Corredoira, M. 2010, Astron. J., 139, 540.

Ly, C., et al. 2007, Astrophys. J., 657, 738.

Majewski, S. R., Skrutskie, M. F., Weinberg, M. D., & Ostheimer, J. C. 2003, Astrophys. J., 599, 1082.

Mandelbaum, R, McDonald, P., Seljak, U., Cen., R. 2003, Mon. Not. R. Astron. Soc., 344, 776

Marchesini, D., et al. 2007, Astrophys. J., 656, 42.

Marinoni, C., et al. 2005, Astron. Astrophys., 442, 801.

Marshall, P. J., et al. 2007, Astrophys. J., 671, 1196

Martin, C. L., & Sawicki, M. 2004, Astrophys. J., 603, 414.

Martinez-Delgado, D., et al. 2010, arXiv:1003.4860.

Martini, P., & Weinberg, D. H. 2001, Astrophys. J., 547, 12.

Matthews, D.J. & Newman, J.A. 2010, Astrophys. J., in press.

McDonald, P., et al. 2005, Astrophys. J., 635, 761.

McDonald, P. 2006, Phys. Rev. D, 74, 103512.

McDonald, P., et al. 2006, Astrophys. J. Supp., 163, 80

McDonald, P. 2008, Phys. Rev. D, 78, 123519.

McDonald, P. & Eisenstein, D. (2007), Phys. Rev. D, 76, 063009

McDonald, P. & Roy, A. 209, J. Cosmology Astropart. Phys., 08, 020.

McDonald, P. & Seljak, U. 2009, J. Cosmology Astropart. Phys., 10, 007.

Miknaitis, G., et al. 2007, Astrophys. J., 666, 674

Milosavljević, M. & Merritt, D. 2001, Astrophys. J., 563, 34.

Milosavljević, M., Merritt, D., Rest, A. & van den Bosch, F. C. 2002, Mon. Not. R. Astron. Soc., 331, L51.

MiniBooNE Collaboration 2010, arxiv:1007.1150

Minos Collaboration 2008, Phys. Rev. Lett., 101, 131802

Montero-Dorta, A. D., et al. 2009, Mon. Not. R. Astron. Soc., 392, 125.

Montesano, F., Sáchez, A. G., & Phleps, S. 2010, arXiv:1007.0755

Morrison, H. L., Olszewski, E. W., Mateo, M., Norris, J. E., Harding, P., Dohm-Palmer, R. C., & Freeman, K. C. 2001, Astron. J., 121, 283.

Moustakas, J., Kennicutt, Jr., R. C. & Tremonti, C. A. 2006, Astrophys. J., 642, 775.

Mukadam, A. S., et al. 2004, Astrophys. J., 607, 982.

Myers, A. D., Brunner, R. J., Richards, G. T., Nichol, R. C., Schneider, D. P., & Bahcall, N. A. 2007, Astrophys. J., 658, 99.

Newman, J. A. 2008, Astrophys. J., 684, 88.

Nilson, P. 1973, Acta Universitatis Upsaliensis. Nova Acta Regiae Societatis Scientiarum Upsaliensis - Uppsala Astronomiska Observatoriums Annaler, Uppsala: Astronomiska Observatorium.

Noxon, J. F. 1978, Planet. Space Sci., 26, 191

Padmanabhan, N., et al. 2007, Mon. Not. R. Astron. Soc., 378, 852.

Padmanabhan, N., et al. 2008, Astrophys. J., 674, 1217.

Padmanabhan, N. & White, M. 2009, Phys. Rev. D, 80, 063508.

Palanque-Delabrouille et al. 2010, in preparation.

PanSTARRS Survey Website 2010, http://pan-starrs.ifa.hawaii.edu/public.

Papovich, C., Rudnick, G., Rigby, J. R., Willmer, C. N. A., Smith, J.-D. T., Finkelstein, S. L., Egami, E., & Rieke, M. 2009, Astrophys. J., 704, 1506.

Parkinson, D., et al. 2010, Mon. Not. R. Astron. Soc., 401, 2169.

Pasquali, A., van den Bosch, F. C., Mo, H. J., Yang, X., & Somerville, R. 2009, Mon. Not. R. Astron. Soc., 394, 38.

Peebles, P. J. E. 1980, The large scale structure of galaxies, Princeton University Press.

Pen, U. 2004, Mon. Not. R. Astron. Soc., 350, 1445.

Percival, W. J. & White, M. 2009, Mon. Not. R. Astron. Soc., 393, 297

Percival, W. J., et al. 2010, Mon. Not. R. Astron. Soc., 401, 2148.

Perlmutter, S., et al. 1999, Astrophys. J., 517, 565.

Pettini, M., Steidel, C. C., Adelberger, K. L., Dickinson, M. & Giavalisco, M. 2000, Astrophys. J., 528, 96.

Phillips, M. M., Lira, P., Suntzeff, N. B., Schommer, R. A., Hamuy, M., & Maza, J. 1999, Astron. J., 118, 1766.

Phillips, M. M. 1993, Astrophys. J. Lett., 413, L105.

Pradhan, A. K., et al. 2006, Mon. Not. R. Astron. Soc., 366, L6.

Prochaska, J. X., & Wolfe, A. M. 2009, Astrophys. J., 696, 1543

Rau, A., et al. 2009, Proc. Astron. Soc. Pacific, 121, 1334

Rengstorf, A. W., et al. 2004, Astrophys. J., 617,184.

Reyes, R., Mandelbaum, R., Seljak, U., Baldauf, T., Gunn, J., Lombriser, L. & Smith, R. 2010, Nature 464, 256.

Rhoads, J. E., Malhotra, S., Dey, A., Stern, D., Spinrad, H., & Jannuzi, B. T. 2000, Astrophys. J. Lett., 545, L85

Richards, G. T., et al. 2002, Astron. J., 123, 2945.

Richards, G. T., et al. 2006, Astron. J., 131, 2766.

Richards, G. T., et al. 2009, Astrophys. J. Supp., 180, 67.

Riess, A. G., et al. 1998, Astron. J., 116, 1009.

Riess, A. G., et al. 1998, Astron. J., 116, 1009.

Riess, A. G., et al. (2010), Astrophys. J., 699, 539

Rodriguez, C., Taylor, G. B., Zavala, R. T., Peck, A. B., Pollack, L. K. & Romani, R. W. 2006, Astrophys. J., 646, 49.

Ross, N. P., et al. 2009, Astrophys. J., 697, 1634.

Rubin, V. C., Ford, W. K. J. & Thonnard, N. 1980, Astrophys. J., 238, 471.

Rubin, V. C. 1983, Science, 220, 1339.

Rubin, V. C., Burstein, D., Ford, W. K., Jr., & Thonnard, N. 1985, Astrophys. J., 289, 81.

Sawicki, M. 2002, Astron. J., 124, 3050.

Schiavon, R. P., et al. 2006, Astrophys. J. Lett., 651, L93.

See, for example, D. Schlegel, M. White, and D. Eisenstein, arXiv:0902:4680.

Schmidt, B. P., et al. 1998, Astrophys. J., 507, 46.

Schmidt, G. D., et al. 2003, Astrophys. J., 595, 1101.

Schmidt, K. B., et al. 2010, Astrophys. J., 714,1194.

Schneider, D. P., et al. 2010, Astron. J., 139, 2360.

Schulz, A. & White, M. 2006, Astropart. Phys., 25, 172.

Scoccimarro, R. 2004, Phys. Rev. D, 70, 083007.

Seljak, U., et al. 2005, Phys. Rev. D, 71, 3515.

Seljak, U., Slosar A., McDonald, P, J. Cosmology Astropart. Phys.06, 10.

Seljak, U., Hamaus, N. & Desjacques, V. 2009, Phys. Rev. Lett., 103, 091303.

Seljak, U. 2009, Phys. Rev. Lett., 102, 021302.

Senatore, L., Smith, K. M. & Zaldarriaga, M. 2010, J. Cosmology Astropart. Phys., 01, 028.

Seo, H.-J. & Eisenstein, D. J. 2007, Astrophys. J., 664, 675.

Seo, H.-J. et al. (2010), Astrophys. J., 720, 1650

Shankar, F., Weinberg, D. H., & Shen, Y. 2010, Mon. Not. R. Astron. Soc., 406, 1959.

Shen, Y., et al. 2007, Astron. J., 133, 2222.

Shen, Y., et al. 2009, Astrophys. J., 697, 1656.

Shen, Y. 2009, Astrophys. J., 704, 89.

Silvestri, N. M., et al. 2006, Astron. J., 131, 1674.

Simpson, F. & Peacock, J. A. 2010, Phys. Rev. D, 81, 043512

Slosar, A., Hirata, C., Seljak, U., Ho, S. & Padmanabhan, N. 2008, J. Cosmology Astropart. Phys., 08, 031.

Slosar, A., Ho, S., White, M., & Louis, T. (2009), J. Cosmology Astropart. Phys., 10, 019

Smith, K. L., Shields, G. A., Bonning, E. W., McMullen, C. C., Rosario, D. J. & Salviander, S. 2010, Astrophys. J., 716, 866.

Song, Y.-S. & Percival, W. J. 2009, J. Cosmology Astropart. Phys., 10, 004.

South Galactic Cap U-Band Sky Survey, 2010, http://batc.bao.ac.cn/Uband.

Springel, V., Di Matteo, T. & Hernquist, L. 2005, Mon. Not. R. Astron. Soc., 361, 776.

Steidel, C. C., Erb, D. K., Shapley, A. E., Pettini, M., Reddy, N., Bogosavljević, M., Rudie, G. C., & Rakic, O. 2010, Astrophys. J., 717, 289.

Stern, D., et al. 2005, Astrophys. J., 631, 163.

Stril, A., Cahn, R. N. & Linder, E. V. 2010, Mon. Not. R. Astron. Soc., 404, 239.

Sumiyoshi, M., et al. 2009, arXiv:0902.2064.

Swanson, M., et al. 2008, Mon. Not. R. Astron. Soc., 387, 1391.

Takahashi, M. I., et al. 2007, Astrophys. J. Supp., 172, 456.

Tegmark, M., et al. 2006, Phys. Rev. D, 74, 123507.

Thomas, S.A., Abdalall, F. B., L̂ahav, O. 2010, Phys. Rev. Lett., 105, 031301

Tinker, J. L. 2007, Mon. Not. R. Astron. Soc., 374, 477

Tinker, J. L. , Weinberg, D. H. , & Zheng, Z. 2006, Mon. Not. R. Astron. Soc., 368, 85

van Dokkum, P. G., et al. 2009, Proc. Astron. Soc. Pacific, 121, 2.

van Waerbeke, L., White, M., Hoekstra, H. & Heymans, C. 2006, Astropart. Phys., 26, 91.

Verde, L., et al. 2002, Mon. Not. R. Astron. Soc., 335, 432.

Vikhlinin, A,., et al. 2010, Astrophys. J., 692, 1060

Wang, Y. 2008, J. Cosmology Astropart. Phys., 05, 021.

Wang, J., Chen, Y., Hu, C., Mao, W., Zhang, S. & Bian, W. 2009, Astrophys. J. Lett., 705, L76.

Weinberg, D. H., Katz, N., & Hernquist, L. 1998, Origins, 148, 21, astro-ph/9708213

Weiner, B. J., et al. 2005, Astrophys. J., 620, 595.

White, M. 2003, arXiv:astro-ph/0305474

White, M. 2005, Astropart. Phys., 24, 334.

White, M., et al. 2009, Mon. Not. R. Astron. Soc., 397, 1348.

White, M. 2010, arXiv:1004.0250

Wolfe, A. M., Gawiser, E., & Prochaska, J. X. 2005, Annu. Rev. Astron. Astrophys., 43, 861

Wood-Vasey, W. M., et al. 2007, Astrophys. J., 666, 694.

Wright, E. L., et al. 2010, arXiv:1008.0031v1.

Xue, X. X., et al. 2008, Astrophys. J., 684, 1143.

Yan, R., Newman, J. A., Faber, S. M., Konidaris, N., Koo, D., & Davis, M. 2006, Astrophys. J., 648, 281.

Yang, X., Mo, H. J., van den Bosch, F. C., Pasquali, A., Li, C., & Barden, M. 2007, Astrophys. J., 671, 153.

Yanny, B., et al. 2009, Astrophys. J., 700, 1282.

Yanny, B., et al. 2009, Astron. J., 137, 4377.

Yeche, Ch., et al. 2010, submitted to å, arXiv:0910.3770v1.

Yee, H. K. C., Ellingson, E., Bechtold, J., Carlberg, R. G. & Cuillandre, J.-C. 1996, Astron. J., 111, 1783.

Yee, H. K. C., Gladders, M. D., Gilbank, D. G., Majumdar, S., Hoekstra, H. & Ellingson, E. 2007, Cosmic Frontiers, 379, 103.

Zehavi, I., Zheng, Z., Weinberg, D. H., & the SDSS Collaboration 2010, submitted to Astrophys. J., arXiv:1005.2413v1.

Zhang, P., et al. 2007, Phys. Rev. Lett., 99, 141302.

Zhu, G., Moustakas, J. & Blanton, M. R. 2009, Astrophys. J., 701, 86.

Zwicky, F. 1933, Helvetica Physica Acta, 6, 110.